\documentclass[12pt]{article}
\pdfoutput=1 
\usepackage{jheppub}
\usepackage{soul}
\usepackage{setspace}
\usepackage{slashed}   
\usepackage{graphicx,url}
\usepackage{booktabs}
\usepackage{bbm}
\usepackage{braket}
\usepackage{colortbl}
\usepackage{amsmath,hyperref,amssymb,cancel,stmaryrd}
\usepackage{amsfonts}
\usepackage{slashed}
\usepackage{arydshln}
\usepackage{mathrsfs}
\usepackage{verbatim}
\usepackage[usenames,dvipsnames]{xcolor}
\usepackage{color} 
  
\usepackage{epsfig} 
\def\half{\frac{1}{2}}

\def\a{\alpha}

\def\Or[#1]{{\text{O}}\left({#1}\right)}
\def\dotl[#1,#2]{\left\langle #1, #2 \right\rangle}
\def\dotlb[#1,#2]{[ #1, #2 ]}
\def\dotp[#1,#2]{(#1) \cdot (#2)}
\def\aff[#1,#2]{\hat{#1}(#2)}
\def\n4sym{{\cal N}=4 SYM}
\def\>{\rangle}
\def\<{\langle}
\newcommand   \corr   [1] {\left\langle #1 \right\rangle }
\newcommand   \inner [2] {\< #1 | #2 \>}
\newcommand   \pa [1] {\left( #1 \right)}
\def\weight[#1,#2,#3]{\{(#1),#2,#3\}}
\def\ads[#1]{$\text{AdS}_{#1}$}

\newcommand{\ba}{\begin{eqnarray}}
\newcommand{\ea}{\end{eqnarray}}
\newcommand{\bal}{\begin{aligned}}
\newcommand{\eal}{\end{aligned}}
\newcommand{\be}{\begin{eqnarray}}
\newcommand{\ee}{\end{eqnarray}}
\newcommand{\bq}{\begin{equation}}
\newcommand{\eq}{\end{equation}}
\newcommand{\benn}{\begin{equation*}}
\newcommand{\eenn}{\end{equation*}}
\newcommand{\bi}{\begin{itemize}}  
\newcommand{\ei}{\end{itemize}}
\renewcommand{\d}{\partial}

\newcommand{\Acal}{{\mathcal A}} 

\newcommand{\Ccal}{{\mathcal C}}

\newcommand{\Gcal}{{\mathcal G}}

\newcommand{\Ical}{{\mathcal I}}

\newcommand{\Lcal}{{\mathcal L}}
\newcommand{\Mcal}{{\mathcal M}}
\newcommand{\Ncal}{{\mathcal N}}

\newcommand{\Ocal}{{\mathcal O}}
\newcommand{\cO}{{\mathcal O}}
\newcommand{\Pcal}{{\mathcal P}}

\newcommand{\Tcal}{{\mathcal T}}

\newcommand{\Wcal}{{\mathcal W}}

\newcommand{\CC}{{\cal C}}

\newcommand{\CI}{{\cal I}}
\newcommand{\CL}{{\cal L}}

\newcommand{\CM}{{\cal M}}
\newcommand{\CN}{{\cal N}}
\newcommand{\CO}{{\cal O}}

\newcommand{\nn}{\nonumber}

\newcommand\oo\infty
\newcommand\s\sigma
\newcommand\de\delta
\newcommand\De\Delta
\newcommand{\p}{\partial}
\newcommand\f\phi
\newcommand\g\gamma
\newcommand\x\times

\newcommand{\ra}{\rightarrow}
\newcommand{\lra}{\leftrightarrow}
\newcommand{\fr}{\frac}
\newcommand{\comm}[2]{[#1,#2]}
\newcommand{\acomm}[2]{\{#1,#2\}}

\newcommand{\CFT}{\textrm{CFT}}
\newcommand{\tfr}{\tfrac}
\newcommand{\KL}{K\"{a}ll\'{e}n-Lehmann }

\newcommand{\sgn}{{\rm sgn}}
\newcommand{\eff}{\textrm{eff}}
\newcommand{\gap}{\textrm{gap}}
\newcommand{\ET}{\textrm{ET}}
\newcommand{\LC}{\textrm{LC}}

\newcommand{\Cmax}{\Ccal_{\max}}
\newcommand{\Dmax}{\De_{\max}}
\newcommand\lrpar{\raise .8ex\hbox{$^\leftrightarrow$} \hspace{-9pt}
\partial}

\newcommand{ \sqb }[1]{ \left[ #1 \right]}

\newcommand\G{\Gamma}

\newcommand{\Chat}{\widehat{C}}

\newcommand{\bk}{\boldsymbol{k}}

\newcommand{\bs}{\boldsymbol{\sigma}}
\newcommand{\dagg}{\dagger}
\newcommand{\norm}[1]{\left\lVert#1\right\rVert}
\newcommand{\Tr}{\textrm{Tr}}

\newcommand{\norder}[1]{%
  {:\mathrel{\mspace{1mu}#1\mspace{1mu}}:}%
}

\newcommand{\ft}[1]{\footnotesize #1 \normalsize}
\usepackage{listings}
\usepackage{xcolor}
\usepackage{mdframed}

\usepackage{mmacells} 
\mmaSet{
  leftmargin=2.3em,
  labelsep=-0.8em,
}

\usepackage{afterpage}
\usepackage{bookmark}
\usepackage{pifont}

\newenvironment{monospace}{\ttfamily}{\par}
\newenvironment{code}{\begin{monospace}\begin{tabbing}}{\end{tabbing}\end{monospace}}

\newcommand{\Bvec}{\boldsymbol{\beta}}
\newcommand{\Lvec}{\boldsymbol{\ell}}

\newcommand{\kvec}{\boldsymbol{k}}
\newcommand{\yvec}{\boldsymbol{k}}

\newcommand{\pr}[1]{\left(#1 \right)}
\newcommand{\Psibar}{\overline{\Psi}}
\newcommand{\vac}{| {\rm vac} \>}

\newcommand{\ptl}{\partial}
\newcommand{\Kvec}{\boldsymbol{k}}

\newcounter{mypageno}
\let\oldtableofcontents\tableofcontents
\renewcommand{\tableofcontents}{%
  \oldtableofcontents
  \clearpage
  \setcounter{mypageno}{\value{page}}%
  \pagenumbering{arabic}%
  \setcounter{page}{\value{mypageno}}%
}

\makeatletter
\def\@fpheader{\vspace{-.1cm}}
\makeatother

\title{\Large
Introduction to Lightcone Conformal Truncation: \\ QFT Dynamics from CFT Data 
}

\author[a]{Nikhil Anand,}
\author[b]{A. Liam Fitzpatrick,}
\author[b]{Emanuel Katz,}
\author[b,c]{Zuhair U. Khandker,}
\author[d,e]{Matthew T. Walters,}
\author[b]{Yuan Xin}

\affiliation[a]{Department of Physics, McGill University, Montr\'eal, QC H3A 2T8, Canada}
\affiliation[b]{Department of Physics, Boston University, Boston, MA 02215, USA}
\affiliation[c]{Department of Physics, University of Illinois, Urbana, IL 61801, USA}
\affiliation[d]{Theoretical Physics Department, CERN, 1211 Geneva 23, Switzerland}
\affiliation[e]{Institute of Physics, \'Ecole Polytechnique F\'ed\'erale de Lausanne (EPFL), CH-1015 Lausanne,
Switzerland}

\abstract{
 We both review and augment the lightcone conformal truncation (LCT) method. LCT is a Hamiltonian truncation method for calculating dynamical quantities in QFT in infinite volume. This document is a self-contained, pedagogical introduction and ``how-to'' manual for LCT. We focus on 2D QFTs which have UV descriptions as free CFTs containing scalars, fermions, and gauge fields, providing a rich starting arena for LCT applications. Along our way, we develop several new techniques and innovations that greatly enhance the efficiency and applicability of LCT. These include the development of CFT radial quantization methods for computing Hamiltonian matrix elements and a new SUSY-inspired way of avoiding state-dependent counterterms and maintaining chiral symmetry. We walk readers through the construction of their own basic LCT code, sufficient for small truncation cutoffs. We also provide a more sophisticated and comprehensive set of Mathematica packages and demonstrations that can be used to study a variety of 2D models. We guide the reader through these packages with several examples and illustrate how to obtain QFT observables, such as spectral densities and the Zamolodchikov $C$-function. Specific models considered are finite $N_c$ QCD, scalar $\phi^4$ theory, and Yukawa theory.
}

\arxivnumber{}

\begin{document}
\maketitle

\section{Introduction}

Quantum Field Theory (QFT) is a remarkably versatile framework and has found applications at nearly all length scales.  However, it is notoriously difficult to investigate quantitatively at strong coupling. It is especially challenging at strong coupling to obtain predictions associated with real-time evolution of observables or the precise wavefunctions of excited states. An old dream \cite{doi:10.1142/0543} is to obtain such information by applying variational methods to QFT.  The most successful variational methods in QFT have been Hamiltonian truncation methods, where the variational parameters are the coefficients of states in a basis of a well-chosen subspace of the full Hilbert space.\footnote{One of the first applications of Hamiltonian truncation to nonperturbative QFT was by Yurov and Zamolodchikov~\cite{Yurov:1989yu,Yurov:1991my}. Since then, there has been exciting progress on many fronts. Hamiltonian truncation has been used to investigate a variety of different QFT models and phenomena, e.g., spontaneous symmetry breaking~\cite{Coser:2014lla,Rychkov:2014eea,Rychkov:2015vap,Bajnok:2015bgw}, scattering~\cite{Bajnok:2015bgw,Gabai:2019ryw}, and quench dynamics~\cite{Rakovszky:2016ugs,Hodsagi:2018sul}, to name a few. In addition, there have been crucial recent advancements that have greatly enhanced the scope and predictive power of truncation methods. These developments include the systematic incorporation of renormalization effects from high energy states allowing for high-precision studies of various models~\cite{Elias-Miro:2015bqk,Elias-Miro:2017xxf,Elias-Miro:2017tup}, progress in numerical diagonalization methods for large matrices~\cite{Lee:2000ac,Lee:2000xna}, and formulations of truncation in $d>2$ dimensions~\cite{Hogervorst:2014rta,Hogervorst:2018otc,EliasMiro:2020uvk}. Indeed, we find ourselves in a time of rapid development for truncation methods in QFT. For a recent overview with a comprehensive list of references, see~\cite{2018RPPh...81d6002J}.}  

This review is meant to serve as an introduction and user's guide for a specific kind of Hamiltonian truncation known as Lightcone Conformal Truncation (LCT)~\cite{Katz:2013qua,Katz:2014uoa,Katz:2016hxp,Anand:2017yij,Fitzpatrick:2018ttk,Delacretaz:2018xbn,Fitzpatrick:2018xlz,Anand:2019lkt,Fitzpatrick:2019cif} that works in lightcone quantization for QFTs that have UV CFT fixed points.  As we will describe in detail, lightcone quantization circumvents a number of issues that are typically challenging for Hamiltonian truncation methods in QFT. Chiefly among its virtues is the fact that bubble diagrams vanish, which allows it to be constructed at least formally in infinite volume, reduces the degree of divergence of most UV divergences, and moreover facilitates the calculation of spectral densities for correlation functions. The price to be paid for these simplifications is usually conceptual,  in that lightcone quantization can behave counterintuitively at times and there can be effects related to modes with zero lightcone momenta that are subtle to include. Nevertheless, it can be a powerful tool for studying new regimes of strongly coupled QFT.  

Our focus throughout will be to prepare the reader for doing their own LCT analyses and to lower the barrier-to-entry for users who may be interested in these techniques but find daunting the amount of new code to write and conceptual subtleties to absorb.  For this reason, we will focus entirely on the restricted class of models that live in two spacetime dimensions and can be obtained by starting with a free theory in the UV and deforming it by a relevant operator.  Although the method is more general, this is the class that is best understood and moreover where the numerical methods are most developed and efficient.  There are still many rich and interesting models within this class, so it can provide a very effective ``warm-up'' arena  that is compelling in its own right.

To be as self-contained as possible, we will take the reader from the most basic aspects of LCT to the development of efficient numerical code and finally through the analysis of concrete models.  By the end of the paper, the reader who 
works through all the examples will have  \emph{their own} basic truncation code that will suffice for small bases.  We also provide more sophisticated Mathematica code that is more efficient and can go to much larger bases, and we discuss the main innovations behind the improved code so that the reader can understand how it works and can use it themselves.  

We emphasize that while the details of the implementation are complicated, the basic overall strategy of our approach is simple and involves the following steps:
\begin{enumerate}
\item Start with a CFT. Construct a complete basis of primary operators with conformal Casimir $\Ccal$ below some threshold $\Cmax \sim \Delta_{\rm max}^2$.
\item Deform the CFT by relevant deformations. Compute the lightcone Hamiltonian matrix elements of the deformations for all states in the truncated basis.
\item Diagonalize the truncated lightcone Hamiltonian 
to determine the mass spectrum.
\item Use the resulting eigenstates to compute dynamical observables (such as spectral densities) in the deformed theory.
\end{enumerate}

Although the basic strategy is general, for practical reasons most of this review will deal with the case where the original CFT is a free theory.  Steps 1 and 2 are the most involved, and we will present three different methods for accomplishing them: the ``Fock Space'' method, the ``Wick Contraction'' method, and the ``Radial Quantization'' method.  The Fock space method is the simplest to understand pedagogically, but the least efficient computationally, whereas the Radial Quantization method is our most efficient method and is what we use in our code. The Wick Contraction method falls somewhere between the other two both conceptually and computationally.  

We present three parts, each of which can largely be  read independently.

{\bf Part I} covers the basics of LCT, explaining the conceptual underpinnings and performing some very simple computations.  It introduces all the ideas that are needed in principle to construct the basis of states used by the truncation, and the matrix elements of the Hamiltonian for these states with some simple interactions.  We also walk through explicitly how to implement these ideas in order to build actual code that computes the Hamiltonian, albeit much too inefficiently to be practical for large bases.  Nevertheless, the reader who just reads Part I should come away understanding what LCT is and how it works.  

In {\bf Part II}, our focus becomes more practical and we demonstrate a number of techniques for drastically improving the computational efficiency of the method.  These methods are the ones that we use in our numeric code for computing Hamiltonian matrix elements, and our goal is for the reader to understand them well enough that they not only have a sense of what the code is actually doing but could make their own improvements or generalizations to the publicly available code.  

Finally, in {\bf Part III} we go through two specific models -- a real scalar with a quartic potential, and a real scalar and a Majorana fermion coupled through a Yukawa potential  -- to show how to take the truncated Hamiltonian and compute useful dynamical quantities. The simplest observable is the spectrum and it is of course just the eigenvalues of the Hamiltonian, but diagonalizing the Hamiltonian also gives us its eigenvectors and we show how to use these to compute spectral densities.  Interpreting the results of these computations can at times be subtle, since truncation and lightcone effects must be understood and taken into account in the analysis, so these applications provide us an opportunity to guide the reader through some of the main issues they would encounter when trying to analyze their own model. 

In Table \ref{tab:timingBenchmark:Intro}, we give some benchmarks of the timing on a single CPU for two versions of the code - a pedagogical version of the ``Wick Contraction'' method with simple code given explicitly in the text in {\bf Part I}, and our most efficient ``Radial Quantization'' code developed in {\bf Part II} - for a scalar theory in 2d with a $\phi^2$ mass  term and a $\phi^4$ quartic  term.

\begin{table}[h]
\centering
\begin{tabular}{|c|c||c|c|c||c|c|c|} 
\multicolumn{2}{c||}{} & \multicolumn{3}{c||}{Pedagogical} & \multicolumn{3}{c}{Radial Quantization} \\ 
\hline
 $\Delta_{\max}$ & $\substack{\text{num of}\\\text{states}}$ & basis & mass & quartic & basis & mass & quartic  \\
 \hline
 10 & 42 &  0.19 & 0.26 & 2.36 & 0.02 & 0.06 & 0.07  \\
 \hline
 20 & 627 & 3061 & 170.1 & 5183 & 0.46 & 1.09 & 3.96  \\
 \hline
 30 & 5604 & {\rm weeks?} & {\rm hours?} &  {\rm weeks?}  &  7.88 & 17.93 & 111.9  \\
 \hline
 40 & 37338 & \multicolumn{3}{c||}{{\rm Good luck}} &  231 & 410 & 3579 \\
 \hline
\end{tabular}
\caption{\label{tab:timingBenchmark:Intro}
The timing benchmark of the scalar $\phi^4$ basis and matrix elements using two different methods. 
 The table shows the time used to compute each component of the scalar basis and matrix element data, at different $\Dmax$. 
 Times are in seconds unless otherwise specified. ``?'' indicates a time estimated by extrapolation. 
 }
\end{table}

Compared to previous work, this review introduces a number of developments that significantly improve efficiency and allow a wider range of theories to be studied.  In section \ref{sec:ScalarBasis}, we introduce a much faster way to construct primary operators by applying a result of Penedones \cite{Penedones:2010ue} for recursively building higher particle-number primaries out of lower ones.   In section \ref{sec:Fermions}, we demonstrate how to treat matrix elements for fermions - in particular, mass terms and Yukawa interactions - that were not understood previously in LCT due to the presence of IR divergences. Moreover, we show that once one has a basis for all-scalar primaries and all-fermion primaries, there is an efficient way to combine them to create mixed scalar-fermion primaries.   Essentially all of {\bf Part II} is dedicated to developing a new method for computing matrix elements by using radial quantization techniques to streamline the computation of correlators in position space, before we conformally map them back to infinite volume flat space and Fourier transform them to get the matrix elements in a momentum space basis. 

This review also includes a new prescription for dealing with state-dependent counterterms and restoring chiral symmetry.  One of the complexities introduced by Hamiltonian truncation is the appearance of state-dependent terms in the Hamiltonian in the presence of divergences.  For example, such state-dependent terms are generated in a theory
containing Yukawa interactions.  Moreover, in lightcone quantization, for reasons explained in section \ref{sec:YukawaTheory}, chiral symmetries do not protect the fermion mass, leaving it vulnerable to corrections which do not vanish as the fermion mass goes to zero.  However, we find that one can generate a counterterm using a mechanism from supersymmetry which removes unwanted state-dependent terms and restores chiral symmetry even in non-supersymmetric theories such as a Yukawa theory.  This mechanism is described in section \ref{sec:YukawaTheory}.

The software package accompanying this paper can be found at the code repository \href{https://github.com/andrewliamfitz/LCT}{\tt https://github.com/andrewliamfitz/LCT}. In addition to code that implements all methods introduced in Parts I-III, this repository also contains several ``demo'' Mathematica notebooks as pedagogical tutorials for users interested in working through the various examples presented in this text. The package Readme file contains more details.
 
\subsection{Reader's Guide} This text covers a large number of technical and conceptual topics pertaining to conformal truncation. We understand that most readers will not require all parts of this text to understand LCT methodology, or to begin doing their own analyses using our code. To this end, we have prepared the following table that may help extract the necessary parts of this text. To use it, \underline{we recommend reading section \ref{sec:ReviewOfLCT} first}, and then looking at the table to decide what to read next:

\begin{center}
\resizebox{6.1in}{!}{%
\begin{tabular}{p{0.5\textwidth};{2pt/2pt}p{0.45\textwidth}}
\hline
\textbf{\textbf{If the reader wants to...}} & \textbf{We recommend that they...}  \\ \hline \hline
Just gain a zeroth-order conceptual understanding of LCT methods and work through some explicit examples & Work through sections \ref{sec:FreeLC}, \ref{sec:FockSpace}.  \\ \hline
Begin writing their own basic LCT code & Read section \ref{sec:2dFFT}, and work through section \ref{sec:simplestcode}. \\ \hline
Understand some of the CFT techniques that greatly improve the efficiency of the code & Read Part II, starting at section \ref{sec:RadialScalars}. \\ \hline
Work through small-scale demonstrations and examples in Mathematica / reproduce figures in the text & Run the ``demo'' Mathematica notebooks from the code repository. \\ \hline 
Work through the more major applications in Part III  &  Download our code/data and begin at Part III, with section \ref{sec:ApplicationPhi4}. More details are also in the code Readme. \\ \hline
\end{tabular}
}
\end{center}

\

\noindent If readers would like to jump to our LCT results in specific theories, we refer them to the following figures:\\

\noindent {\bf Finite $N_c$ massless QCD} \\
Fig.~\ref{fig:gauge-d9-spectral-density}.\, Spectral density of the stress tensor $T_{--}$\\
Fig.~\ref{fig:gauge-d9-data}.\, Mass spectrum (parity-even single particles)\\

\noindent {\bf $\phi^4$-theory } \\
Fig.~\ref{fig:Phi4LowEvals}.\, Mass spectrum\\
Fig.~\ref{fig:Phi4SpectralDensities}.\, Spectral density of $\phi^2$ and Zamolodchikov $C$-function\\
Fig.~\ref{fig:Phi4TheoryUniversality}.\, Universality of $\phi^{2n}$ near the critical point\\
Fig.~\ref{fig:Phi4TheoryTrace}.\, Vanishing of the stress tensor trace near the critical point\\
Fig.~\ref{fig:Phi4CFunction}.\, $C$-function near the critical point\\

\noindent {\bf Yukawa theory } \\
Fig.~\ref{fig:YukawaMassSpectrum}.\, Mass spectrum\\
Fig.~\ref{fig:YukawaCFunction}.\, Zamolodchikov $C$-function\\
Fig.~\ref{fig:breit-wigner}.\, Spectral density of $\phi$ and the associated Breit-Wigner $\phi$ resonance\\

\noindent
In some sections, we have included more detailed explanations that the reader may wish to skip on a first pass. These comments are delineated from the main text by horizontal bars and smaller font.

\clearpage

\newpage
\section*{\Large Part I: Foundations}
\addcontentsline{toc}{part}{Part I: Foundations}
\label{sec:PartI}

%%%%%%%%%%%%%%%%%%%%%%%%%%%%%%%%%%%%%%%%%%%%%%%%%%%%%%%%%%%%%%%%%%%%%%%%%%%%%
%%%%%%%%%%%%%%%%%%%%%%%%%%%%%%%%%%%%%%%%%%%%%%%%%%%%%%%%%%%%%%%%%%%%%%%%%%%%%

\section{Review of Lightcone Conformal Truncation}
\label{sec:ReviewOfLCT}

In this section, we provide a concise introduction to the method of lightcone conformal truncation. The goal of this section is to give a sense of the necessary ingredients involved in using LCT to study deformations of general CFTs, as well as a small preview of the calculations performed in specifically studying deformations of free theories in 2d (which is the focus of this paper). To this end, we will somewhat ignore the technical details, which can be found in later sections, and instead focus on the conceptual structure.

%%%%%%%%%%%%%%%%%%%%%%%%%%%%%%%%%%%%%%%%%%%%%%%%%%%%%%%%%%%%%%%%%%%%%%%%%%%%%

\setcounter{subsection}{-1}
\subsection{Hamiltonian Truncation in QM}
\label{subsec:HamiltonianTruncQM}

Before jumping into Hamiltonian truncation in 2d QFT, it is helpful to see how it works in the more familiar context of 1d quantum mechanics, where it was originally developed.  Truncation, also known as  {\it Rayleigh-Ritz}, methods are simply the idea of finding the spectrum of a Hamiltonian using a variational Ansatz  in terms of a finite set of $n_{\rm max}$ basis wavefunctions $\phi_n$:
\be
| \psi \>  = \sum_{n=1}^{n_{\rm max}} c_n | \phi_n \> .
\ee
The basis wavefunctions $\phi_n$ are chosen a priori and the coefficients $c_n$ are the variational parameters.  We will mainly be interested in the case where the Hamiltonian can be separated into a solvable piece $H_0$ and a deformation $V$:
\be
H = H_0 + g V.
\ee
Despite the apparent similarity with perturbation theory, however, truncation methods are {\it nonperturbative} in the deformation parameter $g$, which does not have to be small.  Instead of using $H_0$ to set up an expansion in powers of $g$, we use $H_0$ to motivate a specific choice for the basis of wavefunctions $\phi_n$ -- we simply take the $n_{\rm max}$ eigenvectors of $H_0$ with the smallest overall eigenvalues:
\be
|\phi_n\> \equiv | n \> , \qquad H_0 |n \> = \mu_n | n \> \qquad (n=1,\ldots,n_{\max}).
\ee
The Hamiltonian acting on any state in the subspace spanned by the $\phi_n$'s is then given by the undeformed eigenvalues $\mu_n$ and the matrix elements $V_{nm} \equiv \< n | V | m\>$ of the deformation.

\begin{figure}[t!]
\begin{center}
\includegraphics[width=0.73\textwidth]{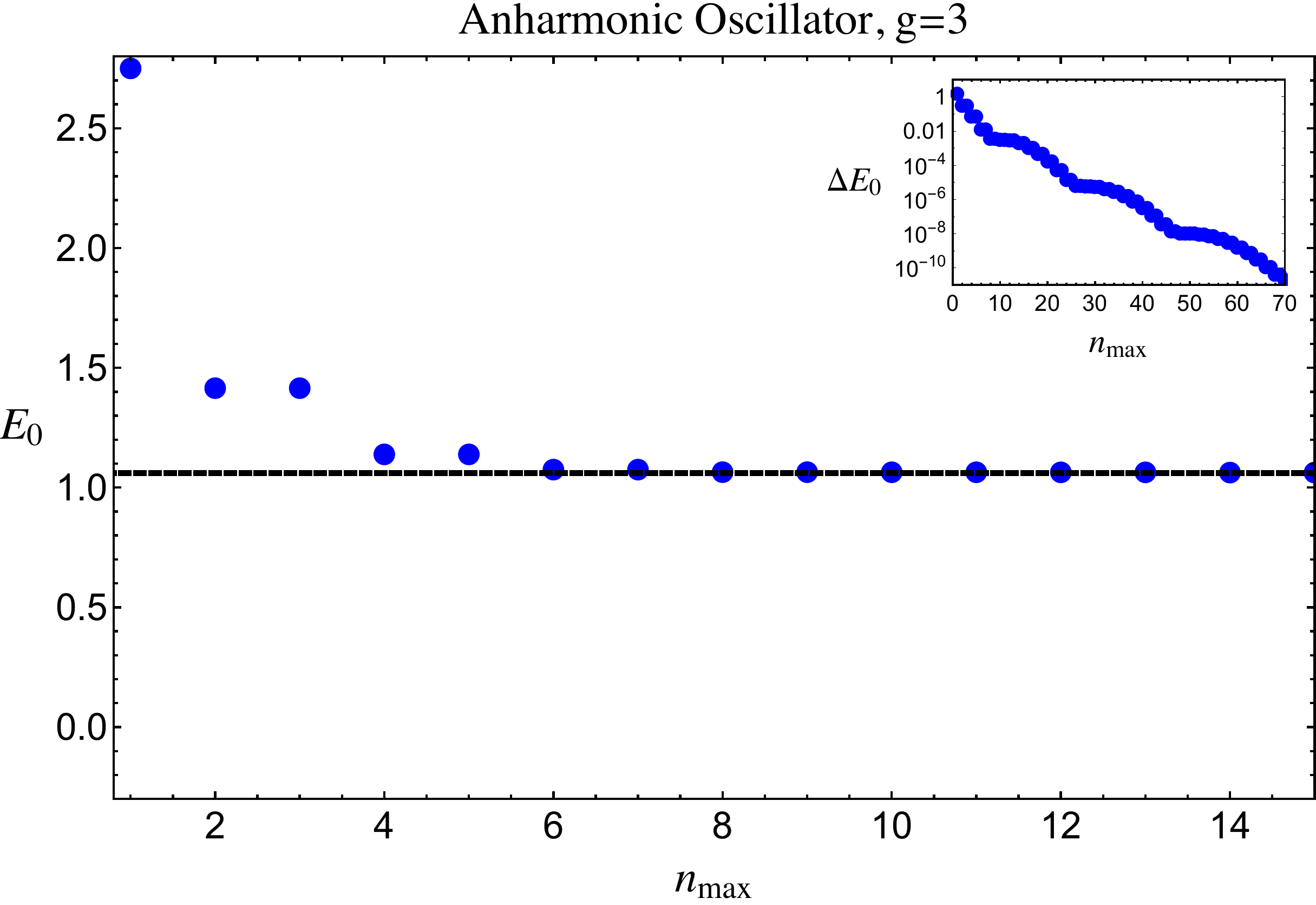}
\caption{Ground state energy $E_0$ and residual $\Delta E_0 \equiv E_0 - E_0^{(n_{\rm max}=\infty)}$ as a function of truncation level $n_{\rm max}$ for the anharmonic oscillator at coupling $g=3$. $E_0^{(n_{\rm max}=\infty)}$ is shown in dashed, black.  The residuals decay exponentially as a function of $n_{\rm max}$ \label{fig:anharmonicE0}}
\end{center}
\end{figure}

One of the simplest applications of this method is the anharmonic oscillator:
\be
H = \frac{1}{2} (p^2 + x^2) + g x^4 = H_{\rm SHO} + g V, \qquad V = x^4 = \frac{1}{4} (a+a^\dagger)^4.
\ee
The eigenvalues of $H_{\rm SHO}$ are simply $\mu_n = n+\frac{1}{2}$, whereas the matrix elements of $V$ in the Simple Harmonic Oscillator (SHO) basis are easy to evaluate, since $\< n | a | m\> = \sqrt{n} \delta_{n, m-1}$.  As a result, in just a few lines of code, one can numerically compute the Hamiltonian in a truncated basis $\{ |n \> \}_{n \le n_{\rm max}}$ for finite $g$ and $n_{\rm max}$; an example is shown in Table \ref{fig:AnharmonicCodeA}.
 We encourage the reader to use code such as that given in the table to experiment with different choices of $n_{\rm max}$ and $g$ themselves.  Numerically diagonalizing this matrix gives results for the lowest energy levels $E_n$ that converge very quickly to the exact result as we increase $n_{\rm max}$, as shown in Fig.~\ref{fig:anharmonicE0}.  By contrast, the perturbative series in $g$ for, say, the ground state energy $E_0$ of $H$ is an asymptotic series, and therefore has an irreducible error $ \sim e^{-\frac{1}{g}}$ for any nonzero coupling.   
For many more results on truncation applied to 1d quantum systems, we refer the reader to \cite{reed1978iv}.

Because Hamiltonian truncation is simply a variational method, the lowest eigenvalue of the truncated matrix for any value of $n_{\max}$ provides an upper bound on the ground state of the full Hamiltonian, as we can see in Fig.~\ref{fig:anharmonicE0}. More powerfully, the $p$-th smallest eigenvalue of the full Hamiltonian is bounded above by the $p$-th smallest eigenvalue of the truncated Hamiltonian \cite{macdonald1933successive}.

\begin{table}[t!]
\begin{tabular}{l>{\columncolor[gray]{0.9}}l}
\verb|X[n_, np_] :=  Switch[n-np,| &  $ x_{n n'} =  \sqrt{\frac{n}{2}}\delta_{n,n'-1} $ \\ 
\verb|                       1,Sqrt[n/2],| &  $ \qquad \qquad + \sqrt{\frac{n'}{2}} \delta_{n+1,n'}$\\
\verb|                       -1,Sqrt[np/2], | & \\
\verb|                       _, 0]| & \\
    & \\
\verb|XMat[nm_]:=Outer[X,#,#]&[Range[0,nm]] | & $({\bf X})_{n n'} = x_{n n'}, \ n, n' \le n_{\rm max}$  \\  & \\
\verb|H[nm_]:=DiagonalMatrix[1/2+Range[0,nm]]| & $H_{n n'} = E_n \delta_{n n'} + g({\bf X}^4)_{nn'}$\\ 
\verb|+g Take[MatrixPower[XMat[nm+2],4],nm+1,nm+1];| \\
\end{tabular}
\caption{Anharmonic Oscillator in Hamiltonian Truncation  \label{fig:AnharmonicCodeA}}
\end{table}

 It is illuminating to see in more detail in this example how Hamiltonian truncation reorganizes perturbation theory from an asymptotic expansion to a convergent one.  Because the interaction $V$ raises/lowers the oscillator number $n$ by at most 4, the perturbation series in $g$ for the ground state energy $E_0$ at $\CO(g^{2n})$ only involves states up to $n$, as one can easily see using time-independent perturbation theory for $E_0$.  By a standard perturbative calculation, $E_0$ up to $\CO(g^2)$ is
 \be
 E_0 = \frac{1}{2} + \frac{3}{4} g- \frac{21}{8} g^2 + \CO(g^3) .
 \ee
 The ground state energy for the few first truncation levels, however, is
 \be
&& n_{\rm max} = 0 : E_0 = \frac{1}{2} + \frac{3}{4} g ,\\
&& n_{\rm max} = 2 : E_0 = \frac{3}{2} + \frac{21}{4} g - \frac{\sqrt{4+36 g+99 g^2}}{2}  \approx \frac{1}{2} + \frac{3}{4} g - \frac{9}{4} g^2 + \CO(g^3) , \nn\\
&& n_{\rm max} = 4 : E_0 = \frac{55 g+10}{4} + W+\frac{4 +60 g+390 g^2}{3 W} \approx \frac{1}{2} + \frac{3}{4} g - \frac{21}{8 } g^2 + \CO(g^3) , \nn
\ee
\begin{equation}   \Big(W=Q^{\frac{1}{2}} R^{\frac{1}{3} }e^{\frac{2 \pi i }{3}}
 \left( 1+ \sqrt{1-R^{-2}}\right)^{\frac{1}{3}}  , \hspace{2mm}  R= \frac{\frac{5575 g^3}{4}+\frac{525 g^2}{2}+8 g}{Q^{\frac{3}{2}}}, \hspace{2mm} Q= 130 g^2+20 g+\frac{4}{3}\Big). \nn
 \end{equation}
Note that at $n_{\rm max}=4$, the series expansion of $E_0$ correctly matches the exact series expansion up to $\CO(g^2)$.  As we increase $n_{\rm max}$, the low-order terms in the series expansion stabilize and stop changing, and in a sense one can think of increasing $n_{\rm max}$ as only adding new higher order terms. However, increasing $n_{\rm max}$ is very unlike perturbation theory in that each time it increases it changes {\it all} the higher order terms, in such a way that the result for finite $g$ converges. Crucially, the truncation results for the ground state energy (past $n_{\rm max} \ge 2$) are non-analytic functions of the coupling, enabling them to capture nonperturbative behavior.

%%%%%%%%%%%%%%%%%%%%%%%%%%%%%%%%%%%%%%%%%%%%%%%%%%%%%%%%%%%%%%%%%%%%%%%%%%%%%

\subsection{General Setup}
\label{subsec:generalsetup}

\begin{figure}[t!]
\begin{center}
\includegraphics[width=0.73\textwidth]{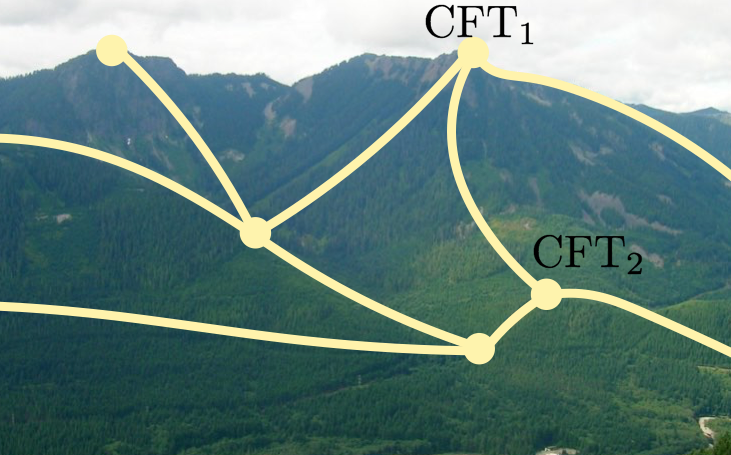}
\caption{Cartoon of  Space of CFTs and the RG flows between them. Not drawn to scale. One of the goals of conformal truncation is to turn this cartoon into a sharp computational tool.}
\label{fig:CFTSpaceCartoon}
\end{center}
\end{figure}

The basic idea of Hamiltonian truncation methods applied to QFT is, at heart, the same as in the 1d quantum mechanics example we have just described.  We begin by separating the Hamiltonian of the theory into a piece $H_0$ that we know how to solve, and a deformation $V$ whose matrix elements we have to calculate.  For {\it conformal truncation} techniques, we further specify that the undeformed Hamiltonian $H_0$ should be that of a CFT.  The motivation for focusing on this class of theories is partly practical and partly philosophical.  The philosophical motivation is that CFTs obey stringent constraints, most notably the conformal bootstrap equation, that can be used to put them on a rigorous mathematical footing.  One can then hope to obtain a rigorous foundation for most or all of QFT as points along the RG flow between CFTs.  A cartoon version of this concept is shown in Fig.~\ref{fig:CFTSpaceCartoon}.  The practical motivation is that in order to apply Hamiltonian truncation to a particular theory, we need to be able to efficiently compute all the matrix elements of the deformation $V$.  Conformal symmetry can often be used to drastically streamline these computations, as we will see. Moreover, in practice we find that in many applications using conformal symmetry to organize the truncated basis leads to rapid convergence to the exact eigenvalues of the full QFT Hamiltonian. It would be very interesting to better understand the nature of this convergence and for which classes of theories this behavior holds.

In more detail, imagine that we have been provided with all of the scaling dimensions and spin quantum numbers $(\Delta_i,\ell_i)$\footnote{Equivalently, the weights $(h_i, \bar{h}_i) = (\frac{\Delta_i+\ell_i}{2}, \frac{\Delta_i - \ell_i}{2})$.} of the primary operators in a CFT, as well as all of the OPE coefficients $C_{ijk}$.  This information is commonly known as the ``CFT data'', since by repeated applications of the OPE it can be used to reconstruct any correlation function of local operators in the theory.  To trigger an RG flow in the theory, we can add a scalar primary operator to the theory:
\be
H = H_{\rm CFT} + g \int d^{d-1} x \, \CO(x).
\ee
If $\CO$ is an irrelevant operator, i.e.~$\Delta_\CO > d$, then in the IR the theory simply flows back to the original CFT.  However, if $\CO$ is relevant, i.e.~$\Delta_\CO<d$, then the RG flow of the theory takes it to a new CFT in the IR.  We will therefore restrict to the case of relevant operator deformations.

To apply Hamiltonian truncation techniques to this setup, we have to choose a basis of states.  Because the UV theory is a CFT by assumption, a natural strategy is to define the basis of states in terms of the primary operators of the UV CFT.  A familiar way to define states in terms of CFT operators is using radial quantization, where states are created by operators acting at the origin on the vacuum, but we can be more general and define basis states as weighted integrals of  primary operators:
\be
|\CO_i, f\> \equiv \int d^d x \, f(x) \, \CO_i(x) | {\rm vac} \>. 
\ee
The matrix elements of the deformation then reduce to weighted integrals over three-point functions of primary operators.  Because the three-point functions in CFTs are determined by conformal invariance in terms of the OPE coefficients $C_{ijk}$ and operator content $(\De_i, \ell_i)$, there is a simple relation between the CFT data and the matrix elements of $H$ in this basis.  

To specify the basis further, we can use the fact that the quadratic Casimir $\CC$ of the conformal algebra,
\be
\CC = D^2 + \frac{1}{2} L_{\mu\nu} L^{\mu\nu} - \frac{1}{2} \{ P_\mu , K^\mu \}, \label{eq:ConfCasimir}
\ee
together with the generators $P_\mu$ of translations are all mutually commuting, and so can be simultaneously diagonalized. We can therefore take our basis of states to be eigenstates of $\CC$ and $P_\mu$.     The quadratic Casimir acting on an irrep whose primary has dimension $\Delta$ and spin $\ell$ is $\CC = \Delta(\Delta-d) + \ell(\ell+d-2)$.  In the absence of additional symmetries, one does not expect two different irreps of the conformal algebra to have the same Casimir, so this prescription would be sufficient to specify the basis.  However, the cases of most interest generally {\it will} have additional symmetries, in order to make it possible to compute the CFT data. In that case, we can easily further label each basis state by choosing a particular primary operator for each irrep.  Primary operators in a CFT are local operators that are annihilated by the special conformal generator $K_\mu$ acting at the origin,
\be
[K_\mu, \CO(0)] = 0 ,
\label{eq:KOPrim}
\ee
and starting from a primary operator, all other operators in the same conformal representation can be obtained by `raising' with $P_\mu$.  So, 
given all the primary operators $\CO_i$ in the theory, we can label our complete basis of states  as follows:\footnote{One might find it surprising that all states on a fixed spacetime slice
can be associated with primary operators, as this sort of statement is often associated with radial quantization. However, we can see that this is the case by considering the K\"all\'en-Lehmann spectral decomposition of any CFT two-point function,
\be
\<\Ocal_i(x) \Ocal_i(0)\> = \sum_\psi \int \fr{d^dp}{(2\pi)^d} \<\Ocal_i(x)|\psi,p_\mu\>\<\psi,p_\mu|\Ocal_i(0)\> = \int \fr{d^dp}{(2\pi)^d} e^{-ip\cdot x} \rho_{\Ocal_i}(p).
\ee
Because in a CFT the two-point functions are diagonal, each primary operator must overlap with a unique linear combination of momentum eigenstates which is orthogonal to all other primaries; this linear combination defines the state $|\Ocal_i,p_\mu\>$. Furthermore, if there were an additional state that had no overlap with any local operators, this state would never contribute to the spectral decomposition of any operator, and so would not affect any correlators in the theory.}
\be
\boxed{| \CO_i, p_\mu \> \equiv \frac{1}{N_{\CO_i}} \int d^d x \, e^{-i p \cdot x} \CO_i(x) | {\rm vac} \>,}
\label{eq:BasisDef}
\ee
where $N_{\Ocal_i}$, which depends on $\De_i$, $\ell_i$, and $p_\mu$, simply ensures that the basis states are properly normalized.  Note that the norm is given simply by the Fourier transform of the two-point function of $\CO_i$, 
\be
\< \CO_i, p_\mu | \CO_j, p'_\mu \> = \frac{1}{N^*_{\CO_i}N_{\CO_j}} (2\pi)^d \delta^d(p-p') \int d^d x \, e^{i p \cdot x}  \< \CO_i(x) \CO_j(0)\>.
\label{eq:NormFromTwoPt}
\ee

To be concrete, consider a free scalar field in $d=3$.  There are two primary operators with Casimir eigenvalue $\CC=-2$, $\phi^2$ and $\phi^4$, so the complete set of eigenstates with $P_\mu=p_\mu$ and $\CC=-2$ are
\begin{equation}
|\phi^2, p_\mu\> \equiv \frac{1}{N_{\phi^2}} \int d^d x \, e^{-i p \cdot x} \phi^2(x) | {\rm vac} \>, \quad |\phi^4, p_\mu\> \equiv \frac{1}{N_{\phi^4}} \int d^d x \, e^{-i p \cdot x} \phi^4(x) | {\rm vac} \>.
\end{equation}
Because we work with eigenstates of $P_\mu$, primaries and all their descendants are automatically packaged together. 
To implement conformal truncation, we truncate this basis to a finite set of states by keeping only the finite set of irreps with conformal Casimir below some maximum value:\footnote{Depending on the context, slight modifications of this truncation can sometimes turn out to be more practical, but in all cases we keep a finite set of irreps with conformal Casimir below some threshold.}
\be
\boxed{\Ccal \le \Ccal_{\rm max}.}
\ee
It is often more intuitive to talk about the truncation as a cutoff $\Delta_{\rm max}$ on the scaling dimension of the primary operator of the irrep.

\subsection{Crash Course on Lightcone Quantization}
\label{subsec:CrashCourseonLC}

To go further, we must choose a time coordinate and a set of spatial surfaces, since this choice determines what we mean by the Hamiltonian. As we will see, there are a number of advantages (and also subtleties) to choosing lightcone (LC) quantization~\cite{Dirac:1949cp,Weinberg:1966jm,Bardakci:1969dv,Kogut:1969xa,Chang:1972xt}, \emph{i.e.}, surfaces of constant ``lightcone'' time $x^+$, where\footnote{It would be perhaps more accurate to refer to $x^+$ as ``lightfront'' time since the surfaces are planes and not cones; in $d=2$, these are equivalent.}
\be
x^{\pm} \equiv \frac{x^0 \pm x^1}{\sqrt{2}} .
\ee
The flat space metric in lightcone coordinates takes the form
\be
ds^2 = 2 dx^+ dx^- - \sum_{i=2}^{d-1} dx_i^2  = 2dx^+ dx^- - d\vec{x}^{\perp2}.
\ee
In this scheme, the Hamiltonian is $P_+ \equiv \fr{1}{\sqrt{2}}(P_0 + P_1)$, the generator of translations in $x^+$.   When we deform by a relevant operator $\CO_R$, we split the Hamiltonian into its undeformed piece $P_+^{(\rm CFT)}$ plus the deformation:
\be
P_+ = P_+^{(\rm CFT)} + \delta P_+, \qquad \delta P_+ = g \int dx^- d^{d-2} x^\perp \, \CO_R(x).
\ee
Because $\delta P_+$ is an integral over space, its matrix elements vanish between states with different $p_-$ or $\vec{p}_\perp$:
\be
\< \CO_i , p_\mu | \delta P_+ | \CO_j, p'_\mu\> = g \, C_{ijR} \, f_{ij}(p, p') \, (2\pi)^{d-1} \delta(p_- - p_-') \delta^{d-2}(\vec{p}_\perp - \vec{p}^{\ \prime}_\perp).
\label{eq:OPpO}
\ee
Here, $C_{ijR}$ is an OPE coefficient and $f_{ij}$ is a purely kinematic function that depends only on the scaling dimensions and spins of $\CO_i$ and $\CO_j$.\footnote{More precisely, when $\CO_i$ and $\CO_j$ have nonzero spin, (\ref{eq:OPpO}) contains a sum over OPE coefficients times kinematic factors for different polarization structures, $\sim \sum_\alpha C^{(\alpha)}_{ijR} \, f^{(\alpha)}_{ij}(p, p')$.}   To see this explicitly, we can write out $\delta P_+$ in the above expression as an integral over the relevant deformation:
\be
\bal
&\<\Ocal_i,p_\mu|\de P_+|\Ocal_j,p'_\mu\> = \frac{g}{N^*_{\CO_i}N_{\CO_j}} \int d^d x \, d^{d-1}y \, d^dz \, e^{i(p \cdot x - p' \cdot z)} \<\Ocal_i(x) \Ocal_R(y) \Ocal_j(z)\> \\[5pt]
& = (2\pi)^{d-1} \delta(p_- - p_-') \delta^{d-2}(\vec{p}_\perp - \vec{p}^{\ \prime}_\perp)\cdot \frac{g}{N^*_{\CO_i}N_{\CO_j}} \int d^d x \, d^d z \, e^{i(p\cdot x-p' \cdot z) } \<\Ocal_i(x) \Ocal_R(0) \Ocal_j(z)\>.
\label{eq:OPpO2}
\eal
\ee
Since $\delta P_+$ does not mix different $p_-$ and $\vec{p}_\perp$, we can restrict to a single ``momentum frame'' with fixed values for them.  Moreover, using boosts and rotations, without loss of generality we can set $p_-=1$ and $\vec{p}_\perp =0$. In this frame, the invariant momentum-squared is
\be
\mu^2 \equiv p^2 = 2 p_+. \label{eq:DefOfMuSq}
\ee

In practice, we are actually interested in the spectrum of the Lorentz invariant mass-squared operator $M^2 \equiv 2P_+ P_- - |\vec{P}_\perp|^2$. In LC quantization, when we deform the UV CFT by adding a relevant operator $\Ocal_R$, only the Hamiltonian $P_+$ is modified, while the generators $P_-$ and $\vec{P}_\perp$ remain unchanged. In our particular choice of frame, the matrix elements of $M^2$ thus take the simple form
\be
\label{eq:HamMatrixElement}
\<\Ocal_i, p |M^2|\Ocal_j,p'\> = \<\Ocal_i,p|(2P_+P_- - |\vec{P}_\perp|^2)|\Ocal_j,p'\> = 2\<\Ocal_i,p|P_+|\Ocal_j,p'\>.
\ee
We see that \emph{diagonalizing the LC Hamiltonian is equivalent to diagonalizing $M^2$, even at finite truncation}. In this work, we will therefore often refer to these two operators interchangeably.

Crucially, diagonalizing the truncated $M^2$ gives us not only its spectrum but also its eigenstates (labeled by their mass eigenvalue $\mu_i^2$) as linear combinations of the primary operators in our basis:
\be
|\mu_i^2,p_-\> = \sum_{\De_j \leq \Dmax} C^{\mu_i^2}_{\Ocal_j} |\Ocal_j,p\>, \qquad (P_+^{(\CFT)} + \de P_+)|\mu_i^2,p_-\> = \fr{\mu_i^2}{2p_-}|\mu_i^2,p_-\>.
\label{eq:FullEigenstates}
\ee
We can use these mass eigenstates to compute dynamical observables. Some of the simplest observables are the spectral densities of local operators $\rho_\Ocal(\mu)$, which encode the decomposition of two-point functions in terms of intermediate mass eigenstates,
\be
\<\Tcal\{\Ocal(x) \Ocal(0)\}\> = \int d\mu^2 \rho_\Ocal(\mu) \int \fr{d^dp}{(2\pi)^d} e^{-ip\cdot x} \fr{i}{p^2 - \mu^2 + i\epsilon}.
\ee
To compute the spectral density for any local operator $\Ocal$, we simply need to compute the overlap of that operator with the eigenstates of the truncated Hamiltonian
\footnote{As we will see in later sections, it is often simpler to study the integrated spectral density
\be
I_\Ocal(\mu) \equiv \int_0^{\mu^2} d\mu'^2 \, \rho_\Ocal(\mu) = \sum_{\mu_i \leq \mu} |\<\Ocal(0)|\mu_i^2,p_-\>|^2.
\ee
}
\be
\rho_\Ocal(\mu) \equiv \sum_i |\<\Ocal(0)|\mu_i^2,p_-\>|^2 \, \de(\mu^2 - \mu_i^2),
\label{eq:SpecDensGeneral}
\ee
which in turn requires us to compute the overlap of the local operator with our original basis states.  Note that by (\ref{eq:FullEigenstates}), these overlaps are simply sums over Fourier transforms of two-point functions of local operators in the original CFT.

There are advantages to LC quantization as well as complications.  One of the main advantages is that the vacuum of the theory remains trivial~\cite{Klauder:1969zz,Maskawa:1975ky,Brodsky:1997de}.  Namely, the vacuum is 
the unique state in the theory with  $p_-=0$, so the LC Hamiltonian $\delta P_+$ does not mix it with any other states.  As a result, there are no vacuum bubble contributions 
in perturbation theory.  This is in contrast to standard or Equal Time (ET) quantization, where both the ground state and the excited states get contributions to their energies that are extensive in the volume of the system, and one thus needs to compute energy differences.  If there are divergences in the theory, the lack of vacuum bubbles can often help 
reduce complexities associated with renormalization.  This is especially true for Hamiltonian truncation methods, where truncation can introduce state-dependent sensitivities to the cutoff that are more challenging to address (e.g., see~\cite{Rutter:2018aog}, as well as \cite{EliasMiro:2020uvk} for a proposed solution).  In addition, the lack of vacuum bubbles also turns off matrix elements in the Hamiltonian where particles are produced
from the vacuum, again simplifying the calculation.  In particular, unlike for ET quantization, in theories where one can take a large-$N$ limit, the calculation can be restricted to the 
``single-trace'' sector~\cite{Fitzpatrick:2018ttk}.

There is another issue to keep in mind with ET quantization, which correlates with the extensivity of all energies with the system volume:  the so-called orthogonality catastrophe~\cite{Anderson:1967zze}.
Roughly, since all states (including low energy states like the new ground state) have large energies, extensive in the volume, they require a large truncation energy to be 
captured by the original basis.  As the volume becomes larger, the truncation energy required must grow accordingly, with 
\be
E_{max} \gtrsim \epsilon(g) L^{d-1}. 
\ee
Here $L^{d-1}$ is the volume and $\epsilon(g)$ is the ground state energy density which depends on the relevant coupling $g$.  As $E_{max}$ grows, more states are involved, with an entropy which grows with the volume.  Hence, the overlap of the new ground state $|\Omega\rangle$ with the original vacuum state is exponentially suppressed
with the volume:
\be
|\langle 0|\Omega\rangle|^2 = e^{-L^{d-1} f(g)}.
\ee
Here $f(g)$ is a model-dependent function.  This exponential suppression applies to overlaps with excited states as well.  Hence, it is more difficult to take the large volume limit.  
However, in known applications, one can find a ``sweet window'' by choosing a volume which is large enough to do physics, but small enough so that the catastrophe is not yet an issue (see Appendix A of~\cite{Elias-Miro:2017tup} for a detailed discussion).  The situation can be further complicated if $f(g)$ is sensitive to the cutoff.  In other words, if the dimension of the relevant operator is $\Delta > (d+1)/2$, then $f(g) = c g^2 E_{max}^{2\Delta-d-1} + \cdots$.  In this case, the naive truncation procedure must be somehow modified in order to proceed, though perhaps by only including effects up to some fixed order in the coupling, similar to the approach of~\cite{EliasMiro:2020uvk}.

While the above are good reasons to use LC quantization, the LC scheme has its own complications.  The main issue involves ``zero modes'', i.e.~modes with $p_-=0$.
Such modes are naively excluded from the quantization scheme.  However, they cannot be simply discarded and must be integrated out properly, inducing new terms in the Hamiltonian.  A procedure for including their contribution was given in \cite{Fitzpatrick:2018ttk} (and summarized in appendix \ref{app:Heff}), where terms in the resulting effective Hamiltonian were computed perturbatively in the coupling.\footnote{For earlier work on LC zero modes, see for example~\cite{Chang:1968bh,Yan:1973qg,Wilson:1994fk,Tsujimaru:1997jt,Hellerman:1997yu,Burkardt2,Yamawaki:1998cy,Rozowsky:2000gy,Heinzl:2003jy,Beane:2013ksa,Herrmann:2015dqa,Hiller:2016itl,Collins:2018aqt}. }

In certain cases, integrating out the zero modes is equivalent to integrating out nondynamical fields, which can generate nonlocal terms, proportional to various relevant couplings squared.  In particular, one chiral half of a massless fermion is nondynamical (has no time derivatives in the action) in lightcone quantization, and we will encounter terms schematically of the form
\be
\sim \psi \frac{1}{\partial} \psi, \qquad \phi \psi \frac{1}{\p} \phi \psi, \qquad \psi \psi \frac{1}{\partial^2} \psi \psi,
\ee
from a fermion mass term, Yukawa interaction, and gauge interaction, respectively. Because these are nonlocal, their Hamiltonian matrix elements are not given by individual OPE coefficients of the UV CFT as we claimed in (\ref{eq:OPpO}).  However, the expression (\ref{eq:OPpO2}) for the matrix elements in terms of three-point functions remains valid.  Moreover,  we will see in Part II how to extract three-point functions of nonlocal operators from higher-point functions of local operators, so the matrix elements are still encoded in the UV CFT data.

A more subtle aspect of the LC zero modes is that they can have contributions at all orders in the coupling which are currently difficult to compute.  However, these contributions often amount simply to changes in the bare parameters of the theory, as was shown in \cite{Fitzpatrick:2018ttk,Fitzpatrick:2018xlz}, and so ignoring them does not change the universality class of the theory.  On the other hand, if the theory undergoes a quantum phase transition, then the expectation is that the universality class of the effective Hamiltonian will change.  Thus, currently, LC quantization can be a useful approach for a generic model as long as one is not interested in observables beyond the phase transition point.\footnote{It is worth noting, however, that it is possible that for certain SUSY theories a description beyond the phase transition point is also possible \cite{Fitzpatrick:2019cif}.}

A final technical challenge for using LCT is that the Hamiltonian matrix element, (\ref{eq:HamMatrixElement}), is IR divergent for relevant operators of dimension $\Delta \leq d/2$.
For free theories in any dimension there is a way of dealing with this divergence by introducing a simple regulator.  Removing the regulator then leads to a straightforward modification of the basis so that all matrix elements remain finite.  See Appendix \ref{app:VertexOps} and section \ref{sec:IRDivergences} for a discussion.  However, for a general CFT it is not currently known how to regulate this IR divergence in a practical manner.  On the other hand, for ET, relevant operators with $\Delta < d/2$ are in fact simple to handle, while for 
$\Delta \geq d/2$ additional challenges arise from the presence of UV divergences.  Hence LCT and ET are nicely complementary here.

%%%%%%%%%%%%%%%%%%%%%%%%%%%%%%%%%%%%%%%%%%%%%%%%%%%%%%%%%%%%%%%%%%%%%%%%%%%%%

\subsection{Overview of Key Steps in 2d}
\label{sec:Calcs}

So far this discussion has been quite general, and can be applied to deformations of CFTs in any number of dimensions. However, now we will turn to the specific focus of this paper: deformations of free field theory in $d=2$. In this section we provide an overview of the key features of LCT in this setting, as well as a small preview of the structure of the necessary calculations.

  Because our goal here is just to give a sense of the steps involved and the simplifications afforded by LC quantization, many results will simply be asserted or justified heuristically; we promise that everything will be derived carefully later. In fact, we will eventually present three different methods for doing the calculations, which we call the ``Fock space'', ``Wick Contraction'', and ``Radial Quantization'' methods.  In this section, we will use the language of the Fock space method, and for various results we will provide references to their derivation using this method in section \ref{sec:2dFFT}.  The Fock space method computes inner products and matrix elements using the mode decomposition in Fourier space directly, whereas the Wick Contraction and Radial Quantization methods work first in position space to compute the two- and three-point functions that appear on the RHS of (\ref{eq:NormFromTwoPt}) and (\ref{eq:OPpO2}), and then Fourier transform the result.  Although the latter two methods are computationally more efficient, we begin our presentation with the Fock space method because it is conceptually the simplest.  In particular, for the most part, the details of the derivations simply involve a lot of bookkeeping, and follow from the mode expansions of fields, e.g.
\be
\phi(x) = \int_0^\infty \frac{d p_-}{(2\pi) \sqrt{2p_-}} \left( e^{- i p \cdot x} a_p + e^{i p \cdot x} a_p^\dagger\right),\qquad
[a_p, a^\dagger_q] = (2\pi) \delta(p_- - q_-),
\label{eq:phimodedecomp0}
\ee
together with the condition $p_->0$ and particle state normalization $|p \> \equiv \sqrt{2p_-} a_p^\dagger | {\rm vac} \>$.

We'll now briefly step through these various steps for the case of scalar field theory, reserving the details and inclusion of fermions for later sections.

\subsubsection*{Basis of Primary Operators}

Because the UV CFT is free, we can organize the primary operators by particle number, constructing each sector separately. Starting with the one-particle sector, we find that there is only one primary operator: the conserved current $\p_\mu\phi(x)$.\footnote{In $d=2$ the scalar field $\phi$ is nonlocal and not a primary operator.} However, as we explain in more detail in section~\ref{sec:2dFFT}, because we are working in lightcone quantization in the sector with $p_- > 0$, we only include the state generated by the left-moving component $\p_-\phi(x)$. Unsurprisingly, this `single-$\phi$' state with momentum $p_-$ is simply proportional to the one-particle Fock space state with momentum $p_-$, 
\be
|\p_-\phi,p\> \propto |p_-\>,
\ee
as one can see explicitly by acting with the Fourier transform of the mode decomposition (\ref{eq:phimodedecomp0}) on the vacuum.

Similarly, for higher particle number states we again only include primary operators that are constructed from $\p_-\phi$, taking the general form \ft{(see section \ref{sec:FreeLC})}
\be
\Ocal_i(x) = \sum_{\kvec} C^{\Ocal_i}_{\kvec} \, \p_-^{k_1} \phi(x) \cdots \p_-^{k_n}\phi(x).
\ee
These operators are all purely \emph{left-moving} (i.e.~holomorphic), with $\De = \ell$, which means the corresponding conformal Casimir eigenvalue $\Ccal$ is uniquely determined by the scaling dimension $\De$,
\be
\Ccal = 2\De(\De-1).
\ee
For 2d free field theory, truncating the basis with respect to the conformal Casimir is therefore equivalent to truncating with respect to the scaling dimension.

To better understand the structure of these basis states, let's briefly consider the two-particle sector. The lowest left-moving primary operator is $(\p_-\phi)^2$, with the corresponding basis state  $|(\p_-\phi)^2,p\> \equiv N_{(\p_-\phi)^2}^{-1} \int dx^- \, e^{-ip_- x^-} (\p_-\phi)^2(x)|\textrm{vac}\>$.  It is often useful to represent these basis states in terms of Fock space states, by computing the overlap
\be
F_{\Ocal_i}(p) \equiv \<p_{1-},\ldots,p_{n-}|\Ocal_i(0)\>.
\ee
For example, for $(\p_-\phi)^2$, the  wavefunction is simply \ft{ (see (\ref{eq:FockWvFns})-(\ref{eq:pphiWvFn}))}
\be
F_{(\p_-\phi)^2}(p) = -2 p_{1-} p_{2-} = -2z(1-z).
\ee
In the second expression, we've used the fact that we're working in the frame $p_- = 1$ and have replaced $p_{1-} \ra z$. The normalization factor for $(\p_-\phi)^2$ can be computed by evaluating the integral \ft{(see (\ref{eq:DPhi2Inner}))}
\bq
|N_{(\p_-\phi)^2}|^2 \propto \int \fr{dp_{1-} dp_{2-}}{p_{1-} p_{2-}} \de(p_- - p_{1-} - p_{2-}) |F_{(\p_-\phi)^2}(p)|^2 \propto \int_0^1 \fr{dz}{z(1-z)} \, z^2(1-z)^2,
\label{eq:SchematicInner}
\eq
where we've suppressed  overall factors to focus on the simple structure of the integral.

For each particle number, we therefore need to construct a complete basis of primary operators up to some scaling dimension $\Dmax$, or equivalently a complete basis of momentum space polynomials $F_\Ocal(p)$ up to some maximum degree. We can then orthonormalize these basis states by evaluating inner products of the same form as eq.~\eqref{eq:SchematicInner}. In section~\ref{sec:2dFFT} we discuss these inner products in more detail, and in section~\ref{sec:RadialScalars} we provide a much more efficient method for evaluating them, motivated by the CFT structure of free field theory.

\subsubsection*{Matrix Elements}

The simplest relevant deformation we can consider is the mass term, which leads to the Hamiltonian contribution
\be
\de P_+ = \fr{m^2}{2} \int dx^- \, \phi^2(x).
\ee
In LC quantization, the mass term is \emph{diagonal} with respect to particle number, such that there are no matrix elements mixing $n$-particle states with $n\pm2$-particle states, and  the one-particle state in the free massless theory is automatically an eigenstate of the mass term, \ft{(see (\ref{eq:HamMassTerm}))}
\be
\de P_+|p_-\> = \fr{m^2}{2p_-}|p_-\>.
\ee
The computation of the new invariant mass is therefore trivial,
\be
M^2 |p_-\> = 2P_-(P_+^{(\CFT)} + \de P_+)|p_-\> = 2p_-\bigg(0 + \fr{m^2}{2p_-}\bigg)|p_-\> = m^2|p_-\>.
\ee
We emphasize that the analogous computation in an equal-time quantization Hamiltonian formulation is much more involved and requires keeping states of arbitrarily high particle number just to calculate the one-particle mass shift.  One way to understand that equal-time should be more complicated is that the energy $\omega_p = \sqrt{p^2+m^2}$ is a nonanalytic function of the mass-squared parameter, and has an infinite Taylor series in $m^2$. By contrast, in lightcone quantization the energy is linear in $m^2$.

At higher particle number, the Fock space states are also eigenstates of the mass term, which makes it straightforward to compute the resulting matrix elements for primary operators. For example, at $n=2$ the action of $M^2$ on a Fock space state is
\be
M^2|p_{1-},p_{2-}\> = 2p_- \bigg( \fr{m^2}{2p_{1-}} + \fr{m^2}{2p_{2-}} \bigg)|p_{1-},p_{2-}\> = \fr{m^2}{z(1-z)} |p_{1-},p_{2-}\>.
\ee
The matrix element for $(\p_-\phi)^2$ can be computed by integrating the wavefunction-squared against this Fock space ``potential'' divided by the normalization factor: \ft{(see (\ref{eq:FirstExampleNorm}) and (\ref{eq:dphisqmat}))}
\bq
\frac{\<(\p_-\phi)^2,p|M^2|(\p_-\phi)^2,p'\> }{{\< (\partial_- \phi)^2,p  | (\partial_- \phi)^2,p'\>} }= \fr{\int_0^1 dz \, z(1-z) \cdot \fr{m^2}{z(1-z)}}{\int_0^1 dz \, z(1-z)} = 6m^2.
\label{eq:SchematicElement}
\eq
This result is above the minimum two-particle mass-squared $M^2 = 4 m^2$, as we expect since we are using a variational method. As we include higher dimension operators in the basis, the lowest eigenvalue of the truncated Hamiltonian will decrease and approach the correct two-particle mass threshold.

For the quartic interaction $\sim \lambda \phi^4$, there are again matrix elements between states with $n$-particles, as well as mixing with $n\pm2$-particle states. In later sections, we will see how to develop efficient methods for computing such matrix elements for all primary operators up to the truncation dimension $\Delta_{\rm max}$. However, conceptually these elements all have the same simple structure as eq.~\eqref{eq:SchematicElement}.

\subsubsection*{Spectral Densities}

Once we have diagonalized the truncated mass-squared $M^2$ matrix and obtained its eigenvalues, by eqs (\ref{eq:FullEigenstates}) and (\ref{eq:SpecDensGeneral}) the last ingredient needed to compute the spectral density of an operator $\Ocal$ is its overlaps $\< \Ocal(0)| \Ocal_j, p\>$ with our basis states.   If $\Ocal$ is one of the primary operators in our basis, this overlap is trivial to compute. However, in this work we will often be interested in operators which are not part of our basis. For example, if we wish to compute the spectral density of $\phi^2$, we need to calculate the overlap of this operator with all of the two-particle states in our basis. We can easily do this by first computing the momentum space wavefunction for $\phi^2$,
\be
F_{\phi^2}(p) \equiv \<\phi^2(0)|p_{1-},p_{2-}\> = 2,
\ee
then computing the resulting overlap with primary operators, such as
\bq
\bal
\<\phi^2(0)|(\p_-\phi)^2,p\> &\propto \int \fr{dp_{1-} dp_{2-}}{p_{1-} p_{2-}} \de(p_- - p_{1-} - p_{2-}) F_{\phi^2}(p) F_{(\p_-\phi)^2}(p) \\
&\propto \int_0^1 \fr{dz}{z(1-z)} \cdot z(1-z).
\eal
\eq

As we demonstrate in this work, the resulting spectral densities can be used to identify second-order phase transitions, discover resonances, compute critical exponents, and study many other properties of the deformed theory. More generally, the wavefunctions of mass eigenstates in terms of primary operators provides a concrete map between parameters in the UV CFT and dynamical observables in the IR.

%%%%%%%%%%%%%%%%%%%%%%%%%%%%%%%%%%%%%%%%%%%%%%%%%%%%%%%%%%%%%%%%%%%%%%%%%%%%%
%%%%%%%%%%%%%%%%%%%%%%%%%%%%%%%%%%%%%%%%%%%%%%%%%%%%%%%%%%%%%%%%%%%%%%%%%%%%%

\section{Free Field Theory in 2D}
\label{sec:2dFFT}

As already mentioned, our focus in this paper will be applying LCT to free CFTs in 2d. Free massless theories are possibly the simplest CFTs, but also very versatile as many QFTs can be described as free theories with relevant deformations.  Moreover, free massless theories are solvable and their CFT data is computable.

In this section, we summarize some basic properties of LCT for 2d free field theory. We begin in section~\ref{sec:FreeLC} by introducing the Fock space mode expansions for free scalars and fermions and the implications for LCT states. Then, we turn to LCT inner products and Hamiltonian matrix elements, which we refer to as the \emph{LCT data}, because once they are computed, one can consider them as fundamental building blocks in their own right from which the rest of the observables in the theory can be obtained.  In section~\ref{sec:FockSpace}, we discuss the ``Fock Space Method," where LCT data are computed as integrals over Fock space momenta. In section~\ref{sec:Jacobi}, we present a related connection between primary operators in free CFTs and Jacobi polynomials. In section~\ref{sec:WickContract}, we discuss the ``Wick Contraction Method," where calculations are instead done by first computing position-space correlators (via Wick contractions)  and then Fourier transforming.  It is often very useful to be able to think about LCT data using both methods.

%%%%%%%%%%%%%%%%%%%%%%%%%%%%%%%%%%%%%%%%%%%%%%%%%%%%%%%%%%%%%%%%%%%%%%%%%%%%%

\subsection{Free Fields on the Lightcone}
\label{sec:FreeLC}

Let us begin with free, massless scalars. The CFT Lagrangian is simply
\be
\CL_{\rm CFT}  = \frac{1}{2} (\partial \phi)^2 = \partial_+ \phi \partial_- \phi.
\ee
The canonical momentum in lightcone quantization is $\pi(x) = \partial_- \phi(x)$, so the canonical commutation relations are
\be
[ \phi(x), \partial_- \phi(y)] = \frac{i}{2} \delta(x_- - y_-).
\label{eq:LCCommReln}
\ee
This unusual commutation relation leads to the following mode decomposition for the field $\phi$:
\be
\phi(x) = \int_0^\infty \frac{d p_-}{(2\pi) \sqrt{2p_-}} \left( e^{- i p \cdot x} a_p + e^{i p \cdot x} a_p^\dagger\right),
\label{eq:phimodedecomp}
\ee
in terms of creation and annihilation operators $a_p, a^\dagger_p$ satisfying
\be
[a_p, a^\dagger_q] = (2\pi) \delta(p_- - q_-).
\ee
One notable feature of (\ref{eq:phimodedecomp}) is that, due to the commutation relation (\ref{eq:LCCommReln}), the usual factor of $\sqrt{2 \omega_p}$ from equal-time quantization has been replaced by $\sqrt{2 p_-}$.

As is well known, $\phi$ itself is not a primary operator in 2d. This is evident at the level of its two-point function, which is logarithmically divergent, 
\be
\langle \phi(x) \phi(0) \rangle = -\frac{1}{4\pi} \log x.
\ee
Instead, to build primaries one needs to utilize the following building blocks:
\be
\p_-\phi, \hspace{5mm} \p_+\phi, \hspace{5mm} e^{i\alpha\phi}.
\ee

As we now explain, only $\p_-\phi$ is needed to construct the LCT basis. The first observation is that, because we are working in momentum space and $p_-  > 0$, the operator $\p_+ \phi$ effectively vanishes by the equations of motion $\p_\mu \p^\mu \phi =0$: 
\be
\p_+ \phi \cong \frac{1}{i p_-} \p_\mu \p^\mu \phi = 0.
\ee
Crucially, we use the fact that  $p_->0$ for each factor of $\phi$ even in multi-$\phi$ operators such as $\p_-^{k_1} \phi \dots \p_-^{k_n} \phi$; a more thorough discussion of $p_-=0$ zero modes is given in appendix \ref{app:Heff}.
 
 Next, consider the vertex operators $V_{\alpha} \equiv e^{i\alpha\phi}$. As explained in appendix~\ref{app:VertexOps}, defining these operators requires the introduction of an IR cutoff $\epsilon$, which then appears in the normalization of $V_\alpha$. However, even after proper normalization, matrix elements of $V_{\alpha}$ diverge as $\epsilon\rightarrow 0$. This means that once we deform the CFT by relevant operators, the vertex operators get lifted out of the spectrum, and hence they can be ignored from the start.\footnote{Vertex operators are lifted from the spectrum specifically because we are considering the deformations $\phi^n$, which break the shift symmetry on $\phi$; see appendix \ref{app:VertexOps}.} Thus, the sole building block for the LCT basis is $\p_- \phi$ (\emph{i.e.}, the basis is holomorphic), and a generic operator is of the form
 \be
\Ocal_i(x) = \sum_{\kvec} C^{\Ocal_i}_{\kvec} \, \p_-^{k_1} \phi(x) \cdots \p_-^{k_n}\phi(x). \label{eq:ScalarGenericOperator}
\ee
Note that in momentum space, all such operators have $p_+ =0$; this follows from the fact explained above that each individual factor $\p_-^k \phi$ has $p_+=0$, and the total momentum is just the sum of these individual momenta.  Therefore, for the 2d free scalar basis, without loss of generality there is only one total momentum $p^\mu$ that we need to consider: $(p_+,p_-) =(0,1)$.  

Let us now turn to fermions. Our starting point is the CFT of a free massless real fermion in 2d,
\be
\Lcal_\CFT = i \psi \p_+ \psi + i \chi \p_- \chi,
\ee
where $\psi$ and $\chi$ are the left- and right-chirality components of the fermion. 
The field $\chi$ is non-dynamical since its time derivative does not appear in the action, so it can be integrated out using its equation of motion.\footnote{In the presence of relevant deformations, integrating out $\chi$ will in general induce nonlocal interactions for $\psi$, as we will see in section~\ref{sec:Fermions}.} The only remaining degree of freedom is $\psi$, which has the mode decomposition
\be
\psi(x) = \int_0^\infty \fr{dp_-}{\sqrt{8\pi^2}} \left( e^{-ip\cdot x} a_p + e^{ip\cdot x} a^\dagger_p \right) ,~~~~
\acomm{a_p}{a^\dagger_q} = (2\pi) \de(p_- - q_-) . 
\ee

Unlike $\phi$, the operator $\psi$ is itself primary and can be used as the building block for other primaries. Nevertheless, it is still true that $\p_+ \psi = 0$, which is just the equation of motion for $\psi$ in the CFT. It follows that $p_+=0$ for every state in the fermion basis, just like for scalars. 

To summarize, the LCT basis for free field theory in 2d consists of states 
\be
\boxed{ \ket{\CO_i, p}_{\text{2d FFT}} \equiv \ket{ \CO_i,\, p_\mu = (p_+,p_-) =(0,p)}, }
\ee
where $\CO_i$ is a primary operator constructed using $\p_-\phi$, $\psi$, and $\p_-$ derivatives.

Note that the scalar and fermion bases are quite similar. The main difference is that scalar primaries are built from $\p_-\phi$, whereas fermion primaries are built from $\psi$ (so fermion primaries can have insertions of $\psi$ without any $\p_-$s attached). In section~\ref{sec:Fermions}, we will see that once we add a mass deformation to the fermion CFT, IR divergences will force us to attach a $\p_-$ to every $\psi$! Moreover, we will see that we can anticipate the effect of the mass deformation by treating $\p_-\psi$ as a new ``primary" operator and making it the basic building block for the fermion basis. In practice, this makes the scalar and fermion bases nearly identical, up to the different scaling dimensions and commuting/anti-commuting properties of $\p_-\phi$ and $\p_-\psi$. \\

\noindent {\bf Notation.} Since the LCT basis for 2d free fields is holomorphic, constructed entirely out of $\p_-$ derivatives acting on the fundamental fields, we will henceforth assume minus subscripts everywhere and for simplicity write
\be
\boxed{\p_- \rightarrow \p, \hspace{10mm} p_- \rightarrow p.} \label{eq:DropSupscriptNotation}
\ee

There are many factors of $i$ and $-1$ that arise in the construction of the basis that are rather annoying and moreover depend on the description being used for the basis itself (for instance, in the Fock space method, it is natural to work with momentum factors $p$, whereas in the Wick Contraction and Radial Quantizion methods it is natural to work with spatial derivatives $\partial$, which differ from $p$ by a factor of $i$).  Worse, these factors obscure the fact that they are mostly overall phase factors in the definition of the basis states themselves and are guaranteed to cancel out in physical results, as we explain in appendix \ref{app:Phases}.  So, we will introduce the following notation, where we replace ``$=$'' by ``$\doteq$'' to indicate that phases have been dropped in a consistent way so that they have no overall effect on physical observables. That is, 
  \be
\boxed{ A \doteq B  \Rightarrow A =  ({\rm phase}) \times B } \label{eq:doteqnotation}
 \ee
 where moreover the relative phase in $A$ and $B$ is composed of factors that are defined in appendix \ref{app:Phases} and that cancel in the final results for the Hamiltonian matrix elements with a particular phase convention for the basis states, so they can be consistently discarded.

%%%%%%%%%%%%%%%%%%%%%%%%%%%%%%%%%%%%%%%%%%%%%%%%%%%%%%%%%%%%%%%%%%%%%%%%%%%%%

\subsection{Fock Space Method}
\label{sec:FockSpace}

Once one has constructed a complete basis of primary operators, the chief computational task of lightcone conformal truncation is to determine their Gram matrix of inner products and their Hamiltonian matrix elements.  As we have mentioned, because this task is so central to applying LCT, over the course of this review we will describe three different methods of increasing sophistication and speed for achieving it.  The first is the ``Fock space method'', where we simply write out the states and operators in terms of their lightcone quantization Fock space creation and annihilation operators, and integrate over momentum space. In this subsection, we will simply do a few example computations with the Fock space approach for the case of scalar fields to show how it works in more detail.

We can express any $n$-particle basis state in terms of Fock space modes by simply inserting the identity as a sum over states,
\bq
\bal
|\Ocal_i,p\> &\equiv \fr{1}{N_{\Ocal_i}} \int dx \, e^{-ipx} \Ocal_i(x)|\textrm{vac}\> \\
&= \fr{1}{N_{\Ocal_i}} \int dx \, e^{-ipx} \fr{1}{n!} \int \fr{dp_1 \cdots dp_n}{(2\pi)^n 2p_1 \cdots 2p_n} |p_1,\ldots,p_n\>\<p_1,\ldots,p_n|\Ocal_i(x)\> \\
&= \fr{1}{n! N_{\Ocal_i}} \int \fr{dp_1 \cdots dp_n}{(2\pi)^n 2p_1 \cdots 2p_n} (2\pi) \de(p-|p|_n) F_{\Ocal_i}(p) |p_1,\ldots,p_n\>,
\label{eq:FockWvFns}
\eal
\eq
where the momentum space wavefunction $F_{\Ocal_i}(p) \equiv \<p_1,\ldots,p_n|\Ocal_i(0)\>$, and we've introduced the useful shorthand notation
\be
|p|_i \equiv \sum_{j=1}^i p_j.
\label{eq:MagDef}
\ee
As a simple concrete example, the resulting expression for $(\p\phi)^2$ is
\bq
\bal
|(\p\phi)^2,p\> &\doteq \fr{1}{2 N_{(\p\phi)^2}} \int \fr{dp_1 \, dp_2}{(2\pi)^2 2p_1 2p_2} (2\pi) \de(p-|p|_2) \, 2p_1 p_2 \, |p_1,p_2\> \\
&= \fr{1}{8\pi N_{(\p\phi)^2}} \int_0^p \fr{dp_1}{p_1 (p-p_1)} p_1 (p-p_1) |p_1,p-p_1\>.
\label{eq:pphiWvFn}
\eal
\eq

We can then use these Fock space representations to easily compute the inner product of states. In studying these inner products, it will be convenient to define the Gram matrix with the momentum-conserving delta function factored out:
\be
\< \CO_i, p| \CO_j, p'\> \equiv 2p(2\pi)\delta(p-p') \, G_{\CO_i \CO_j}. 
\ee
The elements of the Gram matrix then take the general form
\bq
\boxed{G_{\Ocal_i\Ocal_j} = \fr{1}{n! 2p N^*_{\Ocal_i} N_{\Ocal_j}} \int \fr{dp_1 \cdots dp_n}{(2\pi)^n 2p_1 \cdots 2p_n} (2\pi) \de(p-|p|_n) F^*_{\Ocal_i}(p) F_{\Ocal_j}(p).}
\label{eq:GeneralInner}
\eq
For our example of $(\p\phi)^2$, we have the resulting inner product
\begin{equation}
\bal
G_{(\partial \phi)^2,(\partial \phi)^2} &= \fr{1}{8\pi p |N_{(\p\phi)^2}|^2} \int_0^p dp_1 \, p_1 (p-p_1) \\
&= \fr{p^2}{8\pi |N_{(\p\phi)^2}|^2} \int_0^1 dz \, z (1-z) = \fr{p^2}{48\pi |N_{(\p\phi)^2}|^2}.
\eal
\label{eq:DPhi2Inner}
\end{equation}
Next, let's consider the overlap of $(\partial \phi)^2$ with the following operator:
\be
\CO_{(2)}\equiv 6 \partial \phi \partial^3 \phi - 9 (\partial^2 \phi)^2. \label{eq:twopprimaryexample}
\ee
Using the same approach as above, we have
\begin{equation}
G_{(\partial \phi)^2,\CO_{(2)}} = \fr{3p^4}{8\pi N^*_{(\p\phi)^2} N_{\Ocal_{(2)}}} \int_0^1 dz \, z(1-z) \big( z^2 + (1-z)^2 - 3z (1-z) \big) = 0.
\end{equation}
That is, their overlap vanishes.  The underlying reason for this cancellation is that the factors 6 and 9 were chosen so that $\CO_{(2)}$ is a primary operator, and primary operators of different dimensions have no overlap.  Finally, we can compute the norm of $\CO_{(2)}$:
\bq
G_{\CO_{(2)} \CO_{(2)}} = \fr{9p^6}{8\pi |N_{\Ocal_{(2)}}|^2} \int_0^1 dz \, z (1-z) \big( z^2 + (1-z)^2 - 3 z(1-z) \big)^2 = \fr{3p^6}{224\pi |N_{\Ocal_{(2)}}|^2}.
\eq

All together, the $2 \times 2 $ Gram matrix for the operators $(\p \phi)^2$ and  $\CO_{(2)}$  is (in units with $p=1$), 
\be
G_{ij} = \frac{1}{16\pi} \begin{pmatrix} \frac{1}{3|N_{(\p \phi)^2}|^2} & 0 \\ 0 & \frac{3}{14|N_{\CO_{(2)}}|^2} \end{pmatrix}.
\label{eq:GramExample1}
\ee
Because the Gram matrix is already diagonal, to orthonormalize this two-state basis we simply need to set
\be
|N_{(\p\phi)^2}|^2 = \fr{1}{48\pi}, \quad |N_{\Ocal_{(2)}}|^2 = \fr{3}{224\pi}.
\label{eq:FirstExampleNorm}
\ee

Let us do one more example of an inner product, this time for a three-particle state created by an  operator of the form $\p^{\bk} \phi = \p^{k_1} \phi \p^{k_2} \phi \p^{k_3} \phi$.  This ``monomial'' operator is not primary, but the primary operators will all be written as sums over such monomials, making them the building blocks for our basis. 
The monomial's wavefunction is\footnote{In practice, we can often use the fact that all scalar wavefunctions are symmetric under the permutation of any two momenta ($p_i \lra p_j$) to simplify our calculations, replacing the sum over permutations in the wavefunction with $F_{\p^{\bk}\phi}(p) \ra 3! \, p_1^{k_1} p_2^{k_2} p_3^{k_3}$.}
\be
F_{\p^{\bk}\phi}(p) \doteq \sum_{\substack{k'_1,k'_2,k'_3= \\ \textrm{perm}(k_1,k_2,k_3)}} p_1^{k'_1} p_2^{k'_2} p_3^{k'_3},
\label{eq:3ParticleF}
\ee
The norm of the monomial state is
\bq
\bal
G_{\bk\bk } &= \fr{1}{64\pi^2 p |N_{\p^{\bk}\phi}|^2} \int dp_1 \, dp_2 \, dp_3 \, \de(p-|p|_3) \sum_{\substack{k'_1,k'_2,k'_3= \\ \textrm{perm}(k_1,k_2,k_3)}} p_1^{k_1+k'_1-1} p_2^{k_2+k'_2-1} p_3^{k_3+k'_3-1} \\
&= \frac{p^{2(k_1+k_2+k_3)-2}}{64\pi^2 |N_{\p^{\bk} \phi}|^2} \sum_{\substack{k'_1,k'_2,k'_3= \\ \textrm{perm}(k_1,k_2,k_3)}} \frac{\Gamma(k_1 + k_1') \Gamma(k_2+k_2') \Gamma(k_3+k_3')}{\Gamma(k_1+k_1'+k_2+k_2'+k_3+k_3')} .
\eal
\label{eq:3PartFock}
\eq
For the simplest three-particle monomial $\p^{\bk} \phi = (\p \phi)^3$, the above result reduces to
\be
G_{(\d \phi)^3, (\d \phi)^3} = \frac{p^4}{1280\pi^2 |N_{(\p\phi)^3}|^2}.
\label{eq:dphicubednorm}
\ee

Computations of the Hamiltonian matrix elements are quite similar in structure. First, we need to decompose the LC Hamiltonian in terms of Fock space modes. As a simple example, let's consider the (normal-ordered) scalar mass term
\bq
\bal
\delta P_+ &= \frac{m^2}{2}  \int_{-\infty}^\infty dx \, \norder{\phi^2(x)} \\
&= \fr{m^2}{2} \int_0^\infty \frac{dp \, dq}{(2\pi)^2 \sqrt{2p 2q}} \Big[ (2\pi) \delta(p+q) ( a_p^\dagger a_q^\dagger+a_p a_q)  +  2(2\pi) \de(p-q) a_p^\dagger a_q \Big] \\
&= \int_0^\infty \frac{dp}{2\pi} \, \frac{m^2}{2 p}  a_p^\dagger a_p.
\eal
\label{eq:HamMassTerm}
\eq
In the last line, we have used the fact that the integral in the mode decomposition of $\phi$ is only over positive lightcone momenta, so the terms proportional to $\delta(p + q)$  vanish.  This simplification is an example of the generic feature that particles are not pair-produced from the vacuum in LC quantization.

Just like for the inner product, we can define the matrix elements for $M^2$ with the delta function for momentum factored out,
\be
\<\Ocal_i,p|2\de P_+ P_-|\Ocal_j,p'\> \equiv 2p(2\pi)\de(p-p') \, \Mcal^{(\Ocal_R)}_{\Ocal_i\Ocal_j},
\ee
where the superscript indicates the particular relevant deformation $\Ocal_R$ associated with $\de P_+$. For the mass term, we can use the Fock space mode decomposition~\eqref{eq:HamMassTerm} to construct the general matrix element
\bq
\boxed{\Mcal^{(\phi^2)}_{\Ocal_i\Ocal_j} = \fr{1}{n! N^*_{\Ocal_i} N_{\Ocal_j}} \int \fr{dp_1 \cdots dp_n}{(2\pi)^n 2p_1 \cdots 2p_n} (2\pi) \de(p-|p|_n) F^*_{\Ocal_i}(p) F_{\Ocal_j}(p) \sum_{k=1}^n \fr{m^2}{2p_k}.}
\eq
As an example, consider the matrix element for $(\p\phi)^2$,
\begin{equation}
\bal
\Mcal^{(\phi^2)}_{(\p\phi)^2,(\p\phi)^2} &= \fr{1}{4\pi |N_{(\p\phi)^2}|^2} \int_0^p dp_1 \, p_1 (p-p_1) \left( \fr{m^2}{2p_1} + \fr{m^2}{2(p-p_1)} \right) \\
&= \fr{m^2 p^2}{8\pi |N_{(\p\phi)^2}|^2} \int_0^1 dz = \fr{m^2 p^2}{8\pi |N_{(\p\phi)^2}|^2}.
\eal
\label{eq:dphisqmat}
\end{equation}
Using our calculation of $|N_{(\p \phi)^2}|^2 = \fr{1}{48\pi}$ above, we thus produce the result of eq.~\eqref{eq:SchematicElement}. Similarly, we can compute the matrix element of $(\p \phi)^3$:
\bq
\bal
\Mcal^{(\phi^2)}_{(\p \phi)^3, (\p \phi)^3} &= \fr{3!m^2}{64\pi^2 |N_{(\p\phi)^3}|^2} \int dp_1 dp_2 dp_3 \, \de(p-|p|_3) (p_1 p_2 + p_2 p_3 + p_1 p_3) \\
&= \fr{3m^2p^4}{256\pi^2 |N_{(\p\phi)^3}|^2} = 15 m^2.
\eal
\eq

We can apply this same approach to the matrix elements of the quartic interaction, obtaining the Hamiltonian
\bq
\bal
\delta P_+ &= \frac{\lambda}{4!}  \int_{-\infty}^\infty dx \, \norder{\phi^4(x)} \\
&= \fr{\lambda}{4!} \int_0^\infty \frac{dp \, dq \, dk}{(2\pi)^3\sqrt{2p \, 2q \, 2k}} \left( \fr{4 a^\dagger_p a^\dagger_q a^\dagger_k a_{p+q+k}}{\sqrt{2(p+q+k)}} + h.c. + \fr{6 a^\dagger_p a^\dagger_q a_k a_{p+q-k}}{\sqrt{2(p+q-k)}} \right).
\eal
\label{eq:HamPhi4Term}
\eq
The first two terms contribute to mixing between $n$-particle states and $n\pm2$-particle ones, while the third is diagonal with respect to particle number. Note that the $(a^\dagger)^4$ and $a^4$ terms have been removed by the restriction that all particles must have positive momenta, such that there are no matrix elements between $n$- and $n\pm4$-particle states.

As a final example, let's carefully compute the matrix elements of the $\phi^4$ interaction for the three-particle state $(\p \phi)^3$. This matrix element only receives a contribution from the last term in~\eqref{eq:HamPhi4Term}, leading to the expression
\begin{equation}
\bal
\Mcal^{(\phi^4)}_{(\p \phi)^3, (\p \phi)^3} &= \fr{\lambda}{|N_{(\p\phi)^3}|^2} \int \fr{dp_1 dp_2}{(2\pi)^2 2p_1 2p_2 2p_3} p_1 p_2 p_3 \int \fr{dp'_1 dp'_2}{(2\pi)^2 2p'_1 2p'_2 2p'_3} p'_1 p'_2 p'_3 \\
& \qquad \times \sum_{i,j} 2p_i (2\pi) \de(p_i-p'_j),
\eal
\end{equation}
where implicitly $p_3, p_3'$ are fixed by $\sum p_i = \sum p'_i = p$. The top line of this equation comes from the Fock space representation of the external states, while the second line comes from the contractions of the incoming and outgoing Fock space states with the Hamiltonian in~\eqref{eq:HamPhi4Term},
\be
\<p_1,p_2,p_3|\de P_+|p'_1,p'_2,p'_3\> = \lambda (2\pi) \de(p-p') \sum_{i,j} 2p_i (2\pi) \de(p_i-p'_j).
\ee
The delta function $\de(p_i-p'_j)$ corresponds to the remaining `spectator' particle from the in- and out-states that is not contracted with the $\phi^4$ interaction. The final result is
\be
\Mcal^{(\phi^4)}_{(\p \phi)^3, (\p \phi)^3} = \fr{3\lambda p^4}{1024\pi^3|N_{(\p\phi)^3}|^2} = \fr{15\lambda}{4\pi}.
\ee
We leave as an exercise for the reader the following matrix elements that we will encounter later:
\be
\Mcal^{(\phi^4)}_{(\p \phi), (\p \phi)} = 0, \qquad \Mcal^{(\phi^4)}_{(\p \phi), (\p \phi)^3} = \frac{\sqrt{5} \lambda}{4 \pi}.
\label{eq:Phi4Examples}
\ee

%%%%%%%%%%%%%%%%%%%%%%%%%%%%%%%%%%%%%%%%%%%%%%%%%%%%%%%%%%%%%%%%%%%%%%%%%%%%%

\subsection{Primary Operators and Jacobi Polynomials}
\label{sec:Jacobi}

While in the rest of this work we will largely use other methods to construct the basis and evaluate matrix elements, in this subsection we discuss more details of the Fock space representation of primary operators.\footnote{See~\cite{Henning:2019mcv} for a complementary perspective on the construction of primary operators in Fock space, with a natural generalization to higher $d$.}

As we've seen in the previous subsection, constructing a complete basis of primary operators for a free scalar in 2d is equivalent to finding a complete basis of momentum space wavefunctions $F_{\Ocal_i}(p)$ which are orthogonal with respect to the Fock space inner product~\eqref{eq:GeneralInner}. We can organize this basis into eigenfunctions of the conformal Casimir $\Ccal$, which in momentum space maps to the differential operator
\be
\Ccal = -2 \sum_{i<j} p_i p_j \left( \fr{\p}{\p p_i} - \fr{\p}{\p p_j} \right)^2.
\ee

Because this operator is a sum of terms acting only on pairs of particles, we can construct the eigenfunctions recursively in the number of particles, starting with the two-particle Casimir
\be
\Ccal_{12} \equiv -2 p_1 p_2 \left( \fr{\p}{\p p_1} - \fr{\p}{\p p_2} \right)^2.
\ee
The eigenfunctions of this operator take the general form
\be
F_{(\ell_1,\ell_2)}(p_1,p_2) \equiv p_1 p_2 (p_1+p_2)^{\ell_1+\ell_2} P^{(1,1)}_{\ell_1} \bigg(\fr{p_2-p_1}{p_1+p_2}\bigg)
\label{eq:twoparticleJacobis}
\ee
where $P^{(\alpha,\beta)}_\ell$ is the degree-$\ell$ Jacobi polynomial
\be
P^{(\alpha,\beta)}_\ell(x) \equiv \fr{\G(\alpha+\ell+1)}{\ell!\G(\alpha+\beta+\ell+1)} \sum_{m=0}^\ell \binom{\ell}{m} \fr{\G(\alpha+\beta+\ell+m+1)}{\G(\alpha+m+1)} \left( \fr{x-1}{2} \right)^m. \label{eq:JacobiPolyDef}
\ee
The eigenvalues of these two-particle wavefunctions are
\be
\Ccal_{12} \big[F_{(\ell_1,\ell_2)}(p_1,p_2)\big] = 2(\ell_1+1)(\ell_1+2) F_{(\ell_1,\ell_2)}(p_1,p_2),
\ee
which is precisely the Casimir eigenvalue of a holomorphic primary operator with dimension $\De=\ell_1+2$. Note that the Casimir eigenvalue is independent of the second parameter $\ell_2$, which simply controls the overall power of $p = p_1 + p_2$. Eigenfunctions with $\ell_2 > 0$ therefore correspond to \emph{descendants}, so we can restrict to \emph{primaries} by demanding $\ell_2=0$, which is equivalent to requiring that the eigenfunctions are annihilated by the special conformal generator
\be
K = \sum_i p_i \fr{\p^2}{\p p_i^2}.
\ee

As a concrete example, consider the simplest two-particle eigenfunction, with $\ell_1=\ell_2=0$,
\be
F_{(0,0)}(p_1,p_2) = p_1 p_2 P^{(1,1)}_0 \bigg(\fr{p_2-p_1}{p_1+p_2}\bigg) = p_1 p_2.
\ee
This is the momentum space wavefunction for the primary operator $(\p\phi)^2$ (up to an overall normalization factor), which has the conformal Casimir eigenvalue $\Ccal_{(\p\phi)^2} = 4$. Next, we can consider the $(\ell_1,\ell_2)=(2,0)$ eigenfunction,
\be
F_{(2,0)}(p_1,p_2) = p_1 p_2 (p_1+p_2)^2 P^{(1,1)}_2 \bigg(\fr{p_2-p_1}{p_1+p_2}\bigg) = 3 \big(p_1^3 p_2 + p_1 p_2^3\big) - 9 p_1^2 p_2^2,
\ee
which we can recognize as the wavefunction for the primary operator $\Ocal_{(2)}$ in eq.~\eqref{eq:twopprimaryexample}. Jacobi polynomials thus provide an efficient means for constructing primary operators.

We can use these two-particle wavefunctions as building blocks to construct the three-particle Casimir eigenfunctions
\bq
\bal
F_{(\ell_1,\ell_2,\ell_3)}(p_1,p_2,p_3) &\equiv p_1 p_2 p_3 (p_1+p_2)^{\ell_1} P^{(1,1)}_{\ell_1} \bigg(\fr{p_2-p_1}{p_1+p_2}\bigg) \\
& \qquad \times (p_1+p_2+p_3)^{\ell_2+\ell_3} P^{(2\ell_1+3,1)}_{\ell_2} \bigg(\fr{p_3-p_1-p_2}{p_1+p_2+p_3}\bigg).
\eal
\eq
Because the top line clearly corresponds to a two-particle primary operator and the second line is only a function of $p_3$ and $p_1+p_2$, this wavefunction is an eigenfunction of both the two-particle Casimir $\Ccal_{12}$, with eigenvalue $2(\ell_1+1)(\ell_1+2)$, as well as the three-particle Casimir,
\be
\Ccal_{123} \big[F_{(\ell_1,\ell_2,\ell_3)}(p_1,p_2,p_3)\big] = 2(\ell_1+\ell_2+2)(\ell_1+\ell_2+3) F_{(\ell_1,\ell_2,\ell_3)}(p_1,p_2,p_3).
\ee
Similar to before, we can restrict to primary operators by requiring $\ell_3=0$.

For example, consider the eigenfunction with $(\ell_1,\ell_2,\ell_3)=(2,0,0)$,
\bq
\bal
F_{(2,0,0)}(p_1,p_2,p_3) &= p_1 p_2 p_3 (p_1+p_2)^2 P^{(1,1)}_2 \bigg(\fr{p_2-p_1}{p_1+p_2}\bigg) P^{(7,1)}_0 \bigg(\fr{p_3-p_1-p_2}{p_1+p_2+p_3}\bigg) \\
&= \Big(3 \big(p_1^3 p_2 + p_1 p_2^3\big) - 9 p_1^2 p_2^2 \Big) p_3.
\eal
\eq
The first index $\ell_1=2$ thus fixes the two-particle ``building block'' for $p_1$ and $p_2$ to be the wavefunction of $\Ocal_{(2)}$. The second index $\ell_2=0$ then fixes the number of relative derivatives between this building block and the third particle, associated with $p_3$. This wavefunction thus schematically corresponds to the operator
\be
F_{(2,0,0)}(p) \quad \Leftrightarrow \quad \Ocal_{(2)} \p\phi.
\ee

Proceeding with this same recursive construction, we can now write the general $n$-particle Casimir eigenfunction
\be
\boxed{F_{\Lvec}(p_i) \equiv p_1 \cdots p_n |p|_n^{\ell_n} \prod_{i=1}^{n-1} |p|_{i+1}^{\ell_i} P^{(2|\ell|_{i-1}+2i-1,1)}_{\ell_i} \bigg( \fr{p_{i+1}-|p|_i}{|p|_{i+1}} \bigg),}
\label{eq:JacobiBasis}
\ee
which is labeled by the $n$-component index $\Lvec = (\ell_1,\ldots,\ell_n)$, and we've used the notation $|p|_i$ defined in~\eqref{eq:MagDef}. These wavefunctions have the Casimir eigenvalues
\be
\Ccal_{\Lvec} = 2(|\ell|_{n-1}+n) (|\ell|_{n-1}+n-1),
\ee
and can be restricted to primary operators by fixing $\ell_n=0$.

Schematically, these wavefunctions correspond to operators of the form
\be
F_{\Lvec}(p_i) \quad \Leftrightarrow \quad \Ocal_{\Lvec} \sim \p^{\ell_n} \bigg( \p\phi \, \lrpar^{\ell_{n-1}} \Big( \p\phi \cdots \lrpar^{\ell_2} \big( \p\phi \, \lrpar^{\ell_1} \p\phi \big) \Big) \bigg).
\ee
However, the basis of momentum space wavefunctions in eq.~\eqref{eq:JacobiBasis} is actually overcomplete, due to the fact that these states are built from indistinguishable particles. We therefore need to restrict the full space of Jacobi polynomials to only those linear combinations which are symmetric under the exchange of any two particles $p_i \lra p_j$.

While there are some useful tools for improving this symmetrization procedure, which we discuss briefly in appendix~\ref{app:FockSpaceTricks}, in practice we have found that it is more efficient to work directly with the local operators, rather than construct the corresponding symmetric momentum space wavefunctions. In the next subsection, we will present a separate operator construction of basis states and matrix elements, which we will largely use for the remainder of this paper. However, the Fock space representation can often provide a useful, conceptually simple picture when computing matrix elements or comparing results to perturbation theory.

Finally, this construction can easily be generalized to primary operators built from a free fermion $\psi$ (or in fact any holomorphic generalized free field of dimension $\De$),
\be
F^{(\De)}_{\Lvec}(p_i) \equiv (p_1 \cdots p_n)^\De |p|_n^{\ell_n} \prod_{i=1}^{n-1} |p|_{i+1}^{\ell_i} P^{(2|\ell|_{i-1}+2\De i-1,2\De-1)}_{\ell_i} \bigg( \fr{p_{i+1}-|p|_i}{|p|_{i+1}} \bigg).
\ee
Scalar fields thus correspond to the case $\De=1$ (since the operators are built from $\p\phi$), while fermions correspond to $\De=\half$. For bosonic fields we must restrict this basis to \emph{symmetric} wavefunctions, while for fermionic fields we restrict to \emph{antisymmetric} ones.

%%%%%%%%%%%%%%%%%%%%%%%%%%%%%%%%%%%%%%%%%%%%%%%%%%%%%%%%%%%%%%%%%%%%%%%%%%%%%

\subsection{Wick Contraction Method}
\label{sec:WickContract}

In section~\ref{sec:FockSpace}, we learned how to do LCT computations using the ``Fock space method," where one works directly in momentum space and expresses the LCT data as integrals involving Fock space wavefunctions. In this section, we will present a second strategy, where one works in position space until the very last step. 

The main observation is that LCT inner products and matrix elements are, by definition, Fourier transforms of CFT two- and three-point functions, respectively. Recall that our basis states in 2d are given by 
\be
\ket{\CO_i, p} = \frac{1}{N_{\CO_i}} \int dx\,e^{-ipx}\, \CO_i(x) \ket{ {\rm vac} },
\ee
(where $p=p_-$). Let us also recall our notation for the Gram matrix and Hamiltonian matrix elements,
\be
\begin{aligned}
\< \CO_i, p| \CO_j, p'\> &\equiv 2p(2\pi)\delta(p-p') \, G_{\CO_i \CO_j}. \\[5pt]
\<\Ocal_i,p|2\de P_+ P_-|\Ocal_j,p'\> &\equiv 2p(2\pi)\de(p-p') \, \Mcal^{(\Ocal_R)}_{\Ocal_i\Ocal_j},
\end{aligned}
\ee
where $\CO_R$ is the relevant deformation associated with $\delta P_+$. It follows directly from the definition of our basis states that 
\be
\boxed{
\begin{aligned}
G_{\CO_i \CO_j} &= \frac{1}{2p N^*_{\CO_i} N_{\CO_j} } \int dx\,e^{ipx} \< \CO_i(x) \CO_j(0) \>, \\[5pt]
\Mcal^{(\Ocal_R)}_{\Ocal_i\Ocal_j} &= \frac{1}{N^*_{\CO_i} N_{\CO_j} } \int dx \, dz \,e^{ip(x-z)} \< \CO_i(x) \CO_R(0) \CO_j(z) \>,
\end{aligned}
}
\label{eq:LCTDataFT}
\ee
where the correlators in these expressions are \emph{Lorentzian Wightman functions}, with a fixed ordering for the operators to ensure well-defined in- and out-states. Given (\ref{eq:LCTDataFT}), an obvious strategy to compute $G_{\CO_i \CO_j}$ and $\Mcal^{(\Ocal_R)}_{\Ocal_i\Ocal_j}$ is to first work out the position-space correlators appearing on the right and then perform the Fourier transform. 

Fortunately, the Fourier transforms we encounter are known. The formulas we need for two- and three-point functions are, respectively, 
\be
\begin{aligned}
\int dx \, \frac{e^{ipx}}{x^{2\Delta}} &\doteq \frac{2\pi p^{2\Delta-1}}{\Gamma(2\Delta)}, \\[10pt]
\int dx \, dz \, \frac{e^{ip(x-z)}}{x^A z^B (x-z)^C} &\doteq \frac{4\pi^2 \Gamma(A+B-1) p^{A+B+C-2}}{\Gamma(A)\Gamma(B)\Gamma(A+B+C-1)}. 
\end{aligned}
\label{eq:FTFormulas}
\ee
Note that, for simplicity, in these expressions we have suppressed the $i\epsilon$ prescription needed to ensure Wightman ordering of the correlators, step functions enforcing positivity of lightcone momenta, and any resulting overall phases. For a detailed derivation of these formulas, including such additional subtleties, see~\cite{Anand:2019lkt}. At an operational level, though, \eqref{eq:FTFormulas} is all we need. 

With Fourier transform formulas in hand, the task of computing $G_{\CO_i \CO_j}$ and $\Mcal^{(\Ocal_R)}_{\Ocal_i\Ocal_j}$ boils down to computing the position-space correlators appearing on the right-hand side of (\ref{eq:LCTDataFT}). Since we are specifically considering free CFTs in this paper, we can simply use Wick contractions to work out all necessary position-space correlators, starting from the single-particle building blocks:\footnote{Note that we have removed the $x^+$-dependence in the two-point function of $\phi$, which will not affect correlators for primaries built from $\p_-\phi$.}
\be
\<\phi(x) \phi(0)\> = -\fr{\log x}{4\pi}, \qquad \<\psi(x) \psi(0)\> = \fr{-i}{4\pi x}.
\ee
We therefore refer to this general strategy as the ``Wick contraction method" for computing LCT data. In Part II of this work, we will learn how to avoid Wick contractions and instead use radial quantization to compute position-space correlators much more efficiently. 

To see all of these ideas in action, let us revisit the examples of LCT data computed in section~\ref{sec:FockSpace} using the Fock space method and recompute them using the Wick contraction method. In particular, we'll consider the two 2-particle operators $(\p\phi)^2$ and $\CO_{(2)} \equiv 6\p^3\phi\p\phi - 9(\p^2\phi)^2$, as well as the 3-particle operator $(\p\phi)^3$.

First, let us compute the Gram matrix of these operators. Using Wick contractions, it is straightforward to work out the two-point functions of these operators,
\be
\begin{aligned}
&\< (\p\phi)^2(x)\, (\p\phi)^2(0) \> \doteq \frac{2}{(4\pi)^2 x^4}, \hspace{5mm}
&\< \CO_{(2)}(x)\, \CO_{(2)}(0) \> \doteq \frac{1080}{(4\pi)^2 x^8}, \\[5pt]
&\< (\p\phi)^2(x)\, \CO_{(2)}(0) \> = 0, \hspace{5mm}
&\< (\p\phi)^3(x) \, (\p\phi)^3(0) \> \doteq \frac{3!}{(4\pi)^3 x^6}. 
\end{aligned}
\ee
Note that the two-point function of $(\p\phi)^2$ with $\CO_{(2)}$ vanishes by construction, as both operators are primaries with different scaling dimensions. 
Applying (\ref{eq:LCTDataFT})-(\ref{eq:FTFormulas}) then allows us to immediately compute the inner products. For example, the resulting inner product for $(\p\phi)^2$ is
\be
G_{(\p\phi)^2,(\p\phi)^2} = \frac{1}{2p |N_{(\p\phi)^2}|^2} \int dx\,e^{ipx} \< (\p\phi)^2(x) \, (\p\phi)^2(0) \> = \fr{p^2}{48\pi |N_{(\p\phi)^2}|^2},
\ee
exactly reproducing the Fock space calculation in eq.~\eqref{eq:DPhi2Inner}. Filling out the rest of the $2\times 2$ Gram matrix for $(\p\phi)^2$ and $\CO_{(2)}$, we obtain (setting $p=1$)
\be
G_{ij} = \frac{1}{16\pi} \begin{pmatrix}  \frac{1}{3 |N_{(\p \phi)^2}|^2} & 0 \\ 0 & \frac{3}{14 |N_{\CO_{(2)}}|^2} \end{pmatrix},
\label{eq:GramExample1Wick}
\ee
which can be compared with~\eqref{eq:GramExample1}. Finally, we can compute the 3-particle inner product
\be
G_{(\d \phi)^3, (\d \phi)^3} = \frac{p^4}{1280\pi^2 |N_{(\p\phi)^3}|^2}.
\label{eq:NormDPhi3Wick}
\ee
The normalization constants $N_{\Ocal_i}$ are now chosen to set all norms to unity. 

Now, let us turn to some examples of Hamiltonian matrix elements. We start with the mass matrix, which corresponds to the relevant deformation $\CO_R = \frac{1}{2}m^2 \phi^2$. Let us compute some of the diagonal entries of the mass matrix. For instance, starting from these three-point functions,
\be
\begin{aligned}
\< \p\phi(x) \, \phi^2(y) \, \p\phi(z) \> &\doteq \frac{2}{(4\pi)^2(x-y)(y-z)}, \\[5pt]
\< (\p\phi)^2(x) \, \phi^2(y) \, (\p\phi)^2(z) \> &\doteq \frac{8}{(4\pi)^3(x-y)(y-z)(x-z)^2}, \\[5pt]
\< (\p\phi)^3(x) \, \phi^2(y) \, (\p\phi)^3(z) \> &\doteq \frac{36}{(4\pi)^4(x-y)(y-z)(x-z)^4}, \\[5pt]
\end{aligned}
\ee
the formulas (\ref{eq:LCTDataFT})-(\ref{eq:FTFormulas}) yield the following results for mass matrix elements,
\be
\CM_{\p\phi,\p\phi}^{(\phi^2)} = m^2, \hspace{10mm} \CM_{(\p\phi)^2,(\p\phi)^2}^{(\phi^2)} = 6m^2, \hspace{10mm} \CM_{(\p\phi)^3,(\p\phi)^3}^{(\phi^2)} = 15m^2.
\label{eq:WickExamplesPhi2}
\ee

Finally, let us consider some examples involving the quartic interaction, corresponding to the relevant deformation $\CO_R = \frac{1}{4!}\lambda \phi^4$. As examples, the three-point functions
\be
\begin{aligned}
\< \p\phi(x) \, \phi^4(y) \, \p\phi(z) \> &= 0, \\[5pt]
\< \p\phi(x) \, \phi^4(y) \, (\p\phi)^3(z) \> &\doteq \frac{4!}{(4\pi)^4(x-y)(y-z)^3}, \\[5pt]
\< (\p\phi)^3(x) \, \phi^4(y) \, (\p\phi)^3(z) \> &\doteq \frac{3^2\cdot 4!}{(4\pi)^5(x-y)^2(y-z)^2(x-z)^2}, \\[5pt]
\end{aligned}
\ee
yield the following results for $\phi^4$ matrix elements,
\be
\CM_{\p\phi,\p\phi}^{(\phi^4)} = 0, \hspace{10mm} \CM_{\p\phi,(\p\phi)^3}^{(\phi^4)} = \frac{\sqrt{5}\lambda}{4\pi}, \hspace{10mm} \CM_{(\p\phi)^3,(\p\phi)^3}^{(\phi^4)} = \frac{15\lambda}{4\pi}.
\label{eq:WickExamplesPhi4}
\ee

%%%%%%%%%%%%%%%%%%%%%%%%%%%%%%%%%%%%%%%%%%%%%%%%%%%%%%%%%%%%%%%%%%%%%%%%%%%%%
%%%%%%%%%%%%%%%%%%%%%%%%%%%%%%%%%%%%%%%%%%%%%%%%%%%%%%%%%%%%%%%%%%%%%%%%%%%%%

\section{Simplest Possible Scalar Code}
\label{sec:simplestcode}
In this section, we describe in more detail the basic ideas that go into computing the basis and matrix elements in practice when the UV CFT is a free scalar field. As we go, we will build up simple code for each step in the process, and we encourage the reader to write their own version in order to concretely understand how to apply LCT to deformations of free field theories. The emphasis will be on simplicity and conciseness rather than efficiency, so these methods on their own are sufficient only for small bases, and  further improvements in Part II will be needed to go to much larger bases in realistic computation times.

%%%%%%%%%%%%%%%%%%%%%%%%%%%%%%%%%%%%%%%%%%%%%%%%%%%%%%%%%%%%%%%%%%%%%%%%%%%%%

\subsection{Basis of Primary Operators}
\label{sec:ScalarBasis}

We saw in section \ref{sec:2dFFT} that when our UV CFT is a free scalar field and one of the relevant deformations is a mass term $\sim m^2 \phi^2$, then the basis of primary operators we need to consider is spanned by products of $\partial_-$ derivatives of $\phi$. We can denote these operators in the following compact notation:\footnote{As a reminder, we have adopted the convention that $\partial$ derivatives without an index correspond to $\partial_-$ derivatives.}  
\be
\p^{\bk} \phi \equiv \p^{k_1} \phi \cdots \p^{k_n}\phi. \label{eq:ScalarMonoDef}
\ee
Note that any $\bk$ which are related by permutations are equivalent. We will therefore always choose to arrange the vectors such that
\be
k_1 \geq k_2 \geq \cdots \geq k_n.
\ee

By a straightforward generalization of \eqref{eq:3ParticleF} the wavefunctions $F(p)$ of such operators in the Fock space basis are
\be
F_{\p^{\bk} \phi}(p) = \< p_1, \dots, p_n | \p^{\bk} \phi(0)\> \doteq \sum_{\bk' \in \textrm{perm}(\bk)} p_1^{k'_1} \dots p_n^{k'_n}.
\ee
For this reason, we shall refer to the operators $\p^{\bk} \phi$ as ``monomials''.\footnote{Technically, the wavefunctions are monomial symmetric polynomials, but we will use ``monomials'' for short.} Since each insertion of $\phi$ must have at least one derivative acting on it, every $n$-particle monomial must contain at least $n$ derivatives. Because of this, we will define the ``degree'' of a monomial as $|\bk|-n$, where $|\bk| \equiv \sum k_i$. In other words, the degree of a given monomial is the number of \emph{additional} derivatives.

We need to construct primary operators as linear combinations of these monomials.  Primary operators in a CFT are defined as those operators that are annihilated by the special conformal generators $K_\mu$ acting on the operator at the origin $x^\mu=0$. The generator $K_-$ commutes with $P_- \sim \partial_-$, so it automatically annihilates any monomial operator.  Therefore we only need the action of the generator $K\equiv K_+$, which on individual monomials is
\be
\comm{K}{\p^{\bk}\phi(0)} = \sum_{i=1}^n k_i(k_i-1) \p^{k_1}\phi \cdots \p^{k_i-1} \phi \cdots \p^{k_n}\phi(0) .
\ee
To construct a basis of primary operators we therefore need to find the linear combinations of monomials
\be
\CO(x) \equiv \sum_{\bk} C^\Ocal_{\bk} \partial^{\bk} \phi(x),
\label{eq:PrimaryDef}
\ee
which are annihilated by $K$ when acting at the origin. Because primary operators each have a well-defined scaling dimension, we can restrict the sum in eq.~\eqref{eq:PrimaryDef} to monomials with fixed total number of derivatives $|\kvec|=\De$.

In principle, one could construct all the primaries by writing out the action of $K$ on the space of all monomials of a fixed scaling dimension and solving for the kernel of $K$.  However, it is simpler to construct primary operators recursively  by harnessing a result obtained by Penedones in \cite{Penedones:2010ue}.\footnote{See also earlier work by Mikhailov \cite{Mikhailov:2002bp}.} This result states that, given two holomorphic primary operators $A$ and $B$ in a generalized free theory, there is exactly one composite primary operator constructible using $A$ and $B$ for each non-negative integer $\ell$. This composite operator is the double-trace operator 
\be
\left[ AB \right]_\ell \equiv \sum_{m=0}^\ell c^\ell_m(\Delta_A,\Delta_B)\, \p^m A \, \p^{\ell-m} B,
\label{eq:JoaoFormula}
\ee
where the coefficients $c^\ell_m(\Delta_A,\Delta_B)$ are given by the formula
\be
c^\ell_m(\Delta_A,\Delta_B) = \fr{(-1)^m \Gamma(2\Delta_A+\ell) \Gamma(2\Delta_B+\ell)}{m! (\ell-m)! \Gamma(2\Delta_A+m) \Gamma(2\Delta_B + \ell - m)}.
\label{eq:JoaoCoeff}
\ee
This formula allows us to construct primary operators iteratively in particle number $n$ and spin $\ell$ by starting with the simplest primary, $\p\phi$, and successively sewing on additional $\p\phi$'s according to (\ref{eq:JoaoFormula}) to construct new primaries.
\renewcommand{\arraystretch}{1.0}
\begin{table}[t!]
\begin{center}
\begin{tabular}{| l | l | l |}
\hline & & \\[-5pt]
$n$ & General expression & Explicit examples \\
\hline & & \\[-7pt]
1 &  & $\p\phi$ \\
\hline & & \\[-5pt]
2  &   $\Ocal_{(\ell_1)} \equiv \left[ \p\phi \,  \p\phi \right]_{\ell_1} $   &  $\Ocal_{(0)} = \p\phi \p\phi $  \\
    &                                                                                                    & $\Ocal_{(1)}  =0 $ \\
    &                                                                                                    & $\Ocal_{(2)}  = 6 \p^3\phi \p\phi - 9\p^2\phi \p^2\phi$ \\
    &                                                                                                    & $\vdots$ \\
\hline & & \\[-5pt]
3 &    $\Ocal_{(\ell_1,\ell_2)} \equiv \left[ \Ocal_{(\ell_1)}\p\phi \right]_{\ell_2}$   & $\Ocal_{(0,0)} = \p\phi \p\phi \p\phi$ \\
    &                                                                                                    & $\Ocal_{(0,1)}  = \Ocal_{(1,0)} = \Ocal_{(1,1)}  =  0$ \\
    &                                                                                                    & $\Ocal_{(2,0)}  = \fr{3}{8}\Ocal_{(0,2)} = 6\p^3\phi \p\phi \p\phi - 9\p^2\phi \p^2\phi \p\phi$ \\
    &                                                                                                    & $\vdots$ \\
\hline
\end{tabular}
\end{center}
\caption{The first few scalar primaries constructed recursively by starting with $\p\phi$ and successively sewing on additional $\p\phi$'s using (\ref{eq:JoaoFormula}).}
\label{table:recursion}
\end{table}

Table \ref{table:recursion} lists the first few primary operators constructed in this way. Let us unpack this table a bit. For a single particle, $n=1$, the lone primary operator is of course $\p\phi$. At $n=2$, we start with $A=\p\phi$ ($\Delta_A=1)$ and sew on $B=\p\phi$ ($\Delta_B=1$) using~(\ref{eq:JoaoFormula}). We denote the resulting ``double-trace'' operators as $\Ocal_{(\ell_1)} \equiv \left[ \p\phi \,  \p\phi \right]_{\ell_1} $, which have dimension and spin $\ell_1+2$. Explicit expressions for $\Ocal_{(\ell_1)}$ are shown in the table for $\ell_1=0,1,2$. At $n=3$, we can repeat the process starting with any of the operators at $n=2$ and sewing on another $\p\phi$. This time, we denote the resulting operator using two labels $\Ocal_{(\ell_1,\ell_2)} \equiv \left[ \Ocal_{(\ell_1)}\p\phi \right]_{\ell_2}$, with $\ell_1$ indicating which $n=2$ operator was chosen and $\ell_2$ indicating the new ``double-trace'' combination being taken between $\Ocal_{(\ell_1)} $ and $\p\phi$. The table shows several examples. 
Continuing in this way will generate all possible primaries. 

Looking at Table \ref{table:recursion}, we immediately discern several important facts:
\begin{itemize}
\item[(i)] For particle number $n$, the primary operators $\Ocal_{\Lvec}$ are labeled by $n-1$ component vectors $\Lvec=(\ell_1,\dots,\ell_{n-1})$, where each $\ell_i$ specifies which double-trace combination was taken to sew on an additional $\p\phi$. 
\item[(ii)] An operator $\Ocal_{\Lvec}$ is clearly built from $n$ $\phi$'s and $|\Lvec|+n$ derivatives. We will refer to $(n,|\Lvec|) \label{eq:OpLevel}$ as the ``level" of the operator.
\item[(iii)] By construction, the complete list of operators $\{\Ocal_{\Lvec}\}$ spans the space of all primary operators. However, the list is \emph{overcomplete}. This is already evident in the table, where we see that $\Ocal_{(2,0)}$ and $\Ocal_{(0,2)}$ are in fact equal up to an overall constant. More generally, not all of the operators $\Ocal_{\Lvec}$ at a given level $(n,|\Lvec|)$ will be linearly independent. This redundancy is simply a consequence of the fact that the $\phi$'s are indistinguishable.
\end{itemize}

As we have just noted, there are generally linear dependencies amongst the operators  $\Ocal_{\Lvec}$. For small bases, we can simply compute the overcomplete basis and then row reduce to eliminate redundant operators. For greater efficiency, it is actually possible to avoid constructing the redundant operators in the first place by specifying a priori a set of complete but not overcomplete primaries, as detailed below.

\noindent\rule[0.5ex]{\linewidth}{1pt}
\footnotesize

Here we describe the algorithm for choosing a complete and minimal (\emph{i.e.}, not overcomplete) subset of $\Lvec$ vectors. To motivate the algorithm, it is useful to consider partitions of integers. With this in mind, let
\be
P_n(k) \equiv \# \text{ of partitions of $k$ objects into exactly $n$ bins} 
\ee
(\emph{i.e.}, the occupancy of each bin is at least one). The function $P_n(k)$ is related to counts of monomials and primaries in the following way,
\begin{eqnarray}
P_n(k) &=& \# \text{ of monomials with $n$ $\phi$'s and $k$ derivatives}  \nonumber\\[5pt]
\widehat{P}_n(k) &\equiv& P_n(k) - P_n(k-1) \nonumber \\
&=& \# \text{ of primaries with $n$ $\phi$'s and $k$ derivatives}.
\end{eqnarray}
It is straightforward to check that for $\Pcal = P \text{ or } \widehat{P}$ the following recursion relation holds
\be
\Pcal_n(k) = \sum_{j=0}^{\left \lceil{ k/n}\right \rceil } \Pcal_{n-1}(k-jn-1).
\label{eq:PRecursion}
\ee

Now, suppose that we have already selected a complete and minimal list of vectors $\Lvec^\prime$ for $n-1$ particles and want to extend the list to $n$ particles. Specifically, at every level $(n, |\Lvec|)$, we want to find a minimal list of $\Lvec$'s to span that level. We can do so as follows. 

Letting $\mathrm{N}(n, |\Lvec|)$ denote the number of primary operators at level $(n, |\Lvec|)$, the recursion (\ref{eq:PRecursion}) immediately implies
\be
\mathrm{N}(n, |\Lvec|) =  \sum_{j=0}^{\left \lceil{ |\Lvec|/n}\right \rceil +1 } \mathrm{N}(n-1, |\Lvec|-jn).
\label{eq:PHatRecursion}
\ee
This is a very suggestive relation between numbers of primary operators at $n$ and $n-1$ particles. It is suggestive, because for each operator at level $(n-1,|\Lvec|-jn)$ contributing to the right hand side, there is an obvious way to construct an operator at level $(n,|\Lvec|)$: simply sew on an additional $\p\phi$ with $\ell_n = jn$. In other words, for each $\Ocal_{\Lvec^\prime} \in (n-1,|\Lvec|-jn)$, we can construct the operator $[ \Ocal_{\Lvec^\prime}  \, \p\phi ]_{\ell_n=jn} \in (n,|\Lvec|)$. The formula (\ref{eq:PHatRecursion}) strongly suggests that the new operators constructed in this way will be complete at level $(n,|\Lvec|)$.\footnote{We are abusing notation somewhat and writing ``$\Ocal_{\Lvec'} \in (n, |\Lvec|)$'' and $\Lvec' \in (n, |\Lvec|)$ to denote that the level of $\Lvec'$ is $(n,|\Lvec|)$.} 

In practice, this is precisely the algorithm that we use. To state things precisely, given a complete and minimal list of vectors $\Lvec^\prime$ for $n-1$ particles, a complete and minimal list $(n, |\Lvec|)_{\rm minimal}$ of vectors $\Lvec$ for level $(n,|\Lvec|)$ is given recursively by 
\be
\boxed{
(n,|\Lvec|)_{\text{minimal}} = \left\{ (\Lvec', jn) \ \Big| \  0 \le j \le \lceil |\Lvec|/n \rceil+1 \textrm{ and } \Lvec' \in (n-1, |\Lvec|-jn)_{\rm minimal} \right\}.
}
\label{eq:MinLs}
\ee

\normalsize
\noindent\rule[0.5ex]{\linewidth}{1pt}

Based on these observations, we can now write simple Mathematica code for generating a complete basis of linearly independent primary operators of a given particle number $n$ and total degree $|\Lvec|$, which is provided in table~\ref{table:ScalarBasisConstructionCode}. Schematically, this code simply proceeds through the complete, minimal set of vectors $\Lvec$ defined in eq.~\eqref{eq:MinLs}, and for each $\Lvec$, computes the coefficients $C^{\Lvec}_{\bk}$ corresponding to the expansion of that operator in terms of the monomials $\p^{\bk}\phi$, as in~\eqref{eq:PrimaryDef}.

\begin{table}[t!]
\noindent\rule[0.5ex]{\linewidth}{1pt}
\verb|(*| The monomials and number of primaries at each level, and the coefficients  (\ref{eq:JoaoCoeff}) \verb|*)|
\begin{code}
monomialsBoson\verb|[n_,deg_]:=IntegerPartitions[deg+n,{n}]|;\\
\verb|numStates[n_,deg_]|:=\=Length[monomialsBoson[n,deg]]\\
 \> -Length[monomialsBoson[n,deg-1]]; \\
\verb|PrimCoeffs[DA_,DB_,L_,k_]|:=\=\verb|(-1)^k| Gamma[2DA+L]Gamma[2DB+L] \\
 \> /(k!(L-k)!Gamma[2DA+k]Gamma[2DB+L-k]);
\end{code}
\verb|(*| Compute maps that add a $\partial^k \phi$ or take a total derivative, in the monomial basis \verb|*)|
\begin{code}
appendOneScalarMap\=Simp\verb|[n_,deg_,kNew_]|:=Table[\\
 \> If[Reverse[Sort[Append[mon2,kNew]]]==mon1,1,0], \\
 \> \verb|{mon2,monomialsBoson[n,deg]}|, \\
 \> \verb|{mon1,monomialsBoson[n+1,deg+kNew-1]}]|; \\ \\
dBoson\=Simp\verb|[n_,deg_]|:=Table[ \\
 \> Length[C\=ases[ \\
 \>\> Table[temp=mon2;\verb| temp[[i]]|++; Reverse[Sort[temp]],\verb|{i,n}],| \\
 \>\> mon1]], \\
 \> \verb|{mon2,monomialsBoson[n,deg]}|,\\
 \> \verb|{mon1,monomialsBoson[n,deg+1]}]|;
\end{code}
\verb|(*| Make all primary operators at a fixed particle number and degree \verb|*)|
\begin{code}
Pri\=marySetSimp\verb|[n_,deg_]|:=Block[\verb|{dL,vecs,vecsF,res={}}|, \\
 \> If\=[n==1,If[deg==0,res=\verb|{{1}},res={}|], \\
 \>\> Do\=[If[numStates[n-1,degP]!=0,dL=deg-degP; \\
 \>\>\>	vecs=PrimarySetSimp[n-1,degP]; \\
 \>\>\>	vecsF=Table[0, \verb|{Length[vecs]}|, \verb|{Length[monomialsBoson[n,deg]]}|];\\
 \>\>\>	Do\=[vecsF+=\=PrimCoeffs[degP+(n-1),1,dL,k] \\
 \>\>\>\>\>	*Dot[vecs,appendOneScalarMapSimp[n-1,k+degP,dL-k+1]]; \\
 \>\>\>\> vecs=Dot[vecs,dBosonSimp[n-1,k+degP]],\\
 \>\>\>\> \verb|{k,0,dL}]|;\\
 \>\>\>	res=Join[res,vecsF]],\\
 \>\>\> \verb|{degP,deg,0,-n}|]]; \\
 \>\> res];
\end{code}
\noindent\rule[0.5ex]{\linewidth}{1pt}
\caption{Sample Mathematica code for constructing a complete basis of primary operators at fixed particle number $n$ and degree $\De-n$ for a single scalar field.
\label{table:ScalarBasisConstructionCode}}
\end{table}

Our strategy is to compute the coefficients $C^{\Lvec}_{\bk}$ recursively, due to the fact that an $n$-particle primary operator $\Ocal_{\Lvec}$ is simply a ``double-trace'' operator built from an $n-1$-particle primary $\Ocal_{\Lvec/\ell_{n-1}}$ and $\p\phi$,
\be
\Ocal_{\Lvec} = [\Ocal_{\Lvec/\ell_{n-1}} \p\phi]_{\ell_{n-1}},
\ee
where $\Lvec/\ell_{n-1}$ refers to the $n-2$-component vector created by removing the last entry of $\Lvec$,
\be
\Lvec/\ell_{n-1} \equiv (\ell_1,\ldots,\ell_{n-2}). \label{eq:DeleteComponentNotation}
\ee
Suppose that we already know the expansion of $\Ocal_{\Lvec/\ell_{n-1}}$ in terms of $n-1$-particle monomials. We can then compute the monomial expansion of $\Ocal_{\Lvec}$ by using eq.~\eqref{eq:JoaoFormula}, which requires acting with additional derivatives on the expansion of $\Ocal_{\Lvec/\ell_{n-1}}$ and then appending an additional $\p^{k_n}\phi$.

With this in mind, let's slowly work through the code in table~\ref{table:ScalarBasisConstructionCode}, to explain the important steps in more detail. First, we define the function {\tt monomialsBoson[n,deg]}, which simply lists all $n$-particle monomials of a particular degree. For example, we can obtain the list of all two-particle monomials with degree $2$ by entering
\begin{mmaCell}[moredefined={monomialsBoson}]{Input}
  monomialsBoson[2,2]
\end{mmaCell}
The output is the list of all possible $\bk$ vectors:
\begin{mmaCell}{Output}
  \{\{3,1\},\{2,2\}\}
\end{mmaCell}
which in this case correspond to the two monomials $\p^3\phi \p\phi$ and $(\p^2\phi)^2$, respectively.\footnote{Recall that ``degree'' refers to the total number of derivatives minus the number of particles.} We will often express operators as vectors in the space of monomials, using the same ordering of the monomials as the output of {\tt monomialsBoson}. For example, the primary operator $\Ocal_{(2)} = 6\p^3\phi\p\phi - 9(\p^2\phi)^2$ listed in table~\ref{table:recursion} would be represented by the vector {\tt \{6,-9\}}.

The next important function is {\tt appendOneScalarMapSimp[n,deg,kNew]}, which takes the list of monomials of a given particle number and degree and appends an additional $\p^{k_{\textrm{new}}}\phi$ to those monomials. For example, if we start with the one-particle operator $\p\phi$, we can append a factor of $\p^3\phi$ with
\begin{mmaCell}[moredefined={appendOneScalarMapSimp}]{Input}
  appendOneScalarMapSimp[1,0,3]
\end{mmaCell}
The output is a matrix mapping from the space of monomials with particle number $n=1$ and degree $deg=0$ (in this case, the only such monomial is $\p \phi$) to the space of  $n+1$-particle monomials with degree $deg + (k_{\textrm{new}}-1) = 2$ (in this case, $\p^3 \phi \p \phi$ and $\p^2 \phi \p^2 \phi$):
\begin{mmaCell}{Output}
  \{\{1,0\}\}
\end{mmaCell}
For this example,  the  output {\tt \{\{1,0\}\}} is a $2 \times 1$ matrix indicating that appending $\p^3 \phi$ maps $\p \phi$ to the first monomial in the list generated by {\tt monomialsBoson[2,2]}, which of course corresponds to $\p^3\phi\p\phi$.  That is,  {\tt \{\{1,0\}\}}  indicates that
\be
\p^3 \phi \times (\p\phi) = 1 \cdot \p^3\phi\p\phi + 0 \cdot (\p^2\phi)^2.
\ee
 
Next, we define the function {\tt dBosonSimp[n,deg]}, which takes the list of $n$-particle monomials of a given degree and computes the action of a single derivative on each of them. As a simple example, consider the only two-particle, degree-$1$ monomial $\p^2\phi\p\phi$. We can act with a single derivative on this monomial by computing
\begin{mmaCell}[moredefined={dBosonSimp}]{Input}
  dBosonSimp[2,1]
\end{mmaCell}
The output is a matrix from the space of two-particle degree-1 monomials to the space of two-particle degree-$2$ monomials (since acting with a derivative increases the degree by $1$):
\begin{mmaCell}{Output}
  \{\{1,1\}\}
\end{mmaCell}
which indicates that
\be
\p( \p^2\phi\p\phi) = 1 \cdot \p^3\phi\p\phi + 1\cdot (\p^2\phi)^2.
\ee

Finally, we have the function {\tt PrimarySetSimp[n,deg]}, which computes the complete, minimal set of primary operators $\Ocal_{\Lvec}$ of a given particle number and degree, and expresses them as a set of vectors in the space of monomials. Following eq.~\eqref{eq:MinLs}, this function is defined recursively, and uses the set of primaries with one fewer particles (which we can represent schematically as $\Ocal^{(n-1)}_{\Lvec'}$) with degree less than or equal to $|\Lvec|$.\footnote{More precisely, this function uses the set of $n-1$-particle primaries with degree $|\Lvec'| = |\Lvec|-jn$.} It then constructs the ``double-trace'' operators $[\Ocal^{(n-1)}_{\Lvec'} \p\phi]_{|\Lvec|-|\Lvec'|}$ by using {\tt dBosonSimp} to take derivatives of $\Ocal^{(n-1)}_{\Lvec'}$ and {\tt appendOneScalarMapSimp} to sew on the additional $\p\phi$.

As a simple example, we can find all two-particle, degree-$2$ primary operators by entering:
\begin{mmaCell}[moredefined={PrimarySetSimp}]{Input}
  PrimarySetSimp[2,2]
\end{mmaCell}
There is only one such primary, due to the fact that there is only a single one-particle primary for us to construct it from: $\p\phi$. This operator must therefore correspond to
\be
[\p\phi \p\phi]_2 = 6 \p^3\phi\p\phi - 9 (\p^2\phi)^2,
\ee
which is output by {\tt PrimarySetSimp} as a vector in the space of degree-$2$ monomials:
\begin{mmaCell}{Output}
  \{\{6,-9\}\}
\end{mmaCell}

%%%%%%%%%%%%%%%%%%%%%%%%%%%%%%%%%%%%%%%%%%%%%%%%%%%%%%%%%%%%%%%%%%%%%%%%%%%%%

\subsection{Wick Contraction and Orthonormalization}
\label{sec:Ortho}

We now have a general procedure for constructing a complete basis of primary operators for any particle number and scaling dimension. However, we need this basis to be \emph{orthonormal} with respect to the momentum space inner product, such that
\be
\<\Ocal,p|\Ocal',p'\> = 2p (2\pi)\de(p-p') \, \de_{\Ocal\Ocal'}.
\ee
In the previous section, we constructed the set of primary operators $\Ocal$ in \emph{position space}, as linear combinations of monomials
\be
\Ocal(x) = \sum_{\bk} C^{\Ocal}_{\bk} \, \p^{\bk}\phi(x).
\label{eq:PrimaryPositionSpace}
\ee
The first step in orthonormalizing this basis of primary operators is to Fourier transform to momentum space and express the resulting states $|\Ocal,p\>$ as linear combinations of the properly normalized monomial states
\be
|\p^{\bk}\phi,p\> \equiv \fr{1}{N_{\bk}} \int dx \, e^{-ipx} \, \p^{\bk}\phi(x)\vac,
\label{eq:MonoStateDef}
\ee
where the normalization coefficient is defined as
\be
|N_{\bk}|^2 \equiv \fr{1}{2p} \int dx \, e^{ipx} \<\p^{\bk}\phi(x) \p^{\bk}\phi(0)\>. \label{eq:MonoNormalization} 
\ee

Given the position space expansion in eq.~\eqref{eq:PrimaryPositionSpace}, the resulting momentum space representation is clearly
\be
|\Ocal,p\> \equiv \sum_{\bk} \Chat^\Ocal_{\bk} |\p^{\bk}\phi,p\> = \sum_{\bk} \fr{C^\Ocal_{\bk} N_{\bk}}{N_\Ocal} |\p^{\bk}\phi,p\>.
\ee
The inner product between two states can thus be written as a sum of monomial inner products
\be
\<\Ocal,p|\Ocal',p'\> = \sum_{\bk,\bk'} \Chat^{\Ocal*}_{\bk} \Chat^{\Ocal'}_{\bk'} \<\p^{\bk}\phi,p|\p^{\bk'}\phi,p'\>.
\ee
The individual inner products for monomials can be written in the general form
\be
\<\p^{\bk}\phi,p|\p^{\bk'}\phi,p'\> = 2p(2\pi)\de(p-p') G_{\bk\bk'},
\ee
where the Gram matrix $G_{\bk\bk'}$ is defined as
\be
G_{\bk\bk'} = \fr{1}{2p N^*_{\bk} N_{\bk'}} \int dx \, e^{ipx} \<\p^{\bk}\phi(x) \p^{\bk'}\phi(0)\>. \label{eq:DefnMonoGram}
\ee

To orthonormalize our basis of primary operators, we therefore need to first construct the Gram matrix $G_{\bk\bk'}$. As we can see, the coefficients $N_{\bk}$ are defined such that the diagonal elements $G_{\bk\bk} = 1$ by construction. However, for the off-diagonal elements we need to evaluate the Fourier transform of monomial two-point functions.

In section~\ref{sec:2dFFT}, we introduced two different methods for evaluating these inner products: the ``Fock space method'', where we integrate over individual particle momenta weighted by the momentum space wavefunctions for the two monomials, and the ``Wick contraction method'', where we directly compute the position space two-point function, then Fourier transform the resulting expression to momentum space.

In principle, these two approaches are completely equivalent, and we can directly map computations in the Fock space approach to those in terms of Wick contraction. However, there is a key conceptual advantage in phrasing the computation in terms of Fourier transforms of position space correlators, which is to make the CFT structure of the UV theory more manifest. Correlation functions of local operators are the natural set of observables in CFTs, with strong constraints on the precise form of two- and three-point functions. However, this structure is largely obfuscated in the Fock space formulation. We will see the advantage of using CFT techniques even more clearly in Part II, where we introduce a third, much more powerful method for evaluating inner products and matrix elements.

To start, we can write down a general $n$-particle monomial two-point function as a sum over all possible pairings of the incoming and outgoing particles
\be
\<\p^{\bk}\phi(x) \p^{\bk'}\phi(0)\> = \sum_{\bs\in\textrm{perm}(\bk')} \<\p^{k_1}\phi(x)\p^{\sigma_1}\phi(0)\> \cdots \<\p^{k_n}\phi(x)\p^{\sigma_n}\phi(0)\>.
\ee
Using the simple one-particle correlator
\be
\<\p\phi(x)\p\phi(0)\> \doteq \fr{1}{4\pi x^2},
\ee
we can evaluate all these contractions to obtain the general expression
\be
\<\p^{\bk}\phi(x) \p^{\bk'}\phi(0)\> \doteq \fr{A_{\bk\bk'}}{(4\pi)^n x^{|\bk|+|\bk'|}},
\label{eq:2ptWick}
\ee
where the Wick contraction coefficient $A_{\bk\bk'}$ is defined as
\be
A_{\bk\bk'} \equiv \sum_{\bs\in\textrm{perm}(\bk')} \G(k_1+\sigma_1) \cdots \G(k_n+\sigma_n).
\label{eq:WickCoeff}
\ee
It will be useful to compute this expression recursively, expressing $n$-particle coefficients in terms of $n-1$-particle ones,
\be
A_{\bk\bk'} = \sum_{i=1}^n \G(k_n+k'_i) \, A_{\bk/k_n,\bk'/k'_i}, \label{eq:WickContractionCoeff}
\ee
where $\bk/k_i$ is the $n-1$-component vector created by removing the entry $k_i$ from $\bk$.

For example, we can use~\eqref{eq:2ptWick} to compute the following two-particle monomial correlators,
\bq
\bal
\<(\p\phi)^2(x) \, \p^3\phi\p\phi(0)\> &\doteq \fr{\G(1+3)\G(1+1) + \G(1+1) \G(1+3)}{(4\pi)^2 x^6} = \fr{3}{4\pi^2 x^6}, \\
\<(\p\phi)^2(x) \, (\p^2\phi)^2(0)\> &\doteq \fr{\G(1+2)\G(1+2) + \G(1+2) \G(1+2)}{(4\pi)^2 x^6} = \fr{1}{2\pi^2 x^6},
\eal
\eq
to again show that the operators $(\p\phi)^2$ and $\Ocal_{(2)} \equiv 6\p^3\phi\p\phi-9(\p^2\phi)^2$ are orthogonal,
\be
6 \cdot \<(\p\phi)^2(x) \, \p^3\phi\p\phi(0)\> - 9 \cdot \<(\p\phi)^2(x) \, (\p^2\phi)^2(0)\> = 0.
\ee

Next, we need to Fourier transform the monomial two-point function in eq.~\eqref{eq:2ptWick} to momentum space. We can do this with the general integral in eq.~\eqref{eq:FTFormulas}, allowing us to fix the exact value of the normalization coefficients,
\be
|N_{\bk}|^2 = \fr{\pi p^{2|\bk|-2} A_{\bk\bk}}{(4\pi)^n \G(2|\bk|)},
\label{eq:MonoNorm}
\ee
as well as the resulting Gram matrix elements
\be
G_{\bk\bk'} \doteq \fr{\sqrt{\G(2|\bk|)\G(2|\bk'|)}}{\G(|\bk|+|\bk'|)} \cdot \fr{A_{\bk\bk'}}{\sqrt{A_{\bk\bk} A_{\bk'\bk'}}}.
\label{eq:MonoGram}
\ee
As a simple sanity check, we can compare these results with the general three-particle normalization computed via the Fock space method in eq.~\eqref{eq:3PartFock},
\be
|N_{\bk}|^2 =  \fr{1}{64\pi^2} \sum_{\bk' \in \textrm{perm}(\bk)} \fr{\G(k_1+k_1') \G(k_2+k_2') \G(k_3+k_3')}{\G(k_1+k_2+k_3+k_1'+k_2'+k_3')} p^{k_1+k_2+k_3+k_1'+k_2'+k_3'-2}.
\ee

Given this Gram matrix, we can now orthonormalize a set of primary operators with respect to the general inner product
\be
G_{\Ocal\Ocal'} = \sum_{\bk,\bk'} \Chat^{\Ocal*}_{\bk} \Chat^{\Ocal'}_{\bk'} G_{\bk\bk'} \doteq \fr{\pi p^{2\De-2}}{(4\pi)^n \G(2\De) N^*_\Ocal N_{\Ocal'}} \sum_{\bk,\bk'} C^{\Ocal*}_{\bk} C^{\Ocal'}_{\bk'} A_{\bk\bk'}.
\label{eq:PrimaryNorm}
\ee
Table~\ref{table:ScalarBasisOrthogonalizationCode} provides simple Mathematica code which orthonormalizes the primary operators generated by {\tt PrimarySetSimp} in the previous section.

\begin{table}[t!]
\noindent\rule[0.5ex]{\linewidth}{1pt}
\verb|(*| The Wick contraction coefficients~\eqref{eq:WickCoeff} \verb|*)|
\begin{code}
A\verb|[k_,kp_]|:=\=A[k,kp]=If[Length[kp] <= 1, \\
 \> Product[Gamma[k[[i]] + kp[[i]]], \verb|{i,Length[kp]}],| \\
 \> Sum[\=Gamma[k[[-1]] + kp[[i]]] A[Delete[k,-1],Delete[kp,i]], \\
 \>\> \verb|{i,Length[kp]}]];|
\end{code}
\verb|(*| Construct the monomial Gram matrix~\eqref{eq:MonoGram} \verb|*)|
\begin{code}
monoGram\verb|[n_,deg_]|:=Table\=[A[k,kp] / Sqrt[A[k,k] A[kp,kp]], \\
 \> \verb|{k,monomialsBoson[n,deg]},| \\
 \> \verb|{kp,monomialsBoson[n,deg]}];|
\end{code}
\verb|(*| Rescale the coeffs generated by {\tt PrimarySetSimp} by monomial normalizations \verb|*)|
\begin{code}
rescalePrimarySet\verb|[n_,deg_]|:=\=Table\=[Sqrt[A[k,k]], \verb|{i,numStates[n,deg]},| \\
 \>\> \verb|{k,monomialsBoson[n,deg]}]| \\
 \> *PrimarySetSimp[n,deg];
\end{code}
\verb|(*| Orthonormalize all primary operators at a fixed particle number and degree \verb|*)|
\begin{code}
orthoPrimaries\verb|[n_,deg_]|:=Orthogonalize[\=rescalePrimarySet[n,deg], \\
 \> \verb|Dot[#1,monoGram[n,deg],#2]&];|
\end{code}
\noindent\rule[0.5ex]{\linewidth}{1pt}
\caption{Sample Mathematica code for orthonormalizing the basis of primary operators at fixed particle number $n$ and degree $\De-n$ generated by the code in table~\ref{table:ScalarBasisConstructionCode}. To obtain the timing date in table \ref{tab:timingBenchmark:Intro}, we added a ``{\tt //N}'' at the end of {\tt monoGram} and {\tt rescalePrimarySet}.
\label{table:ScalarBasisOrthogonalizationCode}}
\end{table}

The first important function is {\tt monoGram[n,deg]}, which generates the Gram matrix from eq.~\eqref{eq:MonoGram} for all monomials of a given particle number $n$ and degree $|\bk|-n$. For example, we can compute the Gram matrix for all two-particle, degree-$2$ monomials by entering:
\begin{mmaCell}[index=1,moredefined={monoGram}]{Input}
  monoGram[2,2]
\end{mmaCell}
There are two such monomials, $\p^3\phi\p\phi$ and $(\p^2\phi)^2$, so the output is a symmetric $2\times2$ matrix:
\begin{mmaCell}{Output}
  \{\{1,4\mmaSqrt{\mmaFrac{2}{39}}\},\{4\mmaSqrt{\mmaFrac{2}{39}},1\}\}
\end{mmaCell}
The ordering of the monomials labeling the entries in this matrix is the same as that given by {\tt monomialsBoson}. The diagonal elements are all trivially equal to $1$ by construction, but the off-diagonal element tells us that
\be
G_{\p^3\phi\p\phi,(\p^2\phi)^2} = \fr{A_{\p^3\phi\p\phi,(\p^2\phi)^2}}{\sqrt{A_{\p^3\phi\p\phi,\p^3\phi\p\phi} A_{(\p^2\phi)^2,(\p^2\phi)^2}}} = \fr{96}{\sqrt{156 \cdot 72}} = 4 \sqrt{\fr{2}{39}}.
\ee

Next, we have the function {\tt rescalePrimarySet[n,deg]}, which takes the coefficients $C^{\Lvec}_{\bk}$ generated by {\tt PrimarySetSimp} and rescales them by $\sqrt{A_{\bk\bk}}$ to account for the normalization of the momentum space states $|\p^{\bk}\phi,p\>$,
\be
C^{\Lvec}_{\bk} \ra \sqrt{A_{\bk\bk}} \, C^{\Lvec}_{\bk}.
\ee
For example, we saw in the previous subsection that {\tt PrimarySetSimp[2,2] = \{\{6,-9\}\}}, corresponding to the operator $\Ocal_{(2)} = 6\p^3\phi\p\phi - 9(\p^2\phi)^2$. We can rescale the coefficients of this operator by entering:
\begin{mmaCell}[moredefined={rescalePrimarySet}]{Input}
  rescalePrimarySet[2,2]
\end{mmaCell}
The output is now a list of vectors in the space of monomial momentum space states:
\begin{mmaCell}{Output}
  \{\{12\mmaSqrt{39},-54\mmaSqrt{2}\}\}
\end{mmaCell}
We can confirm that this vector is correct by computing the overlap of the resulting state with the degree-$0$ primary $(\p\phi)^2$,
\bq
12\sqrt{39} G_{(\p\phi)^2,\p^3\phi\p\phi} - 54\sqrt{2} G_{(\p\phi)^2,(\p^2\phi)^2} = 12\sqrt{39} \cdot 3\sqrt{\fr{7}{65}} - 54\sqrt{2} \cdot \sqrt{\fr{14}{15}} = 0.
\eq
This linear combination of monomial states thus corresponds to the degree-$2$ primary operator $\Ocal_{(2)}$.

The final function {\tt orthoPrimaries[n,deg]} takes the rescaled vectors of a given particle number and degree generated by {\tt rescalePrimarySet} and orthonormalizes them with respect to the Gram matrix generated by {\tt monoGram}. In our two-particle, degree-$2$ example, there's only one state so there's no need to orthogonalize, but we can still obtain the properly normalized state by evaluating:
\begin{mmaCell}[moredefined={orthoPrimaries}]{Input}
  orthoPrimaries[2,2] // FullSimplify
\end{mmaCell}
where we've simplified the expression just to make the result more readable, obtaining the output:
\begin{mmaCell}{Output}
  \{\{\mmaSqrt{\mmaFrac{26}{5}},-3\mmaSqrt{\mmaFrac{3}{5}}\}\}
\end{mmaCell}
which we can confirm is properly normalized by evaluating
\bq
\bal
&\fr{26}{5} G_{\p^3\phi\p\phi,\p^3\phi\p\phi} + \fr{27}{5} G_{(\p^2\phi)^2,(\p^2\phi)^2} - 2 \cdot \fr{3 \sqrt{78}}{5} G_{\p^3\phi\p\phi,(\p^2\phi)^2} \\
& \hspace{1in} = \fr{26}{5} + \fr{27}{5} - \fr{6 \sqrt{78}}{5} \cdot 4\sqrt{\fr{2}{39}} = 1.
\eal
\eq

\renewcommand{\arraystretch}{1.25}
\begin{table}[t!]
\begin{center}
\footnotesize
\begin{tabular}{| @{}c@{} | @{}c@{} | @{}c@{} | @{}c@{} | @{}c@{} | @{}c@{} |}
\hline & & & & & \\[-9pt]
\hspace{0.1pt} $n$ \hspace{0.1pt} & \hspace{0.1pt} $\De=1$ \hspace{0.1pt} & \hspace{0.1pt} $\De=2$ \hspace{0.1pt} & \hspace{0.1pt} $\De=3$ \hspace{0.1pt} & $\De=4$ & $\De=5$ \\
\hline & & & & & \\[-9pt]
1 & $|\p\phi,p\>$ & & & & \\
\hline & & & & & \\[-14pt]
2 & & $|(\p\phi)^2,p\>$ & & $\sqrt{\tfrac{26}{5}} |\p^3\phi\p\phi,p\> - \sqrt{\tfrac{27}{5}} |(\p^2\phi)^2,p\>$ & \\
\hline & & & & & \\[-9pt]
3 & & & $|(\p\phi)^3,p\>$ & & $\fr{8}{\sqrt{7}} |\p^3\phi(\p\phi)^2,p\> - 3 |(\p^2\phi)^2\p\phi,p\>$ \\
\hline & & & & & \\[-9pt]
4 & & & & $|(\p\phi)^4,p\>$ & \\
\hline & & & & & \\[-9pt]
5 & & & & & $|(\p\phi)^5,p\>$ \\
\hline
\end{tabular}
\normalsize
\renewcommand{\arraystretch}{1.0}
\end{center}
\caption{Orthonormal basis of primary operators for a single scalar field up to $\Dmax=5$, written in terms of the monomial states defined in~\eqref{eq:MonoStateDef}.}
\label{table:LowDmaxBasisScalar}
\end{table}

We now have general code which can construct the complete, orthonormal basis of primary operators built from $\p\phi$ at any particle number $n$ and scaling dimension $\De$. As a simple exercise, we encourage the reader to either use the provided code or write their own to construct the basis up to $\Dmax=5$ and compare with the results in table~\ref{table:LowDmaxBasisScalar}.

%%%%%%%%%%%%%%%%%%%%%%%%%%%%%%%%%%%%%%%%%%%%%%%%%%%%%%%%%%%%%%%%%%%%%%%%%%%%%

\subsection{Scalar Mass Term}
\label{sec:ScalarMassTerm}

Now that we have a basis of primary operators built from the scalar field $\phi$, we can start constructing the Hamiltonian for various relevant deformations of free field theory. The simplest deformation we can consider is a mass term,
\be
\de \Lcal = -\half m^2 \phi^2,
\ee
which gives rise to the Hamiltonian contribution
\be
\de P^{(\phi^2)}_+ = \fr{m^2}{2} \int dx \, \phi^2(x).
\ee
To construct the matrix elements for the mass term, we need to first evaluate the three-point functions
\benn
\<\Ocal(x) \phi^2(y) \Ocal'(z)\>,
\eenn
for the operators in our basis. We then need to Fourier transform these correlators to momentum space to obtain the matrix elements
\bq
\bal
&\<\Ocal,p|2P_-\de P_+|\Ocal',p'\> \equiv 2p(2\pi)\de(p-p') \, \Mcal^{(\phi^2)}_{\Ocal\Ocal'} \\ 
& \qquad = 2p (2\pi)\de(p-p') \cdot \fr{m^2}{2} \cdot \fr{1}{N^*_\Ocal N_{\Ocal'}} \int dx \, dz \, e^{ip(x-z)} \<\Ocal(x) \phi^2(0) \Ocal'(z)\>,
\eal
\eq
which can be evaluated with the general integral given in eq.~\eqref{eq:FTFormulas}.

\begin{figure}[t!]
\begin{center}
\includegraphics[width=0.9\textwidth]{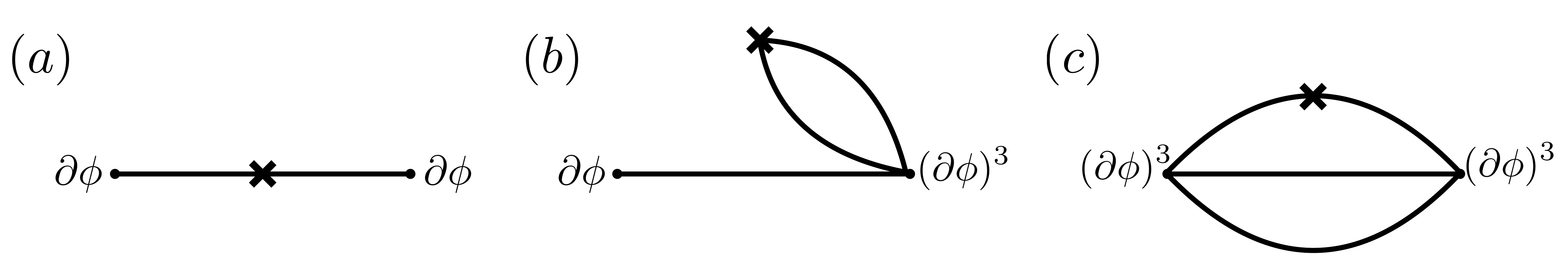}
\caption{Mass term Hamiltonian matrix elements involving $\p\phi$ and $(\p\phi)^3$. The matrix element corresponding to the middle diagram, which involves the creation of particles from the vacuum, vanishes in lightcone quantization.
\label{fig:MassTerm}}
\end{center}
\end{figure}

We briefly saw this procedure in action in section~\ref{sec:WickContract}, where we computed the mass term matrix elements
\be
\CM_{\p\phi,\p\phi}^{(\phi^2)} = m^2, \qquad \CM_{(\p\phi)^3,(\p\phi)^3}^{(\phi^2)} = 15m^2.
\label{eq:MassTermReminder}
\ee
However, what about the off-diagonal matrix element mixing $\p\phi$ and $(\p\phi)^3$ (shown schematically in Fig.~\ref{fig:MassTerm})? We can easily compute the associated three-point functions via Wick contraction, obtaining
\bq
\<\p\phi(x) \, \phi^2(y) \, (\p\phi)^3(z)\> \doteq \fr{12}{(4\pi)^3 (y-z)^2 (x-z)^2}.
\eq

However, if we try to compute the Fourier transform of this correlator, we find that it is \emph{zero},
\be
\Mcal^{(\phi^2)}_{\p\phi,(\p\phi)^3} = \fr{m^2}{2 N^*_{\p\phi} N_{(\p\phi)^3}} \int dx \, dz \, e^{ip(x - z)} \<\p\phi(x) \, \phi^2(0) \, (\p\phi)^3(z)\> = 0.
\ee
Looking carefully at the general integral in eq.~\eqref{eq:FTFormulas}, we see that this occurs because the correlator has $A=0$ (i.e.~there is no factor of $x-y$), such that the gamma function $\G(A)$ in the denominator of~\eqref{eq:FTFormulas} is singular and the expression vanishes.

This behavior is quite general. Any three-point function where one of the external operators does \emph{not} contract with the relevant deformation will either have $A=0$ or $B=0$, and the resulting matrix element will vanish. These matrix elements all involve the creation of particles from the vacuum, as we can see for this example in Fig.~\ref{fig:MassTerm}, and vanish in lightcone quantization.

This restriction leads to a dramatic simplification in the resulting Hamiltonian. For example, the mass term is \emph{diagonal} with respect to particle number. To find the mass eigenstates, we can therefore consider each particle number sector separately. Amazingly, to find the one-particle mass eigenvalue we only need to consider a single matrix element! Indeed, in eq.~\eqref{eq:MassTermReminder} we see that the resulting one-particle matrix element is exactly $m^2$.

Note that, with only one state, we already obtain a reasonable estimate of the lowest three-particle invariant mass, as well. If we added more three-particle states to the basis, we would find that the lowest eigenvalue quickly approaches the correct value of $9m^2$.

Following this simple example, let's now try to develop a general algorithm for computing the mass term matrix elements. In the previous two subsections, we wrote Mathematica code that generated all basis states of a given particle number $n$ and scaling dimension $\De$. These states are expressed as a sum over individual monomials,
\be
|\Ocal,p\> = \sum_{\bk} \Chat_{\bk}^\Ocal |\p^{\bk}\phi,p\>.
\ee
The matrix elements for primary operators can therefore be written as a sum over monomial matrix elements,
\be
\Mcal^{(\phi^2)}_{\Ocal\Ocal'} = \sum_{\bk,\bk'} \Chat_{\bk}^{\Ocal*} \Chat_{\bk'}^{\Ocal'} \Mcal^{(\phi^2)}_{\bk\bk'}.
\ee

First, we need to compute the three-point functions for individual monomials, which we can do via Wick contraction\footnote{Recall that $\bk/k_i$ indicates the vector created by removing the entry $k_i$ from $\bk$.}
\be
\<\p^{\bk}\phi(x) \phi^2(y) \p^{\bk'}\phi(z)\> = \sum_{\substack{k_i \in \bk \\ k'_j \in \bk'}} \<\p^{k_i} \phi(x) \phi^2(y) \p^{k'_j} \phi(z)\> \<\p^{\bk/k_i}\phi(x) \, \p^{\bk'/k'_j} \phi(z)\>,
\ee
We therefore have one ``interacting'' particle, which contracts with $\phi^2$, and $n-1$ ``spectating'' particles.

The spectating piece of this correlator was calculated in the previous subsection in eq.~\eqref{eq:2ptWick}, so we only need to compute the interacting piece,
\be
\<\p^{k_i}\phi(x) \phi^2(y) \p^{k'_j}\phi(z)\> \doteq 2 \cdot \fr{\G(k_i)\G(k'_j)}{(4\pi)^2 (x-y)^{k_i} (y-z)^{k'_j}}.
\ee
We therefore obtain the full correlator
\bq
\bal
&\<\p^{\bk}\phi(x) \phi^2(y) \p^{\bk'}\phi(z)\> \\
& \qquad \quad \doteq \fr{2}{(4\pi)^{n+1} (x-z)^{\De+\De'}} \sum_{\substack{k_i \in \bk \\ k'_j \in \bk'}} \G(k_i)\G(k'_j) A_{\bk/k_i,\bk'/k'_j} \fr{(x-z)^{k_i+k'_j}}{(x-y)^{k_i} (y-z)^{k'_j}}.
\eal
\eq
Using eq.~\eqref{eq:FTFormulas}, we can then Fourier transform this expression, and normalize the result by the coefficients $N_{\bk}$ from~\eqref{eq:MonoNorm} to obtain the monomial matrix elements
\be
\boxed{\Mcal^{(\phi^2)}_{\kvec\kvec'} = \fr{m^2}{\G(|\bk|+|\bk'|-1)} \sqrt{ \fr{\G(2|\bk|)\G(2|\bk'|)}{A_{\bk\bk} \, A_{\bk'\bk'}} } \sum_{\substack{k_i \in \kvec \\ k'_j \in \kvec'}} \G(k_i+k'_j-1) A_{\kvec/k_i,\kvec'/k'_j}.}
\label{eq:MonomialMassTerm}
\ee

Given a set of primary operators, we can then compute the resulting Hamiltonian matrix elements from linear combinations of eq.~\eqref{eq:MonomialMassTerm}. Table~\ref{table:SimpleMassTermCode} shows Mathematica code which computes the $\phi^2$ matrix elements between all $n$-particle primaries of a given incoming scaling dimension $\De_1$ and outgoing dimension $\De_2$ (with the overall factor of $m^2$ removed).

\begin{table}[t!]
\noindent\rule[0.5ex]{\linewidth}{1pt}
\verb|(*| Individual monomial matrix element (\ref{eq:MonomialMassTerm}) \verb|*)|
\begin{code}
monoMass\verb|[k_,kp_]|:=\=Sqrt\=[Gamma[2Total[k]] Gamma[2Total[kp]] \\
 \>\> /(A[k,k] A[kp,kp])] / Gamma[Total[k+kp]-1] \\
 \> *Sum[Gamma[k[[i]]+kp[[j]]-1] \\
 \>\> *A[Delete[k,i],Delete[kp,j]], \\
 \>\> \verb|{i,Length[k]}, {j,Length[kp]}];|
\end{code}
\verb|(*| Construct all mass term matrix elements for primary operators at a fixed particle number for incoming degree {\tt deg1} and outgoing degree {\tt deg2} \verb|*)|
\begin{code}
primary\=MassMatrix\verb|[n_,deg1_,deg2_]|:=If[ \\
 \> numStates[n,deg1]==0 || numStates[n,deg2]==0, \verb|{},| \\
 \> Dot\=[orthoPrimaries[n,deg1], \\
 \>\> Table\=[monoMass[k,kp], \verb|{k,monomialsBoson[n,deg1]},| \\
 \>\>\> \verb|{kp,monomialsBoson[n,deg2]}],| \\
 \>\> Transpose[orthoPrimaries[n,deg2]]]];
\end{code}
\noindent\rule[0.5ex]{\linewidth}{1pt}
\caption{Sample Mathematica code for constructing the $\phi^2$ matrix elements for all $n$-particle primary operators with incoming degree $\De_1-n$ and outgoing degree $\De_2-n$.
\label{table:SimpleMassTermCode}}
\end{table}

The main function is {\tt primaryMassMatrix[n,deg1,deg2]}, which takes the orthonormalized basis states generated by {\tt orthoPrimaries} for two different degrees, {\tt deg1} and {\tt deg2}, and computes all matrix elements between the two sets of primaries. For example, if we want to compute the two-particle matrix element $\Mcal^{(\phi^2)}_{(\p\phi)^2,(\p\phi)^2}$, we can enter:
\begin{mmaCell}[index=1,moredefined={primaryMassMatrix}]{Input}
  primaryMassMatrix[2,0,0]
\end{mmaCell}
The resulting output is a matrix, where the rows correspond to the states with {\tt deg1} and the columns correspond to the states with {\tt deg2}. In this example, there is only one state, so we obtain the single matrix element:
\begin{mmaCell}{Output}
  \{\{6\}\}
\end{mmaCell}
which indicates that
\be
\Mcal^{(\phi^2)}_{(\p\phi)^2,(\p\phi)^2} = 6m^2,
\ee
as we computed in subsection~\ref{sec:WickContract}.

We can use this code to compute the full set of $\phi^2$ matrix elements for the $\Dmax=5$ basis from table~\ref{table:LowDmaxBasisScalar}, with the results shown in table~\ref{table:LowDmaxMassMatrixScalar}. At such low $\Dmax$, we can uniquely identify each primary operator by its particle number $n$ and scaling dimension $\De$, so the rows and columns of this table are labeled by $(n,\De)$ of the corresponding operator. For higher $\Dmax$, there are degeneracies, such that we would need to introduce additional labels to distinguish between operators.

\begin{table}[t!]
\begin{center}
\begin{tabular}{| @{}c@{} | @{}c@{} | @{}c@{} | @{}c@{} | @{}c@{} | @{}c@{} | @{}c@{} | @{}c@{} |}
\hline & & & & & & & \\[-9pt]
\hspace{0.1pt} $(n,\De)$ \hspace{0.1pt} & \hspace{0.1pt} $(1,1)$ \hspace{0.1pt} & \hspace{0.1pt} $(2,2)$ \hspace{0.1pt} & \hspace{0.1pt} $(2,4)$ \hspace{0.1pt} & \hspace{0.1pt} $(3,3)$ \hspace{0.1pt} & \hspace{0.1pt} $(3,5)$ \hspace{0.1pt} & \hspace{0.1pt} $(4,4)$ \hspace{0.1pt} & \hspace{0.1pt} $(5,5)$ \hspace{0.1pt} \\
\hline & & & & & & & \\[-9pt]
$(1,1)$ & $1$ & & & & & & \\
\hline & & & & & & & \\[-9pt]
$(2,2)$ & & $6$ & $\sqrt{14}$ & & & & \\
\hline & & & & & & & \\[-9pt]
$(2,4)$ & & $\sqrt{14}$ & $14$ & & & & \\
\hline & & & & & & & \\[-9pt]
$(3,3)$ & & & & $15$ & $4\sqrt{3}$ & & \\
\hline & & & & & & & \\[-9pt]
$(3,5)$ & & & & $4\sqrt{3}$ & $27$ & & \\
\hline & & & & & & & \\[-9pt]
$(4,4)$ & & & & & & $28$ & \\
\hline & & & & & & & \\[-9pt]
$(5,5)$ & & & & & & & $45$ \\
\hline
\end{tabular}
\end{center}
\caption{Matrix elements of $\fr{m^2}{2} \phi^2$ for the $\Dmax=5$ basis of primary operators shown in table~\ref{table:LowDmaxBasisScalar}, with the overall factor of $m^2$ removed. Each row and column is identified by the particle number and scaling dimension $(n,\De)$ of the corresponding primary operator. Note that in general $n$ and $\De$ are not sufficient to uniquely specify each primary operator, but at this $\Dmax$ there are no degeneracies.}
\label{table:LowDmaxMassMatrixScalar}
\end{table}

%%%%%%%%%%%%%%%%%%%%%%%%%%%%%%%%%%%%%%%%%%%%%%%%%%%%%%%%%%%%%%%%%%%%%%%%%%%%%

\subsection{Adding Interactions}

In addition to the mass term, we can consider self-interactions for the scalar field $\phi$. For simplicity, we will only focus on the case of a quartic interaction,
\be
\de \Lcal = -\fr{1}{4!} \lambda \, \phi^4,
\ee
but this procedure for constructing matrix elements can easily be generalized to other $\phi^n$ interactions or higher-dimensional operators built from derivatives acting on $\phi$.

\begin{figure}[t!]
\begin{center}
\includegraphics[width=0.7\textwidth]{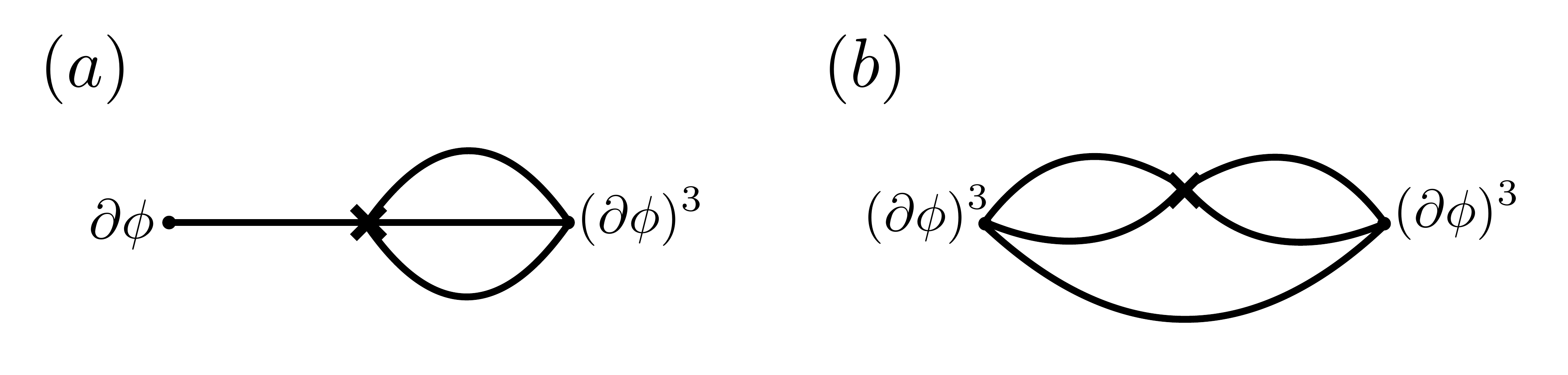}
\caption{Quartic interaction Hamiltonian matrix elements involving $\p\phi$ and $(\p\phi)^3$. The $1 \ra 1$ matrix element has been removed by normal-ordering the $\phi^4$ interaction, and the $1 \ra 5$ matrix element, which involves the creation of particles from the vacuum, vanishes in lightcone quantization.
\label{fig:QuarticTerm}}
\end{center}
\end{figure}

For this interaction, there are naively three classes of matrix elements: $n\ra n$, $n\ra n+2$, and $n\ra n+4$. However, the last type (where particle number changes by $4$) involves the creation of particles from the vacuum, which means the resulting matrix elements vanish in lightcone quantization. We therefore only need to construct two types of matrix elements, which are shown schematically for the external states $\p\phi$ and $(\p\phi)^3$ in Fig.~\ref{fig:QuarticTerm}.

The procedure for constructing these matrix elements is the same as for the mass term. First, we need to evaluate the $\phi^4$ correlation functions for monomial operators, which factorize into an interacting piece and a spectating piece. For example, the $n\ra n$ correlators take the general form
\be
\<\p^{\bk}\phi(x) \phi^4(y) \p^{\bk'}\phi(z)\> = \sum_{\substack{k_{i,j} \in \bk \\ k'_{r,s} \in \bk'}} \<\p^{k_{i,j}} \phi(x) \phi^4(y) \p^{k'_{r,s}} \phi(z)\> \<\p^{\bk/k_{i,j}}\phi(x) \, \p^{\bk'/k'_{r,s}} \phi(z)\>.
\ee
The interacting piece of this correlator we can easily compute to obtain
\be
\<\p^{k_{i,j}}\phi(x) \, \phi^4(y) \, \p^{k'_{r,s}}\phi(z)\> \doteq 4! \cdot \fr{\G(k_i)\G(k_j)\G(k'_r)\G(k'_s)}{(4\pi)^4 (x-y)^{k_i+k_j} (y-z)^{k'_r+k'_s}}.
\ee
Similarly, the interacting part of $n \ra n+2$ correlators takes the form
\be
\<\p^{k_i}\phi(x) \, \phi^4(y) \, \p^{k'_{r,s,t}}\phi(z)\> \doteq 4! \cdot \fr{\G(k_i)\G(k'_r)\G(k'_s)\G(k'_t)}{(4\pi)^4 (x-y)^{k_i} (y-z)^{k'_r+k'_s+k'_t}}.
\ee

We can then combine these correlators with the spectating piece, Fourier transform to momentum space with eq.~\eqref{eq:FTFormulas}, and normalize by the coefficients $N_{\bk}$ in~\eqref{eq:MonoNorm} to obtain the resulting monomial matrix elements. For the $n\ra n$ process, the result is
\bq
\boxed{
\bal
\Mcal^{(\phi^4)_{n\ra n}}_{\kvec\kvec'} &= \fr{\lambda}{4\pi \G(|\bk|+|\bk'|-1)} \sqrt{ \fr{\G(2|\bk|)\G(2|\bk'|)}{A_{\bk\bk} \, A_{\bk'\bk'}} } \\
& \qquad \times \sum_{\substack{k_{i,j} \in \kvec \\ k'_{r,s} \in \kvec'}} \fr{\G(k_i)\G(k_j)\G(k'_r)\G(k'_s)\G(k_i+k_j+k'_r+k'_s-1)}{\G(k_i+k_j)\G(k'_r+k'_s)} A_{\kvec/k_{i,j},\kvec'/k'_{r,s}},
\eal
}
\label{eq:MonomialNtoNTerm}
\eq
while for $n\ra n+2$, we obtain
\bq
\boxed{
\bal
\Mcal^{(\phi^4)_{n\ra n+2}}_{\kvec\kvec'} &= \fr{\lambda}{4\pi \G(|\bk|+|\bk'|-1)} \sqrt{ \fr{\G(2|\bk|)\G(2|\bk'|)}{A_{\bk\bk} \, A_{\bk'\bk'}} } \\
& \qquad \times \sum_{\substack{k_i \in \kvec \\ k'_{r,s,t} \in \kvec'}} \fr{\G(k'_r)\G(k'_s)\G(k'_t)\G(k_i+k'_r+k'_s+k'_t-1)}{\G(k'_r+k'_s+k'_t)} A_{\kvec/k_i,\kvec'/k'_{r,s,t}}.
\eal
}
\label{eq:MonomialNtoN2Term}
\eq

\afterpage{\clearpage}
\begin{table}[p]
\noindent\rule[0.5ex]{\linewidth}{1pt}
\verb|(*| Individual monomial matrix element for $n\ra n$ \eqref{eq:MonomialNtoNTerm} and $n \ra n+2$ \eqref{eq:MonomialNtoN2Term} \verb|*)|
\begin{code}
monoNtoN\verb|[k_,kp_]|:=\=Sqrt\=[Gamma[2Total[k]] Gamma[2Total[kp]] \\
 \>\> /(A[k,k] A[kp,kp])] / Gamma[Total[k+kp]-1] \\
 \> *Sum[Gamma[k[[i]]] Gamma[k[[j]]] \\
 \>\> *Gamma[kp[[r]]] Gamma[kp[[s]]] \\
 \>\> *Gamma[k[[i]]+k[[j]]+kp[[r]]+kp[[s]]-1] \\
 \>\> /(Gamma[k[[i]]+k[[j]]] Gamma[kp[[r]]+kp[[s]]]) \\
 \>\> *A[Delete[k,\verb|{{i},{j}}|],Delete[kp,\verb|{{r},{s}}|]], \\
 \>\> \verb|{i,Length[k]},{j,i-1},{r,Length[kp]},{s,r-1}];| \\ \\
monoNtoNplus2\verb|[k_,kp_]|\=:=S\=qrt[Gamma[2Total[k]] Gamma[2Total[kp]] \\
 \>\> /(A[k,k] A[kp,kp])] / Gamma[Total[k]+Total[kp]-1] \\
 \> *Sum[Gamma[kp[[r]]] Gamma[kp[[s]]] Gamma[kp[[t]]] \\
 \>\> *Gamma[k[[i]]+kp[[r]]+kp[[s]]+kp[[t]]-1] \\
 \>\> /Gamma[kp[[r]]+kp[[s]]+kp[[t]]] \\
 \>\> *A[Delete[k,i],Delete[kp,\verb|{{r},{s},{t}}|]], \\
 \>\> \verb|{i,Length[k]},{r,Length[kp]},{s,r-1},{t,s-1}];|
\end{code}
\verb|(*| Construct all $n \ra n$ matrix elements for primary operators at a fixed particle number for incoming degree {\tt deg1} and outgoing degree {\tt deg2} \verb|*)|
\begin{code}
primary\=NtoNMatrix\verb|[n_,deg1_,deg2_]|:=If[ \\
 \> numStates[n,deg1]==0 || numStates[n,deg2]==0, \verb|{},| \\
 \> Dot\=[orthoPrimaries[n,deg1], \\
 \>\> Table\=[monoMass[k,kp], \verb|{k,monomialsBoson[n,deg1]},| \\
 \>\>\> \verb|{kp,monomialsBoson[n,deg2]}],| \\
 \>\> Transpose[orthoPrimaries[n,deg2]]]];
\end{code}
\verb|(*| Construct all $n \ra n+2$ matrix elements for primary operators at a fixed incoming particle number {\tt n} and degree {\tt deg1} and outgoing particle number {\tt n+2} and degree {\tt deg2} \verb|*)|
\begin{code}
primary\=NtoNplus2Matrix\verb|[n_,deg1_,deg2_]|:=If[ \\
 \> numStates[n,deg1]==0 || numStates[n+2,deg2]==0, \verb|{},| \\
 \> Dot\=[orthoPrimaries[n,deg1], \\
 \>\> Table\=[monoNtoNplus2[k,kp], \verb|{k,monomialsBoson[n,deg1]},| \\
 \>\>\> \verb|{kp,monomialsBoson[n+2,deg2]}],| \\
 \>\> Transpose[orthoPrimaries[n+2,deg2]]]];
\end{code}
\noindent\rule[0.5ex]{\linewidth}{1pt}
\caption{Sample Mathematica code for constructing the $\phi^4$ matrix elements, both $n \ra n$ and $n \ra n+2$, for all primary operators with incoming scaling dimension $\De_1$ and outgoing dimension $\De_2$.
\label{table:SimpleInteractionTermCode}}
\end{table}

We can now write Mathematica code to use eqs.~\eqref{eq:MonomialNtoNTerm} and \eqref{eq:MonomialNtoN2Term} to construct the $\phi^4$ matrix elements for primary operators, shown in table~\ref{table:SimpleInteractionTermCode}. The structure of this code is very similar to that of the mass term in table~\ref{table:SimpleMassTermCode}.

The first important function is {\tt primaryNtoNMatrix[n,deg1,deg2]}, which computes the $n\ra n$ matrix elements between all primaries of incoming degree {\tt deg1} and those of outgoing degree {\tt deg2}, with the overall factor of $\fr{\lambda}{4\pi}$ removed. For example, we can compute the $2 \ra 2$ matrix element $\Mcal^{(\phi^4)}_{(\p\phi)^2,(\p\phi)^2}$ by entering:
\begin{mmaCell}[index=1,moredefined={primaryNtoNMatrix}]{Input}
  primaryNtoNMatrix[2,0,0]
\end{mmaCell}
Just like for the mass term, the output is a matrix with rows corresponding to the states with {\tt deg1} and columns corresponding to the states with {\tt deg2}. In this example, we obtain the single matrix element:
\begin{mmaCell}{Output}
  \{\{3\}\}
\end{mmaCell}
which indicates that
\be
\Mcal^{(\phi^4)}_{(\p\phi)^2,(\p\phi)^2} = \fr{3\lambda}{4\pi}.
\ee

The second main function is {\tt primaryNtoNplus2Matrix[n,deg1,deg2]}, which computes the $n \ra n+2$ matrix elements between all $n$-particle primaries of degree {\tt deg1} and all $n+2$-particle primaries of degree {\tt deg2}. For example, we can compute the $1 \ra 3$ matrix element $\Mcal^{(\phi^4)}_{\p\phi,(\p\phi)^3}$ with:
\begin{mmaCell}[moredefined={primaryNtoNplus2Matrix}]{Input}
  primaryNtoNplus2Matrix[1,0,0]
\end{mmaCell}
The output is again a matrix, though now the rows correspond to $n$-particle states with {\tt deg1} and the columns correspond to $n+2$-particle states with {\tt deg2}. For this example, we obtain:
\begin{mmaCell}{Output}
  \{\{\mmaSqrt{5}\}\}
\end{mmaCell}
which agrees with the result computed previously via the Fock space method in eq.~\eqref{eq:Phi4Examples}
\be
\Mcal^{(\phi^4)}_{(\p\phi),(\p\phi)^3} = \fr{\sqrt{5}\lambda}{4\pi}.
\ee

\begin{table}[t!]
\begin{center}
\begin{tabular}{| @{}c@{} | @{}c@{} | @{}c@{} | @{}c@{} | @{}c@{} | @{}c@{} | @{}c@{} | @{}c@{} |}
\hline & & & & & & & \\[-9pt]
\hspace{0.1pt} $(n,\De)$ \hspace{0.1pt} & \hspace{0.1pt} $(1,1)$ \hspace{0.1pt} & \hspace{0.1pt} $(2,2)$ \hspace{0.1pt} & \hspace{0.1pt} $(2,4)$ \hspace{0.1pt} & \hspace{0.1pt} $(3,3)$ \hspace{0.1pt} & \hspace{0.1pt} $(3,5)$ \hspace{0.1pt} & \hspace{1.2pt} $(4,4)$ \hspace{1.2pt} & \hspace{1.2pt} $(5,5)$ \hspace{1.2pt} \\
\hline & & & & & & & \\[-9pt]
$(1,1)$ & & & & $\sqrt{5}$ & $\tfrac{\sqrt{15}}{2}$ & & \\
\hline & & & & & & & \\[-13pt]
$(2,2)$ & & $3$ & $\sqrt{\tfrac{7}{2}}$ & & & $\sqrt{70}$ & \\
\hline & & & & & & & \\[-13pt]
$(2,4)$ & & $\sqrt{\tfrac{7}{2}}$ & $\tfrac{7}{6}$ & & & & \\
\hline & & & & & & & \\[-9pt]
$(3,3)$ & $\sqrt{5}$ & & & $15$ & $4\sqrt{3}$ & & $2\sqrt{105}$ \\
\hline & & & & & & & \\[-9pt]
$(3,5)$ & $\tfrac{\sqrt{15}}{2}$ & & & $4\sqrt{3}$ & $\frac{33}{2}$ & & \\
\hline & & & & & & & \\[-9pt]
$(4,4)$ & & $\sqrt{70}$ & & & & $42$ & \\
\hline & & & & & & & \\[-9pt]
$(5,5)$ & & & & $2\sqrt{105}$ & & & $90$ \\
\hline
\end{tabular}
\end{center}
\caption{Matrix elements of $\fr{\lambda}{4!} \phi^4$ for the $\Dmax=5$ basis of primary operators shown in table~\ref{table:LowDmaxBasisScalar}, with the overall factor of $\fr{\lambda}{4\pi}$ removed. Each row and column is identified by the particle number and scaling dimension $(n,\De)$ of the corresponding primary operator. Note that in general $n$ and $\De$ are not sufficient to uniquely specify each primary operator, but at this $\Dmax$ there are no degeneracies.}
\label{table:LowDmaxInteractionMatrixScalar}
\end{table}

We encourage the reader to either use this code or write their own to compute the $\phi^4$ matrix elements for all primary operators in the $\Dmax=5$ basis from table~\ref{table:LowDmaxBasisScalar}. The resulting matrix elements are shown in table~\ref{table:LowDmaxInteractionMatrixScalar}, with the rows and columns identified by the particle number and scaling dimension $(n,\De)$ of the corresponding primary operator.

Looking at table~\ref{table:LowDmaxInteractionMatrixScalar}, we see that there is one matrix element, $(2,4) \ra (4,4)$, which is naively allowed but in fact vanishes,
\be
\sqrt{\fr{26}{5}} \Mcal^{(\phi^4)}_{\p^3\phi\p\phi,(\p\phi)^4} - \sqrt{\fr{27}{5}} \Mcal^{(\phi^4)}_{(\p^2\phi)^2,(\p\phi)^4} = 0.
\ee
Note that the corresponding position space three-point function is \emph{not} equal to zero. The expression only vanishes when we Fourier transform to momentum space to obtain the resulting Hamiltonian matrix element.

This structure is actually quite general, such that \emph{all} $n \ra n+2$ matrix elements vanish if the scaling dimension of the $n$-particle primary operator is greater than or equal to the dimension of the $n+2$-particle one,
\be
\Mcal^{(\phi^4)}_{\Ocal_n,\Ocal_{n+2}} = 0 \quad (\De_n \geq \De_{n+2}).
\ee

Because the position space correlators do not vanish, this behavior is not manifest in our current Wick space method, and can only be seen for individual examples after taking the precise linear combinations of monomial matrix elements corresponding to primary operators. However, in section~\ref{sec:RadialScalars} we will introduce a new method for evaluating matrix elements, which will make this selection rule more manifest.

%%%%%%%%%%%%%%%%%%%%%%%%%%%%%%%%%%%%%%%%%%%%%%%%%%%%%%%%%%%%%%%%%%%%%%%%%%%%%

\subsection{Spectrum}
\label{sec:ScalarSpectrum}

Having built up the machinery to do LCT computations for $\phi^4$ theory, let's use it to do some physics! Using the Mathematica code provided in the previous subsections, one can construct the full Hamiltonian for $\phi^4$ theory,
\be
P_+ = \int dx \left( \fr{m^2}{2} \phi^2(x) + \fr{\lambda}{4!} \phi^4(x) \right),
\ee
in the basis of primary operators up to some low value of $\Dmax$ (for example, in tables~\ref{table:LowDmaxMassMatrixScalar} and \ref{table:LowDmaxInteractionMatrixScalar} we have provided the matrix elements up to $\Dmax=5$). This Hamiltonian can be diagonalized numerically (or analytically for low enough $\Dmax$) to obtain an approximation to the mass eigenstates of the full theory. In this subsection, we will focus on the mass eigenvalues, using them to first study the phase structure of 2d $\phi^4$ theory, then discuss the emergent UV and IR scales that arise in conformal truncation.

\subsubsection{Phase Transition in 2d $\phi^4$ Theory}
\label{sec:Phi4PhaseTransition}

For most quantitative questions, we need a sufficiently large basis that the numeric results are at least starting to converge.  However, we can reach some interesting nonperturbative conclusions even with a very small basis by using the fact that Hamiltonian truncation is a variational method, and therefore the smallest energy eigenvalue of the truncated Hamiltonian is an upper bound on the smallest eigenvalue of the full Hamiltonian. 
In LC quantization, this fact is especially powerful, because the vacuum is in its own selection sector $p_-=0$ and the vacuum energy is not renormalized.  Consequently, the smallest eigenvalue of the truncated LC Hamiltonian is an upper bound on the \emph{energy gap} between the first excited state and the vacuum.

\begin{figure}[t!]
\begin{center}
\includegraphics[width=0.95\textwidth]{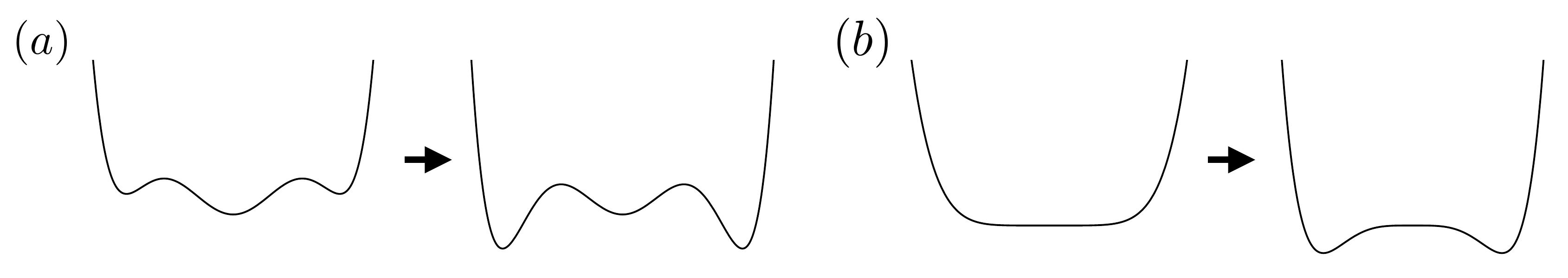}
\caption{Example of a potential $V(\phi)$ with $(a)$ first-order or $(b)$ second-order phase transition as couplings vary in $\phi^6$ theory. In $(a)$, the lowest eigenvalue of the LC Hamiltonian jumps discontinuously from a positive to negative value, indicating the presence of a new global minimum, while in $(b)$, the lowest eigenvalue smoothly crosses zero.}
\label{fig:PhaseTransitions}
\end{center}
\end{figure}

This statement requires an important qualification.  Although the vacuum is not renormalized in LC quantization, it is possible in the case of a phase transition for the location of the true global minimum to change as the parameters of the theory are varied. Consider for instance $\phi^6$ theory, where at zero coupling the theory is in a phase with $\<\phi\>=0$ and the $\mathbb{Z}_2$ symmetry $\phi \rightarrow -\phi$ is unbroken.  One can dial the $\phi^4$ and $\phi^6$ couplings to spontaneously break the $\mathbb{Z}_2$ symmetry with a first-order or second-order phase transition, as shown in Fig.~\ref{fig:PhaseTransitions}. The smallest eigenvalue of the LC Hamiltonian is the gap between the first $p_- > 0$ state and the $\< \phi\>=0$ vacuum; this gap will become \emph{negative} if the $\<\phi\> = 0$ vacuum is no longer the true ground state. If the phase transition is first-order, then the gap never closes, but instead jumps discontinuously from a positive to a negative value. If the phase transition is second-order, however, the gap will smoothly cross zero.

\begin{figure}[t!]
\begin{center}
\includegraphics[width=0.6\textwidth]{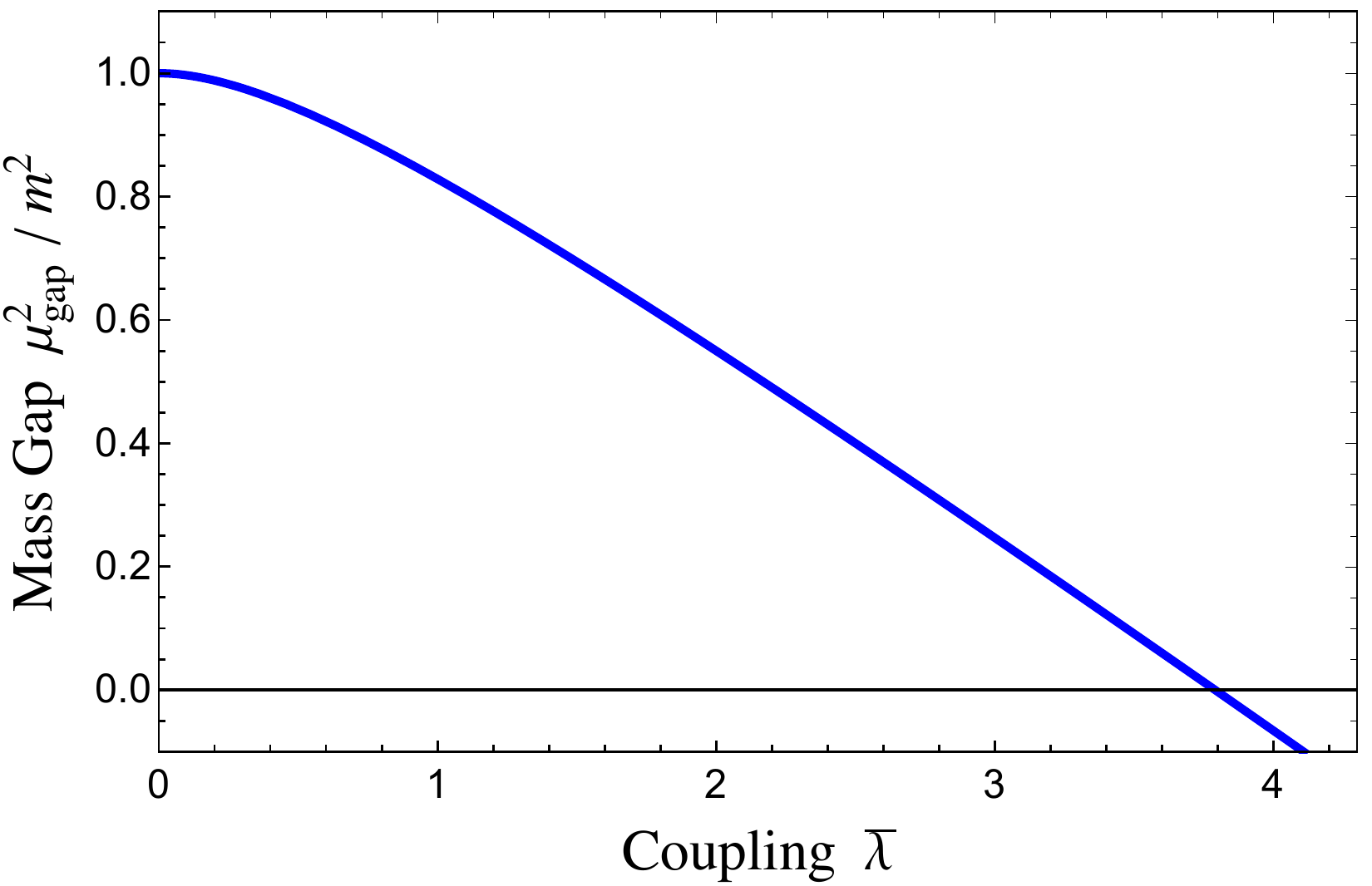}
\caption{Lowest eigenvalue of the truncated matrix $\Mcal$ for $\phi^4$ theory, with only two states in the basis: $\p\phi$ and $(\p\phi)^3$. The eigenvalue crosses zero, proving nonperturbatively that $\phi^4$ theory must undergo a (first- or second-order) phase transition at $\bar{\lambda}_* < 3.8$.}
\label{fig:Phi4LowDmaxGap}
\end{center}
\end{figure}

Therefore, if we find that the smallest eigenvalue of the truncated Hamiltonian passes from positive to negative values as we dial the coupling, then since the true gap is bounded above by the truncated gap, we immediately know the smallest eigenvalue of the full Hamiltonian also must pass from positive (at weak coupling) to negative -- although at finite truncation we cannot say if it did so continuously or discontinuously.  That is to say, we immediately know the theory passes through a phase transition, but we do not know if it is first- or second-order.\footnote{We must compute other observables, such as the spectral density of $T^\mu_{\phantom{\mu}\mu}$, in order to fully determine whether the phase transition is first- or second-order (see section~\ref{sec:ApplicationPhi4}).} With this prelude, let us consider the following extremely simple LCT computation: take the Hamiltonian for $\phi^4$ theory with only two states, 
\be
\CO_1 = \p \phi, \qquad \CO_2 = (\p \phi)^3.
\ee
We have already computed all the matrix elements we need:
\be
\CM = m^2 \begin{pmatrix} 1 & 0 \\ 0 & 15 \end{pmatrix} + \bar{\lambda} \, m^2 \begin{pmatrix} 0 & \sqrt{5} \\ \sqrt{5} & 15 \end{pmatrix},
\ee
where $\bar{\lambda} \equiv \frac{\lambda}{4\pi m^2}$. The determinant of this matrix is 
\be
{\rm det}(\CM) =m^4 \ 5 (3 +3 \bar{\lambda}- \bar{\lambda}^2),
\ee
which clearly crosses from positive to negative as $\bar{\lambda}$ increases from zero. Equivalently, we can analytically solve for the lowest eigenvalue, which clearly becomes negative at large coupling, as shown in Fig.~\ref{fig:Phi4LowDmaxGap}. We have just proven nonperturbatively that 2d $\phi^4$ theory has a phase transition!

Moreover, the value at which this eigenvalue crosses zero ($\bar{\lambda} \approx 3.8$) places an upper bound on the critical coupling $\bar{\lambda}_*$ at which this phase transition must occur. We have therefore also proven that $\bar{\lambda}_* < 3.8$, using a basis of only two states.

\subsubsection{UV and IR Scales in Truncation}
\label{sec:TruncationScales}

To better understand the structure of the mass eigenvalues, let's now consider the mass deformation $m^2\phi^2$ in more detail. Because the mass term conserves particle number in LC quantization, we can focus on the two-particle sector, which will be simple enough that we can take the large truncation limit analytically.  One of the perhaps surprising features that we can see explicitly in this example is that the truncation parameter $\Delta_{\rm max}$ acts not only like a UV cutoff, but also like an IR cutoff.  Roughly, the reason is that with only a finite number of states, there are only a finite set of energy levels that we can cover, and deep enough in the UV \emph{or} the IR we eventually run out of states. 

Using the formulas for the Fock space wavefunctions of two-particle primary operators $[\p\phi \, \p\phi]_\ell$ from section~\ref{sec:Jacobi}, we can compute the mass term matrix elements for two-particle states in closed form:
\begin{equation}
\bal
\Mcal_{\ell \ell'}^{(\phi^2)} &= m^2 \int_0^1 dx \, \widehat{P}_\ell^{(1,1)}(1-2x)\widehat{P}_{\ell'}^{(1,1)}(1-2x) \\
&=2 m^2 \sqrt{\frac{(\min(\ell,\ell')+1)(\min(\ell,\ell')+2)}{(\max(\ell,\ell')+1)(\max(\ell,\ell')+2)} (2\ell+3)(2\ell'+3)},
\eal
\label{eq:TwoPartMassAnalytic}
\end{equation}
where $\ell \geq 0$ takes on even integer values, and we've introduced the normalized Jacobi polynomials
\be
\widehat{P}_{\ell}^{(\alpha,\beta)}(x) = \sqrt{ \fr{\G(\ell+1)\G(\ell+\alpha+\beta+1)\G(2\ell+\alpha+\beta+2)}{\G(\ell+\alpha+1)\G(\ell+\beta+1)\G(2\ell+\alpha+\beta+1)}} \, P_\ell^{(\alpha,\beta)}(x). \label{eq:NormalizedJacobis}
\ee

We can truncate this basis by restricting to $\ell \leq \ell_{\max} \equiv \Dmax-2$. The eigenvalues of the resulting truncated matrix are given by the roots of the characteristic polynomial
\be
{\rm det}\left(\frac{\Mcal^{(\phi^2)}}{m^2}- x\right) = (-x)^{\frac{\Delta_{\rm max}}{2} } P_{\frac{\Delta_{\rm max}}{2}}^{(0,-\frac{1}{2})}\Big(1-\frac{8}{x}\Big),
\label{eq:BosonMassCharPoly}
\ee
where we take $\Delta_{\rm max}$ to be an even integer. We leave it as an exercise to the reader to prove (\ref{eq:BosonMassCharPoly}).\footnote{A hint: note that the inverse of $\Mcal$ is a tridiagonal matrix.}
At large $\Delta_{\rm max}$, the eigenvalues simplify, as one can see by using the following asymptotic formula:
\be
P_k^{(0, -\frac{1}{2} )}(\cos \theta) \approx \frac{\sin \big[(k+\frac{1}{4})\theta + \frac{\pi}{4}  \big] }{\sqrt{ \pi k \sin \frac{\theta}{2} } } + {\cal O}(k^{-\frac{3}{2}}).
\ee
To facilitate the use of the above formula, define $1-\frac{8}{x} = \cos \theta$. At large $\Delta_{\rm max}$, the eigenvalues are then given by $x_j = 4 \csc^2 \frac{\theta_j}{2}$, where $\theta_j$ satisfies 
\be
\left(\frac{\Delta_{\rm max}}{2} + \frac{1}{4}  \right) \theta_j +  \frac{\pi}{4}  = \left( \frac{\Delta_{\rm max}}{2} - j\right) \pi,  \qquad \Big(j=0,1,\ldots,\tfr{\Dmax}{2}-1\Big).
\ee
Consequently, at large $\Delta_{\rm max}$ the eigenvalues of the truncated matrix are approximately
\be
x_j \approx 4 \sec^2 \left( \frac{(2j+1) \pi }{2 \Delta_{\rm max}+1} \right) .
\label{eq:freemassivespectrum}
\ee
The smallest eigenvalues correspond to small values of $j$.  At large $\Delta_{\rm max}$ but fixed $j$, the two-particle spectrum therefore becomes
\be
\mu^2_j = m^2 x_j \approx \left(4  + \frac{\pi^2(2j+1)^2}{\Delta_{\rm max}^2} \right)m^2.
\label{eq:phiphitruncmass}
\ee
This formula shows us two things. First, in the limit of infinite $\Delta_{\rm max}$, the two-particle spectrum does indeed approach a continuum starting at the correct value $(2m)^2$. Second, for large but finite $\Delta_{\rm max}$, the low-energy spectrum is discrete with a level-spacing proportional to $m^2/ \Delta_{\rm max}^2$. In other words, there is an emergent IR scale set by a combination of the truncation parameter $\Delta_{\rm max}$ and the bare dimensionful parameter $m^2$,
\be
\Lambda_{\textrm{IR}}^2 \sim \fr{m^2}{\Dmax^2}.
\ee

In some sense, despite the fact that the theory is formally in infinite volume, truncation effects themselves create IR scales which can be similar to putting the system in a finite-volume box.  In the case of a free theory, we see that the energy spacing would correspond to a circle length of $\Delta_{\rm max}/m$ in the IR.  In the presence of interactions, additional IR scales due to truncation typically emerge, limiting the resolution of LCT at energies far below the scale set by the UV couplings.  An important question is whether such IR effects can be modeled and partially subtracted or perhaps absorbed into renormalizations of the continuum description, but at present it is not understood how to do this in the majority of cases.

We can also use this analytic solution to show that the largest eigenvalues of the truncated matrix behave as $m^2 \Dmax^2$ in the limit of large $\Dmax$. Truncation therefore also generates an emergent UV scale
\be
\Lambda_{\textrm{UV}}^2 \sim m^2 \Dmax^2.
\ee
Schematically, we therefore see that the relevant deformation sets the overall scale (in this case $m^2$), and the truncation parameter $\Dmax$ sets a dynamic range around this scale that we are able to study numerically. It is important to note, however, that the states are not distributed uniformly over this range, but rather are concentrated more heavily in the IR.  In this sense, conformal truncation is not similar to finite volume.  The emergent dynamic range also has implications for the expected convergence of the truncation.
For gapped theories without any scale separation, where the lowest mass eigenstates are near the scale set by the deformation, we therefore naively expect LCT to quickly converge as we increase $\Dmax$. For theories with a large mass hierarchy, or those with a tuned IR fixed point at some critical coupling, we expect the LCT results to converge more slowly and require larger $\Dmax$.

%%%%%%%%%%%%%%%%%%%%%%%%%%%%%%%%%%%%%%%%%%%%%%%%%%%%%%%%%%%%%%%%%%%%%%%%%%%%%

\subsection{Spectral Densities}

Once we've diagonalized the truncated Hamiltonian, we obtain not only the spectrum of mass eigenvalues $\mu_j^2$, but also the corresponding eigenstates, expressed in the UV basis of primary operators:
\be
|\mu_j^2,p\> = \sum_{\De_i \leq \Dmax} C^{\mu_j^2}_{\Ocal_i} |\Ocal_i,p\>.
\ee
We can use these eigenstates to compute observables in the deformed theory. One of the simplest set of observables are the two-point functions of local operators, which can be written in the \KL representation
\be
\<\Ocal(p) \Ocal(-p)\> = \int d^2x \, e^{ipx} \<\Tcal\{\Ocal(x)\Ocal(0)\}\> = \int_0^\infty d\mu^2 \fr{\rho_\Ocal(\mu)}{p^2 - \mu^2 + i\epsilon}. \label{eq:SpectralDecomp}
\ee
The function $\rho_\Ocal(\mu)$ is the \emph{spectral density} of the local operator $\Ocal$, and corresponds to the overlap of $\Ocal$ with the mass eigenstates as a function of their invariant mass,
\be
\rho_\Ocal(\mu) = \sum_j |\<\Ocal(0)|\mu_j,p\>|^2 \de(\mu^2-\mu_j^2).
\ee

We can therefore obtain the spectral density of a particular operator in the deformed theory by computing its overlap with the resulting mass eigenstates,
\be
\<\Ocal(0)|\mu_j,p\> = \sum_{\De_i \leq \Dmax} C^{\mu_j^2}_{\Ocal_i} \<\Ocal(0)|\Ocal_i,p\>.
\ee
If the operator is in our basis, then this overlap is trivial to compute
\be
\<\Ocal_i(0)|\Ocal_j,p\> = 2p N_{\Ocal_i} \de_{ij}.
\ee

For free field theory, there is a particular set of operators which are not in our basis but are useful to study: the scalar operators $\phi^n$. We can compute their overlap with our basis states by first evaluating their overlap with a general $n$-particle monomial
\bq
\bal
\<\phi^n(0)|\p^{\bk}\phi,p\> &= \fr{1}{N_{\bk}} \int dx \, e^{-ipx} \<\phi^n(0) \p^{\bk}\phi(x)\> \\
&\doteq \boxed{\fr{n! \G(k_1) \cdots \G(k_n)}{\G(|\bk|)} \sqrt{ \fr{\G(2|\bk|)}{(4\pi)^{n-1} A_{\bk\bk}} }.}
\eal
\label{eq:MonomialOverlap}
\eq
We can then take linear combinations of these monomial terms to compute the overlap with basis states,
\be
\<\phi^n(0)|\Ocal,p\> = \sum_{\bk} \Chat^\Ocal_{\bk} \<\phi^n(0)|\p^{\bk}\phi,p\>.
\ee

\begin{table}[t!]
\noindent\rule[0.5ex]{\linewidth}{1pt}
\verb|(*| Individual monomial overlap with $\phi^n$ \eqref{eq:MonomialOverlap} \verb|*)|
\begin{code}
monoPhiN\verb|[k_]|:=\=Length[k]!*Product[Gamma[k[[i]]],\verb|{i,Length[k]}|] \\
 \> /Gamma[Total[k]] * \=Sqrt[Gamma[2Total[k]] \\
 \>\> /((4Pi)\verb|^|(Length[k]-1) A[k,k])];
\end{code}
\verb|(*| Compute the overlap of $\phi^n$ with all primary operators at a fixed particle number and degree \verb|*)|
\begin{code}
primaryPhiN\verb|[n_,deg_]|:=\=If[numStates[n,deg]==0, \verb|{}|, \\
 \> Dot[orthoPrimaries[n,deg], \\
 \> Table[monoPhiN[k], \verb|{k,monomialsBoson[n,deg]}|]]];
\end{code}
\noindent\rule[0.5ex]{\linewidth}{1pt}
\caption{Sample Mathematica code for computing the overlap of $\phi^n$ with all $n$-particle primary operators of degree $\De-n$.
\label{table:SimplePhiNOverlap}}
\end{table}

We can write simple Mathematica code to use~\eqref{eq:MonomialOverlap} to compute the overlaps of primary operators with $\phi^n$, shown in table~\ref{table:SimplePhiNOverlap}. The main function is {\tt primaryPhiN[n,deg]}, which computes the overlap of $\phi^n$ with all $n$-particle primaries of a given degree. For example, we can compute the overlap of $\phi^2$ with $(\p\phi)^2$ by entering:
\begin{mmaCell}[index=1,moredefined={primaryPhiN}]{Input}
  primaryPhiN[2,0]
\end{mmaCell}
The output is a list of the overlaps for each primary operator at this particle number and degree, in the same order as the output of {\tt orthoPrimaries}. Because $(\p\phi)^2$ is the unique two-particle, degree-$0$ primary, we thus obtain the single overlap:
\begin{mmaCell}{Output}
  \{\{\mmaSqrt{\mmaFrac{3}{\(\pi\)}}\}\}
\end{mmaCell}
which agrees with our general formula~\eqref{eq:MonomialOverlap}
\bq
\<\phi^2(0)|(\p\phi)^2,p\> \doteq \sqrt{ \fr{3}{\pi} }.
\eq

In principle, we can now use this code to compute spectral densities in $\phi^4$ theory. However, reconstructing the full spectral density for an operator requires many mass eigenstates. We therefore typically need to go to somewhat large values of $\Dmax$ in order to obtain useful results. 
 Fortunately, there is one example we can easily study, which is the free massive theory (i.e., $\lambda =0$). In this case, the Hamiltonian is diagonal with respect to particle number, which means we can reproduce the spectral density of an operator such as $\phi^2$ with only the two-particle primaries, which are efficient to compute.

In fact, for this particular example we can even compute the overlaps for primary operators analytically, using the Fock space approach,
\be
\<\phi^2(0)|[\p\phi \, \p\phi]_\ell,p\> \doteq \fr{1}{\sqrt{2\pi}} \int_0^1 dx \, \widehat{P}^{(1,1)}_\ell(1-2x) = \sqrt{ \fr{2(2\ell+3)}{\pi(\ell+1)(\ell+2)} },
\label{eq:TwoPartOverlapAnalytic}
\ee
allowing us to go to very large values of $\Dmax$. We can compare our LCT results to the known $\phi^2$ spectral density for free massive field theory,
\be
\rho_{\phi^2}(\mu) = \fr{\theta(\mu^2-4m^2)}{\pi\mu\sqrt{\mu^2-4m^2}}.
\label{eq:SFphi2}
\ee

\begin{figure}[t!]
\begin{center}
\includegraphics[width=0.96\textwidth]{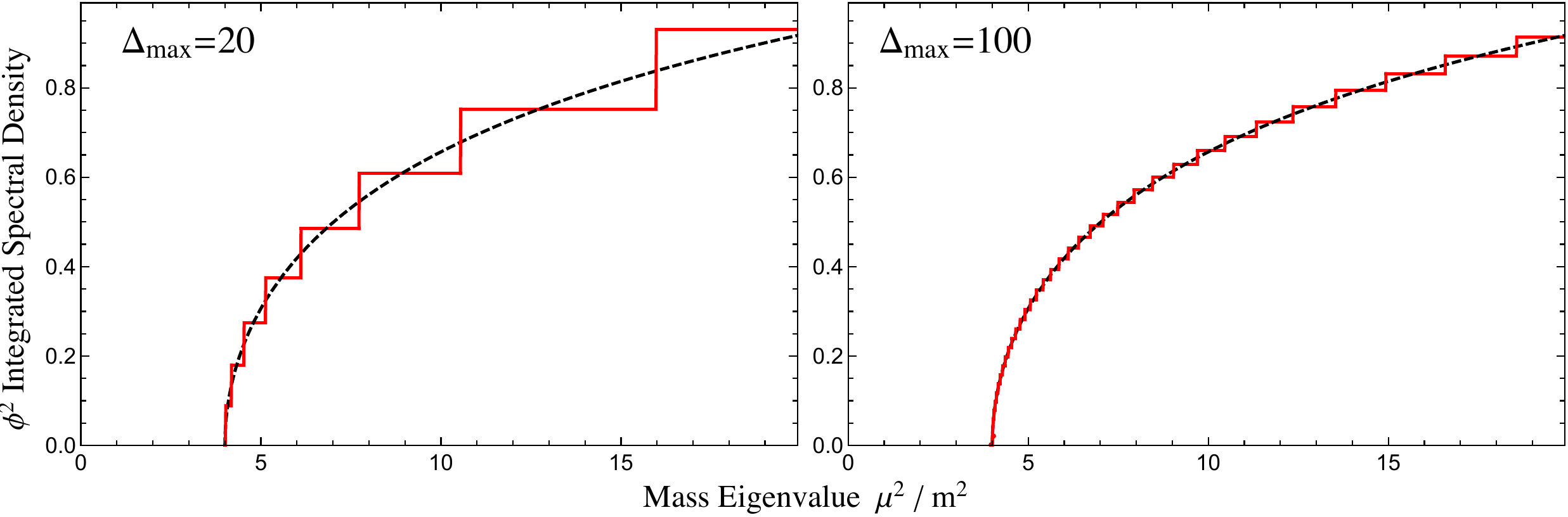}
\caption{Integrated spectral density for $\phi^2$ in free massive theory, from LCT (red, solid) vs the known expression (black, dashed). Left plot shows truncation result at $\Delta_{\rm max}=20$, right shows $\Delta_{\rm max}=100$.}
\label{fig:SFFreeScalar}
\end{center}
\end{figure}

Because the spectral densities are formally sums over delta functions, we can integrate them to obtain piecewise continuous functions that are more suitable for plotting:
\be
I_\Ocal(\mu) \equiv \int_0^{\mu^2} d \mu'^2 \rho_\Ocal(\mu') = \sum_{\mu_j^2 \leq \mu^2} |\<\Ocal(0)|\mu_j^2,p\>|^2. \label{eq:DefnOfIntegratedSpec}
\ee
The integrated spectral density for $\phi^2$ is shown in Fig.~\ref{fig:SFFreeScalar}, for both $\Dmax=20$ (which can be obtained using the provided Mathematica code) and $\Dmax=100$ (which requires the analytic results~\eqref{eq:TwoPartMassAnalytic} and \eqref{eq:TwoPartOverlapAnalytic}). As we can see, the truncation results are in good agreement with the exact expression obtained from integrating~\eqref{eq:SFphi2}.  

%%%%%%%%%%%%%%%%%%%%%%%%%%%%%%%%%%%%%%%%%%%%%%%%%%%%%%%%%%%%%%%%%%%%%%%%%%%%%
%%%%%%%%%%%%%%%%%%%%%%%%%%%%%%%%%%%%%%%%%%%%%%%%%%%%%%%%%%%%%%%%%%%%%%%%%%%%%

\section{Adding Fermions}
\label{sec:Fermions}

In section~\ref{sec:FreeLC}, we saw that \emph{before adding any relevant deformations} to the free CFT, the scalar and fermion LCT bases are qualitatively similar. Both bases are constructed from primary operators, with the difference that $\p\phi$ is the basic scalar primary while $\psi$ is the basic fermion primary. As we will now see, the minute we add a relevant deformation to the free Lagrangian, even just a simple mass term, a fundamental difference arises between scalar and fermionic theories: scalar matrix elements are finite, whereas fermionic matrix elements can have IR divergences. 

In section~\ref{sec:IRDivergences}, we will examine the origin of the IR divergences. In doing so, we will see that the effect of these divergences is to lift out (\emph{i.e.}~make infinitely massive) any states in the fermionic basis that contain a $\psi$ without any derivatives attached to it. At this point, one way to proceed would be to introduce an IR regulator $\epsilon$, which would appear in Hamiltonian matrix elements, and then take $\epsilon\rightarrow 0$ at the end, after diagonalizing the Hamiltonian. As $\epsilon \ra 0$, some mass eigenvalues diverge, but many remain finite and correctly reproduce the low-energy spectrum.

In section~\ref{sec:DirichletBasis}, we present a more efficient strategy for handling IR divergences that avoids a regulator. The idea is to preemptively eliminate 
precisely those linear combinations of states from the fermionic basis that will be lifted out by IR divergences anyway. We call the leftover basis (after re-orthogonalization) the ``Dirichlet" basis, because the momentum space wavefunctions for all states satisfy a particular boundary condition. The actual construction of the Dirichlet basis becomes straightforward with a simple, but crucial observation: the operators in the Dirichlet basis can still be thought of as ``primary" operators, except that they are constructed out of $\p\psi$ instead of $\psi$, as we will explain. Remarkably, because it is built from $\p\psi$, the Dirichlet basis is actually more like the scalar basis than the original fermion basis. 

In section~\ref{sec:FermionMass}, with the Dirichlet basis in hand, we compute the (now finite) matrix elements of the mass term. In section~\ref{sec:MixedScalarFermion}, we explain how to construct the LCT basis for theories with both scalars and fermions. Finally, in section~\ref{sec:YukawaInteraction}, we show how to compute matrix elements for Yukawa theory, in preparation for the applications in Part III of this work.

%%%%%%%%%%%%%%%%%%%%%%%%%%%%%%%%%%%%%%%%%%%%%%%%%%%%%%%%%%%%%%%%%%%%%%%%%%%%%

\subsection{IR Divergences}
\label{sec:IRDivergences}

To illustrate the onset of IR divergences, we deform the free fermion CFT by a mass term,
\be
\CL = \CL_\CFT + \delta \CL = i \bar{\Psi} \slashed{\p} \Psi - m\bar{\Psi}\Psi,
\ee
where $\Psi$ has the left- and right-chirality components $\psi$ and $\chi$:\footnote{The factor of $2^{1/4}$ is chosen to ensure the components satisfy the canonical anticommutation relation $\acomm{\psi(x)}{\psi(y)} = \fr{1}{2} \de(x-y)$.}
\be
\Psi = \fr{1}{2^{1/4}} \binom{\psi}{\chi}.
\ee
In terms of these components, the Lagrangian takes the form
\be
\CL = i \psi \p_+ \psi + i \chi \p_- \chi + \sqrt{2}i m \psi \chi.
\ee
As we alluded to in section~\ref{sec:FreeLC}, the right-chirality component $\chi$ is non-dynamical and can be integrated out using its equations of motion:
\be
\sqrt{2}\p_- \chi = m \psi \, .
\ee
After eliminating $\chi$, the action only involves $\psi$ and the mass term becomes a nonlocal interaction
\be
\Lcal = i \psi \p_+ \psi - \frac{m^2}{2}\psi \frac{1}{i\p_-} \psi.
\ee

It is the matrix elements of this nonlocal mass term that can exhibit IR divergences. To see this, consider a general two-particle monomial operator, $\p^{\bk} \psi = \p^{k_1} \psi \p^{k_2} \psi$ (not necessarily primary). The corresponding LCT state, written as an expansion in Fock space, is
\be
| \p^{\bk} \psi,p\> \doteq \frac{1}{2N_{\p^{\bk} \psi}} \int \frac{dp_1 \, dp_2}{(4 \pi)^2}  (2\pi ) \delta(p-|p|_2) \Big( p_1^{k_1-\frac{1}{2}} p_2^{k_2-\frac{1}{2}} - p_1^{k_2-\frac{1}{2}} p_2^{k_1-\frac{1}{2}} \Big)  | p_1, p_2\> .
\ee
Now consider the mass term matrix elements between these monomials: 
\bq
\Mcal^{(\psi\fr{1}{\p}\psi)}_{\p^{\bk}\psi,\p^{\bk'}\psi} \doteq \frac{1}{N^*_{\p^{\bk} \psi}N_{\p^{\bk'} \psi}} \int \frac{dp_1}{8\pi} \Big( p_1^{k_1+k_1'} p_2^{k_2+k'_2}  - p_1^{k_1+k_2'} p_2^{k_2+k'_1} \Big) \left( \frac{m^2}{2p_1} + \frac{m^2}{2p_2} \right)_{p_2 = p-p_1}.
\label{eq:FermionMassDiv}
\eq
This integral is potentially IR divergent. In particular, the factors of $1/p_i$ come from the $\frac{1}{i\p}$ in the nonlocal mass term and lead to IR divergences at $p_i = 0$ if any of the $k_i$ and $k'_i$ are zero, \emph{i.e.}, if the operator $\p^{\bk} \psi$ has a $\psi$ without a derivative acting on it.  

To understand the effect of these IR divergences, we can put in an IR regulator that removes a region $0 \le p_i \le \epsilon$ in momentum space around $p_i = 0$, so that the limits of integration are from $\epsilon$ to $p-\epsilon$.  For simplicity, consider the mass matrix in the subspace of two-particle monomials with $|\bk|=k_1 + k_2 = 3$.  There are two such monomials:\footnote{These operators actually span the space of all states with $k_1 + k_2 \le 3$. The number of primaries at $k_1+k_2=0,1,2,3$ is $0,1,0,1$, respectively, and descendants of a primary create the same state as the primary in momentum space. }
\be
\CO_1 \equiv \p^3 \psi \psi, \qquad  \CO_2 \equiv \p^2 \psi \p \psi .
\ee
In the limit that we take the IR regulator $\epsilon \rightarrow 0$, we can separate out the Hamiltonian in this two-dimensional subspace into a divergent piece and a finite piece:
\be
\delta P_+ = \log(\epsilon) H_{\rm div} + H_{\rm fin}, \qquad H_{\rm div} \propto \left( \begin{array}{cc} 1 &  0  \\ 0  & 0 \end{array} \right).
\ee
Therefore, in the $\epsilon \rightarrow 0$ limit, the divergent part of the Hamiltonian makes the state $\p^3 \psi \psi$ infinitely heavy and lifts it out of the spectrum, leaving one finite eigenvalue associated with the state $\p^2\psi\p\psi$. 

To keep only the states with finite energy in the $\epsilon \rightarrow 0$ limit, we restrict our basis to the kernel of $H_{\rm div}$.   Clearly in the above example, keeping the kernel of $H_{\rm div}$ means throwing out the state $\CO_1 = \p^3 \psi \psi$.  The reason this state in particular is divergent is that is has a $\psi$ without any derivative acting on it.  A sufficient condition to avoid divergences is for a monomial to have derivatives acting on each insertion of $\psi$, so the wavefunction of its state in the Fock space basis vanishes when any of the momenta $p_i$ vanish, thereby canceling the $1/p_i$ divergence from the mass term. 

In fact, having at least one $\p$ attached to every $\psi$ is also a necessary condition to avoid divergences. To see this, note that due to Fermi statistics, any monomial can have at most one $\psi$ without a derivative acting on it.  For any fixed particle number $n$, we can choose one representative monomial state, say $\p^{n-1} \psi \p^{n-2} \psi \dots \p \psi \psi$, that has exactly one $\psi$ without derivatives.  Any other monomial state with a derivative-free $\psi$ can be reduced via `integration by parts' to this one plus states where all $\psi$s have derivatives.  For instance, 
 \be
 \p^3 \psi \psi = \p (\p^2 \psi \psi) - \p^2 \psi \p \psi = \p ( \p (\p \psi \psi)) - \p^2 \psi \p \psi \cong  \p \psi \psi - \p^2 \psi \p \psi,
 \ee
 where we have used the fact that the derivative of an operator creates the same state as the operator itself.  Therefore at each $n$, in this basis $H_{\rm div}$ is nonzero only for the diagonal entry corresponding to this representative monomial state, implying that only the states with a derivative-free $\psi$ are lifted.

%%%%%%%%%%%%%%%%%%%%%%%%%%%%%%%%%%%%%%%%%%%%%%%%%%%%%%%%%%%%%%%%%%%%%%%%%%%%%

\subsection{Dirichlet Basis}
\label{sec:DirichletBasis}

We have just seen that the IR divergences in the fermion mass matrix remove from the finite-energy spectrum precisely all monomials containing a derivative-free $\psi$. The states that remain and do not get lifted out thus satisfy a Dirichlet boundary condition: their Fock space wavefunctions vanish at $p_i =0$, because each $\psi$ has at least one derivative attached to it. For this reason, we refer to the remnant basis (which is not lifted out and which has finite matrix elements) as the ``Dirichlet" basis. 

Let us now discuss the construction of the Dirichlet basis. Because we have discarded the primary operator $\psi$ but kept its descendant $\p \psi$, it may appear that there is no longer an option of organizing the Dirichlet basis in terms of primary operators. However, the situation is really not very different from the situation with scalars.  For scalars in 2d, the naively primary operator $\phi$ is discarded as a local operator because of IR divergences in its correlators.  Instead, $\p \phi$, naively a descendant, takes on the role of being a primary operator.  Similarly, with $\psi$ discarded, we can attempt to treat $\p \psi$ as a weight $h=\frac{3}{2}$ primary operator, and in fact this works. 

The reason we can pretend that $\p \psi$ is a weight $h=\frac{3}{2}$ primary operator is because the UV CFT is a free theory. In particular, all correlators of $\p \psi$ are simply products of two-point functions, of the form
\be
\< \p \psi(x) \p \psi(y)\> \doteq \fr{1}{2\pi(x-y)^3}, 
\ee
so that the theory of $\p\psi$ is a Generalized Free Theory (GFT).\footnote{Sometimes these are referred to as ``Gaussian'' or ``mean field'' theories.} Correlators of a GFT are indeed conformally covariant, with the scaling dimensions of fields set by their two-point functions.  

We can state this fact algebraically by defining a modified  special conformal generator $\tilde{K}$ that satisfies the conformal algebra and annihilates $\p \psi$,
\begin{align}
	\comm{\tilde K}{\p\psi(0)} = 0 \, ,
\end{align}
and more generally acts on operators $\Ocal$ as if they were made out of a dimension $\Delta=J=\frac{3}{2}$ ``primary operator'' $\p \psi(x)$. Therefore, we can use the same method for constructing the primary operators out of $\p\psi$ that we used for constructing primary operators out of $\p\phi$, where we recursively make primary operators with $n$ particles by sewing $\p \psi$ onto primary operators with $n-1$ particles according to \eqref{eq:JoaoFormula}. The first few ``primary'' operators are shown in table~\ref{table:FermionOperators}. Note that there are fewer independent primary operators than for bosons due to the anticommuting nature of $\p\psi$.

\begin{table}[t!]
\begin{center}
\begin{tabular}{| l | l | l |}
\hline & & \\[-5pt]
$n$ & General expression & Explicit examples \\
\hline & & \\[-7pt]
1 &  & $\p\psi$ \\
\hline & & \\[-5pt]
2  &   $\Ocal_{(\ell_1)} \equiv \left[ \p\psi \,  \p\psi \right]_{\ell_1} $   &  $\Ocal_{(0)} = 0$  \\
    &                                                                                                    & $\Ocal_{(1)}  =6\p^2\psi\p\psi $ \\
    &                                                                                                    & $\Ocal_{(2)}  = 0$ \\
    &                                                                                                    & $\Ocal_{(3)}  = 20\p^4\psi\p\psi - 100\p^3\psi \p^2\psi$ \\
    &                                                                                                    & $\vdots$ \\
\hline & & \\[-5pt]
3 &    $\Ocal_{(\ell_1,\ell_2)} \equiv \left[ \Ocal_{(\ell_1)}\p\psi \right]_{\ell_2}$   & $\Ocal_{(0,0)} = \Ocal_{(0,1)} = \Ocal_{(1,0)} = 0$ \\
    &                                                                                                    & $\Ocal_{(0,2)}  = \Ocal_{(1,1)} = \Ocal_{(2,0)}  =  0$ \\
    &                                                                                                    & $\Ocal_{(0,3)}  = \Ocal_{(2,1)}  =  0$ \\
    &                                                                                                    & $\Ocal_{(1,2)}  = \fr{18}{25}\Ocal_{(3,0)} = 72\p^3\psi\p^2\psi\p\psi$ \\
    &                                                                                                    & $\vdots$ \\
\hline
\end{tabular}
\end{center}
\caption{The first few fermion Dirichlet ``primaries'' constructed recursively by starting with $\p\psi$ and successively sewing on additional $\p\psi$'s using \eqref{eq:JoaoFormula}.}
\label{table:FermionOperators}
\end{table}

\noindent\rule[0.5ex]{\linewidth}{1pt}
\footnotesize

In the scalar case, the number of monomials with $n$ particles and $k$ derivatives was simply the number of partitions $P_n(k)$ of $k$ objects into exactly $n$ bins. However, Pauli exclusion prohibits any two fermions from having the same number of derivatives, such that the number of fermion monomials at a given level is generically less than that of scalars.

Fortunately, there is a simple map between fermion monomials and scalar ones. Given a scalar monomial labeled by $\bk$, we can construct a corresponding fermion monomial by adding it to the ``Fermi surface'' $\bk_{\textrm{F}} \equiv (n-1,n-2,\ldots,1,0)$. The number of fermion monomials with $n$ particles and $k$ derivatives is therefore equivalent to the number of scalar monomials with $k-|\bk_{\textrm{F}}| = k-\fr{n(n-1)}{2}$ derivatives,
\be
P_n^F(k) = P_n(k-|\bk_{\textrm{F}}|).
\ee
Similarly, the number of independent Dirichlet states with $n$ fermions and $k$ derivatives is
\bq
\widehat{P}^F_n(k) \equiv P^F_n(k) - P^F_n(k-1) = \widehat{P}_n(k-|\bk_{\textrm{F}}|).
\eq

We can therefore use the same analysis from section~\ref{sec:ScalarBasis} to obtain a complete, minimal list of vectors $\Lvec$ for the Dirichlet basis.

\normalsize
\noindent\rule[0.5ex]{\linewidth}{1pt}

Once we've used this recursive method to construct a complete basis of Dirichlet operators up to some $\Dmax$,\footnote{Note that a Dirichlet operator is a linear combination of primaries and descendants with some fixed dimension $\De$. When truncating our basis, we therefore still have a well-defined notion of $\Dmax$, preserving the basic conformal structure of our truncation scheme.} we then need to orthonormalize this basis. Just like for the scalar case, we need to compute the monomial Gram matrix
\be
G_{\bk\bk'} = \fr{1}{2p N^*_{\bk} N_{\bk'}} \int dx \, e^{ipx} \<\p^{\bk^\dagger}\psi(x) \p^{\bk'}\psi(x)\>,
\ee
where $\p^{\bk^\dagger}\psi$ indicates a monomial operator in the reverse order (the fermionic field $\psi$ is real),
\be
\p^{\bk^\dagger}\psi \equiv \p^{k_n}\psi \cdots \p^{k_1}\psi.
\label{eq:kdaggdef}
\ee

We can compute the position space two-point function via Wick contraction, obtaining the general expression
\be
\< \p^{\bk^\dagger} \psi(x) \p^{\bk'}\psi(0) \> \doteq
\frac{ \tilde A_{\kvec\kvec'} }{
	(4\pi)^n x^{|\bk|+|\bk'| + n}}.
    \label{eq:FermionMono2pt}
\ee
It is simplest to compute the Wick contraction coefficient $\tilde A_{\kvec\kvec'}$ recursively, by taking the fermion $\p^{k_n}\psi$ from the left monomial and summing over its contractions with each fermion $\p^{k'_i}\psi$ from the right monomial.  Taking into account the signs from anticommuting the fermions, the resulting recursion relation is
\be
\tilde{A}_{\yvec, \yvec^\prime} = \sum_{i=1}^n (-1)^{n - i} 
\G(k_n+k_i'+1) \tilde{A}_{\bk/k_n, \kvec^\prime/k_i^\prime}.
\ee

Using the general integral in eq.~\eqref{eq:FTFormulas}, we can Fourier transform the resulting two-point function to momentum space, allowing us to fix the fermion monomial normalization coefficients
\be
|N_{\bk}|^2 = \fr{\pi p^{2|\bk|+n-2} \tilde{A}_{\bk\bk}}{(4\pi)^n \G(2|\bk|+n)},
\ee
and the monomial Gram matrix elements
\be
G_{\bk\bk'} \doteq \fr{\sqrt{\G(2|\bk|+n)\G(2|\bk'|+n)}}{\G(|\bk|+|\bk'|+n)} \cdot \fr{\tilde{A}_{\bk\bk'}}{\sqrt{\tilde{A}_{\bk\bk} \tilde{A}_{\bk'\bk'}}}.
\ee

With these expressions, we can follow the same procedure as section~\ref{sec:Ortho} to construct an orthonormal basis of Dirichlet states up to some threshold $\Dmax$.

%%%%%%%%%%%%%%%%%%%%%%%%%%%%%%%%%%%%%%%%%%%%%%%%%%%%%%%%%%%%%%%%%%%%%%%%%%%%%

\subsection{Fermion Mass Term}
\label{sec:FermionMass}

In this section, we compute the Hamiltonian matrix elements due to the fermion mass term. Just like for scalars, we focus on the individual monomial matrix elements,
\be
\Mcal^{(\psi\fr{1}{\p}\psi)}_{\bk\bk'} = \fr{m^2}{2N^*_{\bk} N_{\bk'}} \int dx \, dz \, e^{ip(x-z)} \<\p^{\bk^\dagger}\psi(x) \, \psi\fr{1}{i\p}\psi(0) \, \p^{\bk'}\psi(z)\>,
\ee
which can be combined to obtain the matrix elements for the orthonormal basis of Dirichlet ``primaries''.

We can compute the position space three-point function via Wick contraction, taking one ``interacting'' fermion from both the incoming and outgoing states and leaving $n-1$ ``spectating'' fermions,
\bq
\bal
&\<\p^{\bk^\dagger}\psi(x) \, \psi\fr{1}{\p}\psi(y) \, \p^{\bk'}\psi(z)\> \\
& \qquad = \sum_{\substack{k_i \in \bk \\ k'_j \in \bk'}} (-1)^{i-j} \<\p^{k_i} \psi(x) \, \psi\fr{1}{\p}\psi(y) \, \p^{k'_j} \psi(z)\> \<\p^{\bk^\dagger/k_i}\psi(x) \, \p^{\bk'/k'_j} \psi(z)\>.
\eal
\eq
We can evaluate the spectating piece using eq.~\eqref{eq:FermionMono2pt}. However, for the interacting piece, we need to precisely define the nonlocal operator $\frac{1}{\partial}$ in the Hamiltonian. Since $\frac{1}{\partial}$ arose from integrating out the non-dynamical field $\chi$, it is just the $\chi$ propagator.

One foolproof way to determine how to treat the $\chi$ propagator is to compare with equal-time quantization.  Because correlation functions of local operators are independent of the quantization scheme, one can compare the two-point function of $\psi$ in the interacting theory in both schemes and match them.  This matching is made simpler by the fact that LC quantization can be obtained as the infinite-momentum-frame limit of ET, up to additional terms in the Hamiltonian that arise from modes that decouple in the infinite-momentum limit.  The Hamiltonian with these additional terms can be thought of as an ``effective LC Hamiltonian'' $H_{\rm eff}$, and in \cite{Fitzpatrick:2018ttk}, it was shown how to determine them in perturbation theory by comparing the Dyson series of two-point correlators computed in ET and LC quantization.  The key point is that these additional terms arise from delta functions of LC time $\de(x^+)$ in the Dyson series.  Following this prescription, consider the $\chi$ propagator in position space:
\be
\< \chi(x) \chi(0)\> \sim \frac{-i}{x^+ -i \epsilon \, \sgn(x^-)} \sim -\Pcal\frac{i}{x^+} + \pi \delta(x^+) \sgn(x^-), \label{eq:PVPrescription}
\ee
where $\Pcal$ denote the principal value. The coefficient $\sgn(x^-)$ of  $\delta(x^+)$ is the propagator that is actually generated by taking the infinite-momentum limit and decoupling $\chi$. The precise definition of $\frac{1}{\partial}$ in position space is simply the integral over this propagator,
\be
H \supset \int dx^- \, \psi(x) \frac{1}{\partial} \psi(x) \Rightarrow  \int dx^- dy^- \, \psi(x) \sgn(x^-- y^-) \psi(y).
\ee
We discuss this effective LC Hamiltonian approach in more detail in appendix~\ref{app:Heff}.

In practice, the nonlocal inverse derivative $\fr{1}{\p}$ thus just tells us to integrate, \emph{i.e.},
\be
\<\frac{1}{\partial} \psi(y) \p^k \psi(z)\> \Rightarrow \int dy \<\psi(y) \p^k \psi(z)\>,
\ee
where the integration constant is chosen so that the expression vanishes at $y\rightarrow \infty$. Following this procedure, we obtain the ``interacting'' correlator
\bq
\<\p^{k_i} \psi(x) \, \psi\fr{1}{\p}\psi(y) \, \p^{k'_j} \psi(z)\> \doteq \fr{\G(k_i)\G(k'_j)}{(4\pi)^2(x-y)^{k_i} (y-z)^{k'_j}} \left( \fr{k_i}{x-y} + \fr{k'_j}{y-z} \right).
\eq
One can check that this expression is equivalent to our treatment of the mass term in the Fock space description \eqref{eq:FermionMassDiv}.

After combining the interacting and spectating pieces to obtain the full three-point function, we can Fourier transform to momentum space with the help of the general formula~\eqref{eq:FTFormulas}, obtaining the monomial matrix element,
\bq
\boxed{
\bal
\Mcal^{(\psi\fr{1}{\p}\psi)}_{\bk\bk'} &= \fr{m^2}{\G(|\bk|+|\bk'|+n-1)} \sqrt{ \fr{\G(2|\bk|+n)\G(2|\bk'|+n)}{\tilde{A}_{\bk\bk} \, \tilde{A}_{\bk'\bk'}} } \\
& \qquad \qquad \times \sum_{\substack{k_i \in \bk \\ k'_j \in \bk'}} (-1)^{i-j} \G(k_i+k'_j) \tilde{A}_{\kvec/k_i,\kvec'/k'_j}.
\eal
}
\eq
Note that in evaluating this expression we have explicitly assumed that all $k_i,k'_j \geq 1$, since we are restricting ourselves to the Dirichlet basis. We therefore obtain finite expressions for all monomial matrix elements, which can be combined to compute the Hamiltonian matrix elements for Dirichlet states up to some $\Dmax$.

%%%%%%%%%%%%%%%%%%%%%%%%%%%%%%%%%%%%%%%%%%%%%%%%%%%%%%%%%%%%%%%%%%%%%%%%%%%%%

\subsection{Mixed Scalar-Fermion States}
\label{sec:MixedScalarFermion}

So far we have worked out the complete basis of scalar states and fermion states separately.  In a theory with scalars and fermions, we need to be able to make a larger set of primary operators that contain a mixture of $\p \phi$'s and $\p \psi$'s, of the schematic form
\be
\CO = \sum_{\bk} C^{\CO}_{\bk} \p^{\bk_B} \phi \p^{\bk_F} \psi, 
\label{eq:mixonmonomialsdefn}
\ee
where the sum is over monomials with fixed numbers $n_B, n_F$ of bosons and  fermions, and a fixed total  $|\bk_B| + |\bk_F|$.  

Fortunately, such mixed scalar-fermion primary operators can be constructed simply by combining our all-scalar and all-fermion primary operators.  The reason is that our all-scalar and all-fermion primaries already span the Hilbert space of scalar states and fermion states, so instead of building mixed states in a monomial basis, we can build them directly out of products of scalar and fermion primary operators.

More precisely, we first construct the all-scalar primary operators $\{ B_i\}$, and the all-fermion primary operators $\{F_j\}$. By pairing up each $B_i$ with each $F_j$ to make a new primary in all possible ways, we generate a basis for mixed states.  For any choice of $B_i$ and $F_j$, we combine them to make new primary operators using eq.~\eqref{eq:JoaoFormula}, just like we combined $n-1$ scalar primaries with $\p \phi$ to make new $n$-particle scalar primaries in section~\ref{sec:ScalarBasis}.  In fact, this time our task is even simpler, because the states we construct this way using all pairs of $B_i$ and $F_j$ are {\it already orthogonalized} if the bases $\{B_i\}$ and $\{F_j\}$ are separately orthogonal.  In equations, the mixed primary operators are
\begin{equation}
[B_i F_j]_\ell \equiv \sum_m c^\ell_m(\De_{B_i},\De_{F_j}) \, \p^m B_i \p^{\ell-m} F_j,
\label{eq:MixedStatesJoao}
\end{equation}
where the coefficients $c^\ell_m(\De_{B_i},\De_{F_j})$ are defined in eq.~\eqref{eq:JoaoCoeff}. The two-point function between two of these mixed operators is
\begin{equation}
\begin{aligned}
&\<[B_i F_j]_\ell(x) \, [B_r F_s]_{\ell'}(0)\> \\
& \quad = \sum_{m,m'} c^\ell_m(\De_{B_i},\De_{F_j}) c^{\ell'}_{m'}(\De_{B_r},\De_{F_s}) \<\p^m B_i(x) \p^{\ell-m} F_j(x) \p^{m'} B_r(0) \p^{\ell'-m'} F_s(0)\> \\
& \quad = \sum_{m,m'} c^\ell_m(\De_{B_i},\De_{F_j}) c^{\ell'}_{m'}(\De_{B_r},\De_{F_s}) \<\p^m B_i(x) \p^{m'} B_r(0)\> \<\p^{\ell-m} F_j(x) \p^{\ell'-m'} F_s(0)\> \\
& \quad = \de_{ir} \de_{js} \sum_{m,m'} c^\ell_m(\De_{B_i},\De_{F_j}) c^{\ell'}_{m'}(\De_{B_i},\De_{F_j}) \<\p^m B_i(x) \p^{m'} B_i(0)\> \<\p^{\ell-m} F_j(x) \p^{\ell'-m'} F_j(0)\> \\
& \quad = \de_{ir} \de_{js} \<[B_i F_j]_\ell(x) \, [B_i F_j]_{\ell'}(0)\> = \de_{ir} \de_{js} \de_{\ell\ell'} \<[B_i F_j]_\ell(x) \, [B_i F_j]_\ell(0)\>.
\end{aligned}
\label{eq:Mixed2Pt}
\end{equation}
In other words, the composite operators $[B_i F_j]_\ell$ automatically inherit the orthogonality of the building blocks $B_i$ and $F_j$, with no need to reorthogonalize. We thus have a complete, orthogonal basis for mixed states as soon as we put them together according to eq.~\eqref{eq:MixedStatesJoao}. We only need to properly normalize these states, which we can do by evaluating the final two-point function in~\eqref{eq:Mixed2Pt}, and Fourier transforming to momentum space with eq.~\eqref{eq:FTFormulas}.

%%%%%%%%%%%%%%%%%%%%%%%%%%%%%%%%%%%%%%%%%%%%%%%%%%%%%%%%%%%%%%%%%%%%%%%%%%%%%

\subsection{Yukawa Interaction}
\label{sec:YukawaInteraction}

\begin{figure}[t!]
\begin{center}
\includegraphics[width=0.6\textwidth]{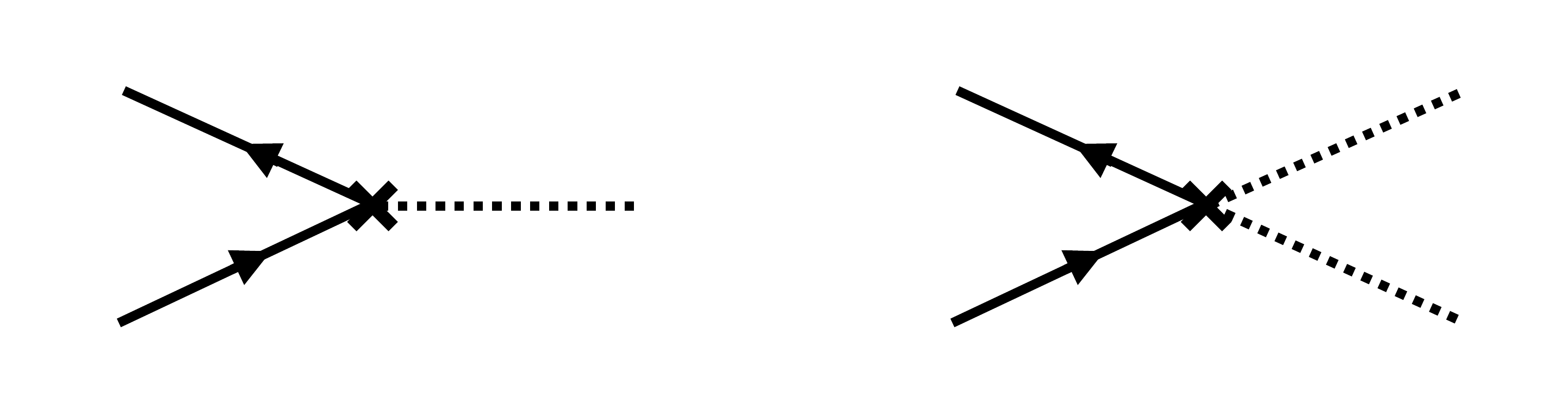}
\caption{Cubic and quartic Yukawa interactions obtained by integrating out the non-dynamical field $\chi$.}
\label{fig:YukawaInteractions} 
\end{center}
\end{figure}

A theory with only a single real fermion has only quadratic relevant terms due to Fermi statistics, so to make the fermion interacting we have to couple it to something else.  If the theory also contains a real scalar, we can add a Yukawa interaction to the Lagrangian:
\be
\Lcal = i\psi\p_+\psi + i \chi\p_- \chi + \sqrt{2}i (m+g \phi) \psi\chi.
\ee
We again must integrate out $\chi$ by using its equations of motion. Since the Lagrangian is quadratic in $\chi$, this is straightforward, and we obtain the new nonlocal Lagrangian:
\be
\Lcal = i\psi\p_+\psi - \fr{m^2}{2} \psi \fr{1}{i\p_-} \psi - mg \, \phi \psi \fr{1}{i\p_-} \psi - \fr{g^2}{2} \phi \psi \fr{1}{i\p_-} \phi \psi.
\label{eq:YukawaLCLag}
\ee
We thus obtain both a cubic and quartic interaction, shown in Fig.~\ref{fig:YukawaInteractions}. We can evaluate matrix elements of these nonlocal interactions either in terms of integrals over momentum in the Fock space approach or by computing matrix elements in position space and Fourier transforming in the Wick contraction approach.

For the Wick contraction method, the factor of $\fr{1}{\p}$ indicates that we need to integrate, just like for the mass term. We discuss this procedure in much more detail in the section on the Yukawa interaction in  appendix \ref{app:GeneratingFunctions}, but here we can consider the simple example of the $2 \ra 1$ interaction
\be
\Mcal_{\bk\bk'}^{(\phi\psi\fr{1}{\p}\psi)} = \fr{mg}{N_{\bk}^* N_{\bk}} \int dx \, dz \, e^{ip(x-z)} \< \p^{k_B}\phi \p^{k_F}\psi(x) \, \phi \psi \fr{1}{i\p} \psi(0) \, \p^{k'_F}\psi(z)\>.
\ee
Evaluating the position space correlator, we obtain
\bq
\<\p^{k_B}\phi \p^{k_F}\psi(x) \, \phi \psi \fr{1}{i\p} \psi(y) \, \p^{k'_F}\psi(z)\> \doteq \fr{\G(k_B) \G(k_F) \G(k'_F)}{(4\pi)^3(x-y)^{k_B+k_F} (y-z)^{k'_F}} \left( \fr{k_F}{x-y} + \fr{k'_F}{y-z} \right).
\eq
If we then Fourier transform to momentum space, we find the resulting matrix element
\be
\Mcal_{\bk\bk'}^{(\phi\psi\fr{1}{\p}\psi)} = mg \sqrt{ \fr{\G(2k_B+2k_F+1)}{\pi\G(2k_B)\G(2k_F+1)} } \, \fr{(k_B+2k_F)\G(k_B) \G(k_F)}{2\G(k_B+k_F+1)}.
\ee 

It is illuminating to also consider how one would treat the $\frac{1}{\partial}$ factor using Fock space modes. For instance, consider the $2 \ra 2$ monomial matrix element:
\bq
\bal
&\Mcal^{(\phi\psi\fr{1}{\p}\phi\psi)}_{\bk\bk'} = \< \p^{k_B} \phi \p^{k_F} \psi | \fr{g^2}{2} \phi \psi \frac{1}{i \p} \phi \psi | \p^{k'_B} \phi \p^{k'_F} \psi\> \\
& \quad \doteq \frac{g^2}{2 N_{\bk}^* N_{\bk'}} \int_0^p \frac{dp_B \, dp_B'}{(4\pi)^4 p_B  p_B'} (2\pi)^2 p_B^{k_B} p_F^{k_F-\frac{1}{2}} p'^{k_B'}_B p'^{k'_F-\frac{1}{2}}_F \< p_B p_F | \phi \psi \frac{1}{i\p} \phi \psi | p_B' p_F'\> \\
& \quad \doteq \frac{g^2}{N_{\bk}^* N_{\bk'}} \int_0^p \frac{dp_B \, dp_B'}{(8\pi)^2} p_B^{k_B-1} p_F^{k_F} p'^{k_B'-1}_B p'^{k'_F}_F \left( \fr{1}{p} + \fr{1}{p_F-p'_B} \right),
\eal
\eq
where the conditions $p_B+p_F=p_B'+p_F'=p$ are implicit. Note that in the last line, the second term has a pole at $p_F=p_B'$.  These poles arise from the on-shell singularities of the $\chi$ propagator. 
The correct way to deal with them is to take their principal value part, which is the real part of the propagator with an  $i \epsilon$:
\be
\frac{1}{p+i \epsilon} = \Pcal \frac{1}{p} - i \pi \delta(p).
\ee
Physically, by taking the real part, we are discarding the $\delta$-function localized part of the spectral weight corresponding to the $\chi$ state, which is not present in lightcone quantization.  The principal value prescription is equivalent to regulating the $p=0$ IR divergence by drilling a hole in the propagator around $p=0$, i.e.~setting it to zero if $-\epsilon < p < \epsilon$, and then taking the limit $\epsilon \rightarrow 0$.  With this prescription, the above matrix element becomes
\be
\Mcal^{(\phi\psi\fr{1}{\p}\phi\psi)}_{\bk\bk'} \doteq \frac{g^2}{N_{\bk}^* N_{\bk'}} \int_0^p \frac{dp_B \, dp_B'}{(8\pi)^2} p_B^{k_B-1} p_F^{k_F} p'^{k_B'-1}_B p'^{k'_F}_F \left( \fr{1}{p} + \Pcal \fr{1}{p_F-p'_B} \right).
\ee
A general integral of this form can be evaluated by using the identity
\be
\int_0^1 dx \, dy \, \Pcal \frac{x^m y^n}{x-y} =\frac{H_m - H_n}{m+n+1},
\ee
where $H_n$ is the $n$-th harmonic number,
\be
H_n \equiv \sum_{s=1}^n \fr{1}{s}.
\ee

%%%%%%%%%%%%%%%%%%%%%%%%%%%%%%%%%%%%%%%%%%%%%%%%%%%%%%%%%%%%%%%%%%%%%%%%%%%%%
%%%%%%%%%%%%%%%%%%%%%%%%%%%%%%%%%%%%%%%%%%%%%%%%%%%%%%%%%%%%%%%%%%%%%%%%%%%%%

\section{Invitation: Two Dimensional QCD}
\label{sec:gaugefields}

As our final application before we move on to more advanced methods to improve computational efficiency, we now show how the techniques used so far can be used to study gauge theories in two dimensions. With the addition of gauge fields, LCT in two dimensions achieves a kind of conceptual completeness  since it encompasses any local 2d Lagrangian with relevant interactions built from products of the fundamental fields.

Gauge fields in two dimensions do not carry propagating degrees of freedom, and so can be integrated out of the theory. If they are coupled to matter, this can generate nonlocal, Coloumb-type interactions between fields, which can easily be accommodated in the methods introduced in sections \ref{sec:simplestcode} and \ref{sec:Fermions}. In this section, we will show how to integrate out gauge fields in practice in LCT, and the resulting form of the interactions. As an application, we will focus on massless two dimensional QCD at low $\Delta_{\textrm{max}}$ and at finite $N_c$, which will serve as a mini-example before we delve into more complicated and detailed applications in Part III. We will defer a more general analysis including massive quarks at finite $N_c$ to an upcoming publication \cite{2dQCDToAppear}.\footnote{The reason we do not consider massive QCD in this work is that there is a technical subtlety that occurs in the massive theory that does not happen in the massless case. It is related to the boundary behavior of the wavefunctions. Consider, for example, the model at large $N_c$ at fixed `t Hooft coupling $\lambda = g^2 N_c$. There, one can show  that the necessary boundary condition for the meson wavefunction $\phi(x)$ (where $x$ is the momentum fraction) near $x=0$ is $\phi(x) \sim x^\beta $ where $\beta$ is a number (not necessarily an integer) related to the quark mass via \cite{tHooft:1974pnl} \begin{equation}
	m^2 = \frac{\lambda^2}{\pi}(1-\pi \beta \cot \pi \beta).
\end{equation}This is somewhat different from the Dirichlet boundary conditions we have considered so far in the context of massive fermions in the previous section (note that those fall off like $\sim x(1-x)$ for two-particle states). For masses in the range $m^2 \gtrsim \frac{\lambda^2}{\pi}$ (corresponding to $1 > \beta \gtrsim \frac{1}{2}$), results obtained using the Dirichlet basis converge relatively quickly. However, in small mass regime $0 < m^2 \lesssim \lambda^2/\pi$ (corresponding to $0 < \beta \lesssim \frac{1}{2}$), there is enhanced sensitivity to the boundary conditions and convergence using the Dirichlet basis is much slower. We will discuss how to improve the convergence in this regime in upcoming work. }

The organization of this section is the following: in \ref{sec:intoutgaugefields}, we explain how integrating out gauge fields works. In \ref{sec:2dmasslessqcd}, we consider 2d QCD in the massless limit as a toy example. In \ref{sec:constructfiniteN} we explain how to build a basis of color singlets. In \ref{sec:gaugeintmatrixelements}, we compute the gauge interaction matrix elements, but  leave technicalities to Appendix \ref{app:appendixgauge}. Finally, in sections \ref{sec:largeNexample} and \ref{sec:finiteNexample} we look at low $\Delta_{\textrm{max}}$ results at both large $N_c$ and finite $N_c$.

%%%%%%%%%%%%%%%%%%%%%%%%%%%%%%%%%%%%%%%%%%%%%%%%%%%%%%%%%%%%%%%%%%%%%%%%%%%%%

\subsection{Integrating out Gauge Fields}\label{sec:intoutgaugefields} We begin with the Yang Mills Lagrangian \begin{equation}
	\Lcal_{\textrm{YM}} = -\frac{1}{2} \Tr F_{\mu\nu} F^{\mu\nu} - A_\mu J^\mu,
\end{equation} where the gauge field is  in the adjoint representation of SU($N_c$) $A_\mu = A_\mu^A T^A$ and the generators are normalized such that $[T^A, T^B] = \half \delta^{AB}$. The field strength is given by $F_{\mu\nu}^A= \p_\mu A_\nu^A - \p_\nu A_\mu^A + g f^{ABC} A_\mu^B A_\nu^C$.

It is particularly convenient to work in   lightcone gauge \begin{equation}
	A_- = 0,
\end{equation}which gives \begin{equation}
	\Lcal_{\textrm{YM}} =  \Tr (\p_- A_+)^2 - A_+ J_- . \label{eq:gaugethrylag}
\end{equation} Clearly, there is no kinetic term associated with $A_+$. The equations of motion, $\ptl^+ F_{+ -}^A = -J_-^A$, give the constraint\footnote{Hereafter, we will drop $-$ subscripts.} \begin{equation}
	\boxed{A_+^A =  \frac{1}{\p_-^2} J_-^A.} \label{eq:AplusConstraint}
\end{equation} We therefore see that the effect of the gauge field was simply to generate a nonlocal Coloumb-type potential between sources. Computing matrix elements involving gauge fields then amounts to computing nonlocal $\frac{1}{\ptl^2}$ matrix elements, similar in spirit to the fermion mass term we encountered in the previous section. Note that the above constraint is quite general in that we have been agnostic about the matter content of the theory residing in $J$. In this sense, it is a generic feature of gauge fields in LCT in two dimensions.

%%%%%%%%%%%%%%%%%%%%%%%%%%%%%%%%%%%%%%%%%%%%%%%%%%%%%%%%%%%%%%%%%%%%%%%%%%%%%

\subsection{2d Massless QCD} \label{sec:2dmasslessqcd}

Our next step is to couple the gauge fields to external, dynamical degrees of freedom. For example, to obtain 2d QCD, we take the current in~\eqref{eq:AplusConstraint} to be the global symmetry current of the quark field $\Psi$ \begin{equation}
	J^A = g \overline{\Psi} \gamma_- T^A \Psi,
\end{equation} such that $\Psi$ transforms in the fundamental of SU($N_c$). The basis for $\gamma$ matrices is chosen to be \begin{equation}
	\gamma^+ = \gamma_- =\begin{pmatrix}
		0 & 0 \\
		\sqrt{2} & 0
	\end{pmatrix}, \quad\quad\quad \gamma^- = \gamma_+ = \begin{pmatrix}
		0 & \sqrt{2}\\
		0 & 0
	\end{pmatrix}, \quad\quad\quad \gamma^0 = \begin{pmatrix}
		0 & 1 \\
		1 & 0
	\end{pmatrix},
\end{equation} which satisfy \begin{equation}
	(\gamma^+)^2 = (\gamma^-)^2 = 0, \quad\quad\quad \{\gamma^+, \gamma^- \} = 2.
\end{equation} We now want to give dynamics to the fermions, with the Lagrangian given by \begin{equation}
	\Lcal_{\textrm{QCD}} = i \Psibar \slashed{D} \Psi - \half \Tr F_{\mu\nu}F^{\mu\nu},
\end{equation} where $D_\mu = \p_\mu - i g A_\mu$. Lightcone conformal truncation was first applied to this model  in \cite{Katz:2014uoa}, although it will be useful to rephrase some of those results in terms of the langauge we have developed in previous sections. To simplify the above equation, we can write $\Psi = \frac{1}{2^{1/4}}\binom{\psi}{\chi}$ in terms of left- and right-moving fields $\psi$ and $\chi$, exactly as we did in section \ref{sec:Fermions}. Since we have already worked out the kinetic term for fermions in section \ref{sec:Fermions}, we will not repeat the procedure here. After integrating out $\chi$ and the gauge field, we obtain the lightcone Hamiltonian \begin{equation}
	\boxed{ P_+^{(\textrm{QCD})} = -\frac{g^2}{2}  \int dx^- \psi^\dagg T^A \psi \frac{1}{\ptl^2} \psi^\dagg T^A \psi. } \label{eq:2dQCDPPlus}
\end{equation}  Note that we have suppressed the vector index on the fermions, but it should be understood that, e.g., $\psi^\dagg T^A \psi =( \psi^\dagg)_{i} (T^A)^{ij} \psi_j$.

Our conventions for the field $\psi$ will mirror those of section \ref{sec:Fermions}, with the inclusion of the complex conjugate of $\psi$: \begin{equation}
		\psi_j(x) = \int \frac{dp}{\sqrt{8\pi^2}} \left[{e^{-ipx} b_{j}(p) + e^{i p x}a_{j}^\dagg(p)}\right], \quad
		\psi_j^\dagg(x) =  \int \frac{dp}{\sqrt{8\pi^2}}\left[ {e^{ipx} b^\dagg_{j}(p) + e^{-i p x}a_{j}(p)} \right], \label{eq:complexfermionmodefunctions}
\end{equation} where the creation and annihilation operators satisfy 
\begin{equation}
	\{ a_i(q), a^\dagg_j(p) \} = \delta_{ij} (2\pi)\delta(q - p) , \quad
		\{ b_i(q), b^\dagg_j(p) \} = \delta_{ij} (2\pi)\delta(q - p).
\end{equation} 
We take the two point function to be \begin{equation}
	\corr{\ptl^{k} \psi_i^\dagg(x) \ptl^{k'} \psi_j(y)} =\frac{\Gamma(k+k'+1)}{4\pi(x-y)^{k+k'+1}}  \delta_{ij} . \label{eq:gen2ptfunctionQCD}
\end{equation}

\subsection{Constructing a Finite $N_c$ Basis} \label{sec:constructfiniteN}

The first step to analyzing 2d QCD is to work out the free fermion basis at finite $N_c$ and the matrix elements corresponding to the mass and interaction terms. Let us start with the basis, where our approach will mirror that of section \ref{sec:Fermions}. However, there are two differences that must be accounted for: 1) the field $\psi$ that appears in~\eqref{eq:2dQCDPPlus} is a \textit{complex} fermion and 2) $\psi$ is in the fundamental of $\textrm{SU}(N_c)$, so our basis states (and consequently, the spectrum) will depend on $N_c$.  We will restrict to the sector of the theory with zero baryon number.  Then, our basis is entirely comprised of color singlet operators of the form \cite{Katz:2014uoa} 
\begin{equation}
\label{eq:qcdMonomial}
	\begin{aligned}
		\ket{\Ocal,p} &\sim \frac{1}{N_{\Ocal}}\int  dx \, e^{-i p x} \\
		&\times \sum_{\Kvec} C_{\Kvec} \left( \ptl^{k_{11}} \psi_{i_1}^\dagg \ptl^{k_{21}} \psi_{i_1} \right) \left( \ptl^{k_{12}} \psi_{i_2}^\dagg \ptl^{k_{22}} \psi_{i_2} \right) \dotsb \left( \ptl^{k_{1n}} \psi_{i_n}^\dagg \ptl^{k_{2n}} \psi_{i_n} \right)(x) \ket{\textrm{vac}},
	\end{aligned}
\end{equation} 
for some coefficients $C_{\Kvec}$, and we use the shorthand notation $\d^{\kvec_1}\psi^\dagger\d^{\kvec_2}\psi$ for the product of $\ptl^{k_{1j}} \psi_{i_j}^\dagg \ptl^{k_{2j}} \psi_{i_j}$. Note that we have restored the SU($N_c$) index on $\psi$ for clarity; every pair of $\psi^\dagg \psi$ contracts these indices amongst themselves.

The $N_c$ dependence of the states can be worked out from the inner product. For example, consider the simplest four particle primary  \begin{equation}
	\ket{( \psi^\dagg  \psi)^2, p} = \frac{1}{N_{( \psi^\dagg  \psi)^2}} \int dx\, e^{-i p x} \left[ \psi^\dagg  \psi\right]^2 (x) \, \ket{\textrm{vac}}.
\end{equation} It is easy to work out the norm of this state; the SU($N_c$) index contractions give \begin{equation}
	\begin{aligned}
		\frac{\inner{( \psi^\dagg  \psi)^2, p'}{( \psi^\dagg  \psi)^2, p}}{2p (2\pi) \delta(p-p')} &= \frac{1}{|N_{( \psi^\dagg  \psi)^2}|^2} \frac{p^2 N_c(N_c-1)}{768\pi^3},
	\end{aligned}
\end{equation} which forces the norm of this state to be (in units where $p = 1$) \begin{equation}
	N_{( \psi^\dagg \psi)^2} = \frac{\sqrt{N_c(N_c-1)}}{16\sqrt{3}\pi^{3/2}}.
\end{equation} Note that when $N_c = 1$, this state is no longer a part of our basis, since fermion statistics would force the state to vanish. In other words, the dimensionality of the LCT basis changes as a function of $N_c$. Determining the complete basis thus requires accounting for the $N_c$ dependence of the inner products. For the simple example above, this was manifest in the overall normalization. However, for higher particle states or higher $\Delta_{\textrm{max}}$ states that are admixtures of monomials, the inner products can be more involved. Consequently, the dependence on $N_c$ will be more nontrivial, but it is straightforward to keep track of the index contractions and generalize the inner product to any number particles.

Now that we have addressed the color dependence, all that remains to construct the basis is to enumerate the primary operators and orthogonalize them. The first step can be accomplished using the method given in section \ref{sec:simplestcode} of conglomerating lower level primaries to form new primaries. We can therefore repeatedly use~\eqref{eq:JoaoFormula} to generate all primaries, starting with the lowest level primaries $A = \psi^\dagg$ and $B =  \psi$. For example, to build four particle operators at level $2$ using~\eqref{eq:JoaoFormula}, we schematically have \begin{equation}
	\begin{aligned}
		\left( \psi^\dagg \overset{\leftrightarrow}{\ptl}^2  \psi \right) \left(  \psi^\dagg  \psi \right), \quad\quad \left(  \psi^\dagg \overset{\leftrightarrow}{\ptl}   \psi \right)  \overset{\leftrightarrow}{\ptl} \left(  \psi^\dagg  \psi \right), \quad\quad  \left(  \psi^\dagg  \psi \right)\overset{\leftrightarrow}{\ptl} \left(  \psi^\dagg \overset{\leftrightarrow}{\ptl}  \psi \right) \nonumber,
	\end{aligned}
\end{equation} where the directional derivative is shorthand for the sum in~\eqref{eq:JoaoFormula}. Finally, in order to orthogonalize these states, we compute the necessary inner products using Wick contractions, as outlined above. For example, the first few orthonormal primaries for $\Delta_{\textrm{max}} \le 3$, a test case that we will use momentarily for matrix elements, are listed in Table \ref{table:LowDmaxFiniteNc}.

\noindent\rule[0.5ex]{\linewidth}{1pt}
\footnotesize

Let us make a brief parenthetical comment about two particle states in this theory. For two particle states, it turns out that the $N_c$ dependence is simple: it only appears in the overall norm of the state and it is $\propto N_c^{-\frac{1}{2}}$. Therefore, it will sometimes be more convenient to represent them as Jacobi polynomials in momentum space, as we did in~\eqref{eq:twoparticleJacobis}. For higher particle states, primaries do not generically factorize into a $N_c$ dependent coefficient and a Jacobi polynomial (though they will be linear combinations of such terms). We can write two particle states as\begin{equation}
\begin{aligned}
	\ket{\cO_\ell, p } &\equiv \frac{1}{N_\ell} \int \frac{dp_1 dp_2}{8\pi^2} (2\pi) \delta(p-p_1-p_2) F_{\ell}(p_1,p_2) b^\dagg_i (p_1) a^\dagg_i (p_2) \ket{\textrm{vac}},
 \nn \\
 F_{\ell}(p_1,p_2) &\equiv \sqrt{2\ell+1} \,
(p_1+p_2)^\ell P_\ell^{(0,0)}
\pr{\frac{p_1-p_2}{p_1+p_2}}, \qquad N_\ell = \frac{ p^{\ell }}{4} \sqrt{\frac{N_c}{\pi}}. \label{eq:2pfiniteNcJacobis}
\end{aligned}
\end{equation} 

\normalsize
\noindent\rule[0.5ex]{\linewidth}{1pt}

\renewcommand{\arraystretch}{1.3}
\begin{table}[t!]
\begin{center}
\begin{tabular}{| @{\hskip .1in}c | c@{\hskip .1in} | c@{\hskip .1in} | c@{\hskip .1in} |}
\hline & & & \\[-9pt]
\hspace{0.1pt} $n$ \hspace{0.1pt} & \hspace{0.1pt} $\De=1$ \hspace{0.1pt} & \hspace{0.1pt} $\De=2$ \hspace{0.1pt} & \hspace{0.1pt} $\De=3$ \hspace{0.1pt}  \\ \hline
2 & $ 4 \sqrt{\frac{\pi}{N_c}}  \psi^\dagg \psi $ & \hspace{0.1pt} $ 4 \sqrt{\frac{3\pi}{N_c}} (\ptl \psi^\dagg \psi - \psi^\dagg \ptl \psi)$ \hspace{0.1pt} & \hspace{0.1pt} $ 4 \sqrt{\frac{5\pi}{N_c}}(\psi^\dagg \ptl^2 \psi - 4 \cdot \ptl \psi^\dagg \ptl \psi + \ptl^2 \psi^\dagg \psi)$ \hspace{0.1pt}\\ \hline
4 & 0 & $\frac{16\sqrt{3} \pi^{3/2}}{\sqrt{N_c(N_c-1)}}(\psi^\dagg \psi)^2 $ & $\frac{32\sqrt{15}\pi^{3/2}}{\sqrt{N_c(N_c-1)}} (\psi^\dagg \psi \psi^\dagg \ptl \psi - \psi^\dagg \psi \ptl \psi^\dagg \psi )$ \\ \hline
6 & 0 & 0 & $ \frac{128 \sqrt{5} \pi^{5/2}}{\sqrt{N_c(N_c-1)(N_c-2)}}(\psi^\dagg \psi)^3$  \\ 
\hline
\end{tabular}
\end{center}
\caption{Orthonormal basis of primary operators (in position space) for finite $N_c$ up to $\Dmax=3$. 
}
\label{table:LowDmaxFiniteNc}
\end{table}

%%%%%%%%%%%%%%%%%%%%%%%%%%%%%%%%%%%%%%%%%%%%%%%%%%%%%%%%%%%%%%%%%%%%%%%%%%%%%

\subsection{Gauge Interaction Matrix Elements}\label{sec:gaugeintmatrixelements} Let us now outline the computation of the gauge interaction matrix elements. The basic idea is quite similar to the $\frac{1}{\ptl}$ matrix elements we computed for the fermion mass term in . Just like with the $\frac{1}{\ptl}$ operator in the mass term, we can define $\frac{1}{\ptl^2}$ in both position space and in momentum space, which equates to computing the matrix elements via Wick contractions or Fock space, respectively. In position space, the definition of $\frac{1}{\ptl^2}$ is that\footnote{Equivalently, we can define it as a double integral \begin{equation}
	\frac{1}{\ptl^2} f(x) = \int^{x} dx' \int^{x'} dx'' f(x''). 
\end{equation}} \begin{equation}
	\int dx f(x) \frac{1}{\ptl^2}g(x) \quad \rightarrow \quad \int dx dy f(x) |x - y| g(y). \label{eqn:1d2kernel}
\end{equation} Intuitively, this is because the Coloumb potential in momentum space gives rise to a linear confining potential in position space. Formally, we can allow for constants of integration, but those constants can be fixed by demanding the appropriate boundary condition, which we will elaborate on later.

With this definition in mind, we can consider interaction matrix elements for generic building block correlators \begin{equation}
	\begin{aligned}
		G_{\kvec\kvec'}(x,y,y',z) = \sum_{a,b,a',b'} \frac{\tilde A_{\Kvec,\Kvec'}^{(a,b,a',b')}}{(x-y)^a (y-z)^b (x-y')^{a'} (y'-z)^{b'}(x-z)^c},
	\end{aligned}
\end{equation} 
where the sum runs through all possible ways to Wick-contracting the spectator fermions and active part that contracts with the middle operator, and the powers $a,b,a',b'$ are determined by the the active fermions.
The power $c$ is not independent and can be fixed in terms of scaling $c = \Delta+\Delta'+2-a-b-a'-b'$, where $\Delta$ and $\Delta'$ are the scaling dimensions of external states. $\tilde A_{\kvec,\kvec^\prime}^{(a,b,a',b')}$ is the product of constants from the two-point functions in~\eqref{eq:gen2ptfunctionQCD}, which consists of signs from permuting fermions past each other, color tensors, $\Gamma$ functions and $4\pi$ factors.
The above correlator can be seen as a building block of the interaction matrix element. For example, the 2-to-2 monomial correlator includes a linear combination of such terms: \begin{equation}
	\begin{aligned}
		&\corr{\ptl^{k_1} \psi^\dagg \ptl^{k_2}\psi(x) \psi^\dagg T^A \psi (y) \psi^\dagg T^A \psi (y') \ptl^{k_1'} \psi^\dagg \ptl^{k_2'}\psi(z) } \\
		&\propto (N_c^2-1) \bigg[\frac{k_1! k_2! k_1'! k_2'!}{(x-y)^{k_1 + 1}(y-z)^{k_2'+1} (x-y')^{k_2+1} (y'-z)^{k_1'+1} }  + \dotsb \bigg], \label{eqn:twototwoactivecorr}
	\end{aligned}
\end{equation} where $\dotsb$ indicates additional terms that arise from Wick contractions and where we made use of the identity \begin{equation}
	(T^A)_{k\ell} (T^A)_{mn} = \half\pa{\delta_{kn}\delta_{\ell m} - \frac{1}{N_c} \delta_{k\ell}\delta_{mn}}. \label{eq:sunidentity}
\end{equation}

We therefore have to consider matrix elements of the form \begin{equation}
	\begin{aligned}
		\frac{\Mcal_{\kvec\kvec'}}{2p(2\pi)\delta(p-p')} \equiv \frac{1}{2p}\int_{-\infty}^{\infty} dx dz e^{i p x - i p' z} \int_{-\infty}^{\infty} dy' |y'| G_{\kvec\kvec'}(x,0,y',z). \label{eq:wickcontractgaugeME}
	\end{aligned}
\end{equation} Here we encounter a technical subtlety: the matrix elements in (\ref{eq:wickcontractgaugeME}) are generally divergent (e.g.~when the external operator is $\psi^\dagg \psi$).

Conceptually, canceling this divergence amounts to accounting for a self-energy contribution to the Hamiltonian which we have naively thrown away. This term is easiest to see in Fock space, where after normal ordering, the interaction term has a contribution \begin{equation}
	\begin{aligned}
		\delta P_+^{(g)} \supset g^2 (T^A)_{ij} (T^A)_{k\ell} \cdot \delta_{i\ell} \int_0^\infty dp \left[ a^\dagg_k (p) a_j(p) + b^\dagg_k(p) b_j (p) \right] C(p),
	\end{aligned}
\end{equation} where \begin{equation}
	C(p) \equiv \int_0^\infty dp' \left[ \frac{1}{(p+p')^2} - \frac{1}{(p-p')^2} \right].
\end{equation} If we use the principal value prescription, $C(p) = -\frac{2}{p}$, so the self energy term can be seen as a renormalization of the bare mass. However, other schemes can give rise to a divergent contribution, such that the divergence cancels the one appearing in the interaction term in~\eqref{eq:wickcontractgaugeME}. The equivalent position space treatment is obtained by Fourier transforming. It is important to note, though, that as long as a particular scheme is fixed for the self-energy term, it cannot enter into physical observables such as the masses of bound states. Therefore, it amounts to a self-consistent prescription to extract finite matrix elements. In this case, the consistency condition is fixed by the existence of chiral symmetry in the massless limit (that is, the existence of massless non-interacting sector in the theory).\footnote{The chiral symmetry in this theory would be anomalous if the chiral symmetry were gauged, and so the presence of a massless particle (even at finite $N_c$) is required by  anomaly matching. }

In the interest of pedagogy, we defer the treatment of this technical subtlety to Appendix \ref{app:appendixgauge}. The resulting expressions for the matrix elements can be found in~\eqref{eq:gaugesummaryeqFirst} through~\eqref{eq:gaugesummaryeqLast}.

%%%%%%%%%%%%%%%%%%%%%%%%%%%%%%%%%%%%%%%%%%%%%%%%%%%%%%%%%%%%%%%%%%%%%%%%%%%%%

\subsection{Example at Large $N_c$}\label{sec:largeNexample} Now that we have the QCD Hamiltonian and matrix elements, let us first study the theory at infinite $N_c$ with fixed  `t Hooft coupling $\lambda \equiv g^2 N_c$. To obtain the single-meson spectrum at infinite $N_c$, we can restrict to LCT two-particle states. To start with, we can choose a small $\Delta_{\textrm{max}} = 4$ cutoff (alternatively, one can use the Jacobi representation in~\eqref{eq:2pfiniteNcJacobis} and restrict to $\ell \le 3$).

In this case, the Hamiltonian takes the form \begin{equation}
	\begin{aligned}
		\Mcal = \frac{\lambda}{\pi} \begin{pmatrix}
			0 & 0 & 0 & 0 \\
			0 & 6 & 0 & \sqrt{\frac{7}{3}} \\
			0 & 0 & 15 & 0 \\
			0 & \sqrt{\frac{7}{3}} & 0 & \frac{77}{3} 
		\end{pmatrix}. \label{eq:LCTLargeNmatrix}
	\end{aligned}
\end{equation} In units of $\lambda/\pi$, the mass matrix has eigenvalues $0$, $\frac{1}{6}(95 \mp \sqrt{3565})$, and $15$. The first corresponds to the aforementioned massless sector (since we are working in the limit $m \to 0$, chiral symmetry is exactly preserved). The fact that this massless mode exists even at finite truncation is nontrivial and reflects the fact that the truncation itself does not break the chiral symmetry. The second eigenvalue is the first massive meson, at $\mu_1^2 \approx 5.882$. 

Let us compare this prediction from LCT to the more familiar large $N_c$ analysis of 2d QCD, originally presented in \cite{tHooft:1974pnl}. As explained in \cite{tHooft:1974pnl}, one obtains a Bethe-Salpeter equation for the meson wavefunction by resumming an infinite set of ladder diagrams that contribute to the meson propagator. The end result is the `t Hooft equation \begin{equation}
	\mu^2 \phi(x) = - \frac{\lambda}{\pi} \int_0^1 dy \frac{\phi(y)-\phi(x)}{(x-y)^2} \label{eq:tHooftequation}
\end{equation} where we have taken the massless limit. $\phi(x)$ is the wavefunction for a meson with momentum fraction $x=p_1/p$, and $\mu$ is the meson mass. Choosing some basis for these wavefunctions $\phi(x) \approx \phi_n(x)$, the Hamiltonian is just given by \begin{equation}
	H_{mn}^{(\textrm{`t Hooft})} = \frac{\lambda}{2\pi} \int_0^1 dx \int_0^1 dy \frac{[\phi_m(x)-\phi_m(y)][\phi_n^*(x)-\phi_n^*(y)]}{(x-y)^2}. \label{eq:thooftHamiltonian}
\end{equation}

\eqref{eq:thooftHamiltonian} can be diagonalized in a number of ways. One canonical choice of basis consists of sine and/or cosine functions. For example, consider a basis of cosines\footnote{Note that these wavefunctions satisfy the correct boundary conditions $\phi_n(x = 0) = \phi_n(x=1)= \textrm{const.}$, as is expected in the massless limit. It is possible to choose a basis of functions with the ``wrong'' boundary condition. However, the answer will not converge; see Appendix \ref{app:appendixgauge} for examples.} \begin{equation}
	\phi_n(x) = \sqrt{2}\cos(n\pi x ), \quad\quad n = 0,1,2 \dots,
\end{equation} normalized such that \begin{equation}
	\int_0^1 dx \phi_m(x) \phi_n(x) = \delta_{m n}.
\end{equation}  The resulting expression for the matrix elements, which can be found in appendix \ref{app:appendixgauge}, can be tabulated for very large truncation parameter. Keeping 1000 odd states with the cosine basis, we find that the lowest eigenvalue is $\mu_1^2 \approx 5.8817 \cdot \lambda/\pi$. Remarkably, this value agrees with the eigenvalue obtained from the just the $ 4 \times 4$ matrix in~\eqref{eq:LCTLargeNmatrix} to within $0.007$ \%! 

We can examine the convergence more closely by computing the LCT matrix elements at larger truncation.  For this infinite $N_c$, two-particle sector, we can independently compute the Hamiltonian matrix elements using~\eqref{eq:thooftHamiltonian} with the LCT momentum space wavefunctions:
\be
\phi_\ell(x) = \sqrt{2\ell+1} P_\ell(1-2x).
\ee
 Then the result for~\eqref{eq:thooftHamiltonian} can be evaluated in closed form.  It is zero when $\ell+\ell'$ is odd, and 
 \begin{equation}
	\begin{aligned}
		H_{\ell\ell'}^{\textrm{`t Hooft, LCT}} = \frac{2\lambda}{\pi}  \sqrt{(2\ell+1)(2\ell'+1)} \left[H_{\frac{\ell_{\textrm{max}}-1}{2}} + H_{\frac{\ell_{\textrm{max}}}{2}} - H_{\frac{\ell_{\textrm{max}}-\ell_{\textrm{min}}-1}{2}} - H_{\frac{\ell+\ell'}{2}} \right], \label{eq:largeNLCTmatrixelements}
	\end{aligned}
\end{equation}
 when $\ell+\ell'$ is even, where $H$ is the harmonic number.

\begin{figure}[t!]
\begin{center}
\includegraphics[width=0.75\textwidth]{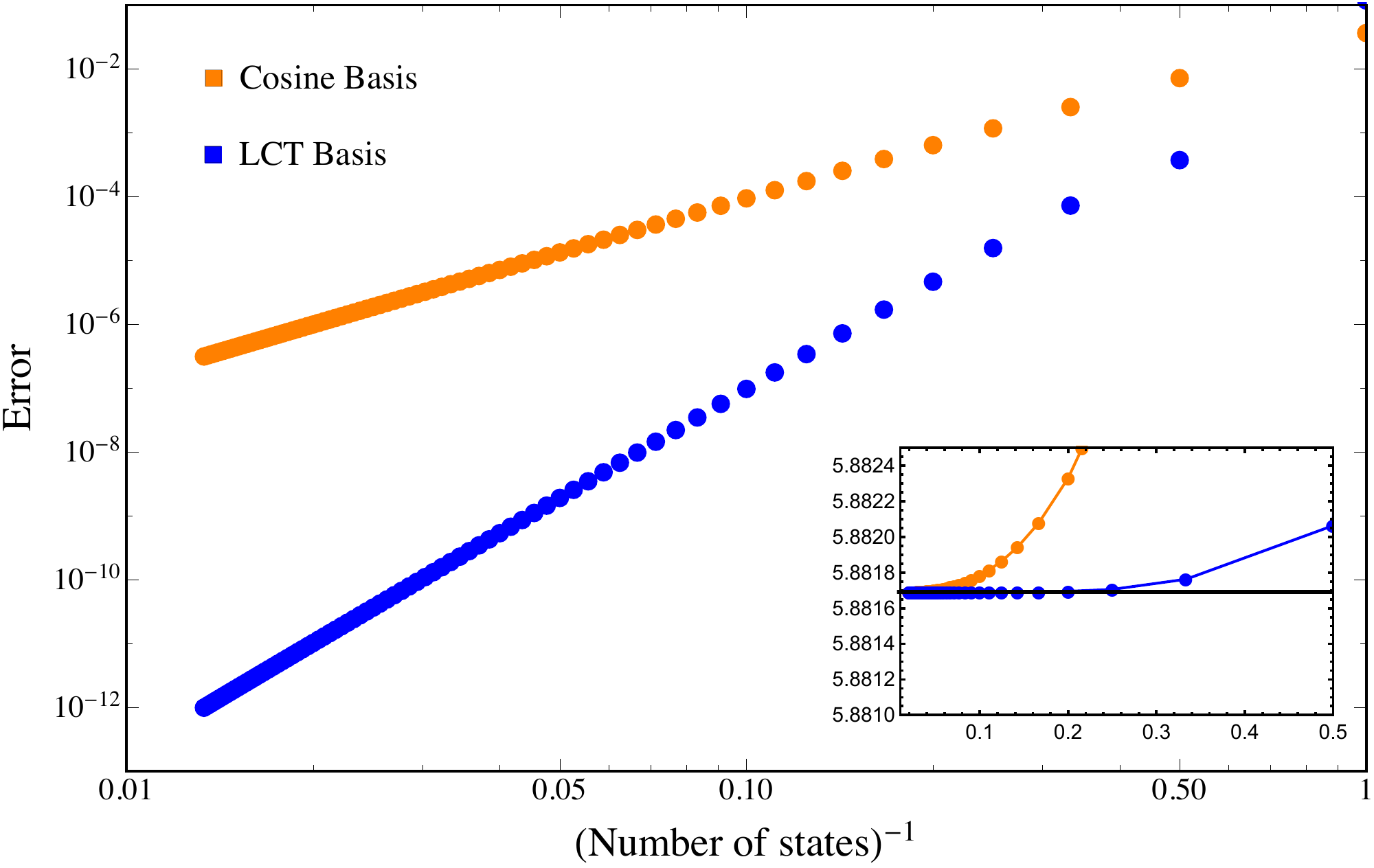}
\caption{The error relative to a baseline eigenvalue, corresponding to the mass of the first excited meson state, as a function of the size of the basis for both cosine (orange) and LCT (blue) bases. The baseline is computed from diagonalizing the mass matrix using the LCT basis with $\ell_{\textrm{max}} = 1000$. For  LCT, the size of the basis corresponds to the maximum scaling dimension of the operators. The inset shows the convergence to the baseline value (black).}
\label{fig:LargeNConvergence} 
\end{center}
\end{figure}

 In Fig.~\ref{fig:LargeNConvergence}, we show the comparison between the cosine basis and the LCT basis for the lowest eigenvalue. We choose the baseline to be the lowest eigenvalue obtained from dialing the LCT truncation to a large value at $\ell_{\textrm{max}} = 1000$. 
 We can plot the convergence to this value, as a function of truncation size using both LCT and the cosine basis. We can see that while both methods converge relatively quickly as a function of the truncation parameter, LCT estimates the lowest eigenvalue with an error of $\sim 10^{-8}$ already at $\ell_{max} = 10$ compared to $\sim 10^{-5}$ for the cosine basis. Moreover, the matrix elements in~\eqref{eq:largeNLCTmatrixelements} have a simple analytic form, while the matrix elements for the cosine basis involve more complicated sine and cosine integral functions.

%%%%%%%%%%%%%%%%%%%%%%%%%%%%%%%%%%%%%%%%%%%%%%%%%%%%%%%%%%%%%%%%%%%%%%%%%%%%%

\subsection{Example at Finite \texorpdfstring{$N_c$}{N_c}}\label{sec:finiteNexample}

\begin{figure}[t!]
\centering
\includegraphics[width=0.75\textwidth]{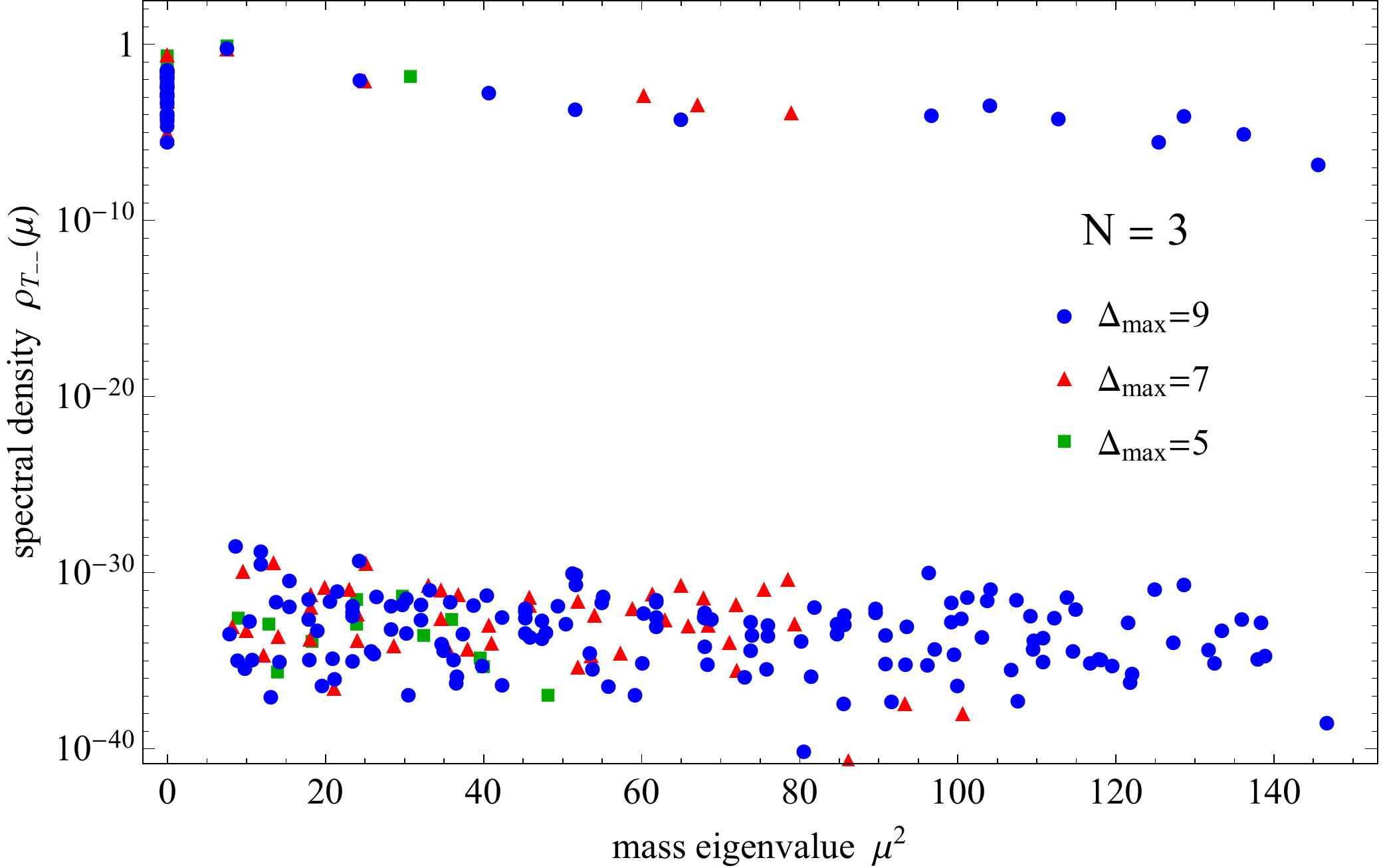}
\caption{\label{fig:gauge-d9-spectral-density} The spectral density of the stress tensor $T_{--}$ at $\Dmax=9,7,5$ and $N_c=3$.
The spectrum of stable massive particles are dressed by massless multiparticle states. The spectral density of $T_{--}$ clearly distinguishes the massive particle spectrum ($\sim 1$ to $\sim 10^{-10}$) from the massless dressing ($\lesssim10^{-30}$). Moreover, focusing on the spectrum of massive particles, we see that the spectrum of the two lowest stable particles have converged.
}
\end{figure}

Let us now turn to an example at finite $N_c$. Using the states in Table \ref{table:LowDmaxFiniteNc} and the matrix elements listed in Appendix \ref{app:appendixgauge}, we find that the gauge interaction matrix is
\begin{align}
\Mcal = \frac{\lambda}{\pi}
\times\frac{N_c^2-1}{N_c^2}\left(
\begin{array}{cccccc}
 0 & 0 & 0 & 0 & 0 & 0 \\
 0 & 6 & 0 & -\frac{6}{\sqrt{N_c-1}} & 0 & 0 \\
 0 & 0 & 15 & 0 & -\frac{5 \sqrt{3}}{\sqrt{N_c-1}} & 0 \\
 0 & -\frac{6}{\sqrt{N_c-1}} & 0 & \frac{6}{N_c-1} & 0 & 0 \\
 0 & 0 & -\frac{5 \sqrt{3}}{\sqrt{N_c-1}} & 0 & 10-\frac{5}{N_c-1} & -\frac{10 \sqrt{3} \sqrt{N_c-2}}{N_c-1} \\
 0 & 0 & 0 & 0 & -\frac{10 \sqrt{3} \sqrt{N_c-2}}{N_c-1} & \frac{30}{N_c-1} \\
\end{array}
\right) \, . \label{eq:gaugeintexamplehamiltonian}
\end{align}
The eigenvalues are $0, 0, 0, 6\pr{1+\frac{1}{N_c}}, 10\pr{1+\frac{1}{N_c}}, 15\pr{1+\frac{1}{N_c}}$ in the unit $\lambda/ \pi$. We can identify the massless pion and two massive mesons with masses $\mu_1^2 = 6\pr{1+\frac{1}{N_c}}$ and $\mu_2^2 = 15\pr{1+\frac{1}{N_c}}$.  The rest of the eigenstates are the multi-meson states. We can already see at this stage the simplicity of LCT and the power of this method to extract observables such as the mass spectrum \textit{as a function of $N_c$}. Furthermore, the matrix in~\eqref{eq:gaugeintexamplehamiltonian} is almost diagonalizable by hand!

We can take the basis at larger $\Dmax$ and compute the meson spectrum numerically. We can take the truncation up to $\Dmax=9$ where the basis contains $\approx 200$ states. The precise number of orthogonal states depends on $N_c$. The spectrum contains the single particle states as well as the continuum of higher particle states, and we would like to identify the low meson mass spectrum. A useful observable is the $T_{--}$ spectral density shown in Fig.~\ref{fig:gauge-d9-spectral-density}. At large $N_c$, the stress tensor creates stable single-meson states with parity even. At finite $N_c$, heavy mesons may decay into light mesons, and $T_{--}$ may create multi-meson states. However, in 2d, massless particles cannot have interactions \cite{Coleman:1973ci}, and so the massless mesons appear in $T_{--}$ only through the free quadratic term $\sim (\p_- \pi)^2$, which doesn't overlap with $\mu > 0$ states.
Hence, states of massless mesons with nonzero center-of-mass energy never contribute to $T_{--}$ spectral density. 
Therefore, the $T_{--}$ spectral density can be used to identify the single-particle light meson states, where they appear as isolated poles in the spectral density until the multi-particle continuum of the lightest {\it massive} states appears.
\begin{figure}[t!]
\centering
\includegraphics[width=0.72\textwidth]{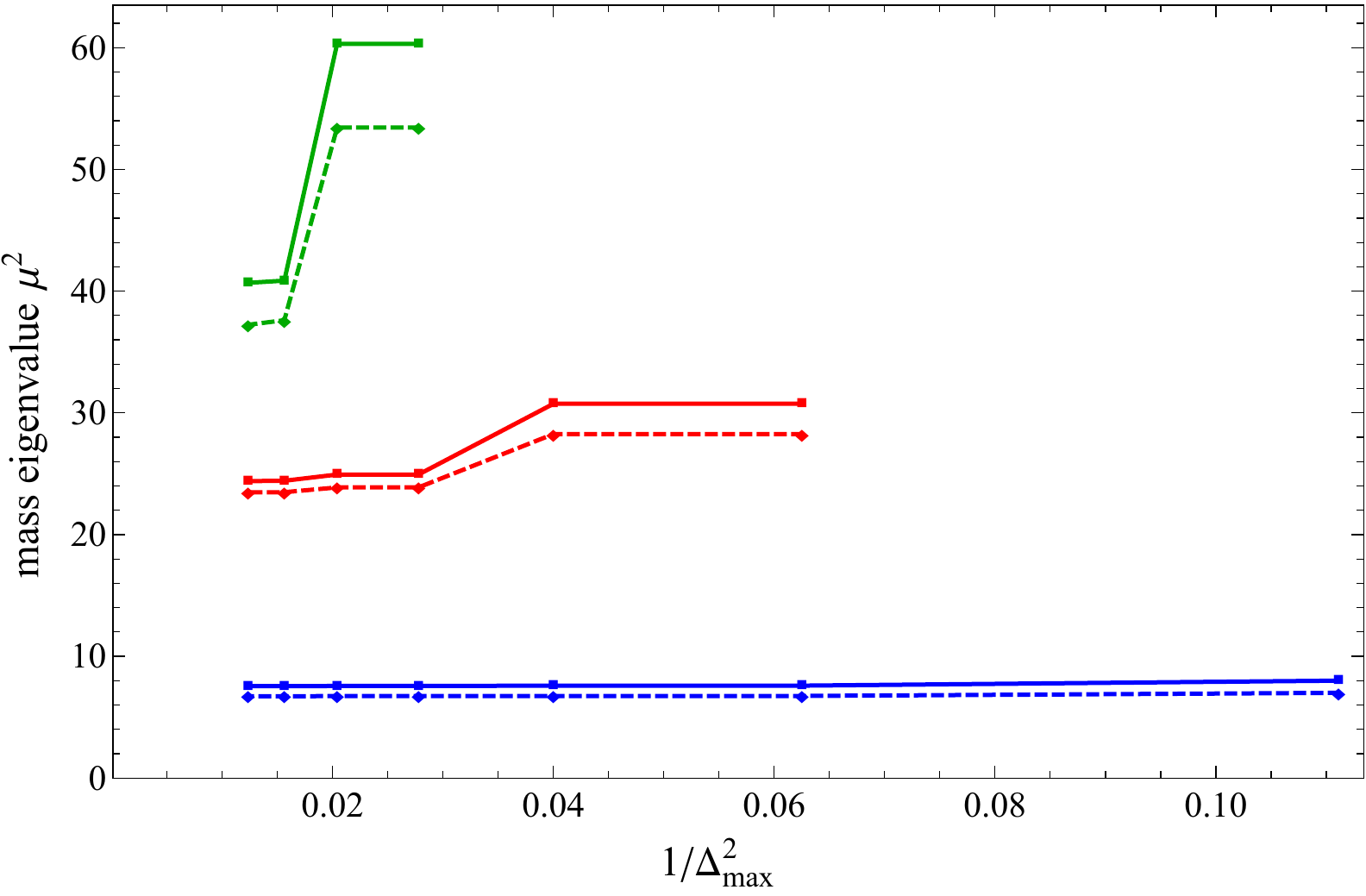}
\caption{\label{fig:gauge-d9-data} The spectrum of parity-even single particles at finite $N$. The solid curve is the $N_c=3$ spectrum and the dashed curve is $N_c=6$. }
\end{figure}
In Fig.~\ref{fig:gauge-d9-data}, we read off the beginning three massive meson states from the $T_{--}$ spectral density, and compare the mass eigenvalues at different $N_c$ and $\Dmax$. The meson mass converges rapidly as $\Dmax$ increases, and the lowest few states have already converged at $\Dmax = 4$.

%%%%%%%%%%%%%%%%%%%%%%%%%%%%%%%%%%%%%%%%%%%%%%%%%%%%%%%%%%%%%%%%%%%%%%%%%%%%%
%%%%%%%%%%%%%%%%%%%%%%%%%%%%%%%%%%%%%%%%%%%%%%%%%%%%%%%%%%%%%%%%%%%%%%%%%%%%%
%%%%%%%%%%%%%%%%%%%%%%%%%%%%%%%%%%%%%%%%%%%%%%%%%%%%%%%%%%%%%%%%%%%%%%%%%%%%%
\clearpage
\section*{\Large Part II: Advanced Improvements} 
\addcontentsline{toc}{part}{Part II: Advanced Improvements}
\label{sec:PartIIAdvImprovements}

%%%%%%%%%%%%%%%%%%%%%%%%%%%%%%%%%%%%%%%%%%%%%%%%%%%%%%%%%%%%%%%%%%%%%%%%%%%%%
%%%%%%%%%%%%%%%%%%%%%%%%%%%%%%%%%%%%%%%%%%%%%%%%%%%%%%%%%%%%%%%%%%%%%%%%%%%%%

\section{Radial Quantization for Scalars}
\label{sec:RadialScalars}

In Part I, we presented all of the steps involved in implementing lightcone conformal truncation and developed simple Mathematica code for applying this method to deformations of free field theory in 2d. 
The goal of Part II is to now significantly improve the computational efficiency of LCT, allowing us to reach higher values of $\Dmax$. To do so, we will capitalize on the CFT structure of the UV basis, making use of radial quantization methods to quickly compute two- and three-point functions. The methods presented here are the ones actually used in the publicly available code released with this paper. 

In this section, we focus on the application of these improved methods to scalar field theory, and in section~\ref{sec:RQFermions} we generalize these methods to include fermions.

%%%%%%%%%%%%%%%%%%%%%%%%%%%%%%%%%%%%%%%%%%%%%%%%%%%%%%%%%%%%%%%%%%%%%%%%%%%%%

\subsection{Motivation}
\label{sec:motivation}

Recall that we consider a QFT Hamiltonian as a deformation of a UV CFT Hamiltonian by some relevant operator(s) $\Ocal_R$, 
\be
H = H_{\text{CFT}} + g \, V = H_{\text{CFT}} + g \int dx\, \Ocal_R(x),
\label{eq:II:HQFT}
\ee
and we evaluate this Hamiltonian in a Hilbert space whose states are defined as Fourier transforms of primary operators from the UV CFT, 
\be
\ket{\cO,p} = \frac{1}{N_{\cO}} \int dx \, e^{-i p x} \, \cO(x) \vac, 
\ee
where $N_{\cO}$ is a normalization constant. It follows that inner products and Hamiltonian matrix elements between conformal truncation states are given by Fourier transforms of CFT two- and three-point functions, respectively,
\begin{align}
\begin{split}
\langle \Ocal, p | \Ocal^\prime, p^\prime \rangle & \equiv 2p(2\pi) \delta(p-p^\prime) \, G_{\cO\cO^\prime}  \label{eq:II:Inner} \\[10pt]
& = \frac{1}{N_{\cO}N_{\cO^\prime}}  \int dx \, dy \, e^{i(px - p^\prime y)} \langle \Ocal(x) \Ocal^\prime(y) \rangle.
\end{split} \\[10pt]
\begin{split}
\langle \Ocal,p | V |  \Ocal^\prime, p^\prime \rangle &\equiv 2p (2\pi) \delta(p-p^\prime)  \, \Mcal^{(\cO_R)}_{\cO\cO^\prime} \label{eq:II:ME} \\[10pt]
& = \frac{1}{N_{\cO}N_{\cO^\prime}}  \int dx \, dy \, dz\, e^{i(px - p^\prime z)} \langle \Ocal(x) \Ocal_R(y) \Ocal^\prime(z) \rangle.
\end{split}
\end{align}
In a CFT, two- and three-point functions of primary operators are completely fixed up to overall coefficients,\footnote{In Lorentzian signature, the phases in these expressions depend on the order of the operators.}
\begin{equation}
\<\Ocal(x_1)\Ocal'(x_2)\> \doteq \fr{\mathfrak{g}_{\Ocal\Ocal'}}{x_{12}^{2\De}}, \quad \<\Ocal(x_1) \Ocal_R(x_2) \Ocal'(x_3)\> \doteq \fr{C_{\Ocal\Ocal'\Ocal_R}}{x_{12}^{\De+\De_R-\De'} x_{23}^{\De'+\De_R-\De} x_{13}^{\De+\De'-\De_R}},
\label{eq:OPEdata}
\end{equation}
where $\mathfrak{g}_{\Ocal\Ocal'}$ is the Zamolodchikov metric and $C_{\Ocal\Ocal'\Ocal_R}$ is the OPE coefficient. 
Naively, we therefore only need to compute the coefficients $\mathfrak{g}_{\Ocal\Ocal'}$ and $C_{\Ocal\Ocal'\Ocal_R}$. However, we will often be interested in the case where $\Ocal_R$ is {\it not} primary; in particular, $\phi^2$ and $\phi^4$ are not primary operators in 2d.  Consequently, the three-point functions do not necessarily take the simple form above, and we must grapple with its more complicated dependence on positions.

Our general strategy for computing the two- and three-point functions is to first compute the position space correlators on the right-hand sides and then apply standard Fourier transform formulas. For free fields,  evaluating these correlation functions by Wick contractions was sufficient for the low truncation levels considered in Part I, but 
the rapid proliferation of contractions for correlators with many fields makes this strategy slow and inefficient for large $\Dmax$.

We can see the source of the inefficiency already at the level of monomial operators $\p^{\bk} \phi$. Recall from section \ref{sec:simplestcode} that monomials are our building blocks for primaries, with general primary operators being written as linear combinations of monomials,
\be
\Ocal(x) = \sum_{\bk} C^\Ocal_{\bk}\, \p^{\bk}\phi(x),
\label{eq:II:OMonExpand}
\ee
for some coefficients $C^\Ocal_{\bk}$. Thus, a two-point function of primaries takes the form
\be
\langle \Ocal(x) \Ocal^\prime(y) \rangle = \sum_{\bk} \sum_{\bk^\prime} C^\Ocal_{\bk} \, C^{\Ocal'}_{\bk'}\, \langle  \p^{\bk}\phi(x)  \, \p^{\bk^\prime}\phi(y) \rangle. 
\label{eq:II:Primary2PF}
\ee
The problem is that all of the monomial two-point functions appearing on the right-hand side are nonzero. That is, inside position space correlators all monomials ``talk" to each other, and each monomial correlator requires Wick contractions to compute. This basic problem only worsens for three-point functions, where additional operators in the middle also need to be Wick contracted. 

It is therefore desirable to have a more efficient method that does not involve this proliferation of contractions.  The method we will present in this section avoids this problem by computing the two- and three-point correlators using radial quantization, i.e.~using the radial direction as the time direction in the mode decomposition of fields.  The basic reason why radial quantization drastically decreases the number of contractions is that monomials $\p^{\bk} \phi$  are simply a tensor product of modes that evolve ``trivially'' from the origin $r=0$ (the infinite past in radial quantization) to $r=\infty$ (the infinite future), so in this scheme the two-point functions are diagonal in a monomial basis, in a sense that will be explained precisely in the next subsection.  Similarly, three-point functions are ``trivial'' except for the radial evolution through the interaction term $\Ocal_R$, and therefore are ``almost diagonal'' in monomial space.

\noindent\rule[0.5ex]{\linewidth}{1pt}
\footnotesize

As discussed in section~\ref{sec:WickContract}, we specifically need \emph{Lorentzian} CFT correlation functions to compute the inner products and matrix elements in eqs.~\eqref{eq:II:Inner} and \eqref{eq:II:ME}. However, when using radial quantization, we are technically computing \emph{Euclidean} correlators. We therefore need to analytically continue the resulting correlation functions to Lorentzian signature by taking the lightcone coordinate $x \ra ix$.\footnote{We must also include an $i\epsilon$ prescription in this analytic continuation to define a particular operator ordering for the resulting Lorentzian correlator, which will affect the Fourier transform to momentum space. These subtleties have already been accounted for in evaluating the Fourier transform formulas~\eqref{eq:FTFormulas}, and are discussed in detail in~\cite{Anand:2019lkt}.} For example, the Euclidean and Lorentzian correlators for the fermion $\psi$ are
\be
\<\psi(x)\psi(0)\>\Big|_{\textrm{Euclidean}} = \fr{1}{4\pi x} \quad \rightarrow \quad \<\psi(x)\psi(0)\>\Big|_{\textrm{Lorentzian}} = \fr{-i}{4\pi x}.
\ee

However, for ``monomial'' correlators involving $\p^{\bk}\phi$ or $\p^{\bk}\psi$, we also need to analytically continue the derivatives $\p \ra -i\p$. These two factors of $i$ for each derivative (one from analytically continuing the position, one from analytically continuing the derivative) cancel, such that the only relative phases between Euclidean and Lorentzian monomial correlators are those in the building block two-point functions with no derivatives.

Concretely, for monomial two-point functions we have the relation
\bq
\bal
\<\p^{\bk}\phi(x) \, \p^{\bk'}\phi(0)\>\Big|_{\textrm{Lorentzian}} &\equiv \<(i\p)^{\bk}\phi(ix) \,(i\p)^{\bk'}\phi(0)\>\Big|_{\textrm{Euclidean}} = \<\p^{\bk}\phi(x) \, \p^{\bk'}\phi(0)\>\Big|_{\textrm{Euclidean}}, \\
\<\p^{\bk^\dagger}\psi(x) \, \p^{\bk'}\psi(0)\>\Big|_{\textrm{Lorentzian}} &\equiv \<(i\p)^{\bk^\dagger}\psi(ix) \, (i\p)^{\bk'}\psi(0)\>\Big|_{\textrm{Euclidean}} = (-i)^n \<\p^{\bk^\dagger}\psi(x) \, \p^{\bk'}\psi(0)\>\Big|_{\textrm{Euclidean}},
\eal
\eq
with a similar relation for three-point functions. There is thus no relative phase for scalar correlators, and a relative factor of $(-i)^n$ for fermion correlators, where $n$ is the number of fermions in the monomial. Fortunately, these phases from analytic continuation simply contribute to the overall phases of basis states, and can safely be ignored, as discussed in appendix~\ref{app:Phases}. These phases are therefore removed in any $\doteq$ equation in the remainder of Part II.

\normalsize
\noindent\rule[0.5ex]{\linewidth}{1pt}

%%%%%%%%%%%%%%%%%%%%%%%%%%%%%%%%%%%%%%%%%%%%%%%%%%%%%%%%%%%%%%%%%%%%%%%%%%%%%

\subsection{Radial Quantization Modes}
\label{eq:RQModes}

Let us begin by quickly recapping some aspects of radial quantization. Recall that in radial quantization (see e.g. \cite{Ginsparg}), we expand $\p\phi$ as\footnote{We have dropped the singular term proportional to $\frac{1}{x}(a_0 + a_0^\dagger)$, since $a_0+a_0^\dagger$ annihilates the vacuum and will never contribute to any of our computations. Sometimes in the literature, $\p \phi$ is defined with a $-i$ on the RHS of (\ref{eq:dphidefRQ}); in the free theory all correlators are invariant under $\phi \rightarrow -\phi$ and so are independent of this choice of sign.} 
\begin{gather}
\p\phi(x) = \frac{i}{\sqrt{4\pi}} \sum_{k=1}^\infty \sqrt{k} \left( x^{-k-1}a_k + x^{k-1}a_k^\dagger    \right), \label{eq:II:RadialPPhi} \\[10pt]
[ a_{k}, a^\dagg_{k^\prime} ] = \delta_{k, k^\prime}.
\label{eq:dphidefRQ}
\end{gather}
Throughout this section, $a_k^\dagger$ and $a_k$ will always denote radial quantization creation and annihilation operators (not to be confused with Fock space modes!). A monomial at the origin acting on the radial quantization vacuum defines an ``in'' state 
\be
\p^{\bk}\phi(0) \vac = \Ncal_{\bk} \, a_{\bk}^\dagger \vac,
\ee
where 
\be
a_{\bk}^\dagger= a_{k_1}^\dagger \cdots a_{k_n}^\dagger, \hspace{10mm} \Ncal_{\bk} \doteq \left( \frac{1}{\sqrt{4\pi}} \right)^n \Gamma(k_1) \cdots \G(k_n) \sqrt{k_1\cdots k_n}.
\label{eq:II:radialscalarNcal}
\ee
A primary operator defines both out and in states via
\begin{eqnarray}
\Ocal(0) \vac &=& \sum_{\bk} C^\Ocal_{\bk} \, \Ncal_{\bk} \, a_{\bk}^\dagger \vac \label{eq:II:ket} \\[5pt]
\bra{{\rm vac}} \Ocal(\infty) &\equiv& \lim_{x\rightarrow 0} x^{-2\Delta} \bra{{\rm vac}} \Ocal(1/x) \doteq \sum_{\bk}  C^\Ocal_{\bk} \, \Ncal_{\bk} \bra{{\rm vac}} a_{\bk}. \label{eq:II:bra}
\end{eqnarray}

It is simple to see now how radial quantization provides a significant speedup in the computation of correlation functions. The key point is that in radial quantization, in and out states are orthogonal,
\be
\langle a_{\bk} \, a^\dagg_{\bk^\prime} \rangle= \norm{\bk}^2 \delta_{\bk, \bk'},
\label{eq:II:RadialInner}
\ee
where we have defined
\be
\norm{\bk}^2 \equiv \prod_{k \in \mathbb{N}} \verb|BC|_k! = \frac{n!}{\textrm{number of permutations of } \bk} , \quad \delta_{\bk ,\bk'} \equiv \delta_{k_1,k_1'} \dotsb \delta_{k_n, k_n'},
\label{eq:II:radialscalarknorm}
\ee
where for each $k$, `$\verb|BC|_k$' is the number, or `bincount', of times $k$ appears in $\bk$.
The Zamolodchikov metric $\mathfrak{g}_{\cO\cO^\prime}$ 
thus has the simple formula
\be
\mathfrak{g}_{\cO\cO^\prime} \equiv \corr{\cO(\infty)\cO^\prime(0)} \doteq \sum_{\bk} C^\Ocal_{\bk}\, C^{\Ocal'}_{\bk} \, \Ncal_{\bk}^2 \norm{\bk}^2,
\label{eq:II:Radial2PF}
\ee
and the general two-point function is given by (\ref{eq:OPEdata}). 
Clearly, evaluating two-point functions 
this way is simpler than using Wick contractions in the double sum (\ref{eq:II:Primary2PF}).

Radial quantization also provides speed ups for three-point functions for basically the same reasons. If we add an operator in the middle of (\ref{eq:II:RadialInner}), the matrix element $ \langle a_{\bk} \,\mathcal{O}_R\, a^\dagg_{\bk^\prime} \rangle$ will no longer be exactly diagonal, but it will still be \emph{almost} diagonal. The reason is that the middle operator will only contribute a handful of its own creation and annihilation operators, such that $\bk$ and $\bk^\prime$ can only differ by the mismatches caused by these new contributions. The general lesson is that matrix elements between radial quantization in and out states tend to be \emph{sparse}. This is the reason why radial quantization matrix elements are much easier to compute than their general position space counterparts. 

Before diving into applications in the following subsections, there is an important subtlety regarding the bra states defined in (\ref{eq:II:bra}) that requires comment. The subtlety is that the second equality in this formula only holds when $\Ocal$ is \emph{primary}, and is generally not true for non-primaries. In particular, it does not hold for individual monomial operators $\p^{\bk}\phi$, which are generally non-primary, \emph{i.e.}, in general 
\be
\bra{{\rm vac}} \p^{\bk} \phi(\infty) \ne \Ncal_{\bk} \bra{{\rm vac}} a_{\bk}.
\label{eq:II:MonSubtle}
\ee
The reason for this is that monomials and other non-primary operators do not transform homogeneously under inversions. Consequently, there will be additional terms on the right-hand side of (\ref{eq:II:MonSubtle}) involving other monomials. The point is that when we take linear combinations of monomials to form a primary $\Ocal$, all of these extraneous terms precisely cancel and we end up with (\ref{eq:II:bra}).\footnote{To see this, start with the invariance of correlation functions under conformal transformations:
\be
\< \CO(x) \dots \> = \left( \Omega^{-\Delta}(x') \dots\right) \< \CO(x') \dots \>,
\ee
where $x'$ is a conformal transformation of $x$, and $\Omega^{-d} (x')\equiv \left| \frac{\partial x'}{\partial x}\right|$ is the Jacobian of the coordinate transformation.  Then, write the primary operator $\CO$ as a sum over monomials:
\be
\sum_{\bk} C^\Ocal_{\bk}\<  \partial^{\bk} \phi(x) \dots \> = \sum_{\bk} C^\Ocal_{\bk} \left( \Omega^{-\Delta}(x') \dots\right)\< \partial^{\bk} \phi(x') \dots \>.
\ee
The individual terms in the sum on the RHS are exactly what we would get if we took the individual terms on the LHS and transformed them {\it as if } the monomials were primary operators themselves. Therefore, as long as we only use the primary operator transformation law on monomials when they appear in linear combinations that form primary operators, we will get the correct answer. 
}
To keep this subtlety manifest in our notation, we will write monomial in and out states in radial quantization as 
\begin{eqnarray}
\p^{\bk}\phi(0) \vac &=& \Ncal_{\bk}\, a_{\bk}^\dagger \vac,\label{monoket} \\[5pt] 
\bra{{\rm vac}} \p^{\bk} \phi(\infty) &\cong& \Ncal_{\bk} \bra{{\rm vac}} a_{\bk}. \label{monobra}
\end{eqnarray}
The first line is always true, while $\cong$ in the second line means that the equation can be used only when applied to all of the terms in a linear combination of monomials that adds up to a primary operator. 

Let us demonstrate some of these ideas with a simple example. Consider the two-particle primary operator that we introduced back in (\ref{eq:twopprimaryexample}): 
\begin{equation}
	\cO_{(2)}(x) = 6 \, \p^{\Kvec_1} \phi(x) - 9 \, \p^{\Kvec_2} \phi(x), \quad\quad\quad \Kvec_1 = (1,3), \quad \Kvec_2 = (2,2).
\end{equation} 
Individually, $\p^{\Kvec_1} \phi(x)$ and $\p^{\Kvec_2} \phi(x)$ are not primary, but as we saw in section \ref{sec:2dFFT}, this specific linear combination of monomials \textit{is} primary. For ket states, we simply have
\be
\begin{aligned}
\p^{\Kvec_1} \phi(0) \vac &= \CN_{\Kvec_1} a_{\Kvec_1}^\dagger \vac, \\[5pt]
\p^{\Kvec_2} \phi(0) \vac &= \CN_{\Kvec_2} a_{\Kvec_2}^\dagger \vac, \\[5pt]
\CO_{(2)}(0) \vac &= ( 6\, \CN_{\Kvec_1} a_{\Kvec_1}^\dagger - 9\, \CN_{\Kvec_2} a_{\Kvec_2}^\dagger  ) \vac.
\end{aligned}
\ee
On the other hand, using the mode expansion (\ref{eq:II:RadialPPhi}) and the definition (\ref{eq:II:bra}) for bra states, it is straightforward to work out that 
\be
\begin{aligned}
\bra{{\rm vac}} \p^{\Kvec_1}\phi(\infty) &\doteq \bra{{\rm vac}} \left( 13\, \CN_{\Kvec_1} a_{\Kvec_1} + 12\, \CN_{\Kvec_2} a_{\Kvec_2}  \right) \\[5pt]
\bra{{\rm vac}} \p^{\Kvec_2}\phi(\infty) &\doteq \bra{{\rm vac}} \left( 8\, \CN_{\Kvec_1} a_{\Kvec_1} + 9\, \CN_{\Kvec_2} a_{\Kvec_2}  \right) \\[5pt]
\bra{{\rm vac}} \CO_{(2)}(\infty) &\doteq \bra{{\rm vac}} \left( 6\, \CN_{\Kvec_1} a_{\Kvec_1} - 9\, \CN_{\Kvec_2} a_{\Kvec_2}  \right) 
\end{aligned}
\ee
We see explicitly that for the individual monomials $\<0|\p^{\bk} \phi(\infty) \ne \Ncal_{\bk} \<0| a_{\bk}$. However, for the primary operator $\CO_{(2)}$, the linear combination of monomials is just right to give the expected bra state. In other words, we can get away with pretending $\<0|\p^{\bk} \phi(\infty)$ is equal to $\Ncal_{\bk} \<0| a_{\bk}$ as long as we are dealing with linear combinations of monomials that are primary. This is the content of (\ref{monobra}). 

We will now use radial quantization to compute conformal truncation inner products and matrix elements in scalar field theory.

%%%%%%%%%%%%%%%%%%%%%%%%%%%%%%%%%%%%%%%%%%%%%%%%%%%%%%%%%%%%%%%%%%%%%%%%%%%%%

\subsection{Inner Products}
\label{sec:radialinnerscalar}

We have in fact already described in the previous subsection essentially all the steps necessary to compute two-point functions. To summarize, with the Zamolodchikov metric $\mathfrak{g}_{\cO\cO^\prime}$ defined in (\ref{eq:II:Radial2PF}), two-point functions of primary operators take the form in~\eqref{eq:OPEdata}. We can Fourier transform this two-point function using the general formula in~\eqref{eq:FTFormulas}. Finally, referring back to (\ref{eq:II:Inner}), the Gram matrix entry for the inner product between states created by $\Ocal$ and $\Ocal^\prime$ is
\be
G_{\cO\cO^\prime} \doteq  \frac{\pi p^{2h-2}  \, \mathfrak{g}_{\cO\cO^\prime}}{\Gamma(2h) N_{\cO} N_{\cO^\prime} }. 
\label{eq:II:InnerFinal}
\ee
Therefore, diagonalizing LCT states is \emph{equivalent} to diagonalizing primary operators according to the standard Zamolodchikov metric.

\subsection{Matrix Elements for $\phi^n$}
\label{sec:radialmassscalar}

Let us now turn to the evaluation of matrix elements. 
In this section, we consider an important class of deformations, $\phi^n$, and work out Hamiltonian matrix elements for these operators. To simplify the presentation, we will focus on the derivation of matrix elements for $\Ocal_R=\tfrac{m^2}{2}\phi^2$, which contains all of the ingredients needed to handle the general case. At the end of this section, we present matrix element formulas for general $\CO_R = \tfrac{\lambda}{n!}\phi^n$, but some of the intermediate steps are presented in appendix \ref{app:GeneratingFunctions}. 

Focusing on $\cO_R=\tfrac{m^2}{2}\phi^2$, we see from the formula (\ref{eq:II:ME}) for computing conformal truncation matrix elements that the position space correlator we need to compute is 
\be
G_{\cO\cO^\prime}^{(\phi^2)}(x,y,z) \equiv \corr{\cO(x) \phi^2(y) \cO^\prime(z)} = \sum_{\bk \bk^\prime} C^\Ocal_{\bk}\, C^{\Ocal'}_{\bk^\prime} \, G_{\bk\bk^\prime}^{(\phi^2)}(x,y,z). 
\label{eq:II:Primary3PF}
\ee
In the second equality above, we have expanded each primary operator in terms of monomials as in (\ref{eq:II:OMonExpand}) and defined the monomial correlator
\be
G_{\bk\bk^\prime}^{(\phi^2)}(x,y,z) \equiv \langle \p^{\bk}\phi(x) \phi^2(y) \p^{\bk^\prime}\phi(z) \rangle.
\label{eq:II:OPhi2O}
\ee

As discussed in detail in section \ref{eq:RQModes}, we can transform the operators $\p^{\bk} \phi$ as if they are primary as long as in our final expressions they appear only in linear combinations that form primary operators. However, the correlator above presents a new problem: $\phi$ (and hence also $\phi^2$) is not a primary operator in 2d. If it were primary, then we could immediately start applying conformal transformations to (\ref{eq:II:OPhi2O}) to map it to a radial quantization matrix element. Since $\phi^2$ is not primary, the mapping to radial quantization cannot be carried out directly for this correlator.

Our strategy to get around the non-primariness of $\phi^2$ in (\ref{eq:II:OPhi2O}) will be to make the replacement
\be
\phi^2(y) \rightarrow \p\phi(y_1)\p\phi(y_2)
\label{eq:II:PhiNTrick}
\ee
and consider the new correlator
\be
G_{\bk\bk^\prime}^{(\p\phi\p\phi)}(x,y_1,y_2,z) \equiv \langle \p^{\bk}\phi(x) \p\phi(y_1) \p\phi(y_2) \p^{\bk^\prime}\phi(z) \rangle.
\label{eq:II:OPPO}
\ee
Since $\p\phi$ is primary, now we \emph{can} apply conformal transformations to this new correlator to map it to a radial quantization matrix element. Moreover, the correlator (\ref{eq:II:OPhi2O}) can be obtained from (\ref{eq:II:OPPO}) after integrating over $y_1$ and $y_2$, as we will demonstrate. The price we have to pay is that (\ref{eq:II:OPPO}) is a four-point function; however, this four-point function is still fixed by conformal symmetry and there is no barrier to computing it in radial quantization. The replacement (\ref{eq:II:PhiNTrick}) is an extremely useful trick that easily generalizes: we can deal with the non-primariness of $\phi^n$ in 2d by promoting it to $\p\phi(y_1)\cdots\p\phi(y_n)$ and then integrating over the $y_i$. 

To compute the correlator (\ref{eq:II:OPPO}), we map it to the following radial quantization matrix element
\be
\begin{aligned}
G_{\bk\bk^\prime}^{(\p\phi\p\phi)}(y_1,y_2) &\equiv \langle \p^{\bk}\phi(\infty) \p\phi(y_1) \p\phi(y_2) \p^{\bk^\prime}\phi(0) \rangle \\[10pt]
& = \Ncal_{\bk} \, \Ncal_{\bk^\prime} \langle a_{\bk} \, \p\phi(y_1) \p\phi(y_2) \,  a_{\bk^\prime}^\dagger \rangle
\end{aligned}
\ee
Specifically, conformal transformations  give
\be
G_{\bk \bk'}^{(\p \phi \p \phi)}(x, y_1,y_2,z)  \cong G_{\bk \bk'}^{(\p \phi \p \phi)}\left(\frac{y_1-z}{x-y_1},\frac{y_2-z}{x-y_2}\right)  \frac{ (x-z)^{2 - \Delta-\Delta'} }{(x-y_1)^2 (x-y_2)^2} ,
\label{eq:II:GkkpFourPoint}
\ee
where we are using the notation $\cong$ introduced in (\ref{monobra}).
So, our strategy will be to compute $G_{\bk\bk'}^{(\p \phi \p \phi)}(y_1,y_2)$, for which we can use radial quantization methods, and then use (\ref{eq:II:GkkpFourPoint}) to infer the full four-point function.  The latter can be integrated in $y_1,y_2$ and then Fourier transformed to obtain the Hamiltonian matrix elements of $\phi^2$ in our basis.

To compute $G_{\bk\bk'}^{(\p \phi \p \phi)}(y_1,y_2)$ in radial quantization, we insert the radial mode decomposition (\ref{eq:II:RadialPPhi}) for the $\partial \phi(y_i)$  insertions. Because the final lightcone matrix elements cannot create particles from the vacuum, we only need to keep the terms where $\p \phi(y_1)$ contributes an $a$ and $\p \phi(y_2)$ contributes an $a^\dagg$, or vice versa; without loss of generality, we can keep only the former case and multiply by a factor of 2.   The result can be written as
\be
G_{\bk\bk'}^{(\p \phi \p \phi)}(y_1,y_2) \doteq \frac{2 \CN_{\kvec} \CN_{\kvec'}}{4\pi} \sum_{\ell, \ell'=1}^\infty  \sqrt{\ell \ell'}y_1^{-\ell -1} y_2^{\ell'-1} \< a_{\kvec} a_\ell  a_{\ell'}^\dagger a_{\kvec'}^\dagger \> .
\ee
 To calculate the expectation value, we move $a_\ell$ to the right and $a^\dagger_{\ell'}$ to the left, and contract with all possible $a_k, a^\dagger_k$s. Note that we can discard the contraction from $[ a_\ell, a_{\ell'}^\dagger]$ since this corresponds to the singularity when $y_1 \rightarrow y_2$ and is subtracted out in the definition of the mass operator $\phi^2$.  Performing these contractions, we obtain
\be
G_{\bk\bk'}^{(\p \phi \p \phi)}(y_1,y_2) &\doteq &\frac{ 2 \CN_{\kvec} \CN_{\kvec'}}{4\pi} \sum_{{\substack{k_i \in \kvec \\ k'_j \in \kvec'}}}  \sqrt{k_i k'_j } y_1^{-k'_j-1}  y_2^{k_i -1} \< a_{\kvec/k_i} a_{\kvec'/k_j'}^\dagger \> .
  \label{eq:MassRadQuantSum}
\ee
The advantage here is that the  sums on $k_i, k_j'$ are sparse, because the inner product $\< a_{\kvec/k_i} a_{\kvec'/k_j'}^\dagger \> $ vanishes unless  $\kvec/k_i$ is the same as $\kvec'/k_j$. In fact, it is obvious that  most $\bk, \bk'$ will differ by more than just one of their $k$s and so the entire sum will vanish.

The next step is to use (\ref{eq:II:GkkpFourPoint}) to reintroduce the positions of the external operators, and then do the $y_{1,2}$ integrations. For conciseness, note that $G_{\bk \bk'}^{(\p \phi \p \phi)}(y_1,y_2)$ above is a sum over terms of the form
\be
G^{(\p \phi \p \phi)}_{k k'}(y_1, y_2) \equiv \sqrt{k k'} y_1^{-k'-1} y_2^{k-1} ,
\ee
with $k - k' = |\bk| - |\bk'| = \Delta-\Delta'$, so the full four-point function is a sum over terms of the form 
\be
 G^{(\p \phi \p \phi)}_{k k'}(x,y_1, y_2,z) \equiv \sqrt{kk'} \left(\frac{y_1-z}{x-y_1}\right)^{-k'-1} \left(\frac{y_2-z}{x-y_2}\right)^{k-1}  \frac{ (x-z)^{2 - \Delta-\Delta'} }{(x-y_1)^2 (x-y_2)^2} .
 \ee 
 We will work with these individual terms, and then sum over them at the end.  First, we do the $dy_1$ and $ dy_2$ integrations to turn the $\p \phi$'s into $\phi$'s. The boundary condition is fixed by demanding the correlator vanish when either $y_1 \rightarrow \infty$  or $y_2 \rightarrow \infty$. We obtain 
\begin{equation}
G_{ k k'}^{(\phi^2)}(x,y_1,y_2,z) \equiv \int d y_1 \, dy_2 \, G^{(\p \phi \p \phi)}_{k k'}(x,y_1,y_2,z)\doteq \frac{
   \left(\left(\frac{x-y_1}{z-y_1}\right){}^{k'}-1\right)\left(\left(\frac{z-y_2}{x-y_2}\right){}^{k}-1\right)}{(-1)^{k'-k}(x-z)^{\Delta+\Delta'}\sqrt{k k'}}.
   \label{eq:Gphiphi}
   \end{equation}
 At this point, we can set $y_1 = y_2 = y$. The combination $\p \phi(y_1) \p \phi(y_2)$ has become the mass term $\phi^2(y)$! 
 
Next, we want to apply the general Fourier transform formula (\ref{eq:LCTDataFT}) for three-point functions to obtain the contribution of each $k,k'$ term to the Hamiltonian matrix elements of $\phi^2$.  
For convenience, define
 \be
 g_k^{(\phi)}(v) \equiv \frac{v^k-1}{\sqrt{k}},
 \label{eq:genfuncPhi}
 \ee
 so that $G_{k k'}^{(\phi^2)}(x,y,z) \doteq (-1)^{k-k'} (x-z)^{-\Delta-\Delta'}g_{k'}^{(\phi)}(\frac{x-y}{z-y}) g_{k}^{(\phi)}(\frac{z-y}{x-y})$. In appendix \ref{app:GeneratingFunctions}, we explain how the integrals over $dx, dy,$ and $dz$ can be reduced to a single contour integral:
 \begin{equation}
\frac{\int dx dy dz \, e^{i(p x-p' z)} \, G_{ k k'}^{(\phi^2)}(x,y,z)}{2 \pi (2p) \delta(p-p')} \doteq (-1)^{k-k'}N_{\rm FT} \int_{\half-i\infty}^{\half+i\infty} \frac{dw}{2\pi i} \, g_{k'}^{(\phi)}(\frac{w-1}{w}) g_k^{(\phi)}(\frac{w}{w-1}),
\label{eq:MassTermContour}
 \end{equation}
 where
 \be
 N_{\rm FT} \equiv 
 \frac{2 \pi^2 p^{\Delta+\Delta'-3}}{\Gamma(\Delta+\Delta'-1)}.
 \label{eq:II:N}
 \ee

The contour integral (\ref{eq:MassTermContour}) is easy to evaluate:\footnote{The integral is symmetric under $k\leftrightarrow k'$, so assume without loss of generality that $k \ge k'$ and evaluate the $dw$ integral by deforming the contour to wrap the pole at $w=0$.
 Since $k\ge k'$,  $ g_{k'}^{(\phi)} \left( \frac{w-1}{w} \right) \left(\frac{w}{w-1}  \right)^{k}$  is regular at $w \sim 0$ and does not contribute. So the only contribution is from the following cross-term:
$\frac{1}{\sqrt{kk'}} \oint \frac{dw}{2\pi i}  \left(1-\frac{1}{w}\right)^{k'}(-1)  = \frac{1}{\sqrt{kk'}} \oint \frac{dz}{2\pi iz^2} (1+z)^{k'}  =\sqrt{\frac{k'}{k}}= \sqrt{\frac{{\rm min}(k,k')}{{\rm max}(k,k')}}.$
}
\be
 \int_{\half-i\infty}^{\half+i\infty} \frac{dw}{2\pi i} \, g_{k'}^{(\phi)}(\frac{w-1}{w}) g_k^{(\phi)}(\frac{w}{w-1}) = \sqrt{\frac{{\rm min}(k,k')}{{\rm max}(k,k')}}.
 \label{eq:basicradialcontourresult}
 \ee.

Finally, let us put everything together. Using the notation of (\ref{eq:II:ME}) our derivation above yields the following formula for matrix elements of $2P_+ P_-$ for $\Ocal_R= \frac{m^2}{2} \phi^2$, 
\begin{equation}
\Mcal_{\cO \cO^\prime}^{(\frac{m^2}{2} \phi^2)} =m^2 \frac{(-1)^{\Delta-\Delta'}N_{\rm FT}}{2\pi N_{\cO}N_{\cO^\prime}} \sum_{\bk, \bk'} \left[   \, C^\Ocal_{\bk} \, \Ncal_{\bk}\, C^{\Ocal'}_{\bk^\prime} \,  \Ncal_{\bk'}  \sum_{\bk/k_i = \bk'/k'_j} \norm{\bk/k_i}^2  \sqrt{\frac{{\rm min}(k_i,k'_j)}{{\rm max}(k_i,k'_j)}} \right]. 
\label{eq:II:ScalarPhi2Final}
\end{equation}
We have used the fact that for each contraction, $k-k'= \Delta-\Delta'$, so the factor $(-1)^{k-k'} = (-1)^{\Delta-\Delta'}$ can be pulled out of the sum. 

For the general case $\cO_R=\frac{\lambda}{n!} \phi^n$, the steps in the derivation are conceptually the same as for $\phi^2$. First, we make the replacement  
\be
\phi^n(y) \rightarrow \p\phi(y_1)\cdots \p\phi(y_n)
\ee
inside correlation functions, thus trading a non-primary three-point function for a primary $(n+2)$-point function. We evaluate this higher-point function using radial quantization methods, integrate with respect to the $y_i$ and set $y_1=\cdots=y_n=y$, and then Fourier transform to obtain Hamiltonian matrix elements. Some of the intermediate steps required to do this are contained in appendix \ref{app:GeneratingFunctions}. Here, we simply state the final result for $\phi^n$ matrix elements:
\be
\Mcal_{\cO \cO^\prime}^{(\frac{\lambda}{n!}\phi^n)} &=&2 \lambda  \frac{(-1)^{\Delta-\Delta'}N_{\rm FT}  }{(4\pi)^{\frac{n}{2}}N_{\cO}N_{\cO^\prime}}  \sum_{\bk, \bk'} \left[   \, C^\Ocal_{\bk} \, \Ncal_{\bk}\, C^{\Ocal'}_{\bk^\prime} \,  \Ncal_{\bk'}  \sum_{\bk/\{k_i\} = \bk'/\{k_j'\} \atop |\{k_i\}| + |\{k_j'\}| = n} 
\norm{\bk/\{k_i\}}^2 \frac{\CI(\{k_i\}, \{k_j'\})  }{\prod_i k_i^{\frac{1}{2}} \prod_j k_j'^{\frac{1}{2}}} \right], \nn\\
&& \CI(\{k_i\}, \{k_j'\}) = \sum_{A_+ \subset \{k_i \} \atop A_- \subset \{ k_j' \}} (-1)^{d(A_+) + d(A_-)} {\rm min}\left(\sum_{k_i \in A_+} k_i, \sum_{k_j' \in A_-} k_j'\right) ,
\label{eq:II:ScalarPhiNFinal}
\ee
where $d(A)$ denotes the number of elements of $A$ and $\bk/\{k_i\}$ indicates the vector $\bk$ with the set of elements $\{k_i\}$ removed.

%%%%%%%%%%%%%%%%%%%%%%%%%%%%%%%%%%%%%%%%%%%%%%%%%%%%%%%%%%%%%%%%%%%%%%%%%%%%%

\subsection{Examples}

In this section, we will revisit some examples of LCT inner products and matrix elements that we computed in Part I using the Fock space (section~\ref{sec:FockSpace}) and Wick contraction (section~\ref{sec:WickContract}) methods. We will now recompute them using the radial quantization formulas presented above. We set $p=1$ throughout.

As a first exercise, let us compute the $2\times 2$ Gram matrix of the following operators: 
\be
\begin{aligned}
\Ocal &=  (\p\phi)^2 \\
\Ocal^\prime &=  6\p^3\phi \p\phi - 9 \p^2\phi \p^2\phi.
\end{aligned}
\ee
In this example, there are three monomials to keep track of, which we denote by $\bk_1 = (1,1)$, $\bk_2=(3,1)$, and $\bk_3=(2,2)$, corresponding respectively to $(\p\phi)^2$, $\p^3\phi \p\phi$, and $(\p^2\phi)^2$. Referring to (\ref{eq:II:OMonExpand}), the monomial expansion coefficients are given by $C^\Ocal_{\bk_1} = 1$, $C^{\Ocal'}_{\bk_2} = 6$, and $C^{\Ocal'}_{\bk_3} = -9$, with all other coefficients vanishing. 

The first step is to compute the Zamolodchikov metric $\mathfrak{g}_{\cO\cO^\prime}$ in (\ref{eq:II:Radial2PF}), where the $\bk$-dependent factors entering the formula were defined in (\ref{eq:II:radialscalarNcal}) and (\ref{eq:II:radialscalarknorm}). In the current example, the factors we need are $\Ncal_{\bk_1} = \tfrac{1}{4\pi}$, $\Ncal_{\bk_2} = \tfrac{\sqrt{3}}{2\pi}$, $\Ncal_{\bk_3} = \tfrac{1}{2\pi}$ and $\norm{\bk_1}^2 = \norm{\bk_3}^2  = 2$, $\norm{\bk_2}^2 = 1$. Plugging the coefficients $C^\Ocal_{\bk}$ along with these factors into (\ref{eq:II:Radial2PF}), we compute the $2\times 2$ Zamolodchikov metric to be  
\be
\mathfrak{g}_{\cO\cO^\prime} = \left( \begin{array}{cc} \frac{1}{8\pi^2} & 0 \\ 0 & \frac{135}{2\pi^2} \end{array}\right).
\ee
Now, we simply plug this into (\ref{eq:II:InnerFinal}), with $h=2$ and $h^\prime = 4$, to get
\be
G_{\cO\cO^\prime} = \frac{1}{16\pi } \left( \begin{array}{cc} \frac{1}{ 3 N_{\cO}^2} & 0 \\ 0 & \frac{3}{14 N^2_{\cO^\prime}}  \end{array}\right). 
\ee
This reproduces~\eqref{eq:GramExample1} and~\eqref{eq:GramExample1Wick} obtained via Fock space and Wick contractions. 

As another inner product example, let us compute the norm of the operator $(\p\phi)^3$. In the language of (\ref{eq:II:OMonExpand}), $\bk=(1,1,1)$ and $C^\Ocal_{\bk}=1$. Moreover, $\Ncal_{\bk} = \tfrac{1}{8\pi^{3/2}}$ and $\norm{\bk}^2 = 6$. Plugging these ingredients into (\ref{eq:II:Radial2PF}), the Zamolodchikov metric is 
\be
\mathfrak{g}_{(\p\phi)^3,(\p\phi)^3} = \frac{3}{32\pi^3}, 
\ee
which we then plug into (\ref{eq:II:InnerFinal}) to get the Gram matrix entry
\be
G_{(\p\phi)^3,(\p\phi)^3} = \frac{1}{1280\pi^2 N_{(\p\phi)^3}^2}.
\ee
This reproduces (\ref{eq:dphicubednorm}) and (\ref{eq:NormDPhi3Wick}).

Now let us turn to mass matrix elements, given by the formula (\ref{eq:II:ScalarPhi2Final}), and reproduce the examples in (\ref{eq:WickExamplesPhi2}). First, consider the case where the external operator is $\p\phi$. In this case, there is only one $\bk$ and one $\bk^\prime$, given by $\bk=\bk^\prime=(1)$, with $C^{\Ocal}_{\bk} = C^{\Ocal'}_{\bk^\prime} =1$. The inner sum in (\ref{eq:II:ScalarPhi2Final}) does not exist in this example. We simply get
\be
\Mcal^{(\frac{m^2}{2}\phi^2)}_{\p\phi, \p\phi} = \frac{m^2}{4 N^2_{\p\phi}} = m^2.
\ee
Next, consider the external operators to both be either $(\p\phi)^2$ or $(\p\phi)^3$. For the case of $(\p\phi)^2$ we have $\bk=\bk^\prime=(1,1)$, while for $(\p\phi)^3$ we have $\bk=\bk^\prime=(1,1,1)$, and in both cases $C^\Ocal_{\bk} = C^{\Ocal'}_{\bk^\prime} =1$. For $(\p\phi)^2$, the inner sum in (\ref{eq:II:ScalarPhi2Final}) yields a factor of $4\cdot  \norm{(1)}^2=4$ (including permutations, there are four ways to delete an entry $k_i$ from $\bk$ and an entry $k_j^\prime$ from $\bk^\prime$ and still have $\bk/k_i = \bk^\prime/k_j^\prime$). For  $(\p\phi)^3$ the inner sum yields a factor of $9\cdot \norm{(1,1)}^2=18$. Putting things together, we reproduce
\be
\Mcal^{(\frac{m^2}{2}\phi^2)}_{(\p\phi)^2, (\p\phi)^2} = \frac{m^2}{8\pi N^2_{(\p\phi)^2}} = 6m^2,  \hspace{10mm} \Mcal^{(\frac{m^2}{2}\phi^2)}_{(\p\phi)^3, (\p\phi)^3} = \frac{3 m^2}{256\pi^2 N^2_{(\p\phi)^3}} = 15m^2.
\ee

Finally, let us consider some $\phi^4$ matrix elements, using the formula~\eqref{eq:II:ScalarPhiNFinal}, and reproduce the examples in (\ref{eq:WickExamplesPhi4}). In the case where both external operators are $\p\phi$, there is no way to satisfy the conditions of the inner sum in (\ref{eq:II:ScalarPhiNFinal}), and so the matrix element trivially vanishes,
\be
\Mcal^{(\frac{\lambda}{4!}\phi^4)}_{(\p\phi),(\p\phi)} = 0.
\ee
Next, consider the case where the in state is $\p\phi$ and the out state is $(\p\phi)^3$, \emph{i.e.}, $\bk=(1)$ and $\bk^\prime=(1,1,1)$. The only contribution to the inner sum in (\ref{eq:II:ScalarPhiNFinal}) comes when $\{k_i\}=(1)$ and $\{k^\prime_j\}=(1,1,1)$. In this case, $\Ical(\{k_i\},\{k^\prime_j\}) $ is computed by
\be
3(-1)^{1+1} \min(1,1) +  3(-1)^{1+2} \min(1,2) +  (-1)^{1+3} \min(1,3) =1 \hspace{10mm}
\ee
Putting things together in (\ref{eq:II:ScalarPhiNFinal}) gives 
\be
\Mcal^{(\frac{\lambda}{4!}\phi^4)}_{(\p\phi),(\p\phi)^3} = \frac{\lambda}{128\pi^2 N_{(\p\phi)}N_{(\p\phi)^3}} = \frac{\sqrt{5}\lambda}{4\pi}.
\ee
The last case to consider is when both in and out states are $(\p\phi)^3$, \emph{i.e.}~$\bk=\bk^\prime = (1,1,1)$. The contribution to the inner sum in (\ref{eq:II:ScalarPhiNFinal}) comes from the nine ways that we can choose $\{k_i\}=\{k^\prime_j\}=(1,1)$, in which case $\Ical(\{k_i\},\{k^\prime_j\})$ is
\be
 4(-1)^{1+1} \min(1,1) +  4(-1)^{1+2} \min(1,2) +  (-1)^{2+2} \min(2,2) = 2 \hspace{10mm}
\ee
Putting things together in (\ref{eq:II:ScalarPhiNFinal}) gives 
\be
\Mcal^{(\frac{\lambda}{4!}\phi^4)}_{(\p\phi)^3,(\p\phi)^3} = \frac{3\lambda}{1024 \pi^3 N^2_{(\p\phi)^3}} = \frac{15\lambda}{4\pi}.
\ee
These matrix element examples all agree with our previous answers.

%%%%%%%%%%%%%%%%%%%%%%%%%%%%%%%%%%%%%%%%%%%%%%%%%%%%%%%%%%%%%%%%%%%%%%%%%%%%%

\subsection{Code Implementation}

We end this section with a discussion of how formulas like (\ref{eq:II:ScalarPhiNFinal}) are actually implemented in the code in order to improve efficiency.  One of the points of the following implementation is that we want to avoid spending a lot of time searching for different ways that $k$s can be subtracted from the `in' and `out' monomials in order for the residuals to match.  Instead, it is significantly faster to construct matrix representations of the creation and annihilation operators acting on monomials, and reduce the computation to linear algebra.  

\begin{figure}[t!]
\centering
$$
\text{\Large
$\sum_{k,k^\prime}$
}
\raisebox{-0.6in}{
	\includegraphics[width=0.44\linewidth]{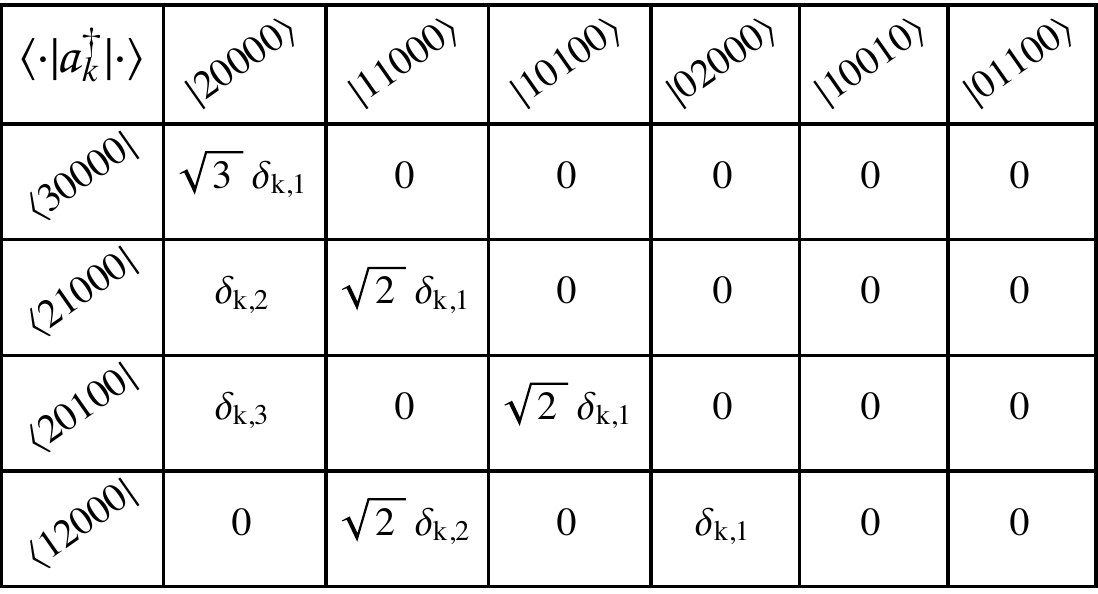}
}
\text{\tiny $^\times$} {\cal I}_{\phi^2}(k,k^\prime) \text{\tiny $^\times$}
\raisebox{-0.6in}{
	\includegraphics[width=0.35\linewidth]{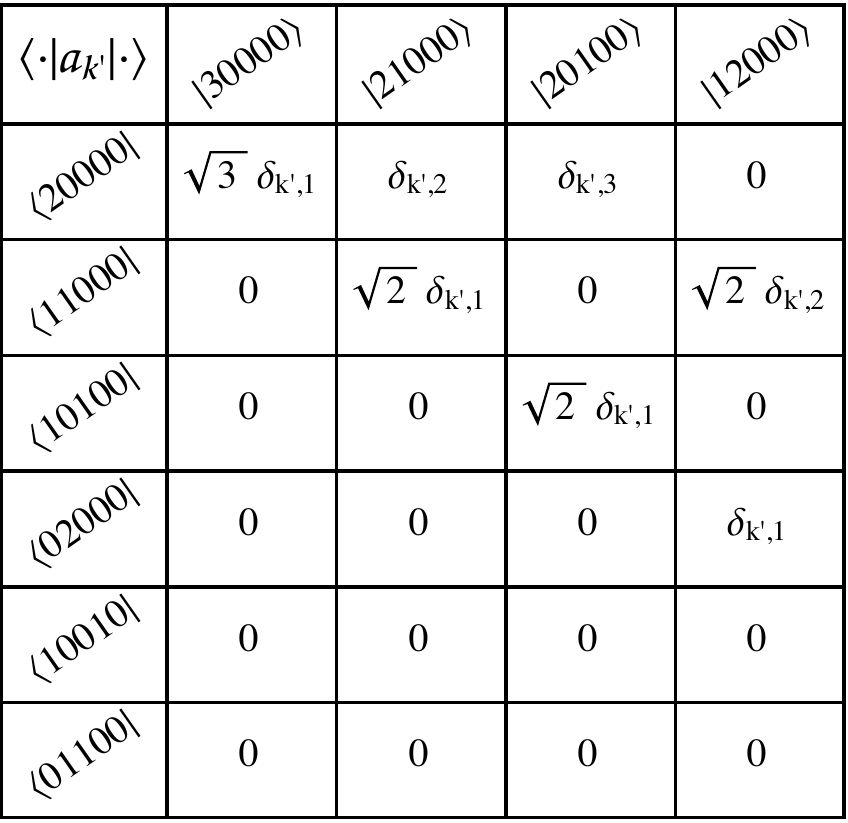}
}
$$
\caption{\label{fig:BosonCodeStructure} The structure of the scalar  mass matrix.}
\end{figure}

Because we are essentially working with separate creation and annihilation operators for each possible value of $k$, it is convenient to work with the occupation number representation of the monomials
\begin{align}
|n_1 n_2 \cdots \> \equiv | (\d \phi)^{n_1} (\d^2 \phi)^{n_1} \cdots \>
\end{align}
where $n_k$ counts the occurrence of $\d^k \phi$ in the monomial.  The actions of an annihilation operator $a_{k'}$  and a creation operator $a^\dagger_k$ are depicted in the right and left matrices, respectively, in Fig. \ref{fig:BosonCodeStructure}.  
 For instance, $a_1$ acting on $|21000\>$ is $\sqrt{2} |11000\>$, since $a |n\> = \sqrt{n} |n-1\>$ for a standard harmonic oscillator.  Constructing these matrix representations is fast, because for any ket state one can quickly enumerate all the possible ways its oscillators can be lowered.  In practice, the Kronecker $\delta$s are carried as additional data for each entry, essentially keeping track of the $k$ that was lowered.  So, continuing with the example ket state $|21000\>$, one immediately can see that only the first two entries in its column in Fig.~\ref{fig:BosonCodeStructure} are nonzero.  

Once $a_k^\dagger$ and $a_{k'}$ have been constructed this way, they can be applied to matrix elements between arbitrary monomials without having to be recomputed each time.  Moreover, for any monomial external states, we essentially only have to take the inner product of a row and a column weighted by $\CI(\{k_i\}, \{k_j'\})$.  For example, consider the mass term $\phi^2$ and take the external monomial states to be $\<12000|$ and $|21000\>$.  Looking  in Fig.~\ref{fig:BosonCodeStructure} at the row in the left matrix associated with $\< 12000|$ and at the column in the right matrix associated with $|21000\>$, the only common entry is $\< 11000| 11000\>$, with $k=2$ and $k'=1$.  We thus find that the only contribution in this case is $\sqrt{2} \cdot \CI_{\phi^2}(2,1) \cdot \sqrt{2}=2$.

%%%%%%%%%%%%%%%%%%%%%%%%%%%%%%%%%%%%%%%%%%%%%%%%%%%%%%%%%%%%%%%%%%%%%%%%%%%%%
%%%%%%%%%%%%%%%%%%%%%%%%%%%%%%%%%%%%%%%%%%%%%%%%%%%%%%%%%%%%%%%%%%%%%%%%%%%%%

\section{Radial Quantization for Fermions}
\label{sec:RQFermions}

In this section, we will introduce radial quantization methods for fermions. The basic strategy is the same as for scalars, but there are
a few added complications that we will cover in this section.

\subsection{Radial Quantization Modes}
\label{sec:radialsetupfermion}

Let us start by recapping what we learned about fermions in section \ref{sec:Fermions}. One of our main lessons was that the building block for the fermion basis is $\p\psi$, where $\psi$ is the left-chirality mode of the real fermion field and we have written $\p \equiv \p_-$ for short. Recall that the reason $\p\psi$ is the basic building block is twofold. First, the right-chirality mode $\chi$ is non-dynamical and can be integrated out, leaving us with $\psi$. Then, adding a mass term (which we always do in this work) introduces IR divergences that lift out from the spectrum any operators that have a $\psi$ without a derivative attached to it. As explained in section \ref{sec:DirichletBasis}, the states that do not get lifted out and remain in the spectrum are the so-called ``Dirichlet states" where all operators are built from $\p\psi$. 

At first pass, it may seem that building operators out of $\p\psi$ raises issues for radial quantization, since $\p\psi$ is strictly-speaking not primary. However, another important lesson from section \ref{sec:Fermions} was that operators built from $\p\psi$ constitute a generalized free theory in which $\p\psi$ can be treated as a primary operator with $h=3/2$. In other words, correlation functions involving $\p\psi$, technically defined via Wick contractions, transform consistently under conformal transformations as if $\p\psi$ were primary. Indeed, we saw that one can even define a shifted special conformal generator $\tilde{K}_\mu$ that annihilates $\p\psi$ and defines a notion of primariness for operators built from $\p\psi$. This is all we need to employ radial quantization. 

Keeping the above lessons in mind, we define the radial quantization mode expansion for $\p\psi$ as follows, 
\begin{gather}
\p\psi(x) = \frac{i}{\sqrt{4\pi}} \sum_{k=1}^\infty \sqrt{k(k+1)} \left( x^{-k-2}b_k + x^{k-1} b_k^\dagger \right),  \label{eq:II:radialfermion} \\[10pt]
\{ b_k, b_{k^\prime}^\dagger \} = \delta_{k,k^\prime}.
\end{gather}
As with scalars, we can define monomial out states for fermions in radial quantization,
\begin{gather}
\p^{\bk}\psi(0) \vac = \Ncal^{(F)}_{\bk} \, b_{\bk}^\dagger \vac, \\[10pt]
b_{\bk}^\dagger= b_{k_1}^\dagger \cdots b_{k_n}^\dagger, \hspace{10mm} \Ncal^{(F)}_{\bk} \doteq \left(\frac{1}{ \sqrt{4\pi} }\right)^n \prod_{i=1}^n \Gamma(k_i) \sqrt{k_i(k_i+1)}. \label{eq:radialfermionoperators}
\end{gather}
Note that we have added a superscript on $\Ncal^{(F)}_{\bk} $ to distinguish it from the normalization constant for scalars. As with scalars, radial quantization inner products are orthogonal, 
\be
 \langle b_{\bk} \, b^\dagg_{\bk^\prime} \rangle= \delta_{\bk, \bk'}, \hspace{10mm} \delta_{\bk ,\bk'} \equiv \delta_{k_1,k_1'} \dotsb \delta_{k_n, k_n'},
\label{eq:II:orthogonalityfermion}
\ee
(for fermions, $\norm{\bk}=1$ due to Pauli exclusion) and again this is the reason behind the efficiency of radial quantization methods. 

A general primary operator\footnote{Recall that primary here means that $\Ocal$ is annihilated by the shifted special conformal generator $\tilde{K}_\mu$ that annihilates $\p\psi$.} $\Ocal$ can be written as a linear combination of monomials
\be
\Ocal(x) = \sum_{\bk} C^\Ocal_{\bk}\, \p^{\bk}\psi(x),
\label{eq:II:primaryfermion}
\ee
and has corresponding radial quantization in and out states given by 
\begin{eqnarray}
\Ocal(0) \vac &=& \sum_{\bk} C^\Ocal_{\bk} \, \Ncal^{(F)}_{\bk} \, b_{\bk}^\dagger \vac \label{eq:II:ketFermions} \\[5pt]
\bra{{\rm vac}} \Ocal(\infty) &\equiv& \lim_{x\rightarrow 0} x^{-2\Delta} \bra{{\rm vac}} \Ocal(1/x) = \sum_{\bk}  \bra{{\rm vac}} C^\Ocal_{\bk} \, \Ncal^{(F)}_{\bk} \, b_{\bk}. \label{eq:II:braFermions}
\end{eqnarray}

The subtlety in defining bra states for individual monomials, which we discussed for scalars in section \ref{eq:RQModes}, persists for fermions. That is, in general
\be
\bra{{\rm vac}} \p^{\bk} \psi(\infty) \ne  \bra{{\rm vac}} \Ncal^{(F)}_{\bk} \, b_{\bk},
\ee
and instead we write
\begin{eqnarray}
\p^{\bk}\psi(0) \vac &=& \Ncal^{(F)}_{\bk}\, b_{\bk}^\dagger \vac, \\[5pt]
\bra{{\rm vac}} \p^{\bk} \psi(\infty) &\cong& \bra{{\rm vac}} \Ncal^{(F)}_{\bk} \, b_{\bk},
\end{eqnarray}
where $\cong$ means that this equation can be used only when applied to all of the terms in a linear combination of monomials that adds up to a primary operator. We are now ready to apply radial quantization to fermions.

\subsection{Inner Product}
\label{sec:radialinnerfermion} 

Inner products are computed in the same way as was done for scalars. To briefly summarize, given two primary operators expanded in terms of monomials as in (\ref{eq:II:primaryfermion}), the Zamolodchikov metric is given by 
\be
\mathfrak{g}^{(F)}_{\Ocal\Ocal^\prime} \equiv \corr{\cO(\infty)\cO^\prime(0)} = \sum_{\bk} C^\Ocal_{\bk}\, C^{\Ocal'}_{\bk} \left(\Ncal^{(F)}_{\bk}\right)^2. \label{eq:ZamoMetricFermions}
\ee
This is a consequence of the orthogonality relation (\ref{eq:II:orthogonalityfermion}). It follows that the general two-point function is 
\be
\corr{\cO(x)\cO^\prime(y)} \doteq \frac{\mathfrak{g}^{(F)}_{\Ocal\Ocal^\prime}}{(x-y)^{2h}},
\ee
where we keep in mind that $\p\psi$ has $h=3/2$. Referring back to  (\ref{eq:II:Inner}), the LCT inner product is given by the Fourier transform of this two-point function, which can be computed using the formulas in (\ref{eq:FTFormulas}). Doing so, we get that the Gram matrix entry between $\CO$ and $\CO^\prime$ is given by (\ref{eq:II:InnerFinal}).

\subsection{Mass Term}
\label{sec:radialmefermion}

Recall from section~\ref{sec:IRDivergences} that with a mass deformation, the Lagrangian for $\psi$ is
\be
\Lcal = i\psi\p_+\psi - \tfrac{1}{2}m^2 \, \psi \frac{1}{i\partial_-} \psi,
\ee
Therefore, we need to work out matrix elements of the relevant deformation $\Ocal_R \equiv \psi \tfrac{1}{\partial} \psi$. Recall that in the case of fermions, we are applying radial quantization in a generalized free theory where $\p\psi$ is a primary operator with $h=3/2$. We can perform conformal transformations on correlators as long as we treat $\p\psi$ as the primary. 

Examining the Lagrangian, we see that the correlator we need to compute is 
\be
G_{\cO\cO^\prime}^{(\psi \frac{1}{\p} \psi)}(x,y,z) \equiv \langle \cO(x)\, \psi \tfrac{1}{\partial} \psi(y) \, \cO^\prime(z) \rangle = \sum_{\bk \bk'} C^\Ocal_{\bk} \, C^{\Ocal'}_{\bk'} \, G_{\bk \bk'}^{(\psi \frac{1}{\p} \psi)}(x,y,z),
\label{eq:II:PsiMassCorr}
\ee
where we have expanded $\cO$ and $\cO^\prime$ in terms of monomials as in (\ref{eq:II:OMonExpand}) and defined the monomial correlator
\be
G_{\bk \bk^\prime}^{(\psi \frac{1}{\p} \psi)}(x,y,z) \equiv \langle \p^{\bk}\psi(x) \, \psi \tfrac{1}{\partial} \psi(y) \,  \p^{\bk'}\psi(z) \rangle. 
\label{eq:II:FermionMassMon3PF}
\ee

We immediately encounter two problems. First, in our generalized free field framework, $\psi$ is not a primary operator, only $\p\psi$ is. Second, the mass term is nonlocal due to the presence of the $1/\p$. Both of these problems can be handled in the same way that we handled the non-primariness of $\phi^n$ for scalars. We make the replacement 
\be 
\psi \tfrac{1}{\partial} \psi(y) \rightarrow \p\psi(y_1) \p\psi(y_2)
\ee 
and consider the new correlator 
\be
G_{\bk \bk^\prime}^{(\p\psi\p\psi)}(x,y_1,y_2,z) \equiv \langle \p^{\bk}\psi(x) \, \p\psi(y_1) \p\psi(y_2) \,  \p^{\bk'}\psi(z) \rangle. 
\label{eq:II:FermionMass4PF}
\ee
In the case of scalars, we turned $\p\phi(y_1)\p\phi(y_2)$ into $\phi^2(y)$ by integrating with respect to $y_1$ and $y_2$ (and then setting $y_1=y_2=y$). The schematic idea is that integration with respect to $y_i$ takes $\phi(y_i) \rightarrow \p^{-1}\phi(y_i)$. In the case of fermions, to turn $\p\psi(y_1) \p\psi(y_2)$ into $ \psi \tfrac{1}{\partial} \psi(y)$ we need to do an extra integration on one of the $\p\psi$'s.

Aside from this extra integration, the derivation of fermion mass matrix elements proceeds very much as in the scalar case. First, one computes (\ref{eq:II:FermionMass4PF}) by mapping it to a radial quantization matrix element, which can be computed by expanding in radial modes. Having computed (\ref{eq:II:FermionMass4PF}), we integrate once with respect to $y_1$ and \emph{twice} with respect to $y_2$ before setting $y_1=y_2=y$ in order to recover (\ref{eq:II:FermionMassMon3PF}). Finally, we Fourier transform to get Hamiltonian matrix elements. The technical details of all of these steps are presented in appendix \ref{app:GeneratingFunctions}.

Here, we just state the final result for the matrix elements of $P^2=2P_+P_-$ for a fermion mass $\CO_R= \frac{m^2}{2} \psi \frac{1}{\p}\psi$: 
\begin{equation}
\Mcal_{\cO \cO^\prime}^{(\frac{m^2}{2}\psi \frac{1}{\partial} \psi)} = m^2 \frac{(-1)^{\Delta-\Delta'}N_{\rm FT}}{4\pi N_\CO N_{\CO'}} \sum_{\bk, \bk'} \left[  2 \, C^\Ocal_{\bk} \, C^{\Ocal'}_{\bk^\prime} \, \Ncal^{(F)}_{\bk}\, \Ncal^{(F)}_{\bk'}  \sum_{\bk/k_i \atop = \bk'/k'_j}\frac{(-1)^{\sigma_{i,j}} }{2} \sqrt{ \frac{k_{\rm min}(k_{\rm min}+1)}{k_{\rm max}(k_{\rm max}+1)}} \right], 
\end{equation}
 where  $k_{\rm min}= {\rm min}(k_i,k'_j)$,  $k_{\rm max} = {\rm max}(k_i, k'_j)$ and $(-1)^{\sigma_{i,j}} = (-1)^{i+j}$ counts the number of permutations required to contract the fermions in $\psi \frac{1}{\partial} \psi$ with the external states. 
 
 For example, consider the simplest primary  $\CO =\p \psi, \bk=(1)$.  From (\ref{eq:ZamoMetricFermions}), we have $\mathfrak{g}_{\CO \CO} = (\CN^{(F)}_{\bk})^2$, so setting $G_{\CO \CO}=1$ in (\ref{eq:II:InnerFinal}) we obtain $N^2_\CO = \frac{\pi (\CN^{(F)}_{\bk})^2}{\Gamma(3)}$.  From (\ref{eq:II:N}), we have $N_{\rm FT} = 2 \pi^2$.  There is only one term in the sum, with $i=j=1$, so the result in this case is
 \be
\Mcal_{\cO \cO^\prime}^{(\psi \frac{1}{\partial} \psi)}= m^2 \frac{2\pi^2}{4\pi \frac{\pi (\CN^{(F)}_{\bk})^2}{\Gamma(3)}} \frac{2 (\CN^{(F)}_{\bk})^2}{2} = m^2,
 \ee
 as expected.

\subsection{Yukawa interaction}
\label{sec:radialfermionyukawa}

Now consider a Yukawa interaction, with Lagrangian given by
\be
\Lcal = i\psi\p_+\psi + i \chi\p_-\chi + \sqrt{2}i (m+g \phi) \psi\chi.
\ee
Recall from section~\ref{sec:YukawaInteraction} that we integrate out $\chi$ to obtain 
\be
\Lcal = i\psi\p_+\psi - \frac{1}{2}(m+g \phi) \psi\frac{1}{i\p_-}(m + g \phi) \psi .
\ee
The two types of interaction terms we need to handle are $\phi\psi\tfrac{1}{\p}\psi$ and $\phi\psi\tfrac{1}{\p}\phi\psi$.

Matrix elements of $\phi\psi\tfrac{1}{\p}\psi$ can be computed using the same technology as for the fermion mass term described in the previous section. In particular, one can use radial quantization methods to work out matrix elements for $\p\phi(y_1) \p\psi(y_2) \p\psi(y_3)$ and then integrate appropriately with respect to the $y_i$. Matrix elements of $\phi\psi\tfrac{1}{\p}\phi\psi$, on the other hand, require some extra care, because we ultimately have to integrate the product $\phi\psi$. The technical details of this integration are covered in appendix \ref{app:GeneratingFunctions}.

Here we state the final result for matrix elements of the two interaction terms. First, for the quartic term $\CO_R=g^2 \phi \psi \frac{1}{\p} \phi \psi$, we have
\be
\Mcal_{\cO \cO^\prime}^{(g^2 \phi \psi \frac{1}{\partial} \phi \psi)} &=&2 g^2 \frac{(-1)^{\tilde{\Delta}-\tilde{\Delta}'}N_{\rm FT}}{(4\pi)^2 N_\CO N_{\CO'}} \nn\\ &\times&  \sum_{\bk, \bk'} \left[   \, C^{\cO}_{\bk} \, C^{\cO'}_{\bk^\prime} \, \Ncal^{(M)}_{\bk}\, \Ncal^{(M)}_{\bk'}  \sum_{\bk/\{k_i| s_i=1\} = \atop  \bk'/\{k_i | s_i =-1 \}}(-1)^{\sigma(\{k_i, s_i\}) } \norm{\bk/\{k_i\}}^2g_{\phi\psi\frac{1}{\p}\phi\psi}(k_i, s_i) \right],  \nn\\
\ee
where $g_{\phi\psi\frac{1}{\p}\phi\psi}(k_i, s_i)$ is given in equation (\ref{yukawa-final-factor}), and the normalization factor $\Ncal^{(M)}_{\bk}$ for monomials is simply the product of the normalization factors  $\Ncal_{\bk_B}$  and $\Ncal^{(F)}_{\bk_F}$ for the boson and fermion parts $\bk_B$ and $\bk_F$ of the monomial (see (\ref{eq:mixonmonomialsdefn})). We have defined $\tilde{\Delta} \equiv \Delta-\frac{1}{2}n_F$, which counts the dimension of operators as if fermions had dimension 0; the factor $(-1)^{\tilde{\Delta}-\tilde{\Delta'}}$ is due to the product of $(-1)^{k'-k}$ factors in the individual monomial contractions. The norm $ \norm{\bk}$ here indicates the norm $\norm{\bk_B}$ of the boson part of $\bk$ (the norm $\norm{\bk_F}$ of the fermion part is 1 due to Fermi statistics).  The notation is somewhat different compared to previous interactions, in order to more compactly include all possible contractions.  Here, the index $i$ always runs from 1 to 4, and the fields in the interaction always are contracted with $k_i$ as follows: $\phi_{k_1} \psi_{k_2} \frac{1}{\partial} \phi_{k_3} \psi_{k_4}$. The $s_i$ label indicates whether the contraction is to the left $(s_i =1)$ or right  $(s_i=-1)$, and should be summed over $+1$ and $-1$.  As before, $\sigma(\{k_i, s_i\})$ counts the number of times that fermion modes must be anticommuted past each other. 

The cubic term $\CO_R = m g \phi \psi \frac{1}{\p} \psi$ is similar,
\be
\Mcal_{\cO \cO^\prime}^{(m g\phi \psi \frac{1}{\partial} \psi)} &=& 2 mg \frac{(-1)^{\tilde{\Delta}-\tilde{\Delta}'}N_{\rm FT}}{(4\pi)^{3/2}N_{\CO} N_{\CO'}} \nn\\
 & \times & \sum_{\bk, \bk'} \left[  C^{\cO}_{\bk} \, C^{\cO'}_{\bk^\prime} \, \Ncal^{(M)}_{\bk}\, \Ncal^{(M)}_{\bk'}  \sum_{\bk/\{k_i| s_i=1\} = \atop  \bk'/\{k_i | s_i =-1 \}}(-1)^{\sigma(\{k_i, s_i\}) }  \norm{\bk/\{k_i\}}^2 g_{\phi\psi\frac{1}{\p}\psi}(k_i, s_i) \right],  \nn\\
\ee
 and also depends on a function $g_{\phi \psi \frac{1}{\p} \psi}$ that is given in an appendix, in equation (\ref{eq:yukawa-cubic-final-factor}).  In this case, the index $i$ runs from $1$ to $3$, and the fields are contracted with $k_i$ in the interaction as follows: $\phi_{k_1} \psi_{k_2} \frac{1}{\p} \psi_{k_3}$. The convention for the signs $s_i$ is the same as above.

%%%%%%%%%%%%%%%%%%%%%%%%%%%%%%%%%%%%%%%%%%%%%%%%%%%%%%%%%%%%%%%%%%%%%%%%%%%%%
%%%%%%%%%%%%%%%%%%%%%%%%%%%%%%%%%%%%%%%%%%%%%%%%%%%%%%%%%%%%%%%%%%%%%%%%%%%%%
%%%%%%%%%%%%%%%%%%%%%%%%%%%%%%%%%%%%%%%%%%%%%%%%%%%%%%%%%%%%%%%%%%%%%%%%%%%%%

\newpage
\section*{\Large Part III: Applications}
\addcontentsline{toc}{part}{Part III: Applications}
\label{sec:PartIII}

%%%%%%%%%%%%%%%%%%%%%%%%%%%%%%%%%%%%%%%%%%%%%%%%%%%%%%%%%%%%%%%%%%%%%%%%%%%%%
%%%%%%%%%%%%%%%%%%%%%%%%%%%%%%%%%%%%%%%%%%%%%%%%%%%%%%%%%%%%%%%%%%%%%%%%%%%%%

\section{Application I: $\phi^4$ Theory}
\label{sec:ApplicationPhi4}
 
In this section we implement our most efficient LCT code to study two-dimensional $\phi^4$ theory. The Lagrangian is
\be
\Lcal = \frac{1}{2}\p^\mu\phi \p_\mu\phi - \frac{1}{2}m^2\phi^2 - \frac{1}{4!}\lambda \phi^4,
\ee
and the corresponding lightcone Hamiltonian is
\be
P_+ = \int dx^- \left(  \frac{1}{2}m^2\phi^2 + \frac{1}{4!}\lambda\phi^4 \right). 
\ee
With $m$ simply setting an overall scale, there is only one physical parameter in the game, which we take to be the dimensionless ratio 
\be
\bar{\lambda} \equiv \frac{\lambda}{4\pi m^2}.
\ee

This theory was studied previously using LCT in \cite{Anand:2017yij}, at a maximum truncation level of $\Dmax=34$, corresponding to 12,310 basis states. That work predated many of the technological developments presented in Part II, which now allow us to reach higher $\Dmax$. Here we will work at $\Dmax=40$, which corresponds to a basis size of 37,338 states. 

We begin in section~\ref{subsec:Phi4Spectrum} by studying the mass spectrum as a function of $\bar{\lambda}$. Recall that in section~\ref{sec:Phi4PhaseTransition}, using a basis of only two states, we proved that 2d $\phi^4$ theory must have a phase transition at some critical coupling $\bar{\lambda}_c$, but we were unable to determine the nature of the transition and could only put a loose upper bound on the value of $\bar{\lambda}_c$. Working at $\Dmax=40$, we will see that the phase transition is second-order and be able to put a much stronger bound on $\bar{\lambda}_c$. 

Next, in section~\ref{subsec:Phi4SpectralDensities}, we compute several examples of nonperturbative spectral densities, which correspond to infinite-volume two-point correlation functions. Among them is the Zamolodchikov $C$-function. Indeed, one of our main messages is that LCT allows one to compute spectral densities at \emph{any} value of $\bar{\lambda}$ in the symmetry preserving phase ($\bar{\lambda} < \bar{\lambda}_c$).\footnote{An analysis of the symmetry-broken phase using LCT is the subject of upcoming work.} In particular, we will see that LCT results for spectral densities converge rapidly with $\Dmax$, especially in the IR. To the best of our knowledge, these spectral density and $C$-function results are \emph{novel predictions} for the nonperturbative dynamics of 2d $\phi^4$ theory and begin to illustrate what LCT has to offer.   

Finally, in section~\ref{subsec:IsingModel}, we focus our attention on dynamics near the critical point $\bar{\lambda}\approx \bar{\lambda}_c$. 
As we will review, in the vicinity of the critical point, the IR physics of $\phi^4$ theory should have an effective description in terms of the 2d Ising CFT with an $\epsilon$ deformation. We compute several spectral densities and show that they exhibit behavior consistent with Ising model predictions. Specifically, we demonstrate the onset of universal behavior in $\phi^n$ spectral densities, the vanishing of the trace of the stress tensor, and the matching of the $C$-function with theoretical predictions. This provides a highly nontrivial check of LCT in the context of $\phi^4$ theory. At this value of $\Dmax$, we are unable to extract the Ising central charge $c_{\text{IR}}=\frac{1}{2}$ right at the critical point, but we explain the barriers involved and discuss this observable as a concrete goal for the future of LCT.  

The Mathematica packages and notebooks used to perform our analysis are included in the supplementary material. In particular, the packages {\tt Basis-Scalar.wl} and {\tt MatrixElements-Scalar.wl} contain the functions needed to generate the scalar basis and Hamiltonian matrix elements. These functions implement the Radial Quantization method presented in section \ref{sec:RadialScalars}. For timing benchmarks see Table \ref{tab:timingBenchmark}. Additionally, the notebook {\tt Phi4Demo.nb} demonstrates how to use these packages and then provides a step-by-step tutorial for generating all of the plots that we present in this section. 

\begin{table}[t!]
\centering
\begin{tabular}{|c|c||c|c|c|c|}
\hline
 $\Delta_{\max}$ & $\substack{\text{num of}\\\text{states}}$ & basis & mass & $n$-to-$n$ & $n$-to-$(n+2)$ \\
 \hline
 10 & 42 & 0.02 & 0.06 & 0.04 & 0.03 \\
 \hline
 20 & 627 & 0.46 & 1.09 & 2.47 & 1.49 \\
 \hline
 30 & 5604 & 7.88 & 17.93 & 65.72 & 46.20 \\
 \hline
 40 & 37338 & 231 & 410 & 1567 & 2012 \\
 \hline
\end{tabular}
\caption{\label{tab:timingBenchmark}
The timing benchmark of the radial quantization scalar $\phi^4$ package. The table shows the time in seconds needed to compute the scalar basis and matrix element data at different $\Dmax$. The timing data is obtained by running the package on a single CPU at machine precision (corresponding to 53 binary digits of precision).
}
\end{table}

\subsection{Spectrum}
\label{subsec:Phi4Spectrum}

\begin{figure}[t!]
\begin{center}
\includegraphics[width=.7\textwidth]{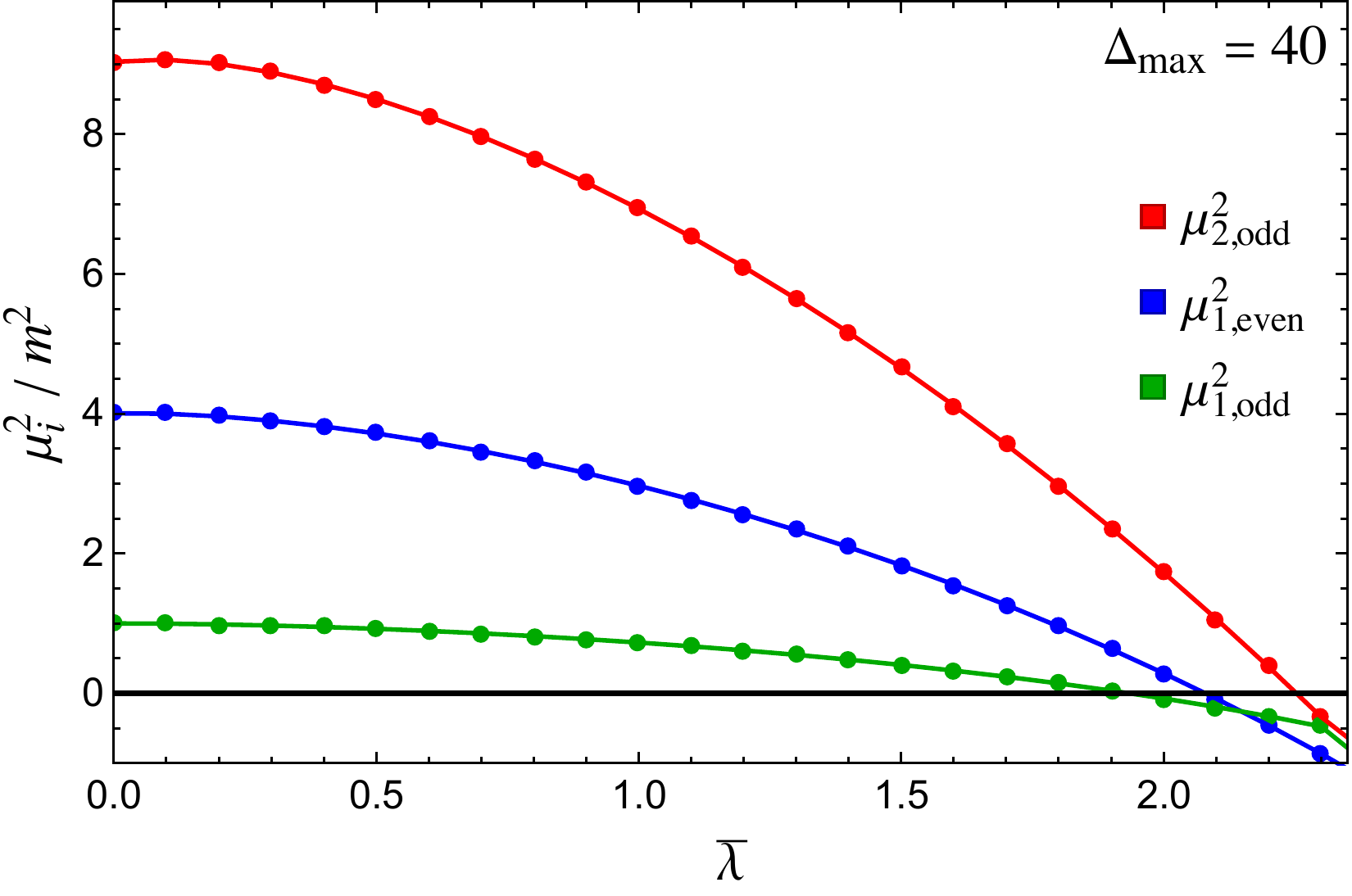}
\caption{Spectrum of $\phi^4$ theory at $\Dmax=40$. $\mu_{1,\text{odd}}^2$ (green) and $\mu_{2,\text{odd}}^2$ (red): lowest and second-lowest eigenvalues, respectively, in the odd-particle-number sector. $\mu_{1,\text{even}}^2$ (blue): lowest eigenvalue in the even-particle-number sector.}
\label{fig:Phi4LowEvals}
\end{center}
\end{figure}

In this section, we examine the $\phi^4$ theory mass spectrum (which are the eigenvalues of the lightcone Hamiltonian) as a function of the dimensionless coupling $\bar{\lambda} \equiv \frac{\lambda}{4\pi m^2}$. Concretely, we vary $\bar{\lambda}$ over a desired range, and for each $\bar{\lambda}$, we diagonalize the full Hamiltonian and record the lowest few eigenvalues. The resulting plot is shown in Fig.~\ref{fig:Phi4LowEvals}. Because this theory is invariant under $\phi \ra -\phi$, we can divide the spectrum into independent odd- and even-particle-number sectors and diagonalize the Hamiltonian in each sector separately. In this plot, the green and red lines denote, respectively, the lowest and second-lowest eigenvalues in the odd-particle-number sector, while the blue line denotes the lowest eigenvalue in the even-particle-number sector. We will use the notation $\mu_{1,\text{odd}}^2$ (green), $\mu_{1,\text{even}}^2$ (blue), and $\mu_{2,\text{odd}}^2$ (red) to refer to these eigenvalues. In the free field limit $\bar{\lambda}=0$, they correspond to the 1-, 2-, and 3-particle mass thresholds. 

In Fig.~\ref{fig:Phi4LowEvals}, we see that as we increase $\bar{\lambda}$, the mass eigenvalues cross zero continuously. This indicates a \emph{second-order} phase transition (see section~\ref{sec:Phi4PhaseTransition}). In the infinite $\Dmax$ limit, all three eigenvalues should cross zero at the same $\bar{\lambda}_c$, consistent with a closing mass gap at a fixed critical coupling. Our plot is consistent with this expectation up to finite $\Dmax$ effects. Note that $\mu_{1,\text{odd}}^2$, $\mu_{1,\text{even}}^2$, and $\mu_{2,\text{odd}}^2$ have an offset in their horizontal intercepts, \emph{i.e.}, they disagree on the value of the critical coupling. This offset is a finite truncation effect that decreases with increasing $\Dmax$, as one can verify. 

Recall that LCT is a variational method and hence allows us to put an upper bound on the value of $\bar{\lambda}_c$. In section~\ref{sec:Phi4PhaseTransition}, using a basis of two states, we derived the loose bound $\bar{\lambda}_c \lesssim 3.8$. At $\Dmax=40$, we of course do significantly better. Reading off the horizontal intercept of the lowest eigenvalue $\mu_{1,\text{odd}}^2$, we obtain the bound\footnote{Our estimate for the critical coupling roughly agrees with other LC quantization results, in particular, the estimate obtained in~\cite{Harindranath:1987db,Harindranath:1988zt} using the method of discretized lightcone quantization (DLCQ)~\cite{Pauli:1985pv,Pauli:1985ps,Brodsky:1997de} and the estimate obtained in~\cite{Burkardt:2016ffk,Chabysheva:2015ynr} using a variant of our conformal basis (see~\cite{Anand:2017yij} for a more detailed comparison). The value of the critical coupling differs in LC and ET quantization.  For  estimates of the ET critical coupling, see ~\cite{Lee:2000ac,Sugihara:2004qr,Schaich:2009jk,Milsted:2013rxa,Bosetti:2015lsa,Rychkov:2014eea}. See \cite{Fitzpatrick:2018xlz} for a discussion of how the ET and LC couplings can be mapped to each other. 
} 
\be
\boxed{ \bar{\lambda}_c \leq 1.94 \hspace{5mm} (\Dmax=40). }
\ee

Given that we can compute the spectrum at different values of $\Dmax$, a natural goal is to try and extrapolate finite $\Dmax$ data to infinite $\Dmax$. In the case of the spectrum, this is straightforward to do in principle. At any fixed $\bar{\lambda}$, we simply track how the spectrum of eigenvalues changes with $\Dmax$ and then try to fit the data and extrapolate. In practice, however, fitting the dependence of the eigenvalues on $\Dmax$ is challenging, because data points are correlated and we are limited by the range of $\Dmax$ we can access.

We can make progress using some additional assumptions. At fixed $\bar{\lambda}$ on the symmetry-preserving side, we suppose the $\Dmax$-dependence of eigenvalues is given by 
\be
\mu_i^2 (\Dmax) = A + \frac{B}{\Dmax^n}, 
\label{eq:Phi4:ExtrapolateModel}
\ee
where the parameters $A$, $B$, and $n$ are $\bar{\lambda}$-dependent. This rough Ansatz follows from an analysis of the $\Dmax$ scaling of individual Hamiltonian matrix elements performed in \cite{Anand:2017yij}. The same analysis also suggests that the exponent $n$ in (\ref{eq:Phi4:ExtrapolateModel}) is constrained by $1\lesssim n \lesssim 2$, with $n\approx 2$ near free field theory and decreasing to $n\approx 1$ as the coupling increases. 

\begin{figure}[t!]
\begin{center}
\includegraphics[width=0.49\textwidth]{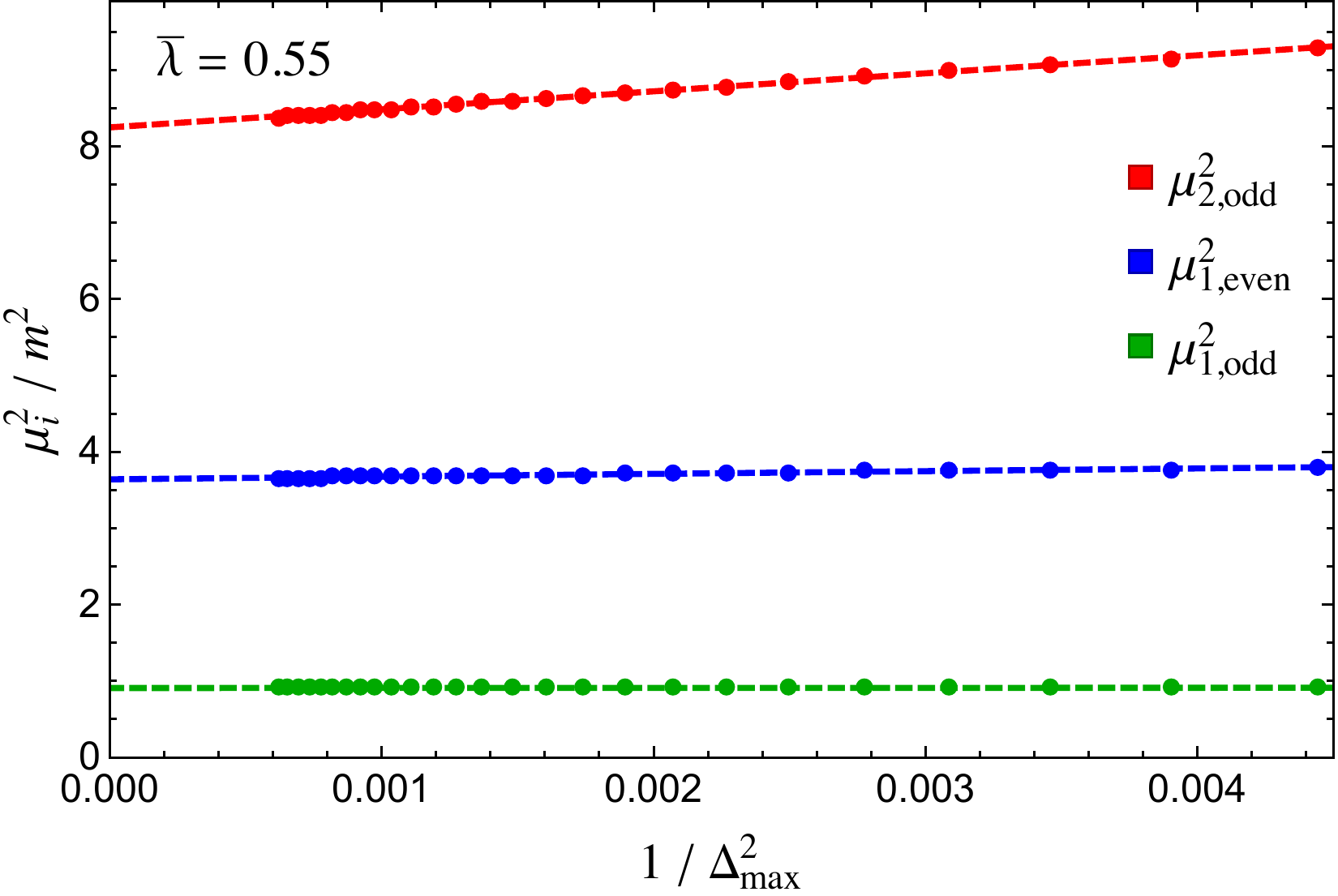}
\includegraphics[width=0.49\textwidth]{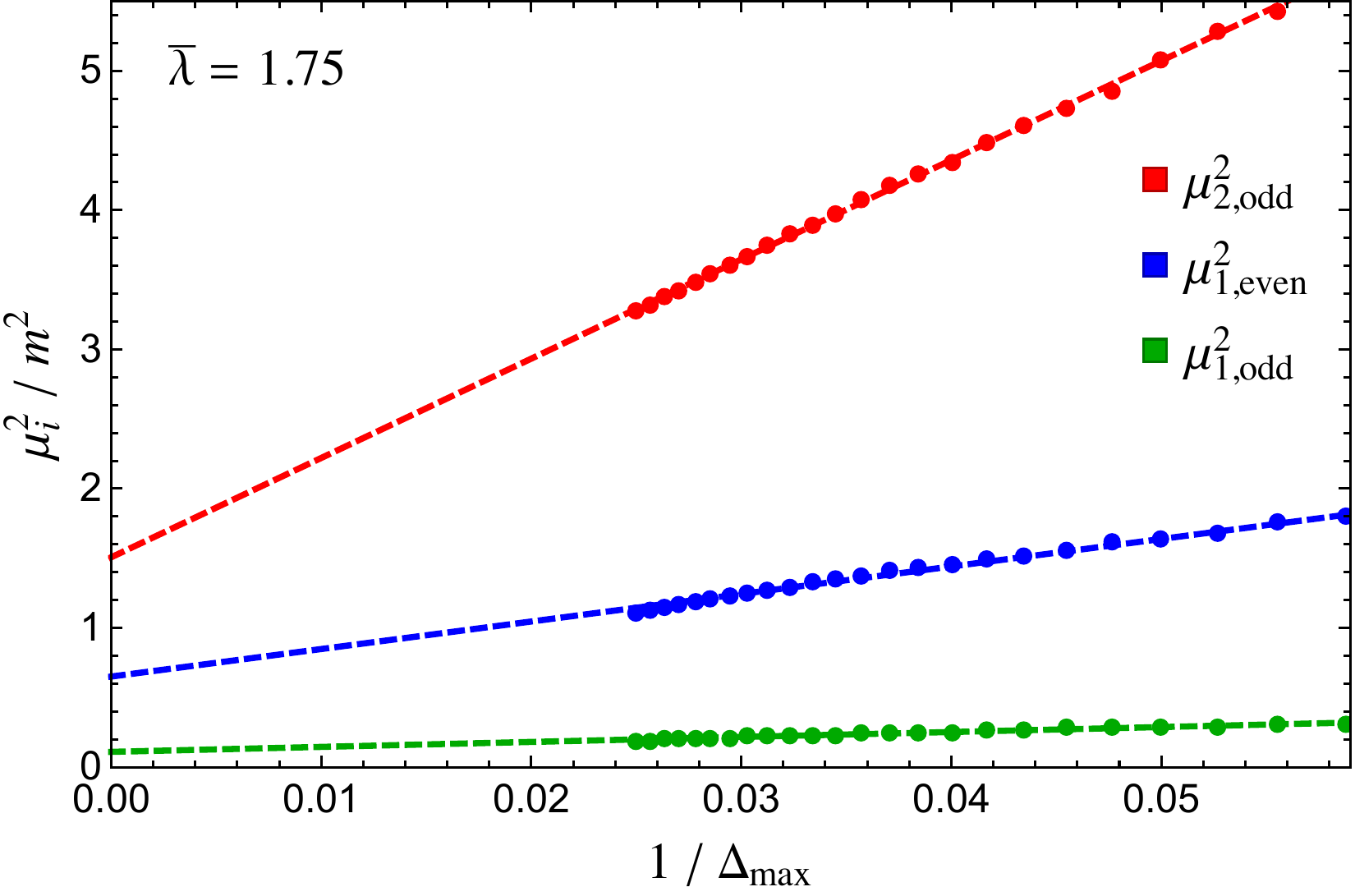}
\caption{Two examples of the dependence of $\mu_{1,\text{odd}}^2$ (green), $\mu_{1,\text{even}}^2$ (blue), and $\mu_{2,\text{odd}}^2$ (red) on $\Dmax$, at fixed $\bar{\lambda}=0.55$ (left) and $\bar{\lambda}=1.75$ (right). The dashed lines are best fits for each $\mu_i^2(\Dmax)$ obtained using the functional form in (\ref{eq:Phi4:ExtrapolateModel}), with the resulting powers $n=2.0$ (left) and $n=1.0$ (right). }
\label{fig:Phi4Extrap}
\end{center}
\end{figure}

Making these assumptions yields extrapolations like the ones shown in Fig.~\ref{fig:Phi4Extrap}. These plots show how $\mu_{1,\text{odd}}^2$, $\mu_{1,\text{even}}^2$, and $\mu_{2,\text{odd}}^2$ vary with $\Dmax$ for two different values of the coupling: in the left figure $\bar{\lambda}=0.55$, corresponding to weak/moderate coupling, and in the right figure $\bar{\lambda}=1.75$, corresponding to strong/nearly-critical coupling. Using extrapolations like these, the authors of \cite{Anand:2017yij} provided evidence for the simultaneous closing of eigenvalues at criticality in the $\Dmax\rightarrow\infty$ limit. 

There remains much to be understood about the behavior of LCT as we vary $\Dmax$. For instance, while the Ansatz~\eqref{eq:Phi4:ExtrapolateModel} roughly matches the eigenvalue behavior in Fig.~\ref{fig:Phi4Extrap}, the fit appears to be much better at weaker and intermediate coupling than near criticality, where additional structure seems very plausible. More generally, we can ask how large $\Dmax$ needs to be in order to establish this scaling behavior, and how to precisely define the uncertainty in infinite $\Dmax$ predictions. By better understanding the corrections as a function of $\Dmax$, we can hope to greatly improve the convergence and resulting extrapolations for LCT.

\subsection{Spectral Densities} 
\label{subsec:Phi4SpectralDensities}

In this section, as one example of the types of observables one can access using LCT, we compute the spectral densities of the operators $\phi^2$  and the stress tensor component $T_{--} \equiv (\p_-\phi)^2$ at strong coupling. 
To the best of our knowledge, these are new results for the nonperturbative dynamics of 2d $\phi^4$ theory. The spectral density $\rho_{T_{--}}(\mu)$ is a particularly important observable in 2d QFTs, because its integral corresponds to the spectral representation of the Zamolodchikov $C$-function~\cite{Zamolodchikov:1986gt,Cappelli:1990yc,Mussardo:2010mgq},
\be
C(\mu) \equiv \frac{12\pi}{p_-^4} \int_0^{\mu^2} d\mu^{\prime 2} \, \rho_{T_{--}}(\mu^\prime).
\label{eq:CFuncDefn}
\ee
As is well known, along an RG flow, $C(\mu)$ monotonically interpolates between the central charges of the UV and IR fixed points and provides a measure of how the number of degrees of freedom in the underlying theory changes with energy scale. 

Before examining the resulting plots, a few explanatory comments are needed. First, we will be comparing spectral density results at different truncation levels in order to study the convergence with $\Dmax$. The naive thing to do would be to \emph{fix the coupling} $\bar{\lambda}$ at a particular value, and then see how our spectral density results behave as we increase $\Dmax$. However, the drawback to doing this is that at a fixed value of the coupling, two things are changing as we increase $\Dmax$: (i) the spectrum itself (including the mass gap), and (ii) the functional form of the spectral density.

We would like to focus our attention here on the functional form of spectral densities. Therefore, it is preferable to \emph{fix the gap}, $m^2_{\text{gap}}$, instead of fixing the coupling $\bar{\lambda}$. That is,  to compare different $\Dmax$, we first fix the value of $m^2_{\text{gap}}$ (in units of the bare mass $m^2$) that we want to study, and for each $\Dmax$ we choose the coupling so that the mass gap is $m^2_{\text{gap}}$. In keeping the mass gap fixed, we are imagining that we are IR observers, where $m^2_{\text{gap}}$ is the physical parameter, and asking how correlation functions change with $\Dmax$. This will be particularly useful in later sections when we compare with theoretical predictions that depend on $m^2_{\text{gap}}$. 

Since we will be keeping $m^2_{\text{gap}}$ fixed, the plots that follow will typically be labeled by a value of $m^2_{\text{gap}}$ rather than a value of $\bar{\lambda}$. The only exception is free field theory, where by definition $\bar{\lambda}=0$. 
In this section, we will only consider spectral densities of operators with even particle number, so the lowest mass eigenvalue corresponds to the two-particle threshold. Thus, in practice we define $m^2_{\text{gap}}$ to be $\frac{1}{4}$ times the lowest eigenvalue in the even-particle sector, \emph{i.e.}, in the notation of the previous subsection 
\be
m^2_{\text{gap}}= \tfrac{1}{4} \mu_{1,\text{even}}^2.
\ee
 Finally, as always, we will be plotting integrated spectral densities
\be
I_{\Ocal}(\mu) \equiv \int_0^{\mu^2} d\mu^{\prime 2} \rho_{\Ocal}(\mu^\prime), 
\ee
which contain the same information as $\rho_{\Ocal}(\mu)$ but are smoother when computed numerically.

\begin{figure}[t!]
\begin{center}
\includegraphics[width=0.95\textwidth]{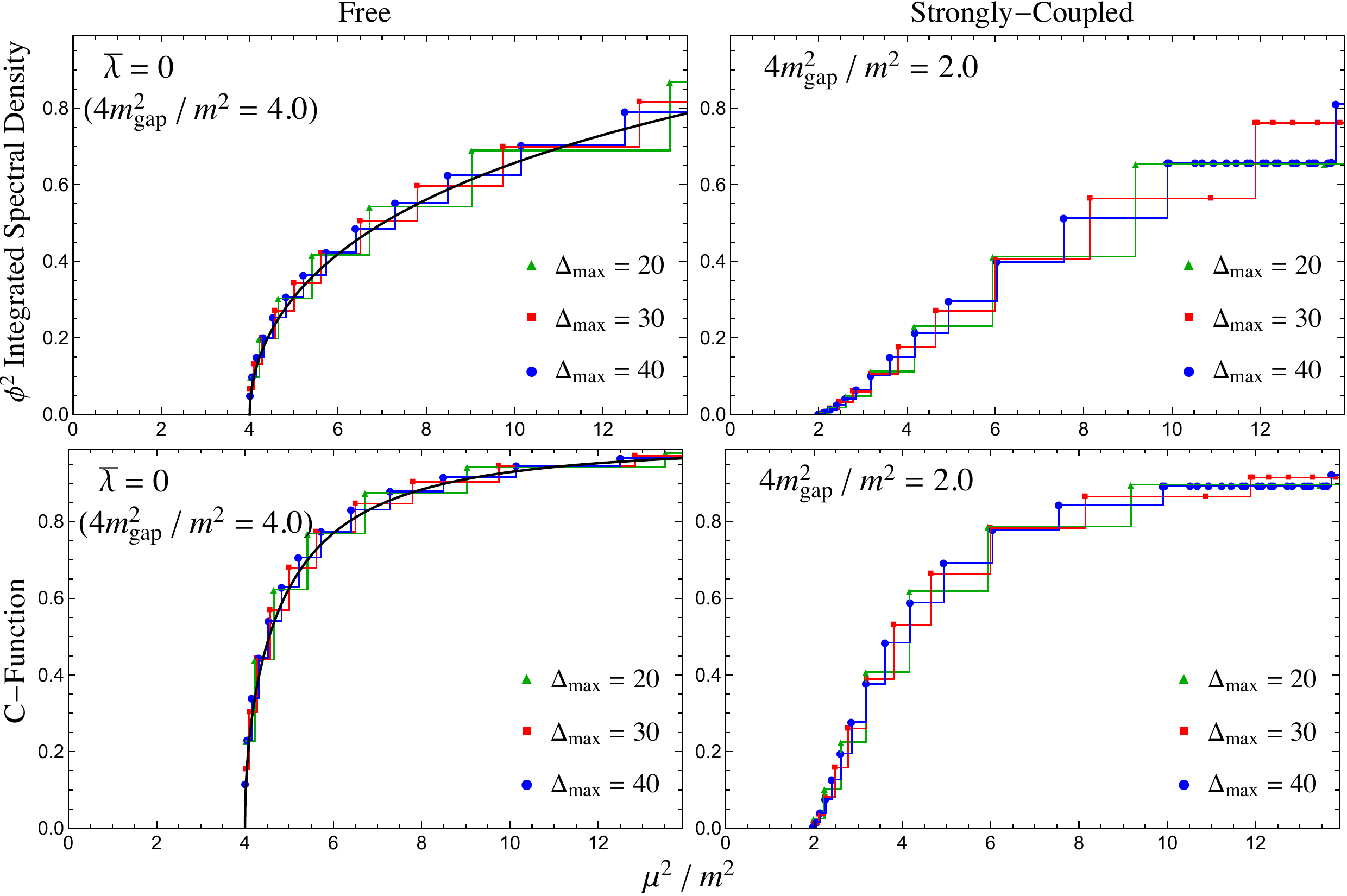}
\caption{Integrated spectral density of the operator $\phi^2$ (first row) and the $C$-function (second row) in 2d $\phi^4$ theory. In each row, the left figure corresponds to free field theory ($\bar{\lambda}=0$), while the right figure corresponds to strong coupling, where $\bar{\lambda}$ has been chosen such that $\fr{4m^2_{\text{gap}}}{m^2}=2.0$ (for $\Dmax=40$ this corresponds to $\bar{\lambda}=1.44$). Each figure shows results for $\Dmax=20$, $30$, and $40$. The black line in the free theory figures (left column) is the exact analytical result, for comparison.} 
\label{fig:Phi4SpectralDensities} 
\end{center}
\end{figure}

Now we are ready to see plots. 
Fig.~\ref{fig:Phi4SpectralDensities} shows the integrated spectral density of the operator $\phi^2$ (first row) and the $C$-function (second row).
 In each row, the left figure corresponds to free field theory ($\bar{\lambda}=0$), which we have included for comparison, while the right figure corresponds to strong coupling, with $\bar{\lambda}$  chosen (see the discussion above) such that $\fr{m^2_{\text{gap}}}{m^2}=0.5$, i.e.~$\fr{4m^2_{\text{gap}}}{m^2}=2$. Each figure shows data for $\Dmax = 20$, $30$, and $40$.
In the free theory figures (left column), the black line denotes the exact analytical result. In each row, we see that there are significant changes to the correlation functions at strong coupling (right column) compared to the non-interacting theory. 

Let us note some features of the $C$-function. On the horizontal axis, the UV regime corresponds to $\mu^2 \rightarrow \infty$. If we were to extend both $C$-function figures to larger $\mu^2$, we would see that in both plots $C(\mu)$ asymptotes to $c_{\text{UV}} = 1$, correctly reproducing the central charge of a free boson. At the opposite end, the IR regime corresponds to $\mu^2 \rightarrow 0$. In both the left and right figures, the theory is gapped. Correspondingly, $C(\mu)$ falls to zero precisely at $\mu^2 = 4m^2_{\text{gap}}$, as it should.

Finally, let us examine convergence with $\Dmax$. We see visually that by $\Dmax=40$, our LCT results are converging rapidly over the range of $\mu^2$ shown. In the free theory, the truncation results correctly reproduce the analytical prediction, even at low values of $\Dmax$, with the resolution improving as $\Dmax$ increases. At strong coupling, we have no predictions to compare with, since these are new results. Nevertheless, by comparing the behavior as we vary $\Dmax$, we see that the results at strong coupling appear to be converging self-consistently. 

There is another extremely important observation to highlight regarding convergence: our LCT results converge \emph{from the IR up}. What we mean by this is that convergence happens most rapidly in the IR, \emph{i.e.}, at small $\mu^2$, and then works its way up to the UV. We see this clearly in Fig.~\ref{fig:Phi4SpectralDensities}, where the agreement between the different $\Dmax$ data is best in the IR. This observation is a strong indication that it is precisely the low-$\Delta$ UV CFT basis states that have the most overlap with the physical IR degrees of freedom, even at strong coupling. This is why truncating in $\Dmax$ seems to be an effective strategy. 

In the next subsection, we will turn to the details of dynamics near the critical point $\bar{\lambda}_c$. As we will see, tuning close enough to the critical point to extract certain Ising model observables like the central charge can be challenging. However, we  emphasize that for generic strong coupling (\emph{i.e.}, not too close to the critical point), LCT results for spectral densities converge rapidly and provide novel predictions for nonperturbative dynamics. This is one of the key messages of this work.

\subsection{Critical Point and the Ising Model}
\label{subsec:IsingModel}

In this section, we compute spectral densities near the critical point $\bar{\lambda}\approx\bar{\lambda}_c$. As is well known, near this critical coupling $\phi^4$ theory is in the same universality class as the 2d Ising model. That is, as $\bar{\lambda} \ra \bar{\lambda}_c$, the spectrum and correlation functions of $\phi^4$ theory in the IR should match those of the Ising model deformed by the $\mathbb{Z}_2$-even operator $\epsilon$:
\be
\Lcal_{\phi^4}(\bar{\lambda}_c) + \fr{1}{4!} 4\pi m^2 (\bar{\lambda}_c-\bar{\lambda}) \phi^4 \quad \overset{\textrm{in IR}}{\Rightarrow}\quad \Lcal_{\textrm{Ising}} - m_{\gap} \epsilon.
\ee
The qualifier ``in the IR" is crucial here; $\phi^4$ theory is \emph{not} the same theory as the 2d Ising model. They are distinct theories with distinct physical observables. Rather, the Ising model is an effective description of $\phi^4$ theory at low energies near criticality. Because the $\epsilon$ deformation of the Ising model is integrable, we can use the analytical Ising results as predictions for the IR behavior of our LCT results near the critical point. The agreement we find provides a highly nontrivial check of LCT for 2d $\phi^4$ theory. After all, we have not input anything about the Ising model. If we correctly reproduce it, it is because we are constructing the full RG flow from the UV to the IR. 

\subsubsection{Universal Behavior}

We start by considering the scalar operators $\phi^n$. Near $\bar{\lambda}_c$, we expect that these operators will all flow in the IR to the lowest-dimension operators in the Ising model with the same quantum numbers (in this case, parity under $\mathbb{Z}_2$), such that
\be
\phi^{2n} \Rightarrow \epsilon + \cdots, \hspace{5mm} \phi^{2n-1} \Rightarrow \sigma + \cdots,
\label{eq:PhiNtoEpsilonSigma}
\ee
where the ellipses denote higher-dimensional operators. For brevity, we will focus on even parity (both parities were considered in~\cite{Anand:2017yij}). For the parity even operators, eq.~(\ref{eq:PhiNtoEpsilonSigma}) implies \emph{universal} IR behavior for the associated spectral densities, in the sense that (up to an overall proportionality constant) they should all match the known spectral density for $\epsilon$ at low mass scales,\footnote{The expression for $\rho_{\epsilon}(\mu)$ can be computed analytically from its decomposition into Fock space states in the free fermion description of the Ising model~\cite{Cappelli:1989yu}.}
\be
\rho_{\phi^{2n}}(\mu) \propto \rho_{\epsilon}(\mu) = \frac{1}{16\pi} \sqrt{1-\frac{4m_{\text{gap}}^2}{\mu^2}}, \hspace{10mm} (\mu^2 \rightarrow 4m_\gap^2).
\label{eq:EpsilonSpectralDensity}
\ee

\begin{figure}[t!]
\begin{center}
\includegraphics[width=.7\textwidth]{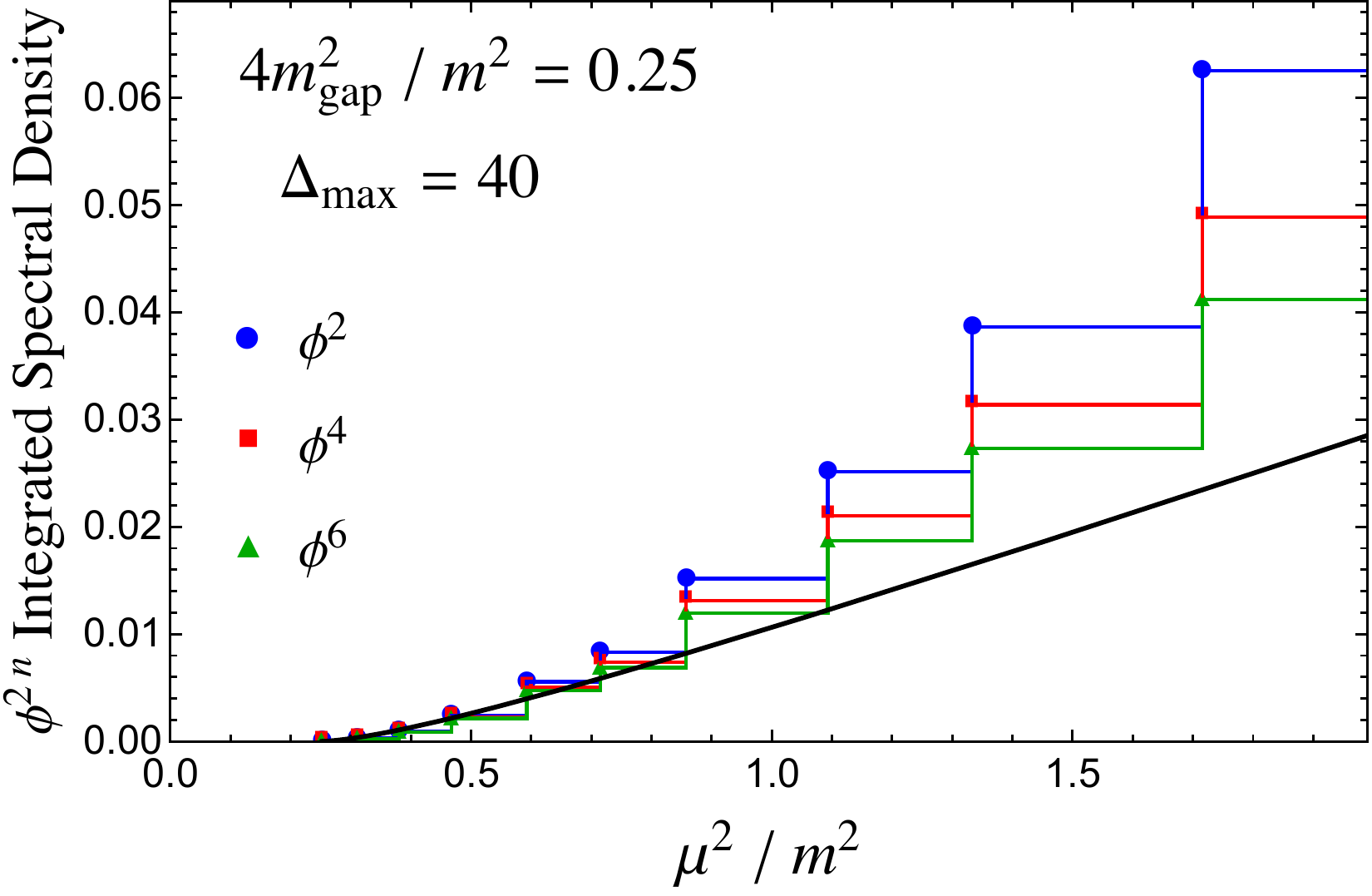}
\caption{Integrated spectral densities for $\phi^2$, $\phi^4$, and $\phi^6$ at $\Dmax=40$ and $\fr{4m_{\text{gap}}^2}{m^2} = 0.25$ (near the critical point). The spectral densities have been rescaled by an overall coefficient such that the first data points match. The black line (Ising prediction) is the integrated spectral density of $\epsilon$.}
\label{fig:Phi4TheoryUniversality}
\end{center}
\end{figure}

Our LCT results indeed reproduce this universal behavior. For example, Fig.~\ref{fig:Phi4TheoryUniversality} shows the spectral densities of $\phi^2$, $\phi^4$, and $\phi^6$ computed at $\Dmax=40$ and $\fr{4m_{\text{gap}}^2}{m^2} = 0.25$, which is close to the critical point ($m_{\text{gap}}^2 = 0$ at criticality). The plot is zoomed in relative to the previous spectral density plots in order to focus on the IR (\emph{i.e.}, small $\mu^2$). The $\phi^{2n}$ spectral densities have been rescaled by an overall coefficient to account for the proportionality constant in~(\ref{eq:EpsilonSpectralDensity}), and the black line is the Ising model prediction for the integrated spectral density of $\epsilon$. In the IR, we see excellent agreement between all three spectral densities and the theoretical prediction. Similar plots can be make for any $m_{\text{gap}}^2 $ close to zero.
 In free field theory, $\phi^{2n}$ for different $n$ are of course distinct operators with completely different spectral densities (for instance in the free theory, $\phi^{2n}$  only has overlap with $2n$-particle states, so its spectral density is exactly zero for $\mu^2 < n^2 m^2$). The onset of universal behavior near the critical point is a sign that we are correctly reconstructing the RG flow to the Ising model.  

Note that the universality we are observing only occurs \emph{in the IR}. In Fig.~\ref{fig:Phi4TheoryUniversality}, for $\frac{\mu^2}{m^2}\gtrsim 1$ the $\phi^{2n}$ spectral densities deviate from each other and from the Ising model prediction. It is worth emphasizing that this is \emph{not} due to truncation error. Although we have not plotted the comparison with other truncation levels in this figure, by $\Dmax=40$ the spectral densities have converged over the range of $\mu^2$ shown. Rather, the deviation between these plots in the UV is physical. As mentioned previously, the Ising model is only an effective IR description of critical $\phi^4$ theory up to some cutoff (set by the UV coupling $\lambda$). In Fig.~\ref{fig:Phi4TheoryUniversality} we are seeing that for $\fr{4m_{\text{gap}}^2}{m^2} = 0.25$, the cutoff of the Ising description is roughly at $\frac{\mu^2}{m^2}\sim 1$. It is worth keeping this lesson in mind, as we will encounter the same behavior in the next two subsections.

\subsubsection{Stress Tensor Trace}

In this section, we study the spectral density of the stress tensor component $T_{+-}$, which corresponds to the trace $T^\mu_{\phantom{\mu}\mu}$ in two dimensions and is another extremely important observable in 2d QFTs. The vanishing of $T^\mu_{\phantom{\mu}\mu}$ signals the onset of conformal symmetry. As we will see in our data, the $T_{+-}$ spectral density goes to zero in the IR as we approach the critical point, indicating that the RG flow is indeed reaching an IR CFT. 

Before we can start, however, we immediately run into a subtlety regarding the proper way to define $T_{+-}$ in the specific setting of 2d $\phi^4$ theory. The subtlety is due to the fact that $\phi$ itself is not a primary operator. By Noether's procedure, we naively should have 
\be
T_{--} = (\p_-\phi)^2, \qquad T_{+-}^{(\text{naive})} = \frac{1}{2}m^2\phi^2 + \frac{1}{4!}\lambda\phi^4,
\label{eq:Tnaive}
\ee
with momentum generators given by
\be
P_- = \oint \frac{dx}{2\pi i} \, T_{--}, \qquad P_+ = \oint  \frac{dx}{2\pi i} \, T_{+-}^{(\text{naive})}.
\label{eq:Pdef}
\ee
In \eqref{eq:Tnaive}, $m$ and $\lambda$ are the bare parameters appearing in the Lagrangian, and in \eqref{eq:Pdef} the contour is a small circle around the origin.\footnote{Here we are working in radial quantization to easily study the OPE between $T_{--}$ and $T_{+-}$.} The problem with this collection of definitions is that the Ward identity is not satisfied. Specifically, 
\be
\left[P_+, T_{--}\right] + [P_-, T_{+-}^{(\text{naive})} ] \neq 0 .
\label{eq:WardProblem}
\ee
As we will now see, this is because $T_{+-}^{(\text{naive})}$ is missing a term. 

One way to demonstrate (\ref{eq:WardProblem}) and understand how to fix it is via the OPE. In the OPE limit, we are working with a free scalar field with two-point function
\be
\langle \phi(x)\phi(y) \rangle = -\frac{1}{4\pi} \log(x-y).
\ee
We can work out OPEs by simply Wick contracting. In particular, it is straightforward to work out the following OPEs, only keeping track of the singular terms,
\bq
\bal
T_{--}(x) \phi^2(y) &\sim \frac{1}{8\pi^2 (x-y)^2} - \frac{1}{2\pi(x-y)} \p\phi^2(y), \\
T_{--}(x) \phi^4(y) &\sim \frac{3}{4\pi^2 (x-y)^2} \phi^2(y) - \frac{1}{2\pi(x-y)} \p\phi^4(y) \\
T_{--}(x)T_{+-}^{(\text{naive})}(y) &\sim \frac{m^2}{16\pi^2 (x-y)^2} + \frac{\lambda}{32\pi^2(x-y)^2}\phi^2(y) \\ 
& \qquad - \, \frac{1}{2\pi(x-y)} \p T_{+-}^{(\text{naive})}(y). 
\eal
\label{eq:OPEs}
\eq
From the first OPE above it follows that 
\be
[ P_-, \phi^2 ] = -\frac{1}{2\pi} \p\phi^2, \label{eq:PMinusPhi2Comm}
\ee
and from the third OPE above it follows that 
\begin{eqnarray}
[ P_-, T_{+-}^{(\text{naive})} ] &=& -\frac{1}{2\pi} \p T_{+-}^{(\text{naive})}, \label{eq:PMinusComm} \\[5pt]
\left[ P_+, T_{--} \right] &=& \frac{\lambda}{32\pi^2} \p \phi^2 + \frac{1}{2\pi} \p T_{+-}^{(\text{naive})}. \label{eq:PPlusComm}
\end{eqnarray}
Summing (\ref{eq:PMinusComm}) and (\ref{eq:PPlusComm}), we see that the Ward Identity (\ref{eq:WardProblem}) does not hold. 

The heart of the problem is that $\phi^n$ is not primary. In any 2d CFT, the OPE of $T_{--}$ with a general scalar primary operator $\Ocal$ should have the form
\be
T_{--}(x)\Ocal(y) \sim \frac{-\Delta_{\Ocal}}{4\pi(x - y)^2} \Ocal(y) - \frac{1}{2\pi(x - y)} \p \Ocal(y).
\ee
The OPEs of $T_{--}$ with $\phi^2$ and $\phi^4$ in (\ref{eq:OPEs}) do not take this form. Note in particular in (\ref{eq:OPEs}) that $\phi^4$ can generate $\phi^2$, such that the distinction between these operators is muddied. This peculiar appearance of $\phi^2$ in the $T_{--} \times \phi^4$ OPE seeps into the OPE of $T_{--} \times T^{\textrm{(naive)}}_{+-}$, resulting in the ``extra" $\p\phi^2$ term in (\ref{eq:PPlusComm}).

Thus, the formulas (\ref{eq:Tnaive}) and (\ref{eq:Pdef}) above are mutually inconsistent because they violate the Ward identity, and something has to give. Fortunately, the commutators in (\ref{eq:PMinusPhi2Comm})-(\ref{eq:PPlusComm}) show us how to fix this inconsistency. A linear combination of these equations does vanish, which amounts to modifying $T_{+-}$ to be 
\be
\boxed{
T_{+-} = T_{+-}^{(\text{naive})} + \frac{1}{16\pi} \lambda \phi^2 = \left( \frac{1}{2}m^2  +  \frac{1}{16\pi} \lambda \right) \phi^2 + \frac{1}{4!}\lambda\phi^4. 
}
\label{eq:T+-New}
\ee
In this way, the Ward Identity leads us to the correct expression for $T_{+-}$. Note that $P_+$ is still given by (\ref{eq:Pdef}). The discrepancy between the true $T_{+-}$ and the integrand of $P_+$ appears to be a strange and unavoidable feature of 2d scalar field theory due to the fact that $\phi$ is not primary. 

\noindent\rule[0.5ex]{\linewidth}{1pt}
\footnotesize

We can confirm the validity of (\ref{eq:T+-New}) using an independent check directly in LC quantization. Ward identities imply relations between Hamiltonian matrix elements and operator overlaps. Using our formulas for matrix elements, we can check that these relations only hold when $T_{+-}$ is defined as in (\ref{eq:T+-New}). Let us see how this works. First, acting with the Ward identity on the vacuum and then acting with an additional $2P_-$ gives the relation
\be
M^2 T_{--}(0) \vac = -2P_-^2 \, T_{+-}(0) \vac.
\label{eq:WardVac}
\ee
The left side can be related to the LCT basis state $|(\p\phi)^2,p\> $,
\be
T_{--}(0) \vac = \int \frac{dp}{2\pi}\, N_{(\p\phi)^2} |(\p\phi)^2,p\>,
\ee
where we previously computed $N_{(\p\phi)^2} \doteq p/ \sqrt{48\pi}$  (e.g., see section~\ref{sec:FockSpace}).
Plug this into (\ref{eq:WardVac}) and then act from the left with any LCT bra state $\bra{\Ocal,p^\prime}$. The left side will become $\Mcal_{ \Ocal, (\p\phi)^2}$, which is the Hamiltonian matrix element of $P^2$ between $\bra{\Ocal,p^\prime}$ and $\ket{(\p\phi)^2,p}$ excluding an overall factor of $2p (2\pi)  \delta(p - p^\prime)$. Meanwhile, the right side is simply an overlap between operators. Consequently the Ward Identity implies
\be
\Mcal_{\Ocal,(\p\phi)^2} \doteq \sqrt{48\pi}\, \langle \Ocal, p | T_{+-}(0) \rangle.  
\ee
Using our formulas for matrix elements and operator overlaps from Part I (or the matrix element and overlap functions in the accompanying notebook {\tt SimpleScalarCode.nb}) one can explicitly check that the relation above is satisified only if $T_{+-}$ is defined as in (\ref{eq:T+-New}). 

\normalsize
\noindent\rule[0.5ex]{\linewidth}{1pt}

\begin{figure}[t!]
\begin{center}
\begin{tabular}{c}
\includegraphics[width=0.95\textwidth]{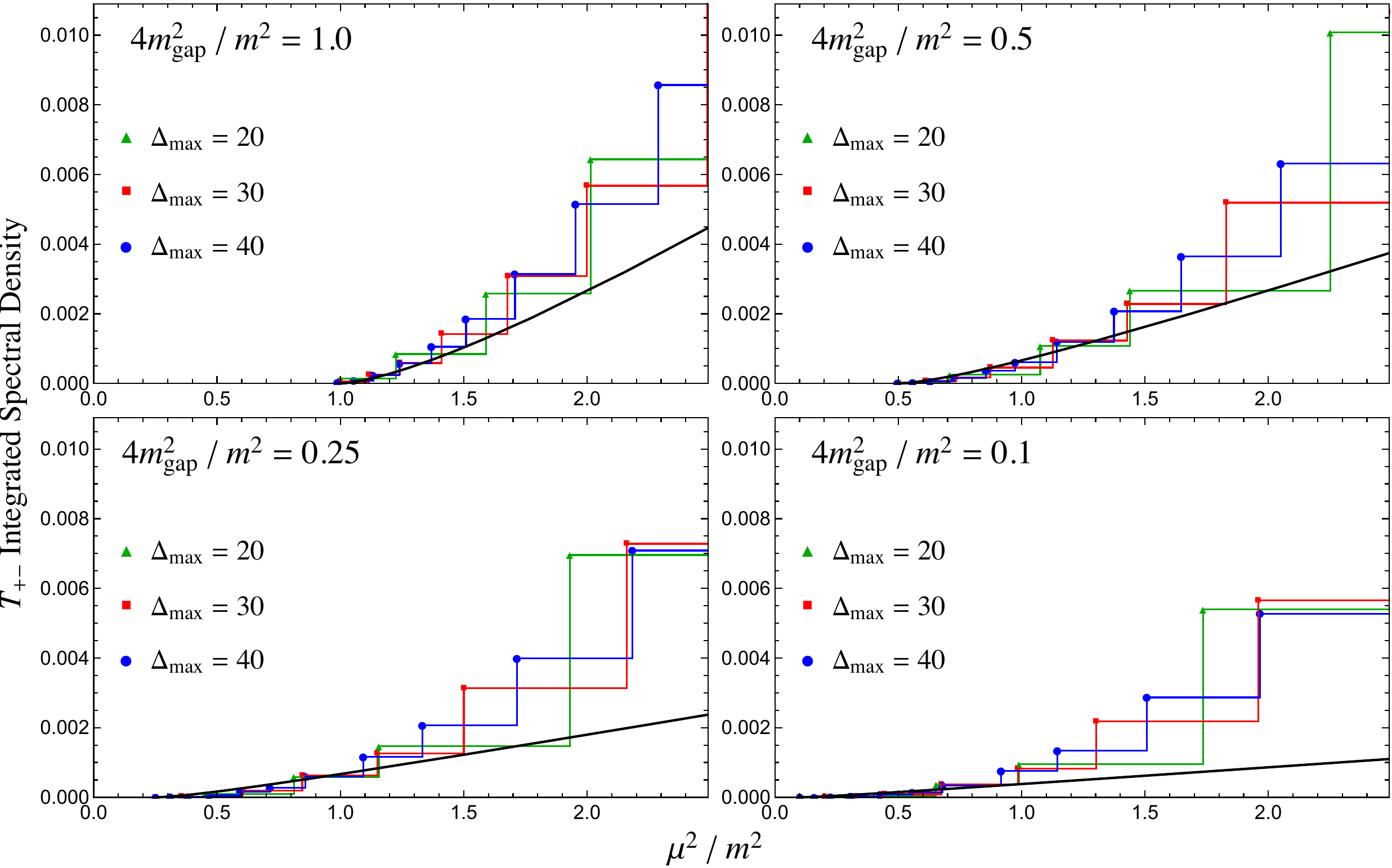} \\
\end{tabular}
\caption{Integrated spectral density of $T_{+-}$ (the trace of the stress tensor) at $\Dmax=20,30,40$ and four different values of $4m_{\text{gap}}^2$ approaching the critical point. The spectral density collapses to zero more and more in the IR as $m_{\text{gap}}^2\rightarrow 0$, signaling the approach to an IR CFT. The black line is the Ising model prediction~\eqref{eq:RhoT+-Ising}.} 
\label{fig:Phi4TheoryTrace} 
\end{center}
\end{figure}

Let us finally turn to our results for $T_{+-}$. 
In free field theory, we see from (\ref{eq:T+-New}) that $T_{+-}$ is simply proportional to the operator $\phi^2$. Thus, in free field theory, the integrated spectral density $I_{T_{+-}}(\mu)$ is proportional to the free field integrated spectral density of $\phi^2$, which is shown in the top left plot in Fig.~\ref{fig:Phi4SpectralDensities}. By comparison, in Fig.~\ref{fig:Phi4TheoryTrace} we plot $I_{T_{+-}}(\mu)$ for four different values of the mass gap: $\fr{4m_{\text{gap}}^2}{m^2} = 1.0$, $0.5$, $0.25$, and $0.1$. The most striking qualitative feature of these plots is that $I_{T_{+-}}(\mu)$ flattens out more and more to zero in the IR as $m_{\text{gap}}^2\rightarrow 0$, \emph{i.e.}, as we approach the critical point. This verifies the onset of conformal symmetry, and signals that we are indeed reaching a CFT in the IR near criticality. 

Of course, we know that the IR CFT we are reaching is the Ising model, so we can compare our data with quantitative predictions. The theoretical prediction from the Ising model is~\cite{Cappelli:1989yu} 
\be
\rho_{T_{+-}}^{\text{(Ising)}}(\mu) = m_{\gap}^2 \rho_\epsilon(\mu) = \frac{m_{\text{gap}}^2}{16\pi} \sqrt{1- \frac{4m_{\text{gap}}^2}{\mu^2} },
\label{eq:RhoT+-Ising}
\ee
which we show as a black line in the plots in Fig.~\ref{fig:Phi4TheoryTrace}. This analytical expression is parametrized solely by $m_{\text{gap}}$, which we take directly from the lowest eigenvalue in the LCT data. In particular, there is no overall proportionality constant that we need to fix. From the figure, we see that in the IR, our numerical results clearly match both the functional form and overall coefficient of the Ising prediction. This provides yet another highly nontrivial check of our numerical results.

\subsubsection{$C$-function}

Now we consider the $C$-function in the vicinity of the critical coupling. Ideally, near a critical point, one would want to use $C(\mu)$ to determine the central charge $c_{\text{IR}}$ of the IR CFT. In practice, however, for IR fixed points that are finely-tuned (as is the case in $\phi^4$ theory) a truncated spectrum will always have a small but nonzero mass gap, $m_{\text{gap}}$. Consequently, $C(\mu)$ will always drop to the trivial value of zero at $\mu^2 = 4m_{\text{gap}}^2$. If $m_{\text{gap}}$ is sufficiently small, there is nevertheless still hope, because $C(\mu)$ will plateau at $c_{\text{IR}}$ before eventually falling to zero. Our ability to extract $c_{\text{IR}}$ is determined by whether we can tune $m_{\text{gap}}$ to be small enough compared with other scales characterizing the RG flow.  

Unfortunately, there are two other scales in the game, which make it difficult to extract the IR central charge. First, we have the IR cutoff in resolution $\Lambda_{\textrm{IR}}$, which is solely a consequence of truncation, as discussed in section~\ref{sec:ScalarSpectrum}. We can roughly think of this IR cutoff as corresponding to the spacing between eigenvalues of the truncated Hamiltonian. The second scale is the UV cutoff of the effective Ising model description in $\phi^4$ theory, $\Lambda_{\textrm{Ising}}$. This physical scale is set by the value of the UV coupling $\bar{\lambda} \sim 2$ near the critical point. In order to read off the Ising central charge $c_{\text{IR}} = \half$ from the $C$-function, we therefore need a large separation between the following three scales:
\be
\Lambda_{\textrm{IR}}^2 \ll m_{\gap}^2 \ll \Lambda_{\textrm{Ising}}^2.
\ee
As we will show experimentally below, for $\Dmax=40$ the IR resolution is not yet small enough to see the IR plateau in the $C$-function. However, we can still test our results by accounting for the corrections to the Ising predictions due to $\Lambda_{\textrm{Ising}}$.

In the language of the Ising EFT, $T_{--}$ receives corrections from higher-dimensional Ising model operators suppressed by the UV cutoff,
\be
T_{--}^{(\phi^4)} \approx T_{--}^{(\text{Ising})} - \frac{\p_-^2\epsilon}{\Lambda_{\textrm{Ising}}} + \cdots,
\label{eq:Teff}
\ee
Here, $T_{--}^{(\phi^4)}$ is the $\phi^4$ theory stress tensor in the IR near criticality, $T_{--}^{(\text{Ising})}$ is the stress tensor of the $\epsilon$-deformed Ising model, $\p_-^2\epsilon$ is the leading irrelevant correction (note that it is suppressed by the cutoff $\Lambda_{\textrm{Ising}}$), and the dots denote other higher-dimensional irrelevant operators. 

Because the $C$-function is the integrated spectral density of $T_{--}$, we thus obtain the IR prediction
\be
C(\mu) \approx C^{(\textrm{Ising})}(\mu) + \de C_{\Lambda_{\textrm{Ising}}}(\mu) + \cdots,
\label{eq:Ceff}
\ee
where $C^{(\textrm{Ising})}(\mu)$ can be fixed in terms of $\rho_\epsilon(\mu)$ by the Ward identity,
\bq
\bal
C^{(\textrm{Ising})}(\mu) &= \fr{12\pi}{p_-^4} \int_0^{\mu^2} d\mu'^2 \rho_{T^{(\textrm{Ising})}_{--}}(\mu') = 48\pi m_{\gap}^2 \int_0^{\mu^2} \fr{d\mu'^2}{\mu'^4} \rho_\epsilon(\mu') \\
&= \half \left( 1 - \fr{4m_\gap^2}{\mu^2} \right)^{\fr{3}{2}},
\eal
\label{eq:IsingCFunc}
\eq
and $\de C_{\Lambda_{\textrm{Ising}}}(\mu)$ is the correction due to $\p_-^2\epsilon$,\footnote{The first term in the integrand comes from the cross term $\<T_{--}^{(\textrm{Ising})} \p_-^2 \epsilon\>$, and the second term comes from $\<\p_-^2 \epsilon \, \p_-^2 \epsilon\>$.}
\be
\de C_{\Lambda_{\textrm{Ising}}}(\mu) = 48\pi \int_0^{\mu^2} \fr{d\mu'^2}{\mu'^2} \left( - 2\fr{m_\gap}{\Lambda_{\textrm{Ising}}} + \fr{\mu'^2}{\Lambda^2_{\textrm{Ising}}} \right)  \rho_{\epsilon}(\mu').
\label{eq:IsingCCorr}
\ee

We do not \emph{a priori} know the value of $\Lambda_{\textrm{Ising}}$. However, we can fix its value by comparing our computation of the $C$-function in $\phi^4$ theory to the Ising prediction at a particular value of $m_\gap$, then use this same value of $\Lambda_{\textrm{Ising}}$ for all other plots near the critical point.

\begin{figure}[t!]
\begin{center}
\includegraphics[width=\textwidth]{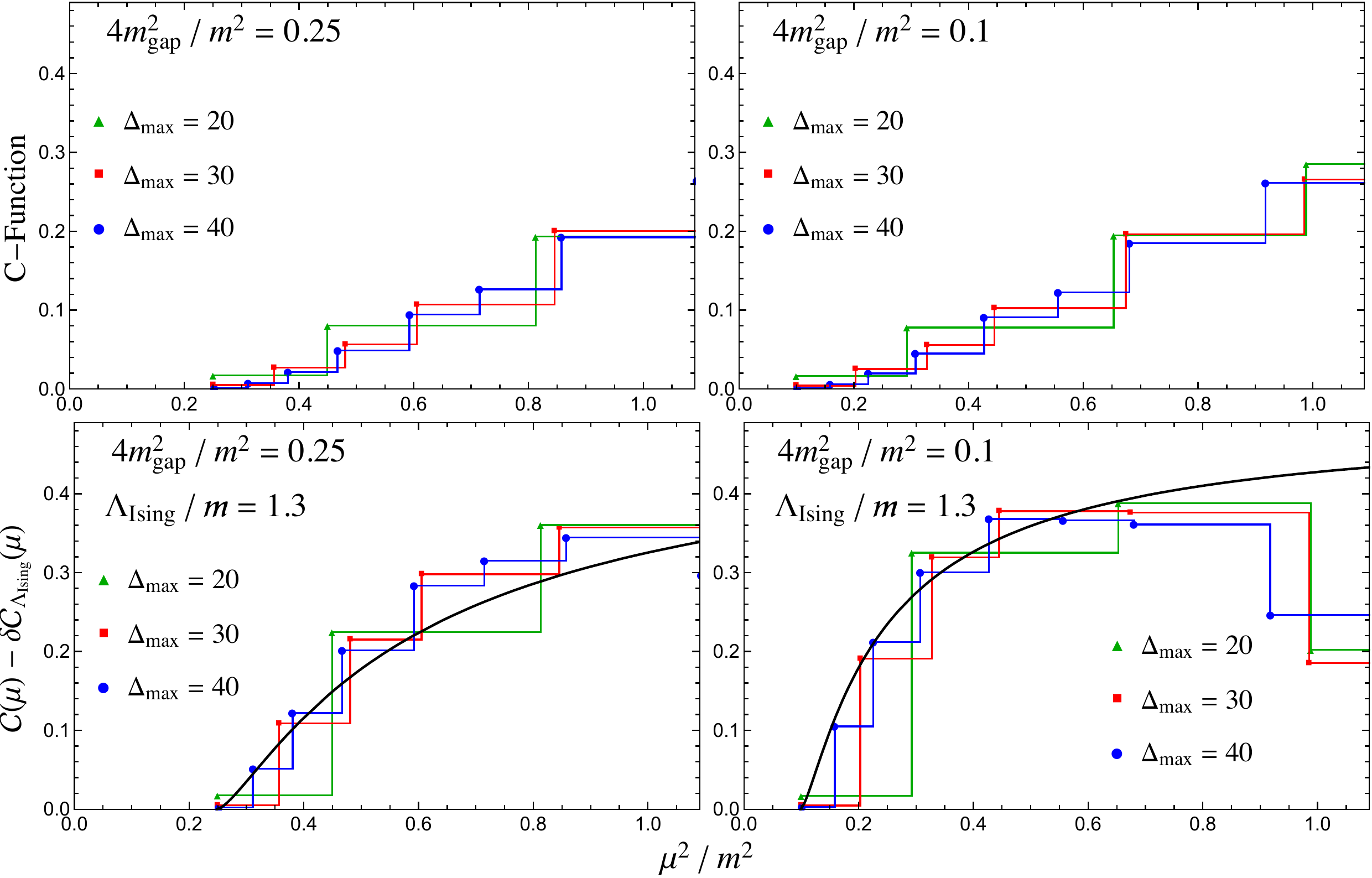}
\caption{\emph{Top:} IR behavior of the $C$-function for $\phi^4$ theory at $\fr{4m_{\text{gap}}^2}{m^2}=0.25$ (left) and $\fr{4m_{\text{gap}}^2}{m^2}=0.1$ (right), for $\Dmax=20$, $30$, and $40$. \emph{Bottom:} The same data, with the $\p_-^2\epsilon$ correction~\eqref{eq:IsingCCorr} subtracted, compared to the Ising model prediction~\eqref{eq:IsingCFunc} (black line). The cutoff parameter $\fr{\Lambda_{\textrm{Ising}}}{m}=1.3$ in~\eqref{eq:IsingCCorr} was fixed using the data in the left plot.}
\label{fig:Phi4CFunction}
\end{center}
\end{figure}

This procedure is demonstrated in Fig.~\ref{fig:Phi4CFunction}. The top row shows our truncation results for the $C$-function at two points near the critical coupling: $\fr{4m_\gap^2}{m^2} = 0.25$ (left) and $0.1$ (right). These plots are zoomed in to small values of $\mu$ in order to focus on the IR regime described by the Ising model. As we can see, the results appear to have largely converged as we vary $\Dmax$; however, we do not see any sign of a plateau at $c_{\textrm{IR}}=\half$, which indicates that these values of $m_\gap$ are not sufficiently small compared to $\Lambda_{\textrm{Ising}}$ to resolve the IR central charge.

It is worth pausing to emphasize that the lack of a plateau for these values of the mass gap is \emph{physical}. Based on the convergence with $\Dmax$, the top row of Fig.~\ref{fig:Phi4CFunction} shows the correct IR behavior of the $C$-function in $\phi^4$ theory. At these values of the mass gap, the corrections due to higher-dimension operators in the Ising description (such as $\p_-^2 \epsilon$) are large enough that they eliminate the plateau, and the only way to suppress these corrections is to go to smaller values of $m_\gap$.

However, if we look at the spacing in eigenvalues for our truncation results, we see that at $\Dmax=40$ our IR cutoff appears to be roughly 
$\fr{\Lambda_{\textrm{IR}}^2}{m^2} \sim 0.05$ (at finite volume, this would be equivalent to a circle of approximate length $30/m$). This means that, for this level of truncation, we cannot accurately reproduce the IR behavior of $\phi^4$ theory for much lower values of $m_\gap$, and therefore cannot directly resolve the value of $c_{\textrm{IR}}$.

In the bottom row of Fig.~\ref{fig:Phi4CFunction}, we subtract the correction $\de C_{\Lambda_{\textrm{Ising}}}(\mu)$ from our truncation results in the top row, in order to directly compare with the Ising model prediction~\eqref{eq:IsingCFunc} at low energies. The value of $\Lambda_{\textrm{Ising}}$ is fixed numerically by matching the truncation data at $\fr{4m_\gap^2}{m^2}=0.25$ (left plot) with the Ising prediction (black line), obtaining the approximate value $\fr{\Lambda_{\textrm{Ising}}}{m} \approx 1.3$. We then use that same value of $\Lambda_{\textrm{Ising}}$ at $\fr{4m_\gap^2}{m^2}=0.1$ (right plot), where our truncation results reproduce the Ising prediction at small $\mu$.

The low extracted value of $\frac{\Lambda_{\textrm{Ising}}}{m} \approx 1.3$ confirms that for these values of the mass gap the corrections from higher-dimensional Ising operators like $\p_-^2\epsilon$ are not very suppressed. To resolve the Ising model central charge, we must therefore be able to push $m_\gap$ far below this cutoff, either by increasing $\Dmax$ or improving the extrapolation of our results. We would like to set this goal, resolving $c_{\textrm{IR}}=\half$ directly from the $C$-function, as an important target for the future of LCT.

As a final remark, let us comment on the deviation between $C(\mu)-\de C_{\Lambda_{\textrm{Ising}}}(\mu)$ and $C^{(\text{Ising})}(\mu)$ above the IR regime in the bottom row of Fig.~\ref{fig:Phi4CFunction}. This deviation is due to  dropping  higher order terms in (\ref{eq:Teff}) and \emph{not} truncation error. To reproduce $C^{(\text{Ising})}(\mu)$ at higher $\mu^2$, one should include the effects of additional irrelevant operators in $\de C_{\Lambda_{\textrm{Ising}}}(\mu)$, or tune to smaller values of $m_\gap$.

\section{Application II: Yukawa Theory} 
\label{sec:YukawaTheory}

For our second application, we will study 2d Yukawa theory, the theory of a real scalar field $\phi$ coupled to a real fermion through a Yukawa interaction $\sim g \phi \psi \chi$.  The lightcone Lagrangian for this theory after integrating out $\chi$ is (see~\eqref{eq:YukawaLCLag}), 
\begin{equation}
\Lcal = \frac{1}{2} (\p \phi)^2 - \frac{1}{2}m_\phi^2 \phi^2 + i\psi\p_+\psi - \frac{1}{2} \psi\frac{m_\psi^2 }{i\p} \psi - m_\psi g \phi \psi\frac{1}{i\p} \psi - \frac{g^2}{2} \phi  \psi\frac{1}{i\p} \phi \psi  .
\label{eq:YukawaLCLag2}
\end{equation}
With the scalar bare mass $m_\phi$ setting the overall scale, we have two physical parameters, the dimensionless ratios:
\benn
\fr{m_\psi}{m_\phi}, \quad \fr{g}{m_\phi}.
\eenn
Using the radial quantization techniques presented in Part II, we will work at $\Dmax=20$, which corresponds to a basis size of $7336$ states.

The overall procedure is the same as the previous application: construct a complete basis of states built from $\phi$ and $\psi$, compute the Hamiltonian matrix elements for both mass terms and the cubic and quartic Yukawa interactions, diagonalize the resulting Hamiltonian to obtain the spectrum at various values for $\fr{m_\psi}{m_\phi}$ and $\fr{g}{m_\phi}$, and use the eigenstates to compute observables such as spectral densities.

However, Yukawa theory has an important new feature which was not present in the previous example of $\phi^4$ theory: UV divergences. In section~\ref{sec:YukawaPert}, we show that at one-loop order there are divergences which cannot be removed simply by normal-ordering terms in the Lagrangian. The UV cutoff regulating these divergences is set by the truncation level $\Dmax$ (see section~\ref{sec:TruncationScales}), such that the spectrum continually shifts as we vary $\Dmax$, even for very large truncation levels. In trying to remove these divergences, we discover a very general feature of Hamiltonian truncation methods: the need for \emph{state-dependent} counterterms. In section~\ref{sec:RegFromQp}, we present a useful trick for constructing such state-dependent counterterms based on supersymmetry, allowing us to easily remove all UV divergences.

In section~\ref{sec:YukawaStrong}, we proceed to study Yukawa theory at strong coupling. As we vary the coupling $g$, we see that the theory has a \emph{first-order} phase transition, with a sharp jump in the spectrum from positive to negative eigenvalues (unlike the smooth transition seen in $\phi^4$ theory). For couplings below this critical point, we compute the Zamolodchikov $C$-function, as well as the integrated spectral density for $\phi$, allowing us to reconstruct the Breit-Wigner resonance for the scalar field (when $m_\phi > 2m_\psi$, i.e.~the scalar is unstable).

\subsection{Perturbation Theory and UV Divergences}
\label{sec:YukawaPert}

Before we launch into numerically diagonalizing the Hamiltonian at strong coupling, it is instructive to compare the truncation to a standard covariant analysis with Feynman diagrams at weak coupling.  There are a number of reasons why such a comparison is not just useful but almost necessary.  The simplest of these reasons is just that it is a strong check of the results -- reproducing the correlation functions at low loop order requires getting the matrix elements right.  Moreover, at low loop order only a small number of particles can participate (in LC, adding particles requires insertions of the interaction vertex), so it is possible to take the truncation level $\Delta_{\rm max}$ quite large and verify the asymptotic limit.  

A much more significant reason, however, is that studying perturbation theory shows us the structure of divergences in the theory.  This fact is especially true in super-renormalizable theories, where all divergences occur at relatively low loop order and so they can all be diagnosed within the perturbative regime.  Because the UV cutoff in lightcone conformal truncation is somewhat unusual, the divergences  can be unfamiliar and subtle, and it is far easier to first understand them in the perturbative regime where many analytic checks are possible.  

The main subtlety we will encounter in divergences in the Yukawa theory is that chiral symmetry no longer protects the mass of the fermion.  The reason for this is simple to see from the Lagrangian (\ref{eq:YukawaLCLag2}): after $\chi$ is integrated out, the mass term is quadratic in $\psi$ and is no longer protected by a $\psi \rightarrow -\psi$ symmetry.  Consequently, if we want to study the theory with a chiral symmetry, then we have to add counterterms tuned as a function of the coupling.  

First, recall the fermion mass shift in a standard covariant approach.  The one-loop correction to the fermion self-energy $\Sigma(\slashed{p})$ is
\be
-i \Sigma(\slashed{q}) &=&(ig)^2 \int \frac{d^2 p}{(2\pi)^2} \frac{i(\slashed{p}+ \slashed{q}+m_\psi)}{(p+q)^2-m_\psi^2 + i \epsilon} \cdot \frac{i}{p^2-m_\phi^2+i \epsilon} \nn\\
 &=& \frac{i g^2}{4\pi} \int_0^1 dx \frac{ (1-x)\slashed{q}+m_\psi}{( x m_\psi^2 +(1-x) m_\phi^2-  x(1-x)q^2 )} .
 \ee
The one-loop mass shift can be evaluated in closed form:
\be
\delta m_\psi^2 = 2 m_\psi \Sigma(m_\psi) =-\frac{g^2}{2\pi} \left(\log \left(\frac{m_{\phi }}{|m_{\psi }|}\right)+\frac{\sqrt{4 m_{\psi
   }^2-m_{\phi }^2}}{m_\phi} \sec ^{-1}\left(\frac{2 |m_{\psi }|}{m_{\phi
   }}\right)\right) .
   \label{eq:FermionYukawaCovariantLoop}
\ee
One can explicitly see that this mass shift is free of UV divergences, and it vanishes at $m_\psi =0$.  

By contrast, consider the mass shift at second order in LCT.  From time-independent perturbation theory, at second order in the interaction the shift is
\be
\delta m_\psi^2 = \sum_k \frac{ |\< \psi | V | \phi \psi\>_k|^2}{m_\psi^2 - \mu^2_k} .
\label{eq:TIPT2}
\ee 
Here, $V$ is the cubic Yukawa term $\sim \phi \psi \frac{1}{\partial} \psi$, and $|\phi \psi\>_k $ denotes the $k$-th mass term eigenstate (with eigenvalue $\mu_k^2$) within the sector of states with one $\phi$ and one $\psi$.   At finite truncation $\Delta_{\rm max}$, the eigenstates $| \phi \psi\>_k$ and eigenvalues $\mu_k^2$ are found by numerically diagonalizing the mass term part of the Hamiltonian.  To see the origin of the divergence at large $\Delta_{\rm max}$, we can start by summing over states in a Fock space basis:\footnote{The factors in this integrand have the following origins: the $\frac{1}{x(1-x)^2}$ is from the inverse of the inner product norm of the $|\phi \psi\>$ states, the $(2-x)^2$ is from the square of the $\< \psi | \phi \psi \frac{1}{\p} \psi|\phi \psi\> \sim (1-x)( 1 + \frac{1}{1-x})$ matrix element, and the denominator is from the difference in the mass term eigenvalues $m_\psi^2$ for the $|\psi\>$ state and $\frac{m_\psi^2}{1-x} + \frac{m_\phi^2}{x}$ for the $|\phi \psi\>$ state. The variable $x$ is the momentum fraction of the $\phi$ particle in the $|\phi\psi\>$ state.}
\be
\delta m_\psi^2 \approx g^2 m_\psi^2 \int_0^1 \frac{dx}{4\pi x(1-x)^2} \frac{(2-x)^2}{m_\psi^2 - \left( \frac{m^2_\psi}{1-x} + \frac{m^2_\phi}{x} \right)} .  
\ee
The integral is logarithmically divergent near $x \sim 1$.  A cut-off  of $\Lambda^2$ on the mass-squared $\frac{m^2_\psi}{1-x} + \frac{m_\phi^2}{x}$ of the intermediate state puts a cut-off of $1-x \gtrsim \frac{m_\psi^2}{\Lambda^2}$, so we conclude\footnote{We can evaluate the integral over $x$ to get the LCT one-loop mass shift more precisely:
\be
\delta m_\psi^2 \approx -\frac{g^2}{2\pi} \left(\log (\gamma \Delta_{\rm max}) +\frac{\sqrt{4 m_{\psi
   }^2-m_{\phi }^2}}{m_\phi} \sec ^{-1}\left(\frac{2 |m_{\psi }|}{m_{\phi
   }}\right)\right) ,
   \label{eq:LCTFermionShift}
\ee
where we have put a cut-off on the $x$ integral of $1-x \gtrsim (\gamma \Delta_{\rm max})^{-2}$, with the intention of fitting the parameter $\gamma$ to numeric results.
}
\be
\delta m_\psi^2 \sim -g^2 \log \frac{\Lambda}{m_\psi} \sim -g^2 \log \Delta_{\rm max}.
\ee
The reason for the last relation is that (see e.g.~eq.~(\ref{eq:phiphitruncmass})) the truncation sets a UV cutoff on the highest mass eigenvalues of the Hamiltonian mass term.  The key point is that the divergence is not suppressed by the mass term at small $m_\psi$.

\begin{figure}[t!]
\begin{center}
\includegraphics[width=0.5\textwidth]{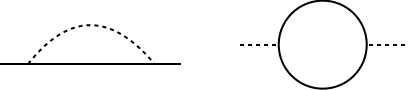}
\caption{One-loop corrections to fermion and scalar self-energies in the Yukawa theory.}
\label{fig:YukOneLoop}
\end{center}
\end{figure}

\begin{figure}[t!]
\begin{center}
\includegraphics[width=0.48\textwidth]{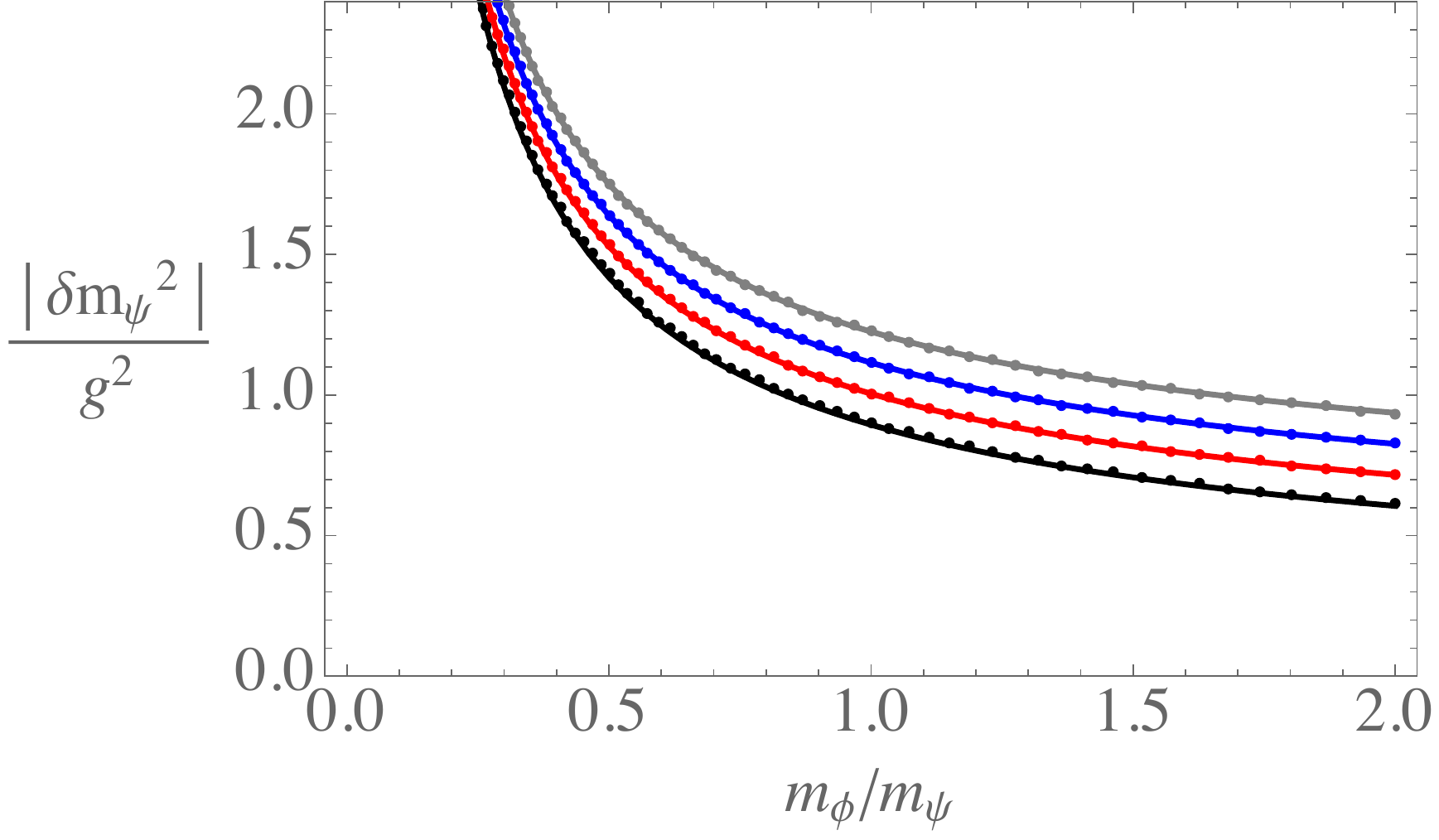}
\includegraphics[width=0.48\textwidth]{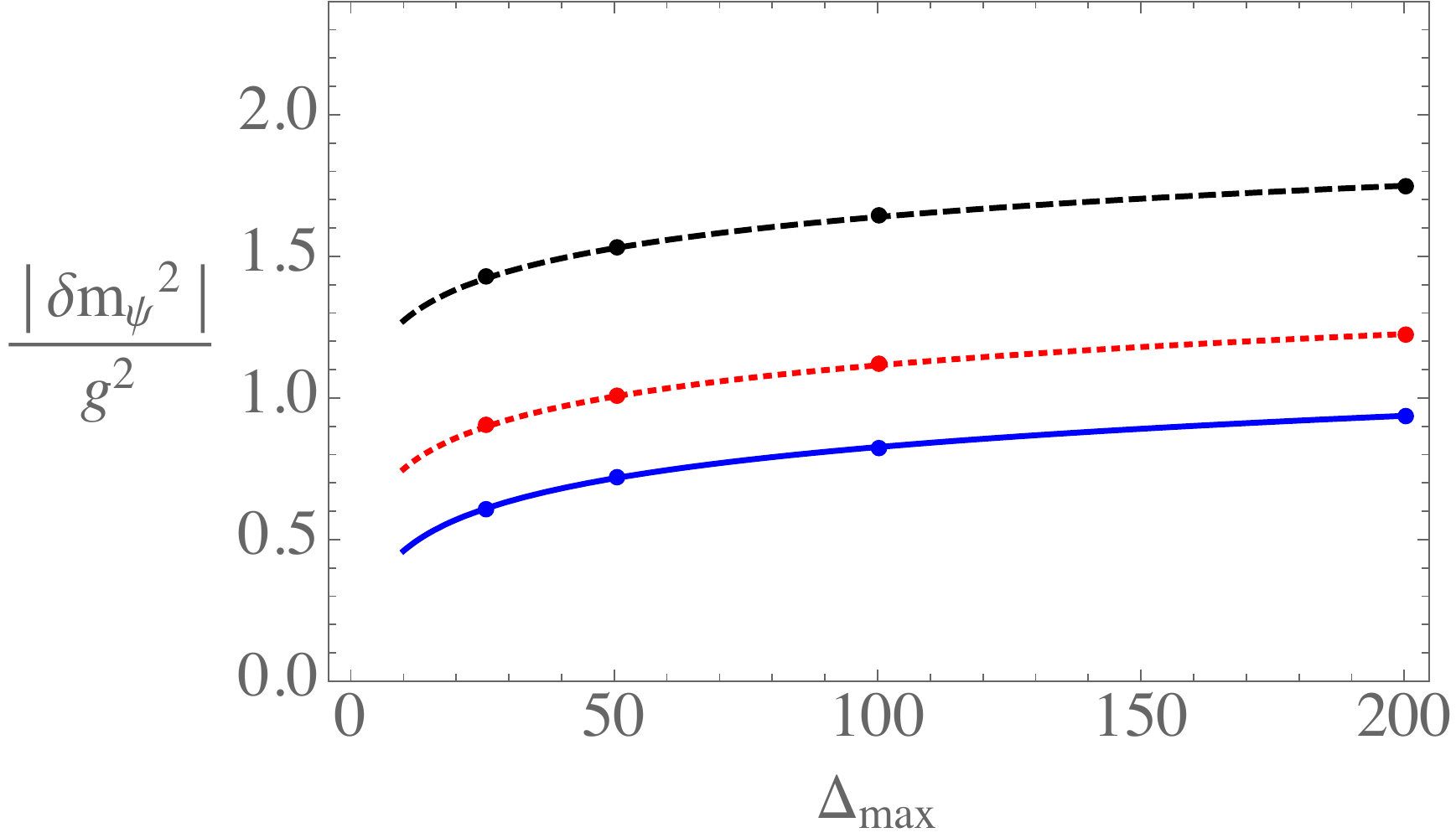}
\caption{{\it Left:} One-loop fermion mass shift in LCT as function of $m_\phi/m_\psi$, for $\Delta_{\rm max}=25.5,50.5,100.5,200.5$, from bottom to top (black, red, blue, gray, respectively).  {\it Right:} One-loop fermion mass shift in LCT as function of $\Delta_{\rm max}$, for $m_\phi/m_\psi = 2, 1, 0.5$ from bottom to top (blue solid, red dotted, black dashed,  respectively).  Points are numeric results, lines are eq. (\ref{eq:LCTFermionShift}) with $\gamma=1.8$ extracted by fitting. The key point is that spectrum is still changing as $\Delta_{\rm max}$ increases even at very large $\Delta_{\rm max}$, but this dependence is completely captured by the $\log(\Delta_{\rm max})$ in  (\ref{eq:LCTFermionShift}).}
\label{fig:FermionMassShiftOneLoop}
\end{center}
\end{figure}

Now that we have seen analytically what we expect for the fermion mass in LCT at one loop, let's compute it numerically.  As before, we will use the second-order time-independent perturbation theory formula (\ref{eq:TIPT2}), but this time using our truncation basis and matrix elements.  We must work to all orders in the masses, so first we diagonalize the mass term matrix elements numerically. The mass term matrix elements for the $[\p\phi \p\psi]_\ell$ basis states are
\be
 \Mcal_{\ell \ell'} &=&  \int_0^1 dx x (1-x)^2 \left( \frac{m_\phi^2}{x} +\frac{ m_\psi^2 }{1-x} \right) \hat{P}_\ell^{(1,2)}(1-2x)\hat{P}_{\ell'}^{(1,2)}(1-2x) \nn\\
   &=&  \left(2 m_\phi^2 + m_\psi^2 (-1)^{\ell+\ell'}\left(\frac{\ell_{\rm min}+2}{\ell_{\rm max}+2}\right) \right) 
\left(  \frac{(\ell_{\rm min}+1)(\ell_{\rm min}+3)}{(\ell_{\rm max}+1)(\ell_{\rm max}+3)} (\ell+2)(\ell'+2)\right)^{\frac{1}{2}}.  \nn\\
 \label{eq:MphiphiMphipsi}
 \ee
 Here, $\ell_{\rm min}, \ell_{\rm max}$ denote ${\rm min}(\ell, \ell')$ and ${\rm max}(\ell, \ell')$, respectively.  The cubic Yukawa term matrix elements between $|\p\psi\>$ and the $|[\p\phi \p\psi]_\ell\>$ states are proportional to
 \begin{equation}
 \bal
 \< \p\psi | \phi \psi \frac{1}{\p} \psi | [ \p\phi \p\psi]_\ell\> &\propto \int_0^1 dx (2-x) \hat{P}_\ell^{(1,2)}(1-2x)\\
 &= (-1)^\ell \sqrt{\frac{2}{(\ell+1)_3}} \left\{ \begin{array}{cc} (\ell + 3)^2 & \ell \textrm{ even} \\ (\ell+ 1)^2 & \ell \textrm{ odd} \end{array} \right\} .
 \eal
\end{equation}
Substituting these into the second order time-independent perturbation theory formula, we can compute the fermion mass shift at $\CO(g^2)$ up to fairly large $\Delta_{\rm max}$.  The results are shown in Fig.~\ref{fig:FermionMassShiftOneLoop}, where they are seen to agree very well with the analytic expression (\ref{eq:LCTFermionShift}). 

\begin{figure}[t!]
\begin{center}
\includegraphics[width=\textwidth]{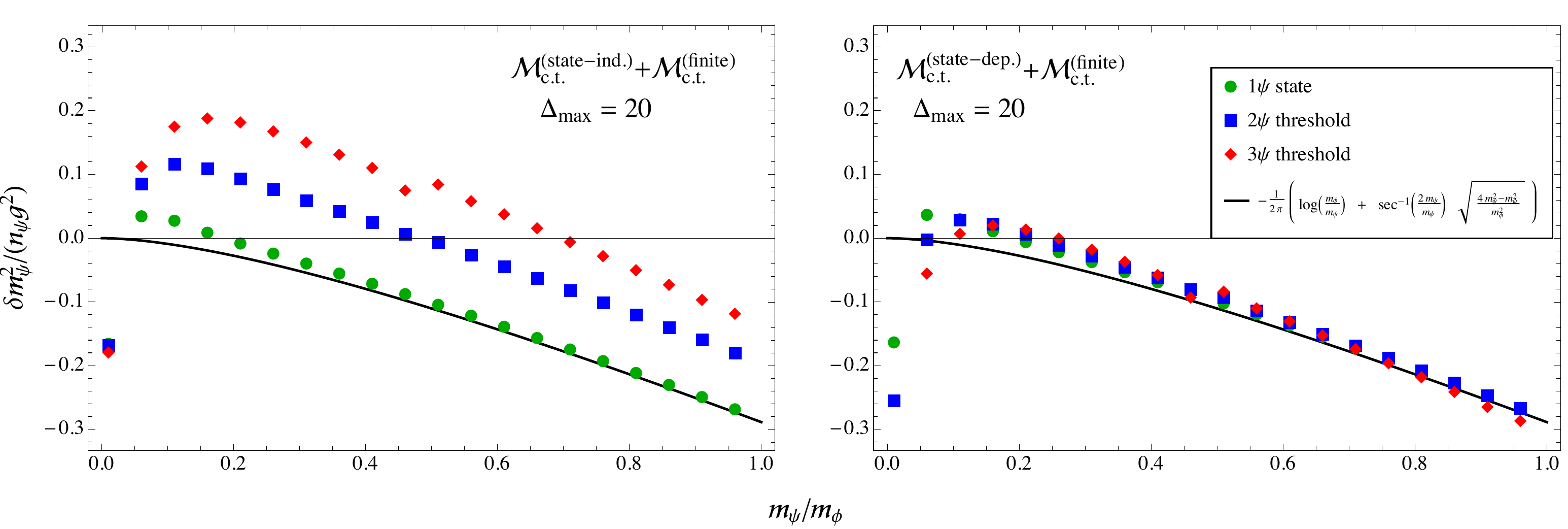}
\caption{ 
The leading order $\CO(g^2)$ shift in the one-, two-, and three-fermion thresholds, normalized by number of fermions squared, as a function of the bare fermion mass $m_\psi$.  
{\it Left:} The mass shifts computed with a state-independent counterterm (\ref{eq:yukawa-local-counterterm}) added to remove the logarithmically divergent term in (\ref{eq:LCTFermionShift}), plus a finite mass shift (\ref{eq:yukawa-finite-counterterm}) to make the one-fermion threshold (green dots) match the covariant result (\ref{eq:FermionYukawaCovariantLoop}) (solid black line).
If the one-loop correction were state-independent, then all three sets of data points would be identical. 
{\it  Right:} Same mass shifts, but now replacing (\ref{eq:yukawa-local-counterterm}) with a state-dependent counterterm (\ref{eq:YukNonLocal}) to remove the logarithmically divergent piece. The residual effect after the subtraction of the divergence is canceled by the same finite mass shift (\ref{eq:FermionYukawaCovariantLoop}). The subtraction removes the state-dependence, as can be seen by the agreement between all three thresholds. 
}
\label{fig:YukawaLocalCounterTermFails2}
\end{center}
\end{figure}

Naively, restoring the chiral symmetry just requires shifting the bare fermion mass to tune back to the chiral point.  Unfortunately, the loop-generated contribution to the fermion mass cannot be canceled simply by a standard, state-independent mass term.  The reason is that when we consider the one-loop divergent contribution to the energy of a multi-particle state,  the value of the cut-off seen by the loop depends on all the particles in the state -- that is, some of the ``$\Delta_{\rm max}$ budget'' is eaten up by the other particles, and the fermion-boson loop effectively sees a reduced value of $\Delta_{\rm max}$.  Canceling the divergence is therefore not as simple as adding $\sim g^2 \log \Delta_{\rm max} \, \psi \frac{1}{\partial} \psi$ to the Hamiltonian.  To see this fact explicitly, we have computed the masses of not only the single-fermion state but also the lowest two- and three-fermion states as a function of coupling, and have attempted to cancel off the shift in the fermion mass by adding a state-independent fermion mass term
\begin{equation}\label{eq:yukawa-local-counterterm}
\CM_{\rm c.t.}^{\rm (state-ind.)} = 
\frac{g^2}{2\pi} \log ( \gamma \Delta_{\rm max} ) \, \psi \frac{1}{i\partial} \psi 
\qquad \gamma = 1.8
\, .
\end{equation}
We still have to fix the finite part of the fermion mass shift, which we do by demanding that the $\CO(g^2)$ correction to the single-fermion mass agrees with the covariant calculation (\ref{eq:FermionYukawaCovariantLoop}).\footnote{Because the LC Hamiltonian (\ref{eq:YukawaLCLag2}) has three deformations -- the mass term, the cubic Yukawa, and the quartic Yukawa -- but only two underlying parameters -- $m_\psi$ and $g$ -- the fermion mass is no longer a true free parameter once the two Yukawa deformations are specified.  }
We can analytically derive the magnitude of the finite piece from comparing 
(\ref{eq:LCTFermionShift}) and (\ref{eq:FermionYukawaCovariantLoop}), and extract the difference in the finite piece at $\Dmax \rightarrow \infty$
\begin{equation}\label{eq:yukawa-finite-counterterm}
\CM_{\rm c.t.}^{\rm (finite)} 
=-\frac{g^2}{2\pi} \log \left(\frac{m_{\phi }}{|m_{\psi }|}\right) \psi \frac{1}{i\partial} \psi .
\end{equation}
The left plot in Fig.~\ref{fig:YukawaLocalCounterTermFails2} shows that, although the shift in the single-fermion state is canceled by this counterterm, the shift in the multi-particle states are $\CO(1)$.

The appropriate fermion mass counterterm must somehow correctly encode a reduced $\Delta_{\rm max}$ for multi-particle states, in order to match the behavior of the divergence. This is a rather general feature of Hamiltonian truncation methods, where we impose a cutoff on the total energy of intermediate states, rather than on the individual particles in loops. Each multi-particle state therefore sees a different \emph{effective} cutoff, such that a simple state-independent shift in the fermion mass cannot correctly remove all UV divergences. Fortunately, as we describe in the next subsection, we can construct a counterterm with the needed state-dependence using a trick inspired by supersymmetry.

\subsection{Matrix Elements and Regulators from $Q_+$}
\label{sec:RegFromQp}

In this subsection, we make a brief digression to discuss how supersymmetric interactions may be implemented  by using the supersymmetry algebra.  Moreover, we will discuss how even in a non-supersymmetric theory, one can use supersymmetry to define a counterterm that cancels state-dependent divergences.\footnote{For more discussion of SUSY in 2d LCT, see \cite{Fitzpatrick:2019cif}. }

To begin, recall that  in $d=2$, ${\cal N}=(1,1)$ SUSY, there are two supercharges $Q_\pm$, and (in a specific convention for their normalization) they satisfy
\be
Q_\pm^2 =  P_\pm .
\ee
Therefore, in a SUSY theory, we can compute $Q_+$ in our truncation and square it to get $P_+$. This approach has several advantages; one obvious one  is that $Q_+$ is structurally much simpler than $P_+$. For a theory of a real scalar superfield $\Phi$ with a superpotential $W(\Phi)$, the supercharge is
\be
 Q_+ = \sqrt{2} \int d x^- W'(\phi) \psi.
 \label{eq:QpGen}
\ee
Now, imagine if we worked with a theory with a cubic superpotential $W(\Phi) = \frac{m}{2} \Phi^2 + \frac{g}{6} \Phi^3$.  Then, in addition to the Yukawa interaction, we would have a cubic and quartic $\phi^3$ and $\phi^4$ interaction, and the fermion and scalar masses would be the same.  Crucially, in such a theory, the divergence in the fermion mass term would be absent by supersymmetry (the mass term would be related by supersymmetry to interaction terms, which are manifestly finite). 

\begin{figure}[t!]
\begin{center}
\includegraphics[width=0.6\textwidth]{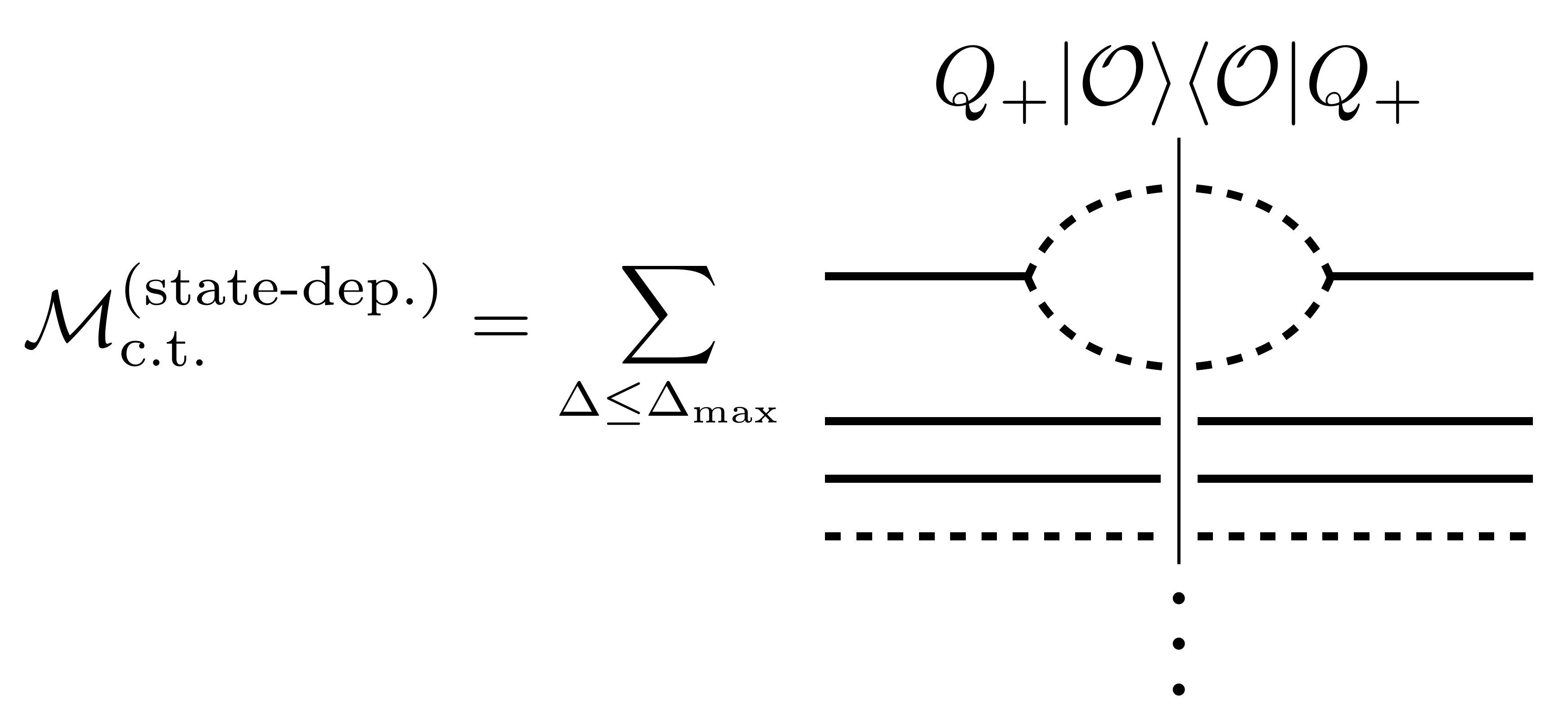}
\caption{Schematic representation of the SUSY-inspired, state-dependent counterterm \eqref{eq:YukNonLocal}. Because this counterterm is constructed from a truncated sum over intermediate states, it correctly reproduces the reduced effective cutoff seen by multi-particle states.}
\label{fig:StateDepCT}
\end{center}
\end{figure}

The key point for our non-supersymmetric Yukawa theory is that the SUSY construction contains two divergent contributions to the fermion mass term that cancel each other, both of which contain state-dependent pieces.  The first contribution is just the one-loop diagram (left diagram in Fig.~\ref{fig:YukOneLoop}) that we have been discussing in the previous subsection.  The second divergent contribution is not exactly a loop correction but rather is a term  generated directly in the Hamiltonian $P_+$ by taking $Q_+^2$, i.e.
\bq
   2 \sum_{\De''\leq \Dmax} \< \CO|  Q_+ | \Ocal''\> \< \Ocal'' |  Q_+ | \CO'\> \supset \< \CO | \Mcal_{\textrm{c.t.}}^{\rm (state-dep.)} | \CO'\>, \quad Q_+ \equiv \frac{g}{\sqrt{2}} \int dx^- (\psi \phi^2)(x^-) ,
\qquad
\label{eq:YukNonLocal}
\eq
where the new divergent piece comes from terms in the matrix product of $Q_+$s where the $\phi$s contract ``inwards'', on the intermediate states $|\Ocal''\>$, and the $\psi$s contract ``outwards'', on the external states $|\CO\>, |\CO'\>$.
Note that this new term is {\it itself} produced by summing over intermediate states and thus mimics the one-loop structure of the original divergence that it cancels.   Now, instead of making our whole theory supersymmetric, we can simply grab this new divergent piece from the above construction, since it is responsible for canceling the state-dependent divergence of the one-loop Yukawa diagram.   Then, we define a new \emph{state-dependent} counterterm (\ref{eq:YukNonLocal}) that is just this second term, restricted as described above to the contractions that generate a fermion bilinear. The schematic structure of this new counterterm is shown in Fig.~\ref{fig:StateDepCT}.

 In a Fock space basis, the shift to the fermion mass in a given multi-particle state from this SUSY-inspired counterterm is
\be
\delta m_\psi^2 \sim g^2 \int \frac{dx}{x(1-x)} \sim g^2 \log ( \widehat{\gamma} \Delta_{\rm max}),
\ee
where we have put a cutoff on $x, 1-x \gtrsim (\widehat{\gamma} \Delta_{\rm max})^{-2}$ near its limits of integration, resulting from the truncation on intermediate states in~\eqref{eq:YukNonLocal}. We see that our new counterterm has the required logarithmic dependence on $\Delta_{\rm max}$. Crucially, though, the resulting coefficient $\widehat{\gamma}$ is now state-dependent.

The residual fermion mass eigenvalues after subtracting the state-dependent divergence still have a contribution from the finite part of the loop correction. The finite piece does not suffer from the state-dependence of the $\Delta_{\rm max}$ cutoff, and so can be captured by the simple fermion mass shift (\ref{eq:yukawa-finite-counterterm}).

In the right plot of Fig.~\ref{fig:YukawaLocalCounterTermFails2}, we show the
one-, two- and three-fermion threshold
with this state-dependent fermion mass counterterm.  Unlike for the previous state-independent counterterm (left plot), the divergent loop contribution is now canceled in the single- and multi-particle states simultaneously. Note that there will still be some state-dependence in the fermion mass shifts at finite $\Dmax$, due simply to truncation effects, but these effects are unrelated to the presence of a UV divergence and vanish as $\Dmax \rightarrow \infty$, such that the LCT results correctly match the covariant calculation.

In the above discussion, we focused on the regime where the coupling $g$ is small but the divergence $g^2 \log\Delta_{\rm max}$ was large.  The issue is particularly clear in this regime because at small coupling one can identify the multi-particle thresholds unambiguously.  For completeness, 
we also quickly discuss the one-loop divergent contribution to the scalar mass. By a standard one-loop computation, the scalar one-loop mass shift from Feynman diagrams is
\be
\delta m_\phi^2 = -\frac{g^2}{2\pi} \left( \log \left( \frac{\Lambda}{|m_\psi|}\right) - \frac{\sqrt{4 m_\psi^2 -m_\phi^2}}{m_\phi} {\rm csc}^{-1}\left( \frac{2 |m_\psi|}{m_\phi} \right) \right),
\label{eq:mphiOneLoopYuk}
\ee
where $\Lambda$ is a hard UV cutoff.  In LCT, the same divergence appears through the truncation cut-off, with $\Lambda \sim \Delta_{\rm max} m_\psi$.  As with the fermion mass divergence, $\log \Delta_{\rm max}$ becomes a state-dependent divergence when it appears as a subdiagram of a contribution to multi-particle states. We can use the same trick we used for the fermion mass to construct a state-dependent counterterm that removes it.  In this case, we again use (\ref{eq:YukNonLocal}), but keep only the contraction where one $\phi$ contracts ``outwards'' and one $\phi$ and one $\psi$ contract ``inwards''.  The resulting counterterm subtracts off the log divergence in (\ref{eq:mphiOneLoopYuk}) and replaces it with a finite constant; we have chosen not to add any additional finite mass shift beyond the counterterm constructed from $Q_+^2$, in which case it turns out that this constant is 2. There are no additional divergences at higher orders in $g$, so we have now fixed all necessary counterterms.

Finally, we end this subsection with a more general comment about implementing these SUSY-inspired tricks in a more elegant way.  Our strategy above was to start with a non-SUSY theory and just extract the divergent counterterms we need from a SUSY formulation in order to regulate and renormalize the theory.  However, it should also be possible to achieve the same result more efficiently by working in the opposite direction: start with a SUSY theory and add local soft SUSY-breaking terms to obtain the non-SUSY theory.  More explicitly, a SUSY theory of a real superfield $\Phi$ with the superpotential $W(\Phi) = \frac{m}{2} \Phi^2 + \frac{g}{6} \Phi^3$ has a Lagrangian containing 
\be
\CL \supset \sqrt{2} i W''(\phi) \psi \chi - \tfrac{1}{2} (W'(\phi))^2 = \sqrt{2} i (m + g \phi) \psi \chi - \frac{1}{2} (m \phi + \frac{g}{2} \phi^2)^2.
\ee
Therefore, we can obtain our Yukawa theory by subtracting the $\phi^3$ and $\phi^4$ interactions and adding a $\phi^2$ mass shift to detune the scalar and fermion masses.  Aside from being more elegant, such an approach would likely be more efficient computationally and allow one to go to much higher truncation $\Delta_{\rm max}$, since the only matrix elements one would have to compute would be those of $Q_+$, $\phi^2, \phi^3$, and $\phi^4$, which are local even in the lightcone formulation.

\subsection{Strong Coupling}
\label{sec:YukawaStrong}

\begin{figure}[t!]
\begin{center}
\includegraphics[width=\textwidth]{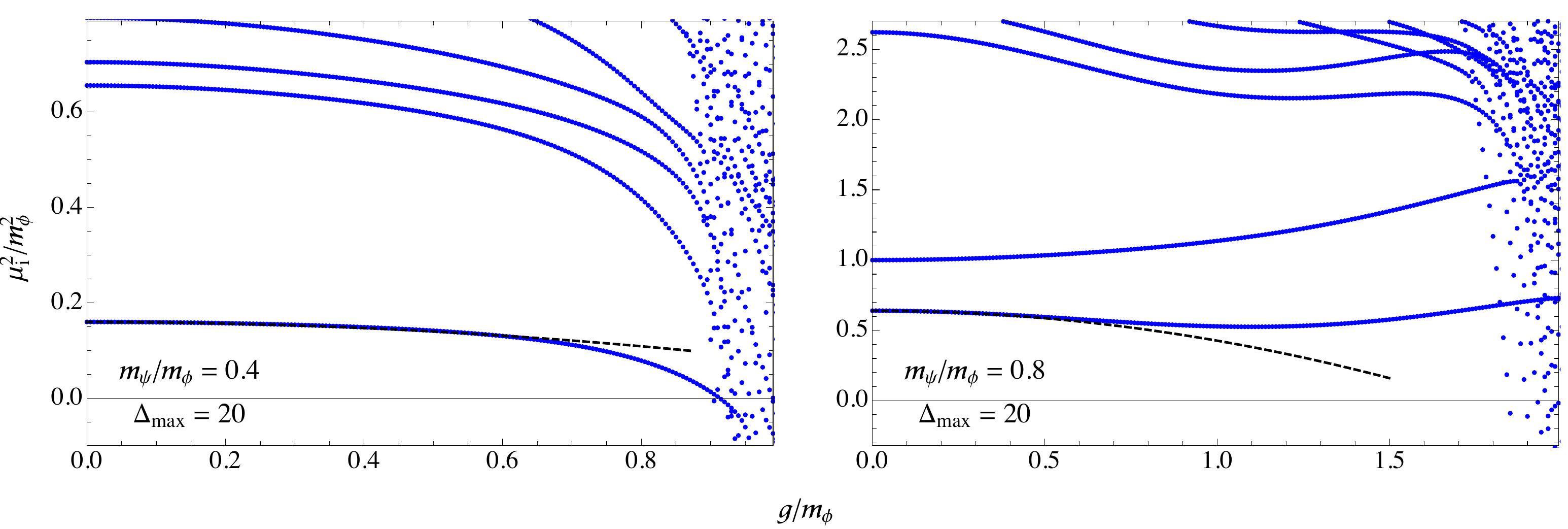}
\caption{Spectrum of mass eigenvalues in Yukawa theory as a function of coupling, at $\Delta_{\rm max}=20$, for $\frac{m_\psi}{m_\phi} = 0.4$ and $0.8$. The blue dots are the mass eigenvalues, respectively. The black dashed curves are the one-loop fermion mass shift (\ref{eq:FermionYukawaCovariantLoop}).
}
\label{fig:YukawaMassSpectrum}
\end{center}
\end{figure}

Now that we have set up the Hamiltonian and counterterms, we are ready to analyze the theory at strong coupling.  The simplest observable is the spectrum of eigenvalues of the Hamiltonian.  We show the mass spectrum for the theory at $\Delta_{\rm max}=20$ as a function of coupling, with the counterterms added, in Fig.~\ref{fig:YukawaMassSpectrum}. In both plots (with $\fr{m_\psi}{m_\phi} = 0.4$ and $0.8$) we see that the theory experiences a first-order phase transition at some critical coupling. In LCT, a first-order phase transition manifests itself as a rapid transition from positive to negative eigenvalues (recall the discussion in section \ref{sec:Phi4PhaseTransition}), as the lowest eigenvector tries to reconstruct a new lower-energy vacuum beyond the critical point.  As $\Delta_{\rm max}$ increases, the transition becomes sharper, and should approach a discontinuous jump in the infinite $\Delta_{\rm max}$ limit. These qualitative features are most obvious at smaller fermion mass, where the phase transition is more strongly first-order.  

\begin{figure}[t!]
\begin{center}
\includegraphics[width=1\textwidth]{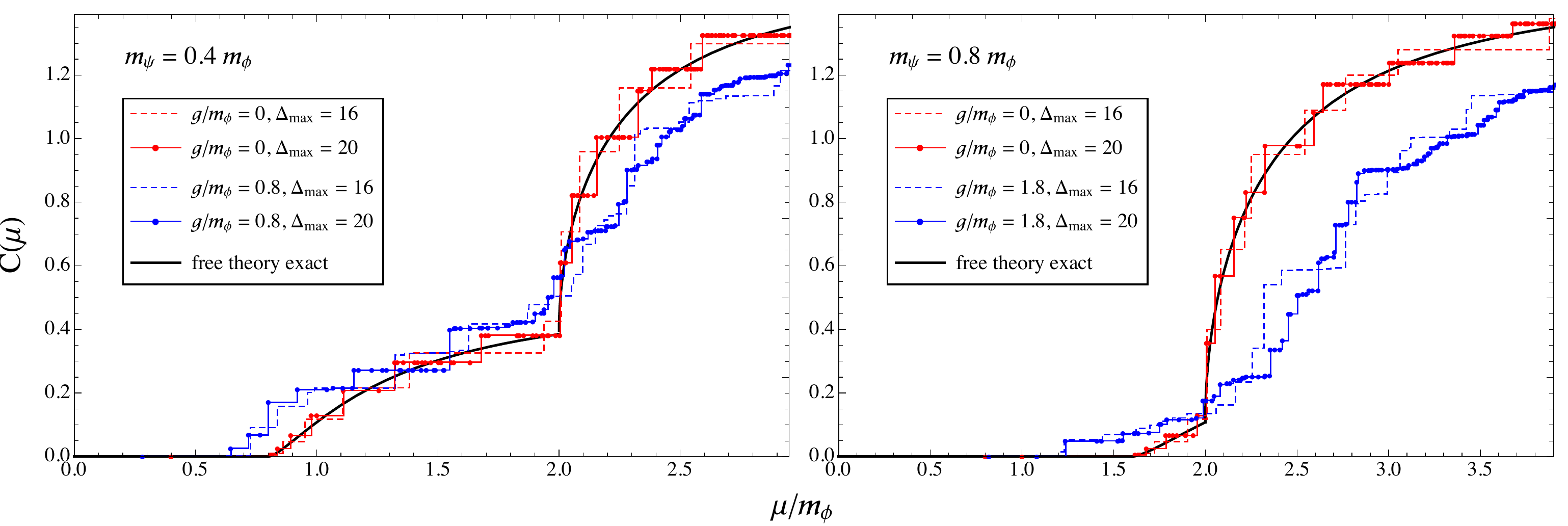}
\caption{Zamolodchikov $C$-function in  Yukawa theory as a function of scale $\mu$, at various $\Delta_{\rm max}$ and couplings $g/m_\phi$, for $\frac{m_\psi}{m_\phi} = 0.4$ and $0.8$ (left and right plots, respectively).  }
\label{fig:YukawaCFunction}
\end{center}
\end{figure}

We also show the Zamolodchikov $C$-function in Fig.~\ref{fig:YukawaCFunction}. Because the stress tensor couples only to parity-even states, it does not see the fermion or scalar until the scale $\mu$ is at least twice their respective mass.  Consequently, the most significant qualitative feature in Fig.~\ref{fig:YukawaCFunction} is that $C(\mu)$ vanishes at small $\mu$, where there are no degrees of freedom due to the mass gap, rises to $\sim \frac{1}{2}$ for a weakly coupled fermion at around $\mu \sim 2 m_\psi$, and finally rises to $\sim \frac{3}{2}$ for a weakly coupled scalar and fermion at around $\mu \sim 2 m_\phi$.  

\begin{figure}[t!]
\includegraphics[width=0.48\linewidth]{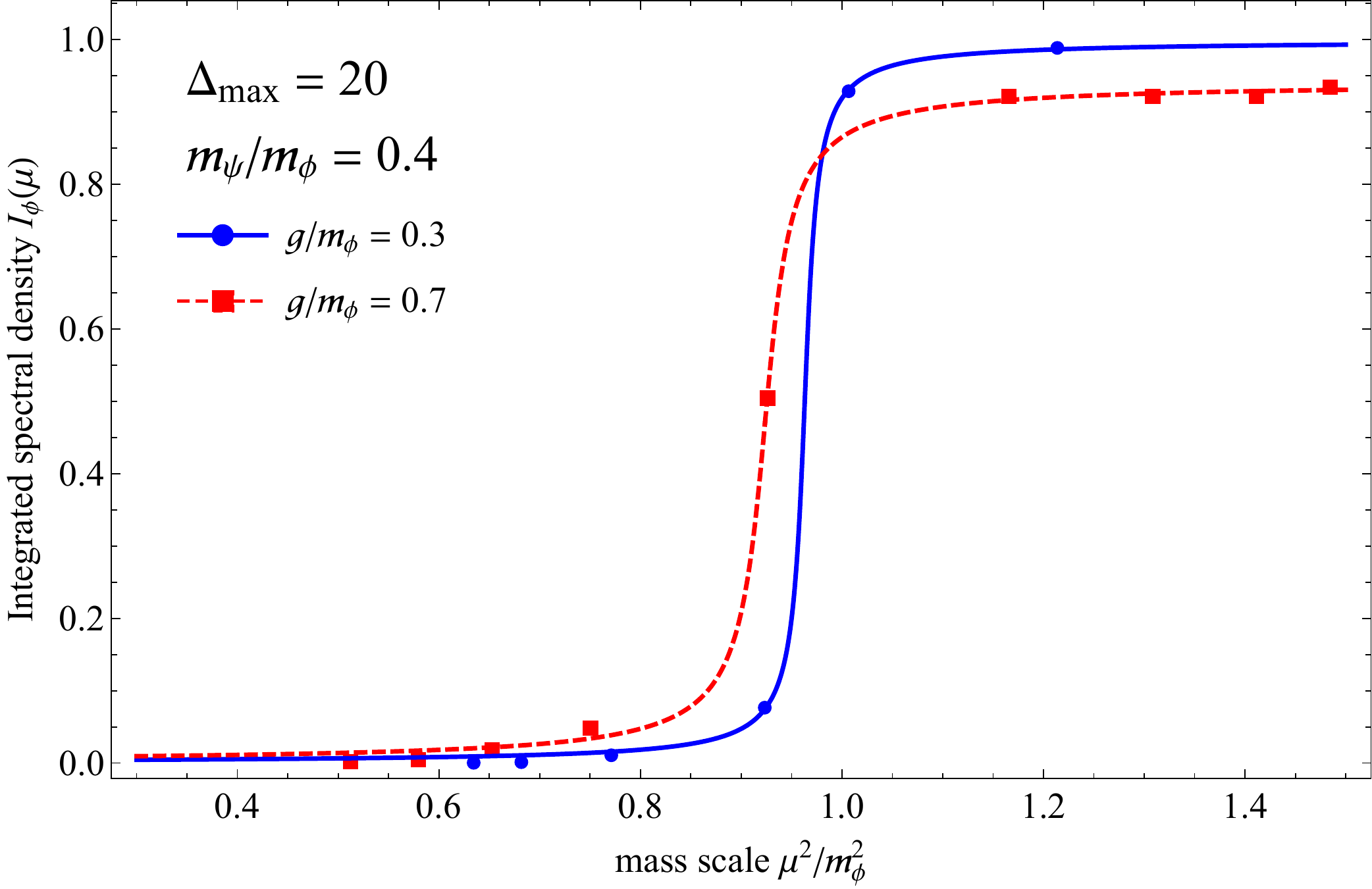}
\includegraphics[width=0.48\linewidth]{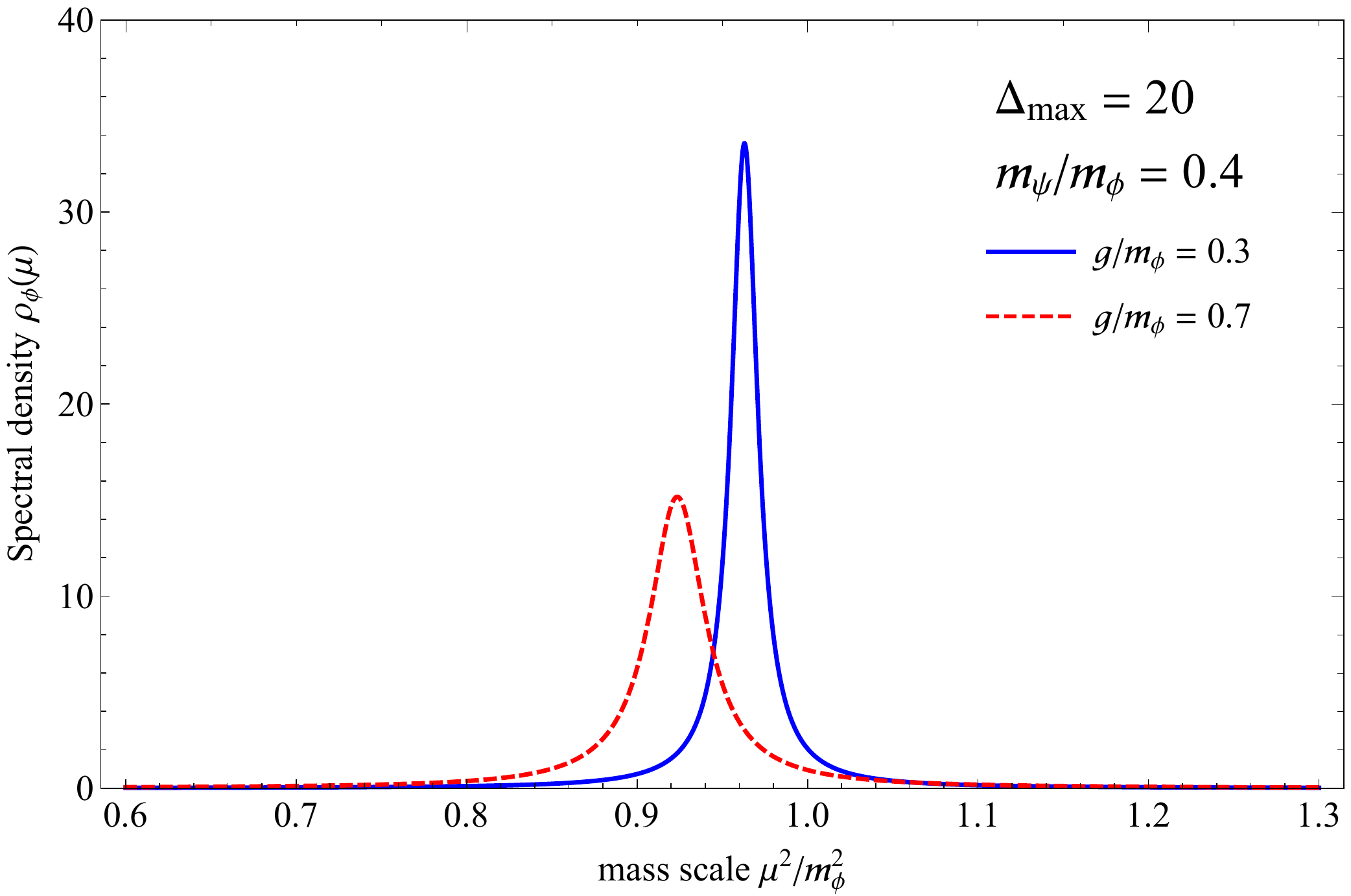}
\caption{\emph{Left:} The $\phi(x)$ integrated spectral density $I_{\phi}(\mu)$ at different couplings. The blue circle and red square are the numerical data for the integrated spectral density at coupling $\frac{g}{m_\phi} = 0.3$ and 0.7, respectively. The blue solid curve and red dashed curve are the best fit curves of (\ref{eq:breit-wigner-integrated}). 
\emph{Right:} The Breit-Wigner resonance (\ref{eq:breit-wigner}) for the scalar field, using the best-fit parameters in the left plot.
The spectral density is computed at $\Dmax = 20$, $\frac{m_\psi}{m_\phi} = 0.4$.
} 
\label{fig:breit-wigner}
\end{figure}

An important useful fact about spectral densities is that they provide a well-defined simple observable that one can use to probe states that are not strictly speaking asymptotic states in the theory and so would be difficult to probe using the spectrum alone.  In particular, in the Yukawa theory with $m_\psi < m_\phi/2$, the $\phi$ particle is unstable and decays to fermions, and  trying to identify a ``$\phi$'' state amidst the two-fermion continuum of states is ambiguous.  Instead, we can compute the spectral density for the $\< \phi \phi\>$ two-point function and look for a resonance with a finite width, similarly to what one would do with an S-matrix.
One can formulate the spectral density as an S-matrix amplitude  by weakly coupling $\phi$ to an external probe. The spectral density of a resonance is a Breit-Wigner bump
\begin{equation}\label{eq:breit-wigner}
\rho_\phi(\mu) = \frac{\textrm{const.}}{(\mu^2 - \mu_\phi^2)^2 + \mu_\phi^2\Gamma^2} \, ,
\end{equation}
where $\mu_\phi$ and $\Gamma$ are the physical mass and width of the scalar resonance.  The $\<\phi\phi \>$ spectral density is shown in Fig.~\ref{fig:breit-wigner}. We first show the integrated spectral density (left plot) near the resonance for the theory at $\Delta_{\rm max}=20$, $\frac{m_\psi}{m_\phi} = 0.4$. The integrated spectral density rises sharply from $\sim 0$ to $\sim 1$ at the energy scale $\mu \sim m_\phi$, which matches our expectation of a resonance. 
The integrated spectral density data fits well to the integrated Breit-Wigner distribution,
\begin{equation}\label{eq:breit-wigner-integrated}
I_\phi(\mu) = \int_0^{\mu^2} d{\mu^\prime}^2 \, \rho_\phi(\mu^\prime) \, .
\end{equation}
On the right, we then show the plot of the resonances  (\ref{eq:breit-wigner}) at different coupling using the best fit parameters $(\mu_\phi, \Gamma)$. As expected, the resonance is narrower at smaller coupling $\frac{g}{m_\phi} = 0.3$ and wider at stronger coupling $\frac{g}{m_\phi} = 0.7$. 

\bookmarksetup{startatroot}
\section{Future Directions and List of Projects}
\label{sec:FutureDir} 

Our hope is that this document and the accompanying code will offer many readers an opportunity to get involved in using and developing the methods of LCT.  We have focused on the simplest class of theories possible that we think nevertheless indicates the potential breadth of applications and addresses many of the fundamental issues and challenges involved in getting started. In this final section, we will discuss several potential avenues for future work. 

We first mention a few areas that we are currently investigating.  In section \ref{sec:gaugefields}, we described how to treat 2d QCD in the absence of a fermion mass; in upcoming work, we will describe how to deal with subtleties involved with adding a mass term when there are gauge interactions.  We also have work in progress studying the broken phase of 2d $\phi^4$ theory, by putting in a $\phi^3$ term so that we expand directly around the symmetry-breaking vacuum.  Although this work has focused on free UV CFTs, part of the philosophical motivation of LCT is to study deformations of any UV CFT, and as a simple example we are exploring deformations of the non-trivial UV CFT given by the critical point of the 2d tricritical Ising model.  To see how global symmetries can be efficiently included, we are studying the generalization of 2d $\phi^4$ to a complex scalar field in 2d.  Finally, a detailed analysis of the generalization of $\phi^4$ theory to $d=3$ will be the subject of forthcoming work.

Beyond these, let us list several other questions that we think would be fruitful to explore in the context of LCT.  Perhaps the simplest extension of these methods would be to the multicritical theory with a $\phi^2, \phi^4$, and $\phi^6$ interaction, which should include the tricritical Ising model in its phase diagram.  It would also be interesting to redo the analysis in section \ref{sec:YukawaTheory} by the strategy suggested at the end of subsection \ref{sec:RegFromQp},  i.e. by starting with a fully supersymmetric theory and deforming by soft SUSY-breaking terms; more generally, it would be useful to know if such a strategy could be widely used  to handle the nonlocal interactions that arise in lightcone quantization from integrating out nondynamical fields.  
There are also important generalizations one would like to make to the kinds of models that can be studied.  Allowing additional symmetry is probably the most obvious such extension.  Going beyond the limited supersymmetry application that we have briefly touched on here,  there are many conjectured dualities in supersymmetric gauge theories in 2d (e.g. \cite{Gukov1,Gukov2}) and it would be nice to be able to test these with LCT.   Another example, the 2d theory of a $U(1)$ gauge field coupled to a charged scalar with a $\theta$ term (i.e. 2d scalar QED) is expected to have a non-trivial phase structure as a function of its parameters.  At infinite $N_c$, 2d QCD with fermions in the adjoint representation is expected to be supersymmetric when the fermion mass is tuned to a particular value \cite{Antonuccio:1998zp,Kutasov:1993gq,Bhanot:1993xp,Demeterfi:1993rs, dalley1993string}. 
% As is discussed in [X], one might speed-up the computation of LCT using quantum devices. It will be useful to establish a publicly available toolbox to link the LCT code and the quantum algorithms, which will provide valuable quantum computing examples of solving field-theories instead of spin-chains. 
As is discussed in \cite{liu2020quantum}, one might speed-up the computation of LCT using quantum devices. It will be useful to establish a publicly available toolbox to link the LCT code and the quantum algorithms, which will provide valuable quantum computing examples of solving field-theories instead of spin-chains.

We think it is unlikely that one could study irrelevant deformations in this framework, but optimistically one might hope to allow marginally relevant or exactly marginal interactions.  The theory of $N$ complex fermions coupled to a heavy scalar field provides a well-defined setting in which to study such deformations, by coupling the fermions to the heavy scalar with a Yukawa interaction. Then, in the UV, the theory is free, but below the mass of the scalar it can be integrated out and one finds a four-fermion interaction that is exactly marginal and integrable (of the form $J \bar{J}$, with anomalous dimensions that are easily calculated in the bosonized description) for $N=1$, and marginally relevant for $N>1$ \cite{Wetzel}. Ideally, this would provide guidance in understanding how to describe the effective theory below the scalar mass directly in LCT.

We also mention a few questions that are further afield.  Generalized free theories provide another class of solvable CFTs, and have interesting RG flows (e.g the flow to the long-distance Ising model \cite{fisher1972critical,Behan:2017emf}).  Our approach to IR divergences in this work was to construct the `Dirichlet' basis, but this construction relied on a free field description and it would be good to understand how to deal with IR divergences more generally.  In principle, finite temperature systems might be addressed with LCT simply by performing a Boltzmann sum over states obtained by diagonalizing the Hamiltonian, though lightcone quantization loses much of its advantage with the reappearance of thermal vacuum bubbles in this case.  One might also study Renyi entropies in an excited state $|E\>$ in 2d by using the fact that they can be formulated as correlators $\< E | \sigma_n(x) \sigma_{-n}(y)| E \>$  of twist fields $\sigma_n$, whose OPEs can be extracted from the OPE data of the CFT itself.

Lastly, in this and previous work, we have taken a practical approach to questions about the rate of convergence of results with $\Delta_{\rm max}$ by simply looking at numeric results, but it would be very interesting and useful to have a more principled understanding of the convergence rate, or even a rigorous proof that the results converge in the limit of infinite truncation. In other Hamiltonian truncation frameworks, valuable work has been done along these lines that also improved convergence by including `renormalization' effects due to changing the truncation level \cite{Rychkov:2014eea,Elias-Miro:2017xxf}. Unfortunately, such methods use the large energy of the heavy states (above the truncation) in the {\it undeformed} Hamiltonian $H_0$ as an expansion parameter, but in LCT the high-dimension states do not have large lightcone energies (in particular, in the free 2d theories in this work, $P_+^{(\rm CFT)} =0$!).  It would likely provide a significant improvement if such renormalization techniques could nevertheless be suitably modified so that they could be applied to LCT.  Efficiency might also be gained with a better {\it a priori} understanding of which states are the most important for the low energy spectrum, especially since the size of our Hamiltonians in this work is approaching the limit of matrices that can be exactly numerically diagonalized on a computer.  In the construction in this work, the energies of the states are spread over a range controlled by the truncation (see e.g. (\ref{eq:freemassivespectrum})), with many states ``wasted'' at high energies when one would like them to more densely concentrated at low energies in order to probe the deep IR. This and other IR truncation effects  deserve further study.

%%%%%%%%%%%%%%%%%%%%%%%%%%%%%%%%%%%%%%%%%%%%%%%%%%%%%%%%%%%%%%%%%%%%%%%%%%%%%
%%%%%%%%%%%%%%%%%%%%%%%%%%%%%%%%%%%%%%%%%%%%%%%%%%%%%%%%%%%%%%%%%%%%%%%%%%%%%
%%%%%%%%%%%%%%%%%%%%%%%%%%%%%%%%%%%%%%%%%%%%%%%%%%%%%%%%%%%%%%%%%%%%%%%%%%%%%

\section*{Acknowledgments}

We thank  Luca D\'elacretaz, Brian Henning, Matthijs Hogervorst,  and Junyu Liu for valuable feedback on the draft and public code. We also especially thank Vincent Genest for collaboration on unpublished early versions of these methods, some of which are discussed in appendix \ref{app:FockSpaceTricks}.    ALF, EK, and YX were supported in part by the US Department of Energy Office of Science under Award Number DE-SC0015845, and ALF in part by a Sloan Foundation fellowship.  MW is partly supported by the National Centre of Competence in Research SwissMAP funded by the Swiss National Science Foundation.  ZK is also supported by the DARPA, YFA Grant D15AP00108. The authors were also supported in part by the Simons Collaboration Grant on the Non-Perturbative Bootstrap, grant \#488649.

%%%%%%%%%%%%%%%%%%%%%%%%%%%%%%%%%%%%%%%%%%%%%%%%%%%%%%%%%%%%%%%%%%%%%%%%%%%%%
%%%%%%%%%%%%%%%%%%%%%%%%%%%%%%%%%%%%%%%%%%%%%%%%%%%%%%%%%%%%%%%%%%%%%%%%%%%%%

\clearpage

\appendix

\section{Notation, Conventions, and Reference Formulae}
In this section, we summarize potentially unfamiliar notation that we have introduced over the course of this text. We also provide our choice of various conventions and an index for frequently used reference formulae. 

\vspace{-.08in}
\subsection*{Notation} Note that the tables below omit notation that is standard in the literature (e.g.~$\Delta$ is scaling dimension, $\cO(x)$ is a local operator, etc.)

\begin{table}[h!]
	\label{tab:NotationTableGeneral}
\resizebox{6.1in}{!}{%

	\begin{tabular}{l@{\hskip .577in}l}

	\hline \hline
	\multicolumn{2}{l}{\textbf{General} } \\ \hline \hline
	$\doteq$ & Equal up to removable phases, p.~\pageref{eq:doteqnotation},  see also Appendix \ref{app:Phases} \\
	$n$ & Number of particles  ($n_F$ fermions, $n_B$ bosons if ambiguous), pp.~\pageref{eq:ScalarGenericOperator}, ~\pageref{eq:mixonmonomialsdefn} \\
	$|v|_i$ & Magnitude of a vector $v$ up to the $i$'th element, p. ~\pageref{eq:MagDef} \\
	$\boldsymbol{v}/v_i$ & Vector $\boldsymbol{v}$ with $i$'th element removed, p.~\pageref{eq:WickContractionCoeff} \\
	$\boldsymbol{v}/\{v_i\}$ & Vector $\boldsymbol{v}$ with set of elements $\{v_i\}$ removed, p.~\pageref{eq:II:ScalarPhiNFinal} \\
	$p$ & Equal to $p_-$, p.~\pageref{eq:DropSupscriptNotation} \\
	$\ptl$ & Equal to $\ptl_-$, p.~\pageref{eq:DropSupscriptNotation}\\ 
	$\rho_{\cO}(\mu)$ & Spectral density of operator $\cO$, p.~\pageref{eq:SpectralDecomp} \\
$I_{\cO}(\mu)$ & Integrated spectral density of operator $\cO$, p.~\pageref{eq:DefnOfIntegratedSpec}\\
${\Pcal}$ & Principal value prescription, pp.~\pageref{eq:PVPrescription}, ~\pageref{eq:PVdef} \\\hline \hline 

	\multicolumn{2}{l}{\textbf{Basis and Matrix Elements}} \\ \hline \hline
	$\ket{\cO, p} \;\;\;\;\;\;$	 &   LCT basis state, p.~\pageref{eq:BasisDef} \\
	$N_{\cO}$ & Normalization of LCT basis state, p.~\pageref{eq:NormFromTwoPt} \\
	$\Ccal$ & Conformal Casimir, p.~\pageref{eq:ConfCasimir} \\
	$\mu^2$ or $\mu_i^2 \;\;\;\;\;\;$ & Mass-squared eigenvalue or $i$-th mass-squared eigenvalue, pp.~\pageref{eq:DefOfMuSq}, \pageref{eq:FullEigenstates} \\
	$\ket{\mu_i^2, p} \;\;\;\;\;\;$ & Mass-squared eigenstate, p.~\pageref{eq:FullEigenstates} \\
	$F_{\cO_i}(p) \;\;\;\;\;\;$ & Momentum space wavefunction/overlap with basis state $\cO_i$, p.~\pageref{eq:FockWvFns} \\
	$G_{\cO_i \cO_j}$ & Basis Gram matrix, p.~\pageref{eq:GeneralInner} \\
	$\Mcal_{\cO_i \cO_j}^{(\cO_R)}$ & LCT matrix element of relevant operator $\cO_R$, p.~\pageref{eq:LCTDataFT} \\
	$F_{\Lvec}(p)$ & Momentum space Casimir eigenfunctions, p.~\pageref{eq:JacobiBasis} \\
	$P_\ell^{(\alpha,\beta)}$ & Jacobi polynomial, p.~\pageref{eq:JacobiPolyDef} \\
$\widehat{P}_\ell^{(\alpha,\beta)}$ & Normalized Jacobi polynomial, p.~\pageref{eq:NormalizedJacobis} \\
$[AB]_\ell$ & Double-trace operator built from $A$ and $B$, p.~\pageref{eq:JoaoFormula} \\
$\cO_{\Lvec}$ & Primary operator built out of double trace combination given by $\Lvec$, p.~\pageref{eq:JoaoFormula} \\
$(n, \Lvec)$ & Level of operator, built out of $n$ field insertions and $|\Lvec|$ derivatives, p.~\pageref{eq:OpLevel} \\ \hline \hline 
	\end{tabular}
	}
\end{table} 
\begin{table}[t]
	\label{tab:NotationTableMonomial}
\resizebox{6.1in}{!}{%

	\begin{tabular}{l@{\hskip .77in}l}

	\hline \hline
	\multicolumn{2}{l}{\textbf{Monomials}} \\ \hline \hline
	$\Kvec$ & List of monomial powers, p.~\pageref{eq:ScalarMonoDef} \\
	$\Kvec^\dagg$ & List of monomial powers in reverse order, p.~\pageref{eq:kdaggdef} \\
	$\ptl^{\Kvec} \phi(x) \;\;\;\;\;\;$ & Monomial (same definition with $\phi \ra \psi$), p.~\pageref{eq:ScalarMonoDef} \\
$N_{\Kvec}$ & Normalization of monomial, p.~\pageref{eq:MonoNormalization} \\
$C_{\Kvec}^{\Ocal}$ & Expansion coefficients of operator $\Ocal$ in terms of monomials, p.~\pageref{eq:ScalarGenericOperator} \\
$G_{\Kvec \Kvec'}$ & Monomial Gram matrix, p.~\pageref{eq:DefnMonoGram} \\
$A_{\Kvec \Kvec'}$ & Wick contraction coefficient for scalars, p.~\pageref{eq:WickContractionCoeff} \\
$\tilde{A}_{\Kvec \Kvec'}$ & Wick contraction coefficient for fermions, p.~\pageref{eq:FermionMono2pt} \\ \hline \hline
	\multicolumn{2}{l}{\textbf{Radial Quantization}} \\ \hline \hline
	$\cong$ & Equal for linear combinations that sum to a primary, p.~\pageref{monobra} \\
	$a_{\Kvec}^\dagg$, $a_{\Kvec}$ & Radial quantization scalar creation and annihilation operators, p.~\pageref{eq:II:radialscalarNcal} \\
$b_{\Kvec}^\dagg$, $b_{\Kvec}$ & Radial quantization fermion creation and annihilation operators, p.~\pageref{eq:radialfermionoperators} \\
$\Ncal_{\Kvec}$ & Radial quantization scalar normalization, p.~\pageref{eq:II:radialscalarNcal} \\
$\Ncal_{\Kvec}^{(F)}$ & Radial quantization fermion normalization, p.~\pageref{eq:radialfermionoperators} \\
$N_{\textrm{FT}}$ & Normalization factors arising from Fourier transform p.~\pageref{eq:II:N} \\
$\norm{\boldsymbol{k}}$ & Radial quantization normalization for a vector $\Kvec$, p.~\pageref{eq:II:radialscalarknorm} \\
$\mathfrak{g}_{\cO \cO'}$ & Scalar Zamolodchikov metric, p.~\pageref{eq:II:Radial2PF}\\
$\mathfrak{g}_{\cO \cO'}^{(F)}$ & Fermion Zamolodchikov metric, p.~\pageref{eq:ZamoMetricFermions} \\
$G_{\Kvec \Kvec'}^{(\cO)}$ & Monomial three-point function, p.~\pageref{eq:II:Primary3PF} \\ \hline \hline
	\end{tabular}
	}
\end{table}

\vfill

\subsection*{Conventions} Here we list various conventions, organized roughly by their category. Rather than reference each equation to pages of the text (as many of these conventions can be found within the same section/page), we point to the general section or part of the text in which they can be found. Fermion conventions that are not listed here follow scalar conventions but with appropriate replacements (e.g.~$[a_p, a^\dagg_q]$ $\rightarrow$ $\{a_p, a^\dagg_q\}$.).

\renewcommand{\arraystretch}{1.5}
\begin{table}[h!]
	\label{tab:LCCoventions}
\resizebox{6.1in}{!}{%

	\begin{tabular}{l@{\hskip 1.8in}c}

	\hline \hline
	\multicolumn{2}{l}{\textbf{Lightcone Kinematics} (See section \ref{sec:ReviewOfLCT}) } \\ \hline \hline
	Metric and signature & \(\displaystyle ds^2 = dt^2-dx^2 \) \\
	Lightcone coordinates & \(\displaystyle x^{\pm} \equiv \frac{t \pm x}{\sqrt{2}} \), $x^+ =$ ``time'' \\
	Metric in lightcone coordinates & \(\displaystyle ds^2 = 2dx^+dx^- \)  \\
	Lightcone momenta & \(\displaystyle p_{\pm} = \frac{1}{\sqrt{2}}(p_0 \pm p_1) \) \\
	Generators of spacetime translations & \(\displaystyle P_{\pm} \equiv \frac{1}{\sqrt{2}}(P_0 \pm P_1) \) \\
	Hamiltonian & $P_+$  \\
	Invariant mass-squared operator &$ M^2 = 2P_+ P_-$ \\
	 \hline \hline 

	\multicolumn{2}{l}{\textbf{Free field theory} (See section \ref{sec:2dFFT}, specifically \ref{sec:FreeLC}, \ref{sec:FockSpace}) } \\ \hline \hline
	Free scalar Lagrangian & \(\displaystyle \Lcal = \frac{1}{2}(\ptl \phi)^2 = \ptl_+ \phi \ptl_- \phi \) \\
	Canonical commuator & \([\phi(x),\ptl_- \phi(y)] = \frac{i}{2}\delta(x^--y^-)\) \\
	Scalar mode expansion & \(\displaystyle \phi(x) = \int_0^\infty \frac{dp_-}{(2\pi) \sqrt{2p_-}} \left(e^{-i p \cdot x} a_p + e^{i p \cdot x} a^\dagg_p \right) \)  \\
	Creation/annihilation commuator & \(\displaystyle [a_p, a^\dagg_q] = (2\pi) \delta(p_- - q_-) \) \\
	Scalar two-point function (Lorentzian) & \(\displaystyle \corr{\phi(x)\phi(0)} = -\frac{\log x}{4\pi} \) \\ 
	Normalization of 1-particle state & \(\displaystyle \ket{p} = \sqrt{2p_-} a^\dagg_p \ket{\textrm{vac}} \) \\
	Chiral components of fermion & \(\displaystyle \Psi = \frac{1}{2^{1/4}} \binom{\psi}{\chi} \) \\
	Fermion Lagrangian & \(\displaystyle \Lcal = i \psi \ptl_+ \psi + i \chi \ptl_- \chi \)  \\
	Free fermion mode expansion & \(\displaystyle \psi(x) = \int_0^\infty \frac{dp_-}{\sqrt{8\pi^2}} \left(e^{-i p \cdot x} a_p + e^{i p \cdot x} a^\dagg_p \right) \) \\
	Fermion two-point function (Lorentzian) & \(\displaystyle \corr{\psi(x)\psi(0)} = -\frac{i}{4\pi x} \) \\ 
	Gamma matrices & \( \displaystyle \gamma^+ = \Bigg(\begin{matrix}
		0 & 0 \\
		\sqrt{2} & 0
	\end{matrix}\Bigg), \quad\gamma^- = \Bigg(\begin{matrix}
		0 & \sqrt{2}\\
		0 & 0
	\end{matrix}\Bigg), \quad\gamma^0 = \Bigg(\begin{matrix}
		0 & 1 \\
		1 & 0
	\end{matrix}\Bigg) \) \\
	LCT free field basis state & \(\displaystyle \ket{\Ocal_i}_{\textrm{2d FFT}} = \ket{\Ocal_i, p_\mu = (p_+,p_-) = (0,1)} \) \\
	Resolution of identity & \(\displaystyle \boldsymbol{1} = \sum_n \frac{1}{n!} \int \frac{dp_1 \dotsb dp_n}{(2\pi)^n 2p_1 \dotsb 2p_n} \ket{p_1, \dots, p_n}\bra{p_1,\dots,p_n} \) \\
	 \hline \hline

	\multicolumn{2}{l}{\textbf{Radial Quantization} (See \hyperref[sec:PartIIAdvImprovements]{Part II}) } \\ \hline \hline
	Scalar mode expansion & \(\displaystyle \ptl \phi(x) = \frac{i}{\sqrt{4\pi}}\sum_{k=1}^\infty \sqrt{k}\left(x^{-k-1} a_k + x^{k-1}a^\dagg_k \right) \) \\
	Scalar operator commutator & \(\displaystyle [a_k, a^\dagg_{k'}] = \delta_{k,k'} \) \\
	Fermion mode expansion & \(\displaystyle \ptl \psi(x) = \frac{i}{\sqrt{4\pi}} \sum_{k=1}^\infty \sqrt{\half k(k+1)}\left(x^{-k-2} b_k + x^{k-1}b^\dagg_k \right) \) \\
	Fermion operator anticommutator & \(\displaystyle \{b_k, b^\dagg_k \} = \delta_{k,k'} \) \\
	 \hline \hline
	\end{tabular}
	}
\end{table}

\clearpage

\subsection*{Reference Formulae} For ease of reference, we list here the most frequently used formulae in this text. \renewcommand{\arraystretch}{2.1}
\begin{table}[h!]
	\label{tab:ReferenceFormulaeGen}
\resizebox{6.1in}{!}{%

	\begin{tabular}{l@{\hskip 2.32in}c}

	\multicolumn{2}{l}{\textbf{General} } \\ \hline \hline
	Spectral decomposition of 2-pt function  & p.~\pageref{eq:SpecDensGeneral} \\
	\multicolumn{2}{c}{ \(\displaystyle \<\Tcal\{\Ocal(x) \Ocal(0)\}\> = \int d\mu^2 \rho_\Ocal(\mu) \int \fr{d^dp}{(2\pi)^d} e^{-ip\cdot x} \fr{i}{p^2 - \mu^2 + i\epsilon} \)} \\
	\multicolumn{2}{c}{ \(\displaystyle \rho_\Ocal(\mu) \equiv \sum_i |\<\Ocal(0)|\mu_i^2,p_-\>|^2 \, \de(\mu^2 - \mu_i^2) \)} \\
	\multicolumn{2}{c}{ \(\displaystyle I_\Ocal(\mu) \equiv \int_0^{\mu^2} d\mu'^2 \, \rho_\Ocal(\mu) = \sum_{\mu_i \leq \mu} |\<\Ocal(0)|\mu_i^2,p_-\>|^2 \)} \\
	 \hline
	Fourier transforms of 2- and 3-pt functions & p.~\pageref{eq:FTFormulas}\\
	\multicolumn{2}{c}{ \(\displaystyle  \int dx \, \frac{e^{ipx}}{x^{2\Delta}} = \frac{2\pi  e^{i\pi\De} p^{2\Delta-1}}{\Gamma(2\Delta)} \)} \\
	\multicolumn{2}{c}{ \(\displaystyle  \int dx \, dz \, \frac{e^{ip(x-z)}}{x^A (-z)^B (x-z)^C} = \frac{4\pi^2 e^{\fr{i\pi}{2}(A+B+C)} \Gamma(A+B-1) p^{A+B+C-2}}{\Gamma(A)\Gamma(B)\Gamma(A+B+C-1)} \)} \\
	For the derivation of these equations, see \cite{Anand:2019lkt}. \\
	 \hline \hline

	\multicolumn{2}{l}{\textbf{Basis and Matrix Elements} } \\ \hline \hline
	Double-trace built out of $A$ and $B$  & p.~\pageref{eq:JoaoFormula} \\
	\multicolumn{2}{c}{ \(\displaystyle \left[ AB \right]_\ell \equiv \sum_{m=0}^\ell c^\ell_m(\Delta_A,\Delta_B)\, \p^m A \, \p^{\ell-m} B \)} \\ 
	\multicolumn{2}{c}{ \(\displaystyle  c^\ell_m(\Delta_A,\Delta_B) = \fr{(-1)^m \Gamma(2\Delta_A+\ell) \Gamma(2\Delta_B+\ell)}{m! (\ell-m)! \Gamma(2\Delta_A+m) \Gamma(2\Delta_B + \ell - m)} \)} \\ \hline
	LCT data & p.~\pageref{eq:LCTDataFT} \\
	\multicolumn{2}{c}{ \(\displaystyle  G_{\CO_i \CO_j} = \frac{1}{2p N^*_{\CO_i} N_{\CO_j} } \int dx\,e^{ipx} \< \CO_i(x) \CO_j(0) \> \)} \\
	\multicolumn{2}{c}{ \(\displaystyle \Mcal^{(\Ocal_R)}_{\Ocal_i\Ocal_j} = \frac{1}{N^*_{\CO_i} N_{\CO_j} } \int dx \, dz \,e^{ip(x-z)} \< \CO_i(x) \CO_R(0) \CO_j(z) \> \)} \\
	 \hline \hline
	\end{tabular}
	}
\end{table}

\clearpage

%%%%%%%%%%%%%%%%%%%%%%%%%%%%%%%%%%%%%%%%%%%%%%%%%%%%%%%%%%%%%%%%%%%%%%%%%%%%%
%%%%%%%%%%%%%%%%%%%%%%%%%%%%%%%%%%%%%%%%%%%%%%%%%%%%%%%%%%%%%%%%%%%%%%%%%%%%%

\section{Zero Modes, $H_{\eff}$, and the Infinite Momentum Limit}
\label{app:Heff}

In this appendix, we review the construction of the effective LC Hamiltonian $H_\eff$ to include the effects of non-dynamical ``zero modes'' (i.e.~particles with LC momentum $p_-=0$). This prescription for $H_\eff$ was initially presented in~\cite{Fitzpatrick:2018ttk}, where interested readers can find a much more thorough discussion of the effects of zero modes, but here we present a brief summary of the need for an effective Hamiltonian and the motivation for our prescription.

The overall goal of conformal truncation is to obtain the eigenvalues and eigenstates of the invariant mass operator $M^2$ for any QFT obtained by deforming a CFT by one or more relevant operators $\Ocal_R$. In this work, we have focused on constructing the operator $M^2$ from the lightcone Hamiltonian $P_+$, which is obtained by integrating the relevant deformation over a slice of fixed lightcone time $x^+ \equiv \fr{1}{\sqrt{2}}(t+x)$,
\be
M^2_{\LC} = 2P_+ P_-, \qquad P_+ \equiv P_{+\CFT} + \lambda \int dx^- \, \Ocal_R(x^+,x^-).
\label{eq:NaiveMLC}
\ee
An alternative (and perhaps more familiar) approach would be to instead construct $M^2$ from the equal-time Hamiltonian $H$, obtained by integrating the deformation over a slice of fixed time $t$,
\be
M^2_{\ET} = H^2 - P_x^2, \qquad H \equiv H_{\CFT} + \lambda \int dx \, \Ocal_R(t,x).
\ee

In both approaches, we then compute the matrix elements of $M^2$ between momentum eigenstates created by primary operators, which can be written in the general form
\bq
\bal
\<\Ocal,p_\mu|M^2_{\LC}|\Ocal',p'_\mu\> &\equiv 2p_- (2\pi) \de(p_- - p_-') \Mcal_{\Ocal\Ocal'}^{(\LC)}(p,p'), \\
\<\Ocal,p_\mu|M^2_{\ET}|\Ocal',p'_\mu\> &\equiv \sqrt{4p_0p'_0} (2\pi) \de(p_x - p_x') \Mcal_{\Ocal\Ocal'}^{(\ET)}(p,p').
\eal
\eq
The dynamical information is all contained within the functions $\Mcal_{\Ocal\Ocal'}(p,p')$, while 
the overall prefactors are set by the normalization of our basis states in lightcone and equal-time quantization, respectively.

These two approaches must agree as the truncation level $\Dmax \ra \infty$, since the eigenvalues of $M^2$ should be independent of the quantization scheme. In fact, this equivalence appears to be quite manifest, as it was shown in~\cite{Fitzpatrick:2018ttk} that the matrix elements of $M^2_{\LC}$ can be obtained by taking the infinite momentum limit of the matrix elements of $M^2_{\ET}$,\footnote{Note that the LHS of eq.~\eqref{eq:LimDefLC} is \emph{independent} of $p_-$, so the matrix elements of $M^2_\LC$ in \emph{any frame} correspond to the infinite momentum limit of $M^2_\ET$.}
\be
\Mcal_{\Ocal\Ocal'}^{(\LC)}(p,p') = \lim_{|p_x|\ra\infty} \Mcal_{\Ocal\Ocal'}^{(\ET)}(p,p').
\label{eq:LimDefLC}
\ee
We can understand this relation kinematically by looking at the difference in $p_-$ at large $p_x$,
\be
p_- - p'_- = \fr{1}{\sqrt{2}} \Big( \sqrt{\mu^2+p_x^2} - p_x \Big) - \fr{1}{\sqrt{2}} \Big( \sqrt{\mu'^2+p_x^2} - p_x \Big) \sim \frac{\mu^2 - \mu'^2}{2\sqrt{2} |p_x|},
\ee
so the LC momentum $p_-$ is conserved in the infinite momentum limit, just as it is in LC quantization.

Naively, it thus appears that we can think of conformal truncation in LC quantization as simply the infinite momentum limit of ET quantization. However, there is an important subtlety, which is most easily seen by considering old-fashioned perturbation theory with respect to the relevant deformation. From eq.~\eqref{eq:LimDefLC}, it is clear that the equivalence between LC and ET holds to leading order in $\lambda$. However, the quadratic and higher terms do not necessarily agree. For instance, there are multiple examples where
\be
\lim_{|p_x| \rightarrow \infty} \sum_{\Ocal'\!,\mu'} \frac{ \big|\de \Mcal^{(\ET)}_{\Ocal\Ocal'}\big|^2}{\mu^2-\mu'^2} \ne \sum_{\Ocal'\!,\mu'} \frac{ \big|\de \Mcal^{(\LC)}_{\Ocal\Ocal'}\big|^2}{\mu^2-\mu'^2},
\ee
where $\de \Mcal$ is the correction to $M^2$ due to the relevant deformation $\Ocal_R$. So, although the individual matrix elements of $\Mcal_{\Ocal\Ocal'}^{(\LC)}$ and $\Mcal_{\Ocal\Ocal'}^{(\ET)}$ match  in the $p_x \rightarrow \infty $ limit, their eigenvalues do not.

How can this be? The problem is that in summing over intermediate states, we often need to impose a cutoff $\Lambda$ on the invariant mass. In many cases, however, the limit of taking this cutoff to infinity and taking the infinite momentum limit \emph{do not commute}:
\be
\lim_{|p_x| \rightarrow \infty} \lim_{\rm \Lambda \rightarrow \infty} \sum_{\Ocal'\!,\mu'\leq\Lambda} \frac{ \big|\de \Mcal^{(\ET)}_{\Ocal\Ocal'}\big|^2}{\mu^2-\mu'^2} \ne  \lim_{\rm \Lambda \rightarrow \infty} \lim_{|p_x| \rightarrow \infty} \sum_{\Ocal'\!,\mu'\leq\Lambda} \frac{ \big|\de \Mcal^{(\ET)}_{\Ocal\Ocal'}\big|^2}{\mu^2-\mu'^2}.
\ee
This noncommutativity arises due to intermediate states whose invariant mass becomes infinite as $|p_x| \ra \infty$, but whose cumulative contribution in perturbation theory remains \emph{finite} as $\Lambda \ra \infty$. As a consequence, \emph{the eigenvalues of the naive $M_\LC^2$ in eq.~\eqref{eq:NaiveMLC} do not always match the eigenvalues of $M_\ET^2$.} We therefore need to add a correction to $M_\LC^2$ to include the contributions that are removed in the infinite momentum limit.

As a simple example, consider a free massive fermion in 2D. In the original undeformed CFT, there are two independent massless components, which obey the equations of motion
\be
\p_+\psi = 0, \quad \p_-\chi = 0.
\ee
In the massive theory, the equal-time Hamiltonian receives the correction
\be
V_\ET \equiv -i\sqrt{2} m \int dx \, \psi(x) \chi(x).
\ee
The full invariant mass operator is thus given by
\be
M^2_\ET = (H_{\CFT} + V_\ET)^2 - P_x^2 = M_\CFT^2 + \acomm{H_\CFT}{V_\ET} + V_\ET^2.
\ee
In the infinite momentum limit, the contribution due to $V_\ET^2$ vanishes~\cite{Fitzpatrick:2018ttk}, which means we can focus solely on the contribution from the linear term.

\begin{figure}[t!]
\begin{center}
\includegraphics[width=0.4\textwidth]{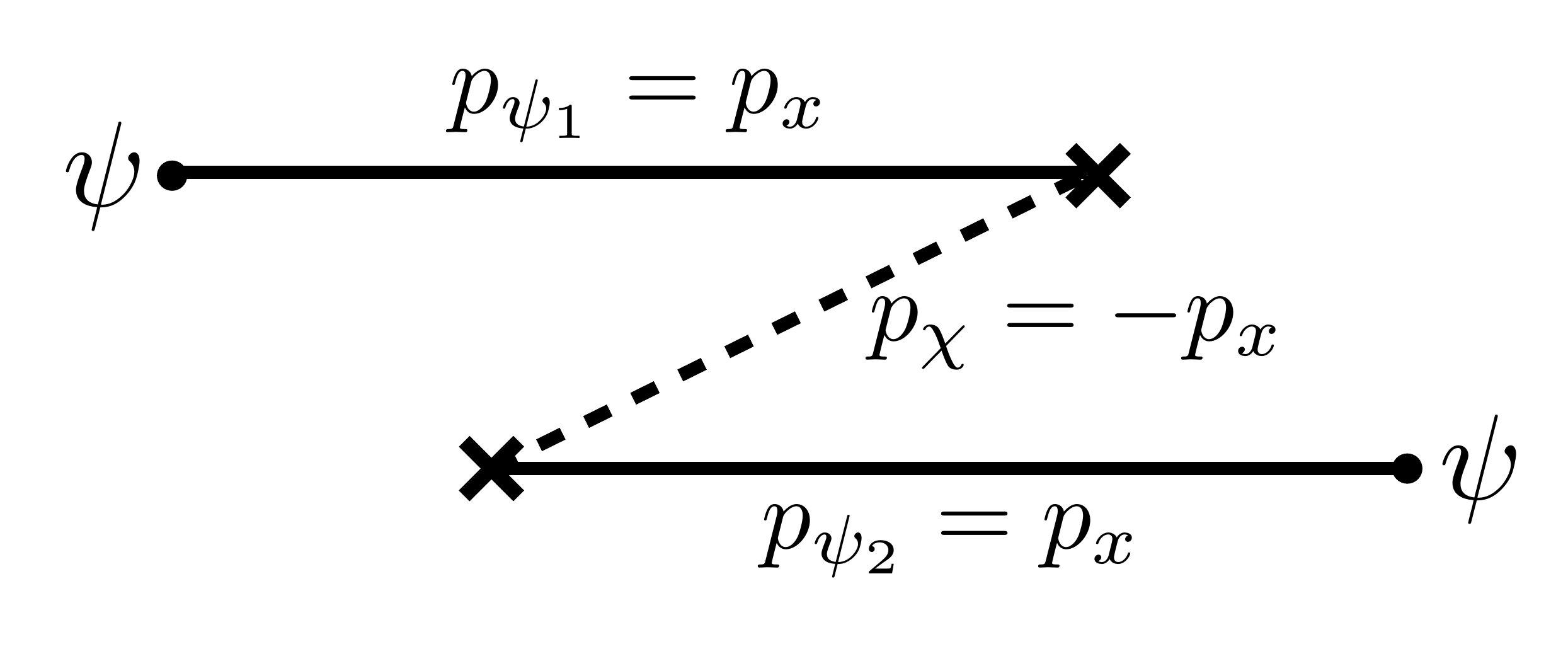}
\caption{Second-order contribution to $m_\psi^2$ due to mixing with three-particle states. In the infinite momentum limit, the invariant mass of the intermediate state $\mu'^2 \ra \infty$, lifting this state above any UV cutoff and naively removing this contribution in LC quantization.
\label{fig:FermionHeff}}
\end{center}
\end{figure}

Let's consider the resulting invariant mass for the one-particle state created by $\psi$. Because $\psi$ is strictly left-moving, it must have $p_x \leq 0$. It therefore cannot directly mix with the right-moving $\chi$, which must have $p_x \geq 0$,
\be
\<\psi,p_x|V_\ET|\chi,p_x'\> = 0.
\ee
The leading contribution to the invariant mass is therefore due to mixing with three-particle states containing two $\psi$ and one $\chi$, shown in Fig.~\ref{fig:FermionHeff}. Writing the sum over intermediate states in terms of Fock space states, we thus have\footnote{The first line of eq.~\eqref{eq:MassShift} simply correponds to a sum over intermediate three-particle states between two insertions of $V_\ET$. The numerator of the second line arises from the factors of $H_\CFT$ in $M^2_\ET \supset \acomm{H_\CFT}{V_\ET}$, which gives the sum of the energies of the external and intermediate states, while the denominator is simply the difference between the invariant mass of the external state (which in this case is zero) and that of the intermediate state, $\mu'^2$.}
\bq
\bal
\de m_\psi^2 = -\int \fr{dp_{\psi_1} \, dp_{\psi_2} \, dp_\chi}{(2\pi)^3 2|p_{\psi_1}| 2|p_{\psi_2}| 2|p_\chi|} &\fr{\<\psi,p_x|V_\ET|p_{\psi_1},p_{\psi_2},p_\chi\>\<p_{\psi_1},p_{\psi_2},p_\chi|V_\ET|\psi,p_x'\>}{\<\psi,p_x|\psi,p'_x\>} \\
& \times \fr{(|p_x|+\sqrt{\mu'^2+p_x^2})^2}{\mu'^2},
\eal
\label{eq:MassShift}
\eq
where the intermediate invariant mass is given by
\be
\mu'^2 = (|p_{\psi_1}|+|p_{\psi_2}|+|p_\chi|)^2 - (p_{\psi_1} + p_{\psi_2} + p_\chi)^2.
\ee
Due to conservation of momentum, the momenta of the intermediate particles are all fixed in terms of the incoming momentum $p_x$,
\be
p_{\psi_1} = p_{\psi_2} = p_x, \quad p_\chi = -p_x.
\ee
We can then evaluate the intermediate matrix elements and rewrite the overall integral into the simpler form
\be
\de m_\psi^2 = \fr{m^2}{2}\int_0^{\Lambda^2} d\mu'^2 \, \de(\mu'^2 - 8p_x^2) \fr{(|p_x|+\sqrt{\mu'^2+p_x^2})^2}{\mu'^2},
\ee
where we've explicitly introduced the cutoff on the invariant mass of the intermediate state.

As we can see, the mass eigenvalue for $\psi$ comes specifically from an intermediate state with mass $\mu'^2 = 8p_x^2$. If we take $|p_x| \ra \infty$ with fixed cutoff $\Lambda$, this state is therefore lifted above our cutoff, such that we lose its contribution. In other words, the naive $M_\LC^2$ in eq.~\eqref{eq:NaiveMLC}, which is equivalent to taking the infinite momentum limit of $M_\ET^2$, has no matrix element mixing $\psi$ with an intermediate three-particle state and is therefore missing this contribution to $m_\psi^2$.

Before discussing how to correct $M_\LC^2$ to include this contribution, let's first understand \emph{why} this intermediate state is removed in the infinite momentum limit. In this example, the intermediate three-particle state has total momentum $p_x \leq 0$, due to the fact that the incoming state is created by the left-moving $\psi$. However, this intermediate state contains at least one $\chi$ particle, which must have $p_\chi \geq 0$, due to the fact that $\chi$ is strictly right-moving (i.e.~has $p_- = 0$). In the limit $p_x \ra -\infty$, the right-moving $\chi$ must therefore have infinite relative momentum with respect to the other left-moving particles, such that the total invariant mass $\mu'^2 \ra \infty$.

This behavior is quite general, such that \emph{all} states involving $\chi$ become infinitely heavy in the limit $p_x \ra -\infty$.\footnote{The decision to send $p_x \ra -\infty$ simply follows from our convention of defining the LC Hamiltonian on slices of fixed $x^+$. If we instead took $p_x \ra +\infty$, we would obtain an equivalent LC Hamiltonian on slices of fixed $x^-$, with the roles of $\psi$ and $\chi$ swapped.} This is simply a manifestation of the right-moving $\chi$ becoming \emph{non-dynamical} in LC quantization. From our discussion in section~\ref{sec:Fermions}, we know that we therefore need to integrate out $\chi$ to obtain an effective Hamiltonian $H_\eff$ for the remaining left-moving degrees of freedom created by $\psi$, to include the corrections that are naively removed in the infinite momentum limit.

More generally, any state involving particles with $p_- = 0$, whether they correspond to a purely right-moving field $\chi$ or a zero mode of a left-moving field, become infinitely heavy as $p_x \ra -\infty$. While in this particular example we know how to use the equation of motion for $\chi$ to obtain $H_\eff$, let's discuss a more general approach, which can be used to include the effects of zero modes in the deformation of any CFT.

This approach, initially proposed in~\cite{Fitzpatrick:2018ttk}, involves first constructing the LC time-evolution operator
\be
U_\LC(x^+,0) \equiv \Tcal\Big\{ e^{-i\int_0^{x^+} dx^{+\prime} [P_{+\CFT} + V_\LC(x^{+\prime})]} \Big\},
\ee
where $V_\LC$ is the naive LC Hamiltonian in eq.~\eqref{eq:NaiveMLC}. We can then define an effective LC Hamiltonian $H_\eff$ as all contributions to this operator which are linear in $x^+$,
\be
\boxed{H_\eff \equiv \lim_{x^+\ra0} i\fr{\p}{\p x^+} U_\LC(x^+,0).}
\label{eq:HeffDef}
\ee
Naively, this definition would simply recover the original Hamiltonian $P_{+\CFT} + V_\LC$. However, as we'll now demonstrate, there are additional contributions, coming precisely from states which are lifted from the Hilbert space in the infinite momentum limit.

Returning to our 2D fermion example, let's use this prescription to compute the $H_\eff$ matrix element for $\psi$,
\be
\<\psi,p_-|H_\eff|\psi,p'_-\> \equiv \lim_{x^+\ra0} i\fr{\p}{\p x^+} \<\psi,p_-|U_\LC(x^+,0)|\psi,p'_-\>.
\ee
We can evaluate the RHS of this expression by expanding the time-evolution operator as the Dyson series
\bq
\bal
U_\LC(x^+,0) &= 1 - i \int_0^{x^+} dx^+_1 [P_{+\CFT} + V_\LC(x_1^+)] \\
& \quad - \half \int_0^{x^+} dx^+_1 dx^+_2 \Tcal\{ [P_{+\CFT} + V_\LC(x_1^+)] [P_{+\CFT} + V_\LC(x_2^+)]\} + \ldots
\eal
\eq
Because $\psi$ is purely left-moving, it is annhilated by the undeformed $P_{+\CFT}$, such that we only need to consider the contributions from $V_\LC$. The first few terms in the expansion are therefore
\bq
\bal
\<\psi,p_-|U_\LC(x^+,0)|\psi,p'_-\> &= \<\psi,p_-|\psi,p'_-\> - i \int_0^{x^+} dx^+_1 \<\psi,p_-| V_\LC(x_1^+)|\psi,p'_-\> \\
& \quad - \half \int_0^{x^+} dx^+_1 dx^+_2 \<\psi,p_-|\Tcal\{ V_\LC(x_1^+) V_\LC(x_2^+) \}|\psi,p'_-\> + \ldots
\eal
\eq
Let's now look at each of these terms more carefully. The first term, while nonzero, will vanish when we take a derivative with respect to $x^+$. The second, linear term is zero, since $V_\LC$ only mixes $\psi$ with $\chi$.

The third term is naively quadratic in $x^+$, which suggests it will vanish when we act with a derivative then take $x^+ \ra 0$. However, if we look more carefully at the four-point function in the integrand, we see that it contains a time-ordered two-point function for $\chi$,
\bq
\bal
&\<\psi,p_-|\Tcal\{ V_\LC(x_1^+) V_\LC(x_2^+) \}|\psi,p'_-\> \\
& \qquad = 2m^2 \int dx_1^- dx_2^- \<\Tcal\{\chi(x_1)\chi(x_2)\}\> \<\psi,p_-|\psi(x_1)\psi(x_2)|\psi,p'_-\>,
\eal
\eq
where we've used the independence of the two fermion modes to factorize this expression into a product of a left-moving correlator and right-moving correlator. The time-ordered $\chi$ two-point function takes the form
\be
\<\Tcal\{\chi(x_1)\chi(x_2)\}\> = \fr{-i}{4\pi(x_{12}^+ - i\epsilon \, \sgn (x_{12}^-))} = \Pcal \left( \fr{-i}{4\pi x_{12}^+} \right) + \fr{1}{4} \de(x_{12}^+) \sgn(x_{12}^-), \, \,
\label{eq:Chi2Pt}
\ee
where $\Pcal$ indicates the principal value. 
This four-point function therefore contains a delta function in $x_{12}^+$, which eliminates one of the integrals, reducing this expression to a term which is \emph{linear} in $x^+$. We thus obtain a nonzero contribution to $H_\eff$ from this second-order term, which reproduces our expectation from integrating out $\chi$ in section~\ref{sec:Fermions},
\be
H_\eff = \fr{m^2}{2} \psi \fr{1}{i\p_-} \psi.
\ee
In fact, our prescription for $H_\eff$ finally explains how to interpret the $\fr{1}{\p_-}$ obtained from the equation of motion for $\chi$: this factor corresponds to the coefficient of $\de(x^+)$ in the $\chi$ propagator~\eqref{eq:Chi2Pt}.

While it may not be immediately apparent, the $\de(x_{12}^+)$ in this four-point function is due to the three-particle intermediate state we considered in old-fasioned perturbation theory, which was removed in the infinite momentum limit. In fact, a factor of $\de(x^+)$ occurs anytime a correlator \emph{loses its spectral decomposition in LC quantization} (i.e.~has finite contributions which are naively removed in the infinite momentum limit). These delta functions in higher-point functions then give rise to contributions to $H_\eff$, reproducing the effects of the infinite mass intermediate states that have been integrated out.

While this discussion has been somewhat technical, the prescription for $H_\eff$ in~\eqref{eq:HeffDef} can be understood as simply demanding that LC quantization reproduce correlation functions in the deformed theory. For example, consider the general two-point function
\benn
\<\widehat{\Ocal}(x^+) \widehat{\Ocal}'(0)\>,
\eenn
where $\widehat{\Ocal}$ indicates that this is a correlator in the deformed theory. We can rewrite this correlator using a general LC time-evolution operator,
\be
\<\widehat{\Ocal}(x^+) \widehat{\Ocal}'(0)\> = \<\Ocal(0) U_\eff(x^+,0) \Ocal'(0)\>, \quad U_\eff(x^+,0) \equiv \Tcal\Big\{ e^{-i\int_0^{x^+} dx^{+\prime} H_\eff(x^{+\prime})} \Big\}. \qquad 
\ee
Expanding this expression as a Dyson series, we can in principle completely fix $H_\eff$ by matching the full correlator to linear order in $x^+$.

However, this requires us to know correlation functions in the deformed theory. Fortunately, correlation functions should be the same in any quantization scheme, so we can also compute this correlator in ET quantization, with the corresponding time-evolution operator
\be
U_\ET(t,0) \equiv \Tcal\Big\{ e^{-i\int_0^t dt' [H_\CFT + V_\ET(t')]} \Big\}.
\ee
We can expand this time-evolution operator as a series in the relevant deformation, computing the correlator perturbatively in $\lambda$. We can then fix $H_\eff$ by matching to all terms in this perturbative expansion that are linear in $x^+$. The prescription in eq.~\eqref{eq:HeffDef} can therefore be thought of as simply a matching procedure between LC and ET quantization.

Note that while in this example we have focused on the case where an entire field becomes non-dynamical in LC quantization, non-trivial contributions to $H_\eff$ can also arise due to zero modes of dynamical fields. For example, in 2D $\phi^4$ theory, zero modes lead to a coupling-dependent shift in the bare mass~\cite{Fitzpatrick:2018ttk,Burkardt,Burkardt2}. However, these effects are still captured by the general prescription in eq.~\eqref{eq:HeffDef}, and must be included to correctly match LC results with those in ET quantization~\cite{Fitzpatrick:2018xlz}.

%%%%%%%%%%%%%%%%%%%%%%%%%%%%%%%%%%%%%%%%%%%%%%%%%%%%%%%%%%%%%%%%%%%%%%%%%%%%%
%%%%%%%%%%%%%%%%%%%%%%%%%%%%%%%%%%%%%%%%%%%%%%%%%%%%%%%%%%%%%%%%%%%%%%%%%%%%%

\section{Technical Details of Gauge Interaction} \label{app:appendixgauge}

In this section, we discuss some technical details glossed over in section \ref{sec:gaugefields}. In \ref{subsec:appgaugematrixele}, we explain how to handle divergences that occurs in the gauge interaction and present the form of the matrix elements. In \ref{subsec:LargeNapp}, we present formulas for the Hamiltonian matrix elements at large $N_c$, for various choices of basis.

\subsection{Matrix elements} \label{subsec:appgaugematrixele}

\subsubsection{Two Particle Warm-up}
Before we jump into the details of multi-particle matrix element, let's warm up with matrix elements of two-particle states. The two particle matrix elements are simple enough to compute in closed form using Fock space methods, and is enough to cover the entire large $N_c$ physics. We will see that the gauge interaction term has IR divergence for certain matrix elements, which is cured by accounting for the self-energy contribution and taking the principal value. 

Recall from (\ref{eq:2pfiniteNcJacobis}), reproduced below for convenience, that we can represent the two-particle states using the creation and annihilation operators
\begin{align}
&\ket{\cO_\ell, p } \equiv \frac{1}{N_\ell} \int \frac{dp_1 dp_2}{8\pi^2} (2\pi) \delta(p-p_1-p_2) F_{\ell}(p_1,p_2)  \ket{\psi^\dagg_i (p_1) \psi^\dagg_i (p_2)},
 \nn \\
 &\ket{\psi^\dagg_i (p_1) \psi^\dagg_i (p_2)} \equiv b_i^\dagg(p_1) a_i^\dagg(p_2) \vac, \qquad  N_\ell = \frac{p^\ell}{4} \sqrt{\frac{N_c}{\pi}}, \label{eq:2pJacobiQCD}
\end{align}
where $F_{\ell}(p_1,p_2)$ is the two-particle wave function in momentum space \begin{equation}
	F_{\ell}(p_1,p_2) \equiv \sqrt{2\ell+1} \,
(p_1+p_2)^\ell P_\ell^{(0,0)}
\pr{\frac{p_1-p_2}{p_1+p_2}}.
\end{equation} In order to compute the gauge interaction matrix element between these states, we must contract the Hamiltonian interaction term with the Fock states. Schematically, we have 
\begin{align}\label{eq:gauge-interaction-fock}
 &\left\< \psi_m^\dagger (p_1) \, \psi_m (p_2) \right|
 \int dy \,
(\psi^\dagger_i T_{ij}^A \psi_j) \frac{1}{\d^2} (\psi^\dagger_k T_{kl}^A \psi_l) (y)
 \left| \psi_n^\dagger (p_1^\prime) \, \psi_n (p_2^\prime) \right\> \nn \\
=~& 0\times 
\raisebox{-.3in}{\includegraphics[width=1.5in]{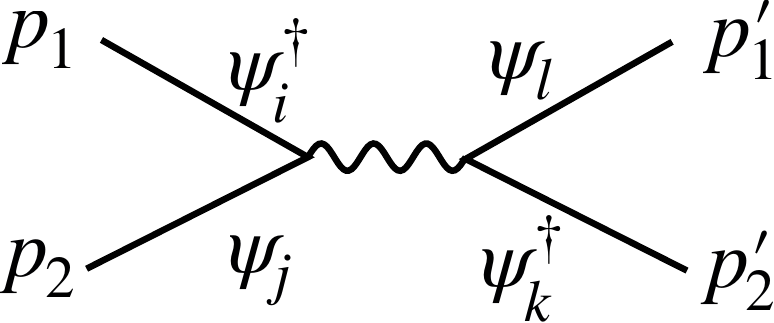} } 
+ (N_c^2-1) \times 
\raisebox{-.40in}{\includegraphics[width=1.3in]{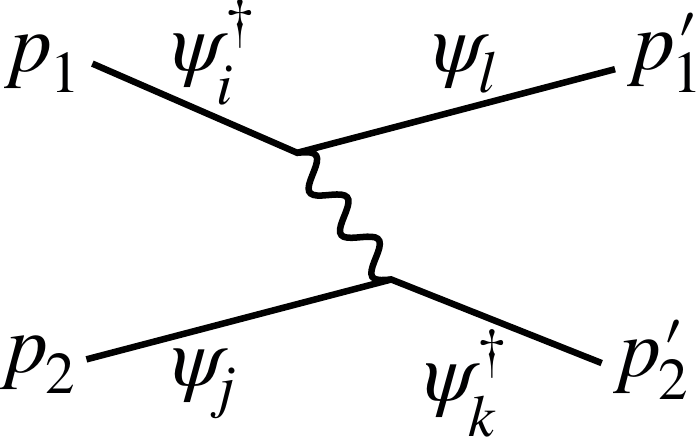} }
~~+ (p\leftrightarrow p^\prime) 
\nn \\
\sim~& 
\frac{1}{(p_1-p_1^\prime)^2},
\end{align}
and then integrate against the wave functions to get the final answer. The color factor is zero for the first diagram and the second diagram gives $(N_c^2-1)$, where we used (\ref{eq:sunidentity}), reproduced here
\begin{align}
(T^A)_{k\ell} (T^A)_{mn} = \half\pa{\delta_{kn}\delta_{\ell m} - \frac{1}{N_c} \delta_{k\ell}\delta_{mn}}.
\end{align}
Thus the 2-to-2 matrix element only gets contribution from t-channel diagram, where the nonlocal potential $1/\d^2$ picks up a factor $1/(p_1-p_1^\prime)^2$. Note that when $p_1\rightarrow p_1^\prime$ we get an IR divergence. This is because by normal ordering the deformation we have thrown away a divergent self-energy term, 
\begin{align}\label{eq:gauge-self-energy}
\raisebox{-.2in}{\includegraphics[width=1.2in]{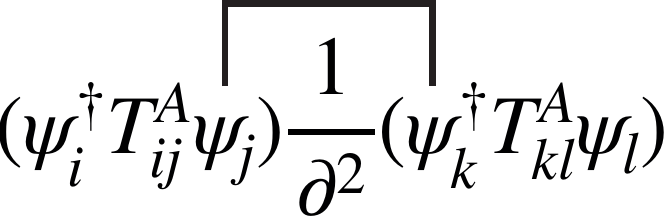} } 
= 
\raisebox{-.3in}{\includegraphics[width=1.4in]{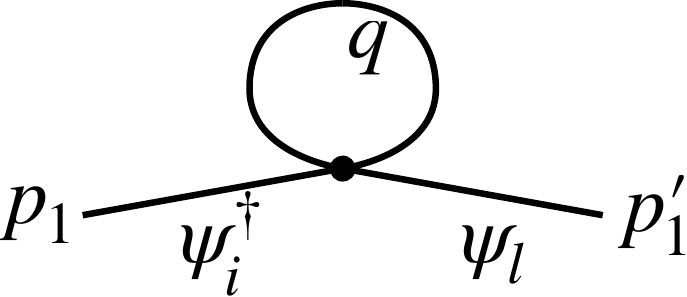} } 
\sim \int \frac{dq}{(p_1-q)^2} \delta(p_1-p_1^\prime) \, .
\end{align}
When we add (\ref{eq:gauge-interaction-fock}) and (\ref{eq:gauge-self-energy}) the divergence will cancel out.
The cancellation is manifest if we take the integrand to be anti-symmetric under $p_1 \leftrightarrow p_1^\prime$
The end result is a modification of the integral
\begin{align}\label{eq:gauge-self-energy-shift}
&\int dp_1dp_1^\prime \,
\frac{
	F_\ell(p_1,p-p_1) F_{\ell^\prime}(p_1^\prime,p-p_1^\prime)
}{(p_1-p_1^\prime)^2} \nn \\
\rightarrow \, & \Pcal \int dp_1dp_1^\prime \, F_\ell(p_1,p-p_1)
\frac{
	F_{\ell^\prime}(p_1^\prime,p-p_1^\prime) - F_{\ell^\prime}(p_1,p-p_1)
}{(p_1-p_1^\prime)^2}  ,
\end{align}
where the integral is finite as a {\it principal value} integral 
\begin{equation}
\Pcal \int \frac{\psi(k)dk}{k^2} \equiv \frac{1}{2} \int 
\frac{\psi(k+i \epsilon)dk}{(k+i\epsilon)^2} +
\frac{1}{2} \int \frac{\psi(k-i \epsilon)dk}{(k-i\epsilon)^2} \, .
\label{eq:PVdef}
\end{equation}
We can thus work out the full formula for the two particle matrix elements, including factors of the coupling in the Hamiltonian
\begin{align}
\CM_{\ell\ell^\prime} &\equiv -\frac{g^2}{2}
	\left\< \CO_{\ell},p \right|
 \int dy \,
(\psi^\dagger_i T_{ij}^A \psi_j) \frac{1}{\d^2} (\psi^\dagger_k T_{kl}^A \psi_l) (y)
 \left| \CO_{\ell^\prime},p^\prime \right\> 
\nn\\
&= 
 g^2 \frac{N_c^2-1 }{N_c}
 \frac{\sqrt{2\ell+1}\sqrt{2\ell^\prime+1}}{\pi}
  \, \nn \\
&~~~~~~ 
\Pcal
\int_{0}^{1} dx_1 dx_2
   P_{\ell}^{(0,0)}\left(1-2
   x_1\right) 
   \frac{
    P_{\ell'}^{(0,0)}\left(1-2 x_1\right)
    - P_{\ell'}^{(0,0)}\left(1-2 x_2\right)
   }{\left(x_1-x_2\right){}^2}
    \, ,
    \label{eq:gauge-integral-2pt}
\end{align}
where we made the substitution $x_1 = p_1/P$, $x_2 = p_1^\prime/P$ and plugged in the expression for $F_{\ell}(p_1,p_2)$ in~\eqref{eq:2pJacobiQCD}. The integrals above converge and the IR divergence has canceled, as promised.

\subsubsection{Higher Particles}
We can use Wick contraction to compute more complicated matrix elements involving more than two particle external states. However, our building blocks will still be the two particle correlators
\begin{align}\label{eq:gauge-building-block}
\left\< \partial^k \psi_i^\dagger(x) \partial^{k^\prime} \psi_j(x^\prime) \right\> = 
\, \frac{\Gamma(k + k^\prime +1)}{4\pi(x-x^\prime)^{k+k^\prime+1}} \cdot \delta_{ij} \, .
\end{align}
In order to compute the matrix element, we must put 
$\CO$
at position $x$,\footnote{Strictly speaking, we mean $\cO^\dagg$ for the outgoing external operator, but we will abuse notation and refer to it as $\cO$.} 
$\CO'$
at position $z$, and the deformation at position $y$, and compute the Fourier transform: 
\begin{align}
\CM_{\CO\CO^\prime} = \frac{1}{N_\CO^* N_{\CO'}}
\int dxdydz \, e^{iPx - iP^\prime z}
\big\< \CO(x)\, 
(\psi^\dagger_i T_{ij}^A \psi_j) \frac{1}{\d^2} (\psi^\dagger_k T_{kl}^A \psi_l) (y)\, 
\CO^\prime(z) \big\> \, .
\end{align}
These matrix elements are nonzero only when the particle number difference is $n - n^\prime = 0$ or $\pm2$ between in and out states. We expand the external operators $\CO$ and $\CO'$ as sums of ``monomials'' defined in 
(\ref{eq:qcdMonomial}),
\begin{align}
\CM_{\CO\CO^\prime} \supset \CM_{\kvec,\kvec'} 
\equiv 
\int& dxdydz \, e^{iPx - iP^\prime z} \times \nn \\
&\big\< \d^{\kvec_1}\psi \d^{\kvec_2}\psi^\dagg (x)\, 
(\psi^\dagger_i T_{ij}^A \psi_j) \frac{1}{\d^2} (\psi^\dagger_k T_{kl}^A \psi_l) (y)\, 
\d^{\kvec_1'}\psi^\dagg \d^{\kvec_2'}\psi(z) \big\> \, .
\end{align}
In coordinate space the nonlocal kernel $1/\d^2$ is an integral defined via
\begin{align}
(\psi_i^\dagger T_{ij}^A \psi_j) \frac{1}{\d^2} (\psi_k^\dagger T_{kl}^A \psi_l) (y)
= (\psi_i^\dagger T_{ij}^A \psi_j)(y) \int^y dy^{\prime} \int^{y^\prime} dy^{\prime\prime} (\psi_k^\dagger T_{kl}^A \psi_l) (y^{\prime\prime}) \, .
\end{align}
We first compute the four-point function $G_{\kvec,\kvec'}(x, y, y^{\prime\prime}, z)$ as an integrand, which has the general form
\begin{equation}
\begin{aligned}\label{eq:gauge-5pt-func}
G_{\kvec,\kvec'}(x, y, y^{\prime\prime}, z) &\equiv \corr{\d^{\kvec_1}\psi \d^{\kvec_2}\psi^\dagg (x) (\psi_i^\dagger T_{ij}^A \psi_j)(y) (\psi_k^\dagger T_{kl}^A \psi_l) (y^{\prime\prime}) \d^{\kvec_1'}\psi^\dagg \d^{\kvec_2'}\psi(z) } \\
 &= \sum_{a,b,a',b'}
\frac{\tilde A_{\kvec,\kvec^\prime}^{(a,b,a',b')}}{
    (x-y)^a (y-z)^b (x-y^{\prime\prime})^{a^\prime}
    (y^{\prime\prime}-z)^{b^\prime}
    (x-z)^{c}
} \, ,
\end{aligned} 
\end{equation}
where $\tilde A_{\kvec,\kvec^\prime}^{(a,b,a',b')}$ is the product of constants from the two-point functions in eq. \ref{eq:gauge-building-block}, which consists of signs from permuting fermions past each other, color tensors, $\Gamma$ functions and $4\pi$ factors.
Then we integrate out $y^{\prime\prime}$ and Fourier transform to get the matrix elements
\begin{align}\label{eq:gauge-matrix-element-compact}
\CM_{\kvec,\kvec^\prime} &\equiv \int dxdydz \, e^{ipx - ip^\prime z}
 \int^y dy^{\prime} \int^{y^\prime} dy^{\prime\prime} G(x, y, y^{\prime\prime}, z) \nn \\
 &= (2\pi) \delta(p-p^\prime)
 \sum_{a,b,a',b'} \frac{
    4\pi^2 i^{\Delta+\Delta^\prime-2}
    P^{\Delta+\Delta^\prime}
    \tilde A_{\kvec, \kvec'}^{(a,b,a',b')}
}{\Gamma(\Delta+\Delta^\prime-1)} I(a,b,a^\prime,b^\prime) \,, 
\end{align}
where we have eliminated $c$ using the relation $a+b+a^\prime+b^\prime+c = \Delta+\Delta^\prime+2$ from dimensional analysis.

The integral above is subject to IR divergences and is sensitive to the boundary condition. The correct treatment is equivalent to taking the self-energy shift and the principal value integral in the momentum space, similar to the two particle case discussed in the previous section. We will discuss the details of this procedure in the following section and how to compute the function $I(a,b,a',b')$.

\subsubsection{Determining the Function \texorpdfstring{$I(a,b,a',b')$}{I} }\label{gauge-momentum-space-appendix}
\paragraph{Wick contraction t-channel}
Starting from (\ref{eq:gauge-5pt-func}), we can Fourier transform each individual spatial factor using 
\begin{align}
\int \frac{e^{i p x}\, dx}{(x-i \epsilon )^a} = \frac{2 \pi  i^a p^{a-1} \theta (p)}{(a-1)!} 
\end{align}
and expand the correlation function in parton momenta
\begin{align}
&\pr{\frac{\d}{\d y^{\prime\prime}}}^{-2}
\frac{1}{
    (x-y)^a (y-z)^b (x-y^{\prime\prime})^{a^\prime}
    (y^{\prime\prime}-z)^{b^\prime}
    (x-z)^{c}
}
\nn \\
\doteq & \int dp_1 dp_2 dp_1^\prime dp_2^\prime dq\, 
e^{-i x (p_1+p_1^\prime+q)} e^{i z (p_2+p_2^\prime+q)}
e^{i y (p_1-p_2)} \pr{ \frac{1}{\d^2} e^{i y^{\prime\prime} (p_1^\prime-p_2^\prime)} }
\nn \\
& ~~~~\times \frac{1
}{
\Gamma(a)\Gamma(b)\Gamma(a^\prime)\Gamma(b^\prime)
\Gamma(c)}
p_1^{a-1} p_2^{b-1} {p_1^\prime}^{a^\prime-1} {p_2^\prime}^{b^\prime-1} q^{c-1}~
\theta (p_1) \theta (p_2) \theta (p_1^\prime) \theta (p_2^\prime) \theta (q) \, ,
\end{align}
where the momenta $p_i$ and $p_i^\prime$ are the momenta of active fermions and $q$ comes from the spectators. 
We take the standard Fourier transformation to obtain the momentum space matrix element
\begin{align}
&\int e^{i p x - i p^\prime z} dxdydz 
\pr{\frac{\d}{\d y^{\prime\prime}}}^{-2} 
\frac{1}{
    (x-y)^a (y-z)^b (x-y^{\prime\prime})^{a^\prime}
    (y^{\prime\prime}-z)^{b^\prime}
    (x-z)^{c}
}
\nn \\
&\doteq ~ (2\pi) \delta(p-p^\prime) 
\int dp_1 dp_2 dp_1^\prime dp_2^\prime dq\, \nn \\
&~~~~ \times (2\pi) \delta( p_1+p_1^\prime+q - p ) (2\pi) \delta( p_2+p_2^\prime+q - p ) \nn \\
&~~~~ \times \frac{1
}{
\Gamma(a)\Gamma(b)\Gamma(a^\prime)\Gamma(b^\prime)
\Gamma(c)}  \times \frac{
p_1^{a-1} p_2^{b-1} {p_1^\prime}^{a^\prime-1} {p_2^\prime}^{b^\prime-1} q^{c-1}
}{(p_1^\prime-p_2^\prime)^2} \nn \\
&\doteq ~ 
(2\pi) \delta(p-p^\prime)\frac{
    4\pi^2 
    p^{\Delta+\Delta^\prime}
}{\Gamma(\Delta+\Delta^\prime-1)} I(a,b,a^\prime,b^\prime),
\end{align}
where the momentum on the denominator comes from acting the $1/\d^2$ on the exponential of $y^{\prime\prime}$. The spatial integral becomes momentum conservation. As usual, we can normalize by the total external momentum, and express the integral in terms of the momentum fractions.
A particulaly convenient substitution is
\begin{align}
p_1 &\equiv x_1 x_2 \nn \\
p_2 &\equiv x_1 x_3 \nn \\
q &\equiv 1-x_1 \nn \\
p_1^\prime &= x_1(1-x_2) \nn \\
p_2^\prime &= x_1(1-x_3) \, ,
\end{align}
which separates the active part and the spectators.
We have thus worked out a general formula to evaluate the gauge interaction matrix elements, term by term from Wick contraction, as a momentum integral
\begin{align}\label{eq:gauge-interaction-momentum}
I(a,b,a^\prime,b^\prime) =& \frac{
\Gamma(\Delta+\Delta^\prime-1)
}
{
\Gamma(a)\Gamma(b)\Gamma(a^\prime)\Gamma(b^\prime)
\Gamma(c)}
\int \, dx_1 x_1^{a+b+a^\prime+b^\prime-4}(1-x_1)^{c-1} \nn \\
&~~\times 
\int dx_2 dx_3 \, 
\frac{x_2^{a-1} (1-x_2)^{a^\prime-1} x_3^{b-1} (1-x_3)^{b^\prime-1} }{(x_2-x_3)^2} \, ,
\end{align}
It is nice that the momentum fraction of the spectators factors out of the principal value integral, making it possible to find a closed form expression for the active part. The integral over the spectators' momentum is
\begin{align}
\int \, dx_1 x_1^{a+b+a^\prime+b^\prime-4}(1-x_1)^{c-1}
= \frac{\Gamma (c) \Gamma \left(a+b+a'+b'-3\right)}{\Gamma \left(a+b+c+a'+b'-3\right)} \, .
\end{align}

Now we are left with the integral of $x_2$ and $x_3$
\begin{align}
I(a,b,a^\prime,b^\prime) &= 
\frac{\Gamma \left(a+b+a'+b'-3\right) }{
    \Gamma(a)\Gamma(a^\prime)\Gamma(b)\Gamma(b^\prime)
}
\int dx_2 dx_3 \, 
\frac{x_2^{a-1} (1-x_2)^{a^\prime-1} x_3^{b-1} (1-x_3)^{b^\prime-1} }{(x_2-x_3)^2}
\nn \\
&= \frac{\Gamma \left(a+b+a'+b'-3\right) }{
    \Gamma(a)\Gamma(a^\prime)\Gamma(b)\Gamma(b^\prime)
} \nn \\
& ~~~~~~~~~~~\times \sum_{m_1}^{a^\prime-1}  \sum_{m_2}^{b^\prime-1}
\binom{a^\prime-1}{m_1} \binom{b^\prime-1}{m_2} \,
I_1(a+m_1-1,b+m_2-1),
\end{align}
where we have defined the general integral
\begin{align}\label{eq:gauge-master-integral-1}
I_1(a,b) &\equiv 
\int dx_2 dx_3 \, \frac{
    x_2^a x_3^b 
}{(x_2-x_3)^2} + (\text{self-energy shift}) \, .
\end{align}
The integral is divergent, and this divergence must be canceled by a self-energy term. Note that this integral looks identical to the two-to-two matrix element in the Fock space, and hence we choose the same scheme for the self-energy shift as (\ref{eq:gauge-self-energy-shift}):
\begin{align}
I_1(a,b)&= \Pcal \int dx_2 dx_3 \, \frac{
        x_2^a x_3^b - x_2^{a+b}
    }{(x_2-x_3)^2} \nn \\
   &= \frac{a H_a+b H_b-1}{a+b} - H_{a+b-1}
   \label{eq:gauge-integral-t-channel}
\end{align}
where the ``$\Pcal$'' stands for taking the  principal value.

\paragraph{The other channel:}
In (\ref{eq:gauge-5pt-func}) one or more of the four numbers ($a$, $b$, $a^\prime$ or $b^\prime$) can vanish and the resulting matrix elements will again diverge, which is a special case that needs to be handled separately.

Without loss of generality, we can set $a=0$, and we have the freedom to integrate with respect to either $y$ or $y^{\prime\prime}$  and get the same answer. 
We can just compute
\begin{align}\label{eq:gauge-5pt-trivial}
\left(
    \int^y dy^{\prime} \int^{y^\prime} dy^{\prime\prime}
    \frac{1}{(y^{\prime\prime}-z)^b}
\right)
\frac{1}{(x-y)^{a^\prime}}
\frac{1}{(y-z)^{b^\prime}}
\frac{1}{(x-z)^{c}} 
\, .
\end{align}
Unless $b=2$, we can be agnostic about the boundary condition of the integral because
the boundary value at $y^{\prime\prime}\rightarrow \infty$ and $y^{\prime}\rightarrow \infty$ vanishes, and
the integral over $y^{\prime\prime}$ is just the naive indefinite integral  
\begin{align}\label{eq:gauge-interaction-momentum-trivial}
\int^y dy^{\prime} \int^{y^\prime} dy^{\prime\prime}\frac{1}{(y-z)^b} = \begin{cases}
\frac{1}{(b-1)(b-2)(y-z)^{b-2}} & b>2\\
\text{depends on the boundary condition} & b=2
\end{cases}
\end{align}
For $b=2$ case, we have to be more careful. Schematically, the correct boundary condition of the coordinate space integral is equivalent to an appropriate principal value prescription in momentum space similar to what we have discussed in (\ref{eq:gauge-integral-2pt}) and (\ref{eq:gauge-integral-t-channel}).
We will discuss this case momentarily. 
For now we can focus on the $b>2$ case where we can perform the usual integral over three-point function as in (\ref{eq:FTFormulas}) and obtain the final result
for $b>2$
\begin{align}
I(0,b,a^\prime,b^\prime) 
=\,& \frac{\Gamma \left(b+a^\prime+b^\prime-3\right)}{(b-2) (b-1) \Gamma \left(a^\prime\right) \Gamma
   \left(b+b^\prime-2\right)}\, \quad\quad\quad (b > 2) .
\end{align}

\paragraph{$b=2$ case:}
The $b\rightarrow2$ limit of (\ref{eq:gauge-interaction-momentum-trivial}) depends on the boundary condition,
so we proceed in the momentum space. 
Like the t-channel case, we write the spatial factors in (\ref{eq:gauge-5pt-trivial}) in momentum space and Fourier transform the overall formula with respect of $x$ and $z$. Note that $p_1$ is missing since $(x-y)$ factor is missing. 
\begin{align}\label{eq:gauge-s-channel-special-case}
& \int e^{i p x - i p^\prime z} dxdydz
\int^y dy^{\prime} \int^{y^\prime} dy^{\prime\prime}
\frac{1}{(y^{\prime\prime}-z)^2}
\frac{1}{(x-y)^{a^\prime}}
\frac{1}{(y-z)^{b^\prime}}
\frac{1}{(x-z)^{c}}   
\nn \\
\doteq &~ (2\pi) \delta(p-p^\prime) 
\int dp_2 dp_1^\prime dp_2^\prime (2\pi)\delta(p_1^\prime+q-p) 
\times (2\pi) \delta(p_2+p_2^\prime +q -p) \nn \\
&~~\times \frac{
1
}{\Gamma(b)\Gamma(c)\Gamma(a^\prime)\Gamma(b^\prime)}
\times \frac{
p_2 (p_1^\prime)^{a^\prime-1}(p_2^\prime)^{b^\prime-1} q^{c-1}} 
{(p_1^\prime - p_2^\prime)^2}
 \nn \\
\doteq &~ (2\pi) \delta(p-p^\prime)\frac{
    4\pi^2 
    p^{\Delta+\Delta^\prime}
}{\Gamma(\Delta+\Delta^\prime-1)} I (0,2,a^\prime,b^\prime,c)
\end{align}
where we used the momentum conservation 
\begin{equation}
 p_2 = p_1^\prime - p_2^\prime
\end{equation}
To proceed, we can further parameterize the momenta as
\begin{align}
p_2 &\equiv x_1 (1-x_2) \nn \\
q &\equiv 1-x_1 \nn \\
p_1^\prime &= x_1 \nn \\
p_2^\prime &= x_1 x_2 \, ,
\end{align}
and compute the integral
\begin{align}
I(0,2,a^\prime,b^\prime,c) = \frac{
    \Gamma(\Delta+\Delta^\prime-1)
}{\Gamma(b)\Gamma(c)\Gamma(a^\prime)\Gamma(b^\prime)} 
\int dx_1 \, x_1^{a^\prime +b^\prime -2} (1-x)^{c-1}
\int dx_2 \frac{x_2^{b^\prime-1}}{1-x_2}.
\end{align}
The $x_1$ integral is finite,
\begin{align}
\int dx_1 \, x_1^{a^\prime +b^\prime -2} (1-x)^{c-1}
= \frac{\Gamma(a^\prime+b^\prime-1)\Gamma(c)}
   {\Gamma(a^\prime+b^\prime+c-1)}~,
\end{align}
while the other integral is divergent. We need to find the scheme for the self-energy regulator. The key is that the wave functions that contract with $\frac{1}{\d^2}\psi^\dagger\psi$ needs to be symmetrized under $\psi^\dagger \leftrightarrow \psi$, i.e.~under $p_1^\prime \leftrightarrow p_2^\prime$. Thus the correct self-energy shift is 
\begin{align}
\frac{ (p_1^\prime)^{a^\prime-1}(p_2^\prime)^{b^\prime-1} }
{(p_1^\prime - p_2^\prime)^2} &\rightarrow
\frac{ (p_1^\prime)^{a^\prime-1}(p_2^\prime)^{b^\prime-1} 
 - (p_2^\prime)^{a^\prime+b^\prime-2}
}
{(p_1^\prime - p_2^\prime)^2} \nn \\
\int dx_2 \frac{x_2^{b^\prime-1}}{1-x_2}
&\rightarrow \Pcal \int dx_2 \frac{
    x_2^{b^\prime-1} - x_2^{a^\prime+b^\prime-2}
}{1-x_2}
= - H_{b-1} + H_{a+b-2}, .
\end{align}
We finally have
\begin{align}
I(0,2,a^\prime,b^\prime) 
=\,& \frac{\Gamma \left(a^\prime+b^\prime-1\right)}{\Gamma
   \left(a^\prime\right) \Gamma \left(b^\prime\right)}
   \left(H_{a^\prime+b^\prime-2}-H_{b^\prime-1}\right) \, .
\end{align}

\subsubsection{Summary} To summarize, let us collect the key equations from above. We defined the gauge interaction matrix elements between generic monomials  \begin{equation}
	\begin{aligned}
		\boxed{\frac{\Mcal_{\Kvec \Kvec'}}{2p(2\pi)\delta(p-p')} = \sum_{a,b,a',b'}  \frac{
    2\pi^2 
    p^{\Delta+\Delta^\prime-1}
    \tilde A_{\kvec,\kvec^\prime}^{(a,b,a',b')}
}{\Gamma(\Delta+\Delta^\prime-1)} I(a,b,a^\prime,b^\prime),} \label{eq:gaugesummaryeqFirst}
	\end{aligned}
\end{equation} where $I(a,b,a',b')$ is determined for the following cases: \begin{itemize}
\item $a,b,a^\prime,b^\prime$ all nonzero:
\begin{align}
I(a,&b,a^\prime,b^\prime) 
=\, 
\frac{\Gamma \left(a+b+a'+b'-3\right) }{
    \Gamma(a)\Gamma(a^\prime)\Gamma(b)\Gamma(b^\prime)
}
\times \sum_{m_1}^{a^\prime-1}  \sum_{m_2}^{b^\prime-1}
\binom{a^\prime-1}{m_1} \binom{b^\prime-1}{m_2} \\
& \times 
\frac{\left(a+m_1-1\right) H_{a+m_1-1}+\left(b+m_2-1\right)
   H_{b+m_2-1}-1}{a+b+m_1+m_2-2}-H_{a+b+m_1+m_2-3} \, , \nn
\end{align} where $H_k$ is the harmonic number.
\item $a=0, b>2$: 
\begin{align}
I(0,b,a^\prime,b^\prime) 
=\,& \frac{\Gamma \left(b+a^\prime+b^\prime-3\right)}{(b-2) (b-1) \Gamma \left(a^\prime\right) \Gamma
   \left(b+b^\prime-2\right)}\, .
\end{align}
\item $a=0, b=2$:
\begin{align}
I(0,2,a^\prime,b^\prime) 
=\,& \frac{\Gamma \left(a^\prime+b^\prime-1\right)}{\Gamma
   \left(a^\prime\right) \Gamma \left(b^\prime\right)}
   \left(H_{a^\prime+b^\prime-2}-H_{b^\prime-1}\right) 
\end{align}
and as a special case,
\begin{align}
I(0,2,a^\prime,0) 
=\,& \frac{\Gamma(a^\prime-1)}{\Gamma(a^\prime)}\, .
\end{align} This function has the symmetry 
\begin{align}
I(a,b,a^\prime,b^\prime) &= I(a^\prime,b^\prime,a,b)\nn \\
I(a,b,a^\prime,b^\prime) &= I(b,a,b^\prime,a^\prime), \label{eq:gaugesummaryeqLast}
\end{align} so the above formulas cover all cases.
\end{itemize}

\subsection{Large \texorpdfstring{$N_c$}{N_c}}\label{subsec:LargeNapp} Next, we present the result to (\ref{eq:thooftHamiltonian}), which are the matrix elements for 2d QCD at large $N_c$, with various bases for the meson wavefunction. The first basis is the cosine basis for $\phi_n(x) = \sqrt{2}\cos(n \pi x)$. The Hamiltonian matrix elements are given by \begin{equation}
 	\begin{aligned}
 		&H_{mn}^{(\textrm{`t Hooft, cos})} = \frac{\lambda}{2\pi} \int_0^1 dx \int_0^1 dy \\
 		& \times \frac{\left(\sqrt{2} \cos (\pi  m x)-\sqrt{2} \cos (\pi  m y)\right) \left(\sqrt{2} \cos
   (\pi  n x)-\sqrt{2} \cos (\pi  n y)\right)}{(x-y)^2}.
 	\end{aligned}
 \end{equation} We can compute the matrix elements by changing variables to $u = \frac{x-y}{\sqrt{2}}$ and $v= \frac{x+y}{\sqrt{2}}$ so that \begin{equation}
	H_{mn}^{(\textrm{` t Hooft, cos})} =  \frac{\lambda}{\pi} \int_0^\frac{\sqrt{2}}{2} du \int_{|u|}^{\sqrt{2}-|u|} dv \frac{4 \sin \left(\frac{\pi  m u}{\sqrt{2}}\right) \sin \left(\frac{\pi  m
   v}{\sqrt{2}}\right) \sin \left(\frac{\pi  n u}{\sqrt{2}}\right) \sin \left(\frac{\pi 
   n v}{\sqrt{2}}\right)}{u^2}. \label{eq:thooftcosinematrixelementint}
\end{equation}The resulting matrix elements are given by {\small \[ H_{mn}^{(\textrm{`t Hooft, cos})} = \frac{\lambda}{\pi} \\
 \begin{cases} 
      2 \left(2 \text{Ci}(n \pi )-\text{Ci}(2 n \pi )+\pi  n \text{Si}(n \pi )+(-1)^n+\log
   \left(\frac{2}{\pi  n}\right)-\gamma_E -1\right), \quad  m = n \\
   \frac{\left((-1)^{m+n}+1\right) }{(m-n)(m+n)} \bigg[\left(n^2-m^2\right) \text{Ci}((m-n) \pi )+m^2
   \left(\log \left(1-\frac{n^2}{m^2}\right)-\text{Ci}((m+n) \pi )\right) \\
   +n^2
   \left(\text{Ci}((m+n) \pi )-2 \text{Ci}(n \pi )-\log
   \left(\frac{m^2}{n^2}-1\right)\right)+2 m^2 \text{Ci}(m \pi )\bigg], \quad\, m\ne n,
   \end{cases}
\]} where $\textrm{Ci}$ and $\textrm{Si}$ indicate the cosine and sine integral functions and $\gamma_E$ is the Euler constant. Other bases, such as sines or complex exponentials, can be obtained in a similar way. For example, for the sine basis, we have $\phi_n(x) = \sqrt{2}\sin(n \pi x)$. However, note that this basis does not have the correct boundary conditions in the massless limit, so it will have extremely poor convergence. In this case, the matrix elements are given by \[ H_{mn}^{(\textrm{`t Hooft, sin})} = \frac{ \lambda}{\pi} \\
 \begin{cases} 
      2 \left(\text{Ci}(2 n \pi )+\pi  n \text{Si}(n \pi )+(-1)^n-\log (2 \pi  n)-\gamma_E
   -1\right), \quad \, \,\,\, m = n \\
   \frac{\left((-1)^{m+n}+1\right) }{(m-n)(m+n)} \bigg[m^2 (-\text{Ci}((m+n) \pi ))+n^2 \text{Ci}((m+n)
   \pi )\\
   -2 m n \text{Ci}(m \pi )+2 m n \text{Ci}(n \pi )+(m-n) (m+n) \text{Ci}((m-n) \pi
   )\\
   +2 m n \log \left(\frac{m}{n}\right)+2 (m-n) (m+n) \tanh
   ^{-1}\left(\frac{n}{m}\right)\bigg], \quad\quad\quad\quad m\ne n.
   \end{cases}
\] 

\begin{comment} For the complex exponential basis, we take $\phi_n(x) = \exp\left[\pi i  n (2x-1)\right]$, so $\int_0^1 dx \phi_m(x) \phi^*_n(x) = \delta_{mn}$. Note that due to the orthogonality constraint, only even frequencies can appear in the eigenfunctions, so one cannot compare odd excited states to the other bases. Abusing notation from before, it is again useful to change variables to $u = \pi(2x-1) $ and $v = \pi(2y-1) $. Then, the matrix elements take the form \begin{equation}
	\begin{aligned}
		H_{mn}^{(\textrm{` t Hooft, exp})} = -\frac{\lambda}{2} \int_{-\pi}^\pi  du \int_{-\pi}^\pi  dv \frac{ e^{-i m (u+v)} \left(e^{i m u}-e^{i m v}\right) \left(e^{i n u}-e^{i n
   v}\right)}{(u-v)^2}.
	\end{aligned}
\end{equation} They can be evaluated in closed form, giving \[ H_{mn}^{(\textrm{`t Hooft, exp})} =  \lambda \\
 \begin{cases} 
     -2 (-\text{Ci}(2 n \pi )-2 \pi  n \text{Si}(2 n \pi )+\log (2 \pi  n)-\cos (2 \pi  n)+\gamma_E +1), \quad \, \,\,\, m = n \\
  \frac{2 (-1)^{m+n}}{m-n} \bigg[(n-m) \text{Ci}(2 (m-n) \pi )-n \left(\text{Ci}(2 n \pi )+\log \left(\frac{m}{n}-1\right)\right) \\
  +m \text{Ci}(2 m \pi )+m \log \left(1-\frac{n}{m}\right)\bigg], \quad\quad\quad\quad m\ne n.
   \end{cases}
\]
\end{comment}

 Finally, we recall the analytic formula (\ref{eq:largeNLCTmatrixelements}) for the 2-to-2 matrix element in the LCT basis (which is the only matrix element necessary at large $N_c$) using the Jacobi representation of the basis states: 
 \begin{equation}
	\begin{aligned}
		H_{\ell\ell'}^{\textrm{`t Hooft, LCT}} = 
		 \frac{2\lambda}{\pi} & \sqrt{(2\ell+1)(2\ell'+1)} \\
		 &\times \left[H_{\frac{\ell_{\textrm{max}}-1}{2}} + H_{\frac{\ell_{\textrm{max}}}{2}} - H_{\frac{\ell_{\textrm{max}}-\ell_{\textrm{min}}-1}{2}} - H_{\frac{\ell+\ell'}{2}} \right] , 
	\end{aligned}
\end{equation}
 when $\ell+\ell'$ is even, and vanishes when $\ell+ \ell'$ is odd.   $H_n$ is the $n$-th harmonic number.

%%%%%%%%%%%%%%%%%%%%%%%%%%%%%%%%%%%%%%%%%%%%%%%%%%%%%%%%%%%%%%%%%%%%%%%%%%%%%
%%%%%%%%%%%%%%%%%%%%%%%%%%%%%%%%%%%%%%%%%%%%%%%%%%%%%%%%%%%%%%%%%%%%%%%%%%%%%

\section{Radial Quantization Method Technical Details}
\label{app:GeneratingFunctions}
In this appendix, we discuss some of the details of the manipulations required to evaluate the matrix elements using radial quantization techniques.  

\subsection{Reduction of Fourier Transform Integral}

First, we argue that the three integrals over positions $x,y,z$ reduce to a single integral, as in (\ref{eq:MassTermContour}).   Start with a general integral of the form
\be
\CI = e^{-\frac{i \pi}{2} a} \int_{-\infty}^\infty dx dy dz e^{i (P x- P' z)} (x-z-i \epsilon)^{-a} F\left(\frac{x-y-i \epsilon}{z-y+ i \epsilon}\right).
\ee
The $i \epsilon$ prescription follows from the operator ordering (see e.g.~\cite{Anand:2019lkt}, eqs. (3.7)-(3.8)).  Change coordinates from $x,y,z$ to $(w, y, z')$ according to
\be
z=  z' w+ y, \quad x = z' (w-1)+y.
\ee
The new form of the integral $\CI$ is 
\be
\CI &=& 2 \pi \delta(P-P')  e^{-\frac{i \pi}{2} a}  \int dw dz' e^{-i P z'} (-z'-i \epsilon)^{-a+1} F\left(  \frac{w-1}{w} \right) \\
&=& 2 \pi (2P) \delta(P-P')  \frac{2\pi^2  P^{a-3}}{\Gamma(a-1)}\int_{\frac{1}{2} - i \infty}^{\frac{1}{2} + i \infty} \frac{dw}{2\pi i} F\left( \frac{w-1}{w} \right).
\label{eq:GenXYZIntegral}
\ee
This formula agrees  with (\ref{eq:MassTermContour}).

We still have to explain how we obtained the specific $w$ contour in the above integral. In our applications,  $F(\frac{w-1}{w})$ as a function of $w$ just has poles at $w=0$ and $w=1$, or branch cuts from $w=-\infty$ to $0$ and $w=1$ to $\infty$. In these cases, the $w$ contour can be determined quickly as follows. Do the change of variables in stages, first eliminating $y$ by a translation, then taking $z=z'+x$, and then taking $x = z'(w-1)$.  The $z$ integral along the real axis becomes the $z'$ integral along the real axis, which has a branch cut starting at $i \epsilon$ and a pole (or branch cut endpoint) at $-i \epsilon$. This pole is turned into a pole in $w$ by the second change in variables, so the $z'$ integral has only the branch cut and its integration is performed explicitly above.  The $x$ integral has a pole (or branch cut endpoint) at $i \epsilon$ and at $-z' -i\epsilon$, so its integral along the real axis runs between these two poles.  When we do the last change of variables, the $x$ integral between these two poles becomes a $w$ integral between the poles (or branch cut endpoints) at $0$ and $1$.

\subsection{Scalar $\phi^n$ Interaction}
\label{app:ScalarPhiN}

Here we will generalize our treatment of the scalar mass term to a $\phi^n$ interaction with any $n$.    As with the mass term, we define
\be
\Gcal_{\bk, \bk'}^{(\partial \phi^n)}(y_i) &\equiv& \< \bk | \p \phi(y_1) \dots \p \phi(y_n) | \bk'\>, \nn\\
\Gcal_{\bk, \bk'}^{(\p \phi^n)}(x, y_i, z) &\equiv& \< \p^{\bk} \phi(x) | \p \phi(y_1) \dots \p \phi(y_n) \p^{\bk'} \phi(z)\>.
\ee
They are related by a conformal transformation that maps $x$ to $\infty$ and $z$ to 0: 
\be
\Gcal_{\bk, \bk'}^{(\p \phi^n)}(x, y_i, z) = \Gcal_{\bk,\bk'}^{(\p \phi^n)}\left(\frac{y_i-z}{x-y_i}\right)  \frac{ (x-z)^{n - \Delta-\Delta'} }{\prod_{i=1}^n (x-y_i)^2 }  .
\label{eq:GkkpNPoint}
\ee

We compute $\Gcal_{\bk,\bk'}^{(\p \phi^n)}$ in radial quantization by inserting the mode decompositions of the monomials and the $\phi(y_i)$s. The result is a sum over terms where the creation/annihilation operators from the $\partial \phi$s contract with the creation/annihilation operators from the external states.  For any contribution, let $s$ the number of annihilation operators coming from $\partial \phi$s and $n-s$ be the number of creation operators.  By symmetrizing the $\phi$s, we can take the annihilation operators to come from $\partial \phi(y_i)$ with $1 \le i \le s$, and the creation operators from $\partial \phi(y_i)$ with $s+1 \le i \le n$, and multiply by $n!$ for the number of different ways of contracting the $\phi$s from the $\phi^n$ interaction:\footnote{The sum $\sum_{ \kvec/\{k_i\} = \kvec'/\{k'_j\}}$ means the sum over all choices of  a subset $\{k_i\}$ of $\bk$ and a subset $\{k'_j\}$ of  $\bk'$, such that $\bk$ and $\bk'$ are the same after removing the subsets, and moreover the the total number of $k$s in  $\{k_i, k'_j\}$ is $n$. For each such choice of subsets,  $s$ is the number of $k$s in $\{ k_i\}$.}
\be
\Gcal_{\bk,\bk'}^{(\p \phi^n)}(y_i) &\doteq&
\left( \frac{1}{\sqrt{4 \pi}} \right)^n \CN_{\kvec} \CN_{\kvec'} \sum_{ \kvec/\{k_i\} = \kvec'/\{k'_j\}}  n! \Gcal^{(\partial \phi^n)}_{\{ k_i\}, \{k_j'\}}(y_i), \nn\\
&& \Gcal^{(\partial \phi^n)}_{\{ k_i\}, \{k_j'\}}(y_i)  \equiv  \left( \prod_{j=s+1}^{n} y_j^{-k'_j-1} \sqrt{k'_j} \right) \left( \prod_{i=1}^s y_i^{k_i-1} \sqrt{k_i} \right) .
\label{eq:InsertingModeExpansionPhiN}
\ee
Restoring the dependence on $x$ and $y$, each of the individual terms $\Gcal^{(\partial \phi^n)}_{\{ k_i\}, \{k_j'\}}(y_i)$ in $\Gcal_{\bk,\bk'}^{(\p \phi^n)}(y_i)$ becomes an individual term in $\Gcal_{\bk,\bk'}^{(\p \phi^n)}(x,y_i,z)$ of the form
 \be
\Gcal^{(\partial \phi^n)}_{\{ k_i\}, \{k_j'\}}(x,y_i,z)   &\doteq & (x-z)^{-\Delta-\Delta'} \prod_{i=1}^n \left[ |a_i|^{\frac{1}{2}} \left( \frac{x-y_i}{y_i-z} \right)^{a_i} \frac{x-z}{(x-y_i)(y_i-z)}\right],
  \ee
where
\be
a_i = \left\{\begin{array}{cc} -k_i & 1 \le i \le s \\
k'_i & s+1 \le i \le n \end{array} \right. .
\label{eq:RadialAiDef}
\ee
Integrating in each $y_i$ and choosing the boundary condition so that the correlator decays like $y_i^{-1}$ at infinity, we find the contribution to the three-point function is 
\be
\Gcal_{\{ k_i \}, \{k'_j\}}^{(\phi^n)}(x,y,z) \doteq  (x-z)^{-\Delta-\Delta'} (-1)^{\sum_i k_i - \sum_j k'_j} \prod_{i=1}^n \left[ \frac{\left(\frac{x-y}{z-y} \right)^{a_i}-1}{ \sqrt{|a_i|} }\right] .
\label{eq:Gphi4AllPositions}
\ee
To obtain the contribution to the Hamiltonian matrix elements, we integrate over $y$ and Fourier transform with respect to $x$ and $z$ using (\ref{eq:GenXYZIntegral}):
\begin{equation}
\frac{1}{N_{\rm FT}} \int dx dy dz e^{i (p x-p' z)} \Gcal_{\{ k_i \}, \{k'_j\}}^{(\phi^n)}(x,y,z)  \doteq (-1)^{\sum_i k_i - \sum_j k'_j} \int_{\half-i\infty}^{\half+i \infty} \frac{dw}{2\pi i} \prod_{i=1}^n \left[ \frac{\left(\frac{w-1}{w} \right)^{a_i}-1}{\sqrt{|a_i|}}\right] .
\label{eq:phiNGenIntegral}
\end{equation}
To obtain the full matrix element for the primary states, we sum over these individual contraction terms.

For any individual $\phi^n$, we can  evaluate the contour integral as a function of the $a_i$s by expanding out the products and grouping them together into a sum of terms of the form (\ref{eq:basicradialcontourresult}) that we encountered for the mass term $\phi^2$. For instance, for $n=4$ with $a_1, a_2>0$ and $a_3, a_4<0$, we can group together the positive $a_i$ terms as 
\be
(v^{a_1}-1)(v^{a_2}-1) = (v^{a_1+a_2}-1) - (v^{a_1}-1) - (v^{a_2}-1),
\ee 
and similarly for the negative $a_i$ terms.  So for this case, the product in (\ref{eq:phiNGenIntegral}) reduces to a sum over nine terms, each of which is of the form that we evaluated in (\ref{eq:basicradialcontourresult}).  A general formula for the contour integral is therefore
\begin{equation}
\int_{\half-i\infty}^{\half+i \infty} \frac{dw}{2\pi i} \prod_{i=1}^n \left[ \left(\frac{w-1}{w} \right)^{a_i}-1\right] = \sum_{A_+ \subset \{ a_i > 0\} \atop A_- \subset \{ a_i < 0 \}} (-1)^{d(A_+) + d(A_-)} {\rm min}\left(\sum_{a_i \in A_+} a_i, \sum_{a_j \in A_-} -a_j\right) ,
\end{equation}
where $d(A)$ denotes the number of elements of $A$.
In words, the above equation says that for every possible subset of the positive $a_i$s and of the negative $a_i$s, take the minimum of the sum over the elements in the positive subset and of  the sum over (minus) the elements in the negative subset, multiply by an overall minus sign if the total number of elements from both subsets combined is odd, and then sum this quantity over all such subsets.

\subsection{Fermion Mass Term}

As discussed in section \ref{sec:RQFermions}, when we apply radial quantization to fermions, we treat $\partial \psi$ as a $h=\frac{3}{2}$ primary operator.  Consequently, we must integrate $\partial \psi$ to obtain $\psi$ in any interaction term.  Additional integrations are typically required because in lightcone, we integrate out the chiral field $\chi \sim (m/\p)\psi$.  As our first example, we consider the fermion mass term $\sim m^2 \psi \p^{-1} \psi$.

As in the scalar case, we begin start with a correlator containing only primary operators. For now, we will allow any even number $n$ of intermediate insertions of $\p \psi$, and later will specialize to $n=2$: 
\be
\Gcal_{\bk, \bk'}^{(\p \psi^n)}(y_i) &\equiv&  \< \bk |  \partial \psi(y_1) \dots \partial \psi(y_n) | \bk' \>, \nn\\
\Gcal_{\bk, \bk'}^{(\p \psi^n)}(x,y_i, z) &\equiv&  \< \partial^{\bk} \psi(x)   \partial \psi(y_1) \dots \partial \psi(y_n) \partial^{\bk'} \psi(z)\>, \nn\\
\Gcal_{\bk, \bk'}^{(\p \psi^n)}(x,y_i, z) &=&\Gcal_{\bk, \bk'}^{(\p \psi^n)} \left( \frac{y_i-z}{x-y_i} \right) \frac{ (x-z)^{\frac{3}{2}n-\Delta-\Delta'} }{  \prod_{i=1}^n (x-y_i)^{3}} .
\ee
The radial mode expansion for $\partial \psi$ is (\ref{eq:II:radialfermion}), reproduced here for convenience:
\begin{equation}
\p\psi(x) = \frac{i}{\sqrt{4\pi}} \sum_{k=1}^\infty \sqrt{k(k+1)} \left( x^{-k-2}b_k + x^{k-1} b_k^\dagger \right).\label{eq:II:radialfermion2} 
\end{equation}
Inserting this mode expansion into $\Gcal_{\bk, \bk'}^{(\p \psi^n)}(y_i)$, we find (with a similar notation to (\ref{eq:InsertingModeExpansionPhiN}))
\be
\Gcal_{\bk, \bk'}^{(\p \psi^n)}(y_i) &\doteq &
\left( \frac{1}{\sqrt{4 \pi}} \right)^n \CN_{\kvec} \CN_{\kvec'} \sum_{ \kvec/\{k_i\} = \kvec'/\{k'_j\}} n!(-1)^{\sigma(\{ k_i\}, \{k_j'\})} \Gcal^{(\partial \psi^n)}_{\{ k_i\}, \{k_j'\}}(y_i), \nn\\
&& \Gcal^{(\partial \psi^n)}_{\{ k_i\}, \{k_j'\}}(y_i)  \equiv  \left( \prod_{j=s+1}^{n} y_j^{-k'_j-2} \sqrt{k'_j(k_j'+1)} \right) \left( \prod_{i=1}^s y_i^{k_i-1} \sqrt{k_i(k_i+1)} \right) . \nn\\
\ee
where $ (-1)^{\sigma(\{ k_i\}, \{k_j'\})} $ keeps track of the number of times we have to anticommute the $b_k, b^\dagger_k$s to the left and right.

Each term $\Gcal^{(\partial \psi^n)}_{\{ k_i\}, \{k_j'\}}(y_i)$ in $\Gcal^{(\partial \psi^n)}_{\bk,\bk'}(y_i)$ becomes a term in $\Gcal_{\bk,\bk'}^{(\p \psi^n)}(x,y_i,z)$ of the form 
 \be
\Gcal^{\partial \psi^n}_{\{ k_i\}, \{k_j'\}}(x,y_i,z)   &\doteq & (x-z)^{-\Delta-\Delta'} \prod_{i=1}^n \left[ \sqrt{ b_i(b_i+1)}\left( \frac{x-y_i}{y_i-z} \right)^{b_i} \frac{(x-z)^{\frac{3}{2}}}{(x-y_i)(y_i-z)^2}\right],\nn\\
  \ee
where 
\be
b_i = \left\{\begin{array}{cc} -k_i -1& 1 \le i \le s \\
k'_i & s+1 \le i \le n \end{array} \right. .
\label{eq:RadialBiDef}
\ee
Each $y_i$ variable can be integrated in closed form to turn the $\partial \psi$s into $\psi$s.  We choose the integration constant so that the correlator decays like $y_i^{-2}$ at $y_i \rightarrow \infty$. This behavior at infinity follows from the fact that we use only Dirichlet basis states for the fermion external operators, and $\< \psi(y) \partial \psi(z)\>$ decays like $y^{-2}$ at large $y$ in the free theory.  The structure of the result is clearer if we define
\be
g_b^{(\psi)}(v) \equiv  \frac{v^b(b (v-1)-1) +1}{\sqrt{ b(b+1)}}.
\label{eq:genfuncPsi}
\ee
Then, integrating all the $y_i$s, we obtain
\be
\Gcal_{\{k_i\}, \{k_j'\}}^{(\psi^n)}(x,y_i, z) \doteq (-1)^{k-k'}(x-z)^{-\frac{n}{2}-\Delta-\Delta'} \prod_{i=1}^n  g_{b_i}^{(\psi)}\left( \frac{x-y_i}{z-y_i} \right)  .
\label{eq:GpsiNAllPositions}
\ee
We can eliminate the strange asymmetry between $k_i$ and $k_j'$ in the definition of $b_i$ by using the identity
\be
g_{-k-1}^{(\psi)}(v) = g_k^{(\psi)}(v^{-1}).
\ee
Using this identity, we can write $\Gcal^{(\psi^n)}_{\{ k_i\}, \{k_j'\}}(x,y_i,z)$ as
\begin{equation}
\Gcal_{\{k_i\}, \{k_j'\}}^{(\psi^n)}(x,y_i, z) \doteq (-1)^{k-k'} (x-z)^{-\frac{n}{2}-\Delta-\Delta'} \prod_{i=1}^s g_{k_i}^{(\psi)}\left( \frac{z-y_i}{x-y_i} \right)  \prod_{i=s+1}^n  g_{k_i}^{(\psi)}\left( \frac{x-y_i}{z-y_i} \right)  .
\label{eq:GpsiNAllPositions2}
\end{equation}
To construct the mass term $\sim \psi \frac{1}{\partial} \psi$, we need to do another integration on one of the $\psi$s.  This integration can also be done in closed form, and $\frac{1}{\partial} \psi$ produces the new function 
\be
g_b^{(\frac{1}{\partial} \psi)}(v) \equiv - \frac{\frac{ \left(v^{b+1}-1\right)}{v-1}-b-1}{\sqrt{ b(b+1)}}, \qquad \partial_w g_b^{(\frac{1}{\partial} \psi)}(1-w^{-1}) = -g_b^{(\psi)}(1-w^{-1}).
\label{eq:genfuncpinvpsi} 
\ee
Note that $g_{-k-1}^{(\frac{1}{\partial} \psi)}(v) = -g_k^{(\frac{1}{\partial} \psi)}(v^{-1})$. The result of integrating one of the $y_i$s to turn $\psi(y_i)$ into $\partial^{-1} \psi(y_i)$ is simply to make the replacement
$g_{b_i}^{(\psi)} \rightarrow -  (x-z) g_{b_i}^{(\frac{1}{\partial} \psi)}$
in (\ref{eq:GpsiNAllPositions}). Equivalently, make the replacement $g_{k_i}^{(\psi)} \rightarrow \pm (x-z)  g_{k_i}^{(\frac{1}{\partial} \psi)}$ in (\ref{eq:GpsiNAllPositions2}), where the sign is $(+)$ for $ i \le s$ and $(-)$ otherwise; this sign is ``removable''  (see \ref{app:Phases}).

Let us apply these results to the fermion mass term. The individual contraction terms are
\be
\Gcal_{k, k'}^{(\psi \frac{1}{\partial} \psi)}(x,y, z) \doteq (-1)^{k-k'}  (x-z)^{-\Delta-\Delta'}   g_{k}^{(\psi)}\left( \frac{z-y}{x-y} \right)  g_{k'}^{(\frac{1}{\partial} \psi)}\left( \frac{x-y}{z-y} \right) .
\label{eq:GpsiMassWithPositions}
\ee
We have taken $y_1=y_2=y$.  By equivalent arguments to those for scalar operators, the integration over $x,y,z$ becomes a single contour integral over $w$:
\begin{equation}
\frac{1}{N_{\rm FT}} \int dx dy dz e^{i(px-p'z)} \Gcal_{k, k'}^{(\psi \frac{1}{\partial} \psi)}(x,y, z)  \doteq (-1)^{k-k'} \int_{\half-i\infty}^{\half+i \infty} \frac{dw}{2\pi i} g_{k'}^{(\frac{1}{\partial} \psi)}(\frac{w-1}{w}) g_k^{(\psi)}(\frac{w}{w-1}).
\label{eq:FermionMassContour}
\end{equation}
This last contour integral can be done explicitly, using a similar argument to the one we used for the scalar mass term,\footnote{Explicitly:
using equation (\ref{eq:genfuncpinvpsi}), we can write the RHS of (\ref{eq:FermionMassContour}) (without the $(-1)^{k-k'}$) as
\be
   \oint \frac{dw}{2\pi i} \frac{ w\left( (\frac{w-1}{w} )^{k'+1} -1\right)  }{\sqrt{2k'(k'+1)}}\partial_w \frac{  (1- w)\left( (\frac{w}{w-1} )^{k+1} -1\right) }{\sqrt{2k(k+1)}} .
 \ee 
 We dropped the $(k'+1)$ and $(k+1)$ terms from the $g^{(\frac{1}{\partial} \psi)}$ functions because they are killed by the derivative $\partial_w$, which can act to the left or to the right using integration by parts. As with the scalar mass term, this last expression is symmetric under $k \leftrightarrow k'$ so we may take $k'\le k$ without loss of generality; then,  all cross-terms are manifestly regular at $w \sim 0$ except for 
 \begin{equation}
   \oint \frac{dw}{2\pi i}  \frac{ w\left( (\frac{w-1}{w} )^{k'+1} \right)  }{\sqrt{k'(k'+1)}} \partial_w \frac{  (1- w)\left( -1\right) }{\sqrt{k(k+1)}}    = \oint \frac{dz}{2\pi i z^2} \frac{(1-z)^{k'+1}}{\sqrt{k(k+1)k'(k'+1)}}  = \frac{1}{2} \sqrt{\frac{k'(k'+1)}{k(k+1)}} .
 \end{equation}
 } 
 with the result
 \be
   \oint \frac{dw}{2\pi i} g_{k'}^{(\frac{1}{\partial} \psi)}(\frac{w-1}{w}) g_k^{(\psi)}(\frac{w}{w-1}) = \frac{1}{2} \sqrt{ \frac{k_{\rm min}(k_{\rm min}+1)}{k_{\rm max}(k_{\rm max}+1)}},
   \ee
   where  $k_{\rm min}= {\rm min}(k,k'),  k_{\rm max} = {\rm max}(k, k')$.

\subsection{Yukawa Interaction}

So far, we have seen how to integrate the radial mode expansions of $\partial \phi$ and $\partial \psi$ to make $\phi$ and $\psi$, as well as $\frac{1}{\partial} \psi$, inside correlators.  A new complication arises when we consider the Yukawa interaction, because we have a term of the form
\be
\phi \psi \frac{1}{\partial} \phi \psi
\ee
where we have to integrate the product $\phi \psi$.  For certain contractions, this integration produces branch cuts as a function of the variable $w$ from (\ref{eq:GenXYZIntegral}), whereas up until now we have only had to deal with poles.  Our strategy will be to separate out the poles from the branch cut, which generally is due to a logarithm, and deal with each separately when we integrate $\int dx dy dz e^{i (px - p'z)}$ to get the LC matrix elements.  

First, we discuss how to do the $\frac{1}{\partial}$ integration in the interaction term itself. We can compute matrix elements of 
\be
\phi(y) \psi(y) \phi(y') \psi(y')
\ee
using the methods in the previous subsections; each individual contraction of radial modes produces a term proportional to
\begin{equation}
(-1)^{\sum_i k_i s_i} g_{k_1}^{(\phi)}\left(\left(\frac{x-y}{z-y} \right)^{s_1}\right) g_{k_2}^{(\psi)}\left(\left(\frac{x-y}{z-y} \right)^{s_2}\right) g_{k_3}^{(\phi)}\left(\left(\frac{x-y'}{z-y'} \right)^{s_3}\right) g_{k_4}^{(\psi)}\left(\left(\frac{x-y'}{z-y'} \right)^{s_4}\right)  ,
 \label{eq:PhiPsiPhiPsiTerm}
\end{equation}
where $s_i = \pm$ depending on whether the contraction was to the left or the right. The factor $(-1)^{\sum_i k_i s_i}$ is equal to $(-1)^{\sum_i k_i}$, but by writing it this way, it is clear that it is equivalent to $(-1)^{(\Delta-\tfrac{1}{2} n_F)- (\Delta'-\tfrac{1}{2} n'_F)}$, where $\Delta, \Delta'$ and $n_F, n_F'$ are the dimension and number of fermions in the external states.  The function $g_k^{(\psi)}$ was given in (\ref{eq:genfuncPsi}) and $g_k^{(\phi)}$ in (\ref{eq:genfuncPhi}). To compute $(\phi \psi \frac{1}{\p} \phi \psi)(y)$, we want to integrate with respect to $y'$ and then set $y'=y$.  The integration constant should be chosen to subtract off the value at $y' \rightarrow \infty$. \footnote{This prescription for the integration constant follows from the origin of $\frac{1}{\partial}$.  It is produced by the $\chi$ propagator when we integrate out $\chi$, and in momentum space we take the propagator to be the principal value part  $\Pcal \frac{1}{p}= \textrm{Re}\frac{1}{p+ i \epsilon}$. Physically, we are removing $\chi$ as a degree of freedom from the theory since the imaginary part of the propagator is exactly the spectral weight due coming from the state $\chi$.  Fourier transforming  $\int dp e^{i p (y-y')} \Pcal \frac{1}{p} \sim {\rm sign}(y-y')$, we see that the $y'$ integral is of the form $\int_{-\infty}^\infty {\rm sign}(y-y')f(y') = F(\infty)+ F(-\infty) - 2 F(y)$, where $F$ is the indefinite integral of $f$.  In most of the cases we will encounter, $F(\infty)=F(-\infty)$.  In some cases, $F(y)$ will contain logs, and one must be more careful about ``the value at infinity''.     }  It is convenient to switch to the variable $w$ that we have been using above,
\be
w \equiv \frac{y-z}{x-z}, \quad w' \equiv \frac{y'-z}{x-z}.
\ee
Write (\ref{eq:PhiPsiPhiPsiTerm})  as
\begin{equation}
 g_{k_1}^{(\phi)}\left(v^{s_1}\right) g_{k_2}^{(\psi)}\left(v^{s_2}\right) g_{k_3}^{(\phi)}\left(v'^{s_3}\right) g_{k_4}^{(\psi)}\left(v'^{s_4}\right)  , \qquad v = 1-w^{-1}.
 \label{eq:PhiPsiPhiPsiTerm2}
\end{equation}
Now we are supposed to integrate with respect to $w'$ and set $w'=w$ after choosing the integration constant so that the integral vanishes at $w' \rightarrow \infty$.  Because $g_{k_3}^{(\phi)}\left(v'^{s_3}\right) g_{k_4}^{(\psi)}\left(v'^{s_4}\right)$ falls off like $w'^{-3}$ at infinity, its integral then decays like $w'^{-2}$ at infinity.

In general, the $w'$ integral can be done in closed form and written in terms of hypergeometric functions, but with a little more work we can beat the integrand into a more useful form where we separate out the power law pieces from the log pieces explicitly.  The basic idea is that $g_{k_3}^{(\phi)}\left(v'^{s_3}\right) g_{k_4}^{(\psi)}\left(v'^{s_4}\right)$ is a sum over a finite number of poles at $w'=0$ and $w'=1$:
\be
g_{k_3}^{(\phi)}\left(v'^{s_3}\right) g_{k_4}^{(\psi)}\left(v'^{s_4}\right) = \sum_{n=1}^{k_3+k_4} \frac{r_{0,n}}{w'^n} + \frac{r_{1,n}}{(1-w')^n}.
\ee
The logarithm comes from the $n=1$ terms $r_{0,1}= r_{1,1}$, where equality follows from the fact that the above vanishes like $w'^{-2}$ at infinity.  The remaining terms integrate to poles, which can be grouped back into a sum over integer powers of $v'$.  Performing this task is tedious but straightforward.  Let
\be
\frac{\hat{g}^{(\frac{1}{\partial} \phi \psi)}(k_3, k_4, s_3, s_4, v)}{\sqrt{k_3 }\sqrt{ k_4 (k_4+1)}} \equiv g^{(\frac{1}{\partial} \phi \psi)}(k_3, k_4, s_3, s_4, v)  \equiv \int dw \hat{g}_{k_3}^{(\phi)}\left(v^{s_3}\right) \hat{g}_{k_4}^{(\psi)}\left(v^{s_4}\right).
\ee
 The result  can be summarized by the following decomposition:\footnote{The sign inside $\log(-v)$ arises from doing the principal value $\frac{1}{\partial}$ integral with the appropriate $i \epsilon$ prescription.}
\begin{align}\label{yukawa-nonlocal-general-final}
   - \hat{g}^{(\frac{1}{\partial} \phi \psi)}(k_3, k_4, s_3, s_4, v)  =& s_3 k_3 \pr{
        g_\ell^{(H)}( v^{s_4 s_k})
        - g_{k_3}^{(H)}( v^{s_3})
    }  \nn \\
    & -s_3 \hat{g}_{k_3}^{(\frac{1}{\p}\psi)} \pr{ v^{s_3} }-s_4 \, \hat{g}_{k_4}^{(\frac{1}{\p}\psi)} \pr{ v^{s_4} } +s_4 s_k \, \hat{g}_{\ell}^{(\frac{1}{\p}\psi)} \pr{ v^{s_4 s_k} } \nn \\
    &- \begin{cases}
        k_3\log(-v) & s_3 s_4 = -1 \text{ and } k_3 \leq k_4 \\
        0 & \text{otherwise}
    \end{cases} \, ,
\end{align}
where $k = s_3 s_4 k_3 + k_4$, $s_k$ and $\ell$ are given by
\begin{align} 
	\text{if } k\geq 0, \text{ then }& s_k=1, \ell=k \\
	\text{ else }& s_k=-1, \ell=-k-1  \, ,
\end{align}
and we have defined 
\begin{equation}
g_k^{(H)}(v) \equiv \sum_{m=1}^k \frac{v^{m}-1}{m}, \qquad \frac{\hat{g}_k^{(\frac{1}{\partial} \psi)}(v)}{ \sqrt{ k(1+k)}} \equiv  g_k^{(\frac{1}{\partial} \psi)}(v). 
\end{equation}
The last step is to integrate $\int dx dy dz e^{i(P x-P'z)}$.  When we multiply $g^{(\frac{1}{\partial} \phi \psi)}(k_3, k_4, s_3, s_4, v)$ by $ g_{k_1}^{(\phi)}\left(v^{s_1}\right) g_{k_2}^{(\psi)}\left(v^{s_2}\right)$ to get the integral, we get a sum over terms that are all either integer powers of $v$ or integer powers times a log, i.e.~$v^k \log v$.  We have already seen how to deal with integer powers, so we just have to understand how to deal with the log terms.  The log term  $\log(-v) = \log(\frac{1-w}{w})$ has a branch cut from $w=0$ to $-\infty$ and a branch cut from $w=1$ to $+\infty$.  The $w$ integration contour passes between these two branch cuts.  The integral of $v^k \log (-v)$ along this contour diverges.  However, we know that the product $ g_{k_1}^{(\phi)}\left(v^{s_1}\right) g_{k_2}^{(\psi)}\left(v^{s_2}\right)$ vanishes at $v \sim 1$ like $(1-v)^2$ or faster.  So, term-by-term, we can replace each power of $v^k$ in $g_{k_1}^{(\phi)}\left(v^{s_1}\right) g_{k_2}^{(\psi)}\left(v^{s_2}\right)$ with $v^k - 1 - k(v-1)$, i.e.~with the first two terms of its series expansion around $v=1$ subtracted off, and we will not change the full sum.  In equations,
\be
\sum_{k=-N}^N c_k v^k = \sum_{k=-N}^N c_k (v^k-1-k(v-1)),
\ee
assuming the LHS vanishes like $\sim (1-v)^2$ or faster at $v\sim 1$. So, we can instead consider the integral 
\be
\Ical(k) \equiv \int_{\frac{1}{2} - i \infty}^{\frac{1}{2} + i \infty}\frac{dw}{2\pi}  (v^k-1-k(v-1))  \log\left( \frac{1-w}{w}\right)  = k(1-H_{|k|}),
\ee
where $H_k$ is the $k$-th harmonic number. 
 For the purposes of evaluating the integrals over the $v^k \log v$ terms, we can therefore apply the rule 
\be
\oint \frac{dw}{2\pi i} v^k \log (-v) \rightarrow \Ical(k).
\ee

Having sorted out the nonlocal piece of Yukawa coupling generating functional, the next step is to put back the rest of the Yukawa term $\phi \psi$, perform the contour integral and obtain a formula in terms of $k_i$'s and $s_i$'s. Given the form of (\ref{yukawa-nonlocal-general-final}), it is efficient to compute the formula of the following building blocks 
\begin{align}
    \label{yukawa-gy1-definition}
    g_{Y,1}(k, k_1, k_2, s_1, s_2) &\equiv \oint \frac{dw}{2\pi i} \, \hat{g}_k^{(\frac{1}{\p}\psi)}(v) \hat{g}_{k_1}^{(\phi)}(v^{s_1}) \hat{g}_{k_2}^{(\psi)} (v^{s_2}) , \\
    \label{yukawa-gy2-definition}
    g_{Y,2}(k, k_1, k_2, s_1, s_2)  &\equiv \oint \frac{dw}{2\pi i} \, g_k^{(H)}(v) \hat{g}_{k_1}^{(\phi)}(v^{s_1}) \hat{g}_{k_2}^{(\psi)} (v^{s_2}) ,  \\
    \label{yukawa-gylog-definition}
    g_{Y,\log}( k_1, k_2, s_1, s_2) &\equiv \oint \frac{dw}{2\pi i} \, \log(-v) \hat{g}_{k_1}^{(\phi)}(v^{s_1}) \hat{g}_{k_2}^{(\psi)} (v^{s_2}),
\end{align}
and recycle these formula for different parts of (\ref{yukawa-nonlocal-general-final}) by substituting combination of $k_i$'s and $s_i$'s. We can fix the sign of power of $v$ in the $\hat{g}_k^{(\frac{1}{\p}\psi)}(v)$ and $g_k^{(H)}(v)$ piece to be always positive, since one can expand the contour to infinity, which is regular, and capture the pole at $w\rightarrow 1$ instead of 1, then redefine $w\rightarrow 1-w$ (thus $v\rightarrow \frac{1}{v}$) to flip the sign of $s_i$
\begin{align}
    \oint \frac{dw}{2\pi i} \, g_{\rm any}(v^{-1}) g_{k_1}^{(\phi)}(v^{s_1}) g_{k_2}^{(\psi)} (v^{s_2})  = \oint \frac{dw}{2\pi i} \, g_{\rm any}(v) g_{k_1}^{(\phi)}(v^{-s_1}) g_{k_2}^{(\psi)} (v^{-s_2}) \, .
\end{align}

Now let's get a formula for each building blocks of (\ref{yukawa-gy1-definition}) - (\ref{yukawa-gylog-definition}):
 
\begin{itemize}
    \item To get $g_{Y,1}$ it is convenient to consider a different elemental integral:
    \begin{align}
        \oint \frac{dw}{2\pi i} \, \frac{v^k}{v-1} = \frac{1}{2} k(k-1) \Theta(k-1) \, .
    \end{align}
    Expand the factor to have common denominator $(v-1)$:
    \begin{align}
        &-(v-1) \hat{g}_k^{(\frac{1}{\p}\psi)}(v) \hat{g}_{k_1}^{(\phi)}(v^{s_1}) \hat{g}_{k_2}^{(\psi)} (v^{s_2}) \nn \\
        =& k v^{k_1 s_1}-k v^{k_1 s_1+1}-v^{k_1 s_1+1}+v^{k_1 s_1+k+1}
        +k v^{k_2 s_2}+k k_2 v^{k_2 s_2}-k v^{k_2 s_2+1}-k k_2 v^{k_2 s_2+1} \nn \\ 
        &-k_2 v^{k_2 s_2+1}-v^{k_2 s_2+1}+k_2 v^{k_2 s_2+k+1}+v^{k_2 s_2+k+1}
        -k v^{k_1 s_1+k_2 s_2}-k k_2 v^{k_1 s_1+k_2 s_2} \nn \\ 
        &+k v^{k_1 s_1+k_2 s_2+1}
        +k k_2 v^{k_1 s_1+k_2 s_2+1}+k_2 v^{k_1 s_1+k_2 s_2+1}+v^{k_1 s_1+k_2 s_2+1}-k_2 v^{k_1 s_1+k_2 s_2+k+1} \nn \\ 
        &-v^{k_1 s_1+k_2 s_2+k+1}-k k_2 v^{k_2 s_2+s_2}+k k_2 v^{k_2 s_2+s_2+1}
        +k_2 v^{k_2 s_2+s_2+1}-k_2 v^{k_2 s_2+k+s_2+1} \nn \\ 
        &+k k_2 v^{k_1 s_1+k_2 s_2+s_2}-k k_2 v^{k_1 s_1+k_2 s_2+s_2+1}
        -k_2 v^{k_1 s_1+k_2 s_2+s_2+1}+k_2 v^{k_1 s_1+k_2 s_2+k+s_2+1} \nn \\ 
        &-v^{k+1}+k v-k+v,
    \end{align}
    and use the elemental integral term by term to get a big conditional expression as follows:\\
  \renewcommand{\arraystretch}{1.5}
    \begin{align}\label{yukawa-gy1}
        \begin{tabular}{|c|c|}
            \hline
            $(s_1,s_2)$& $2\times g_{Y,1}(k, k_1, k_2, s_1, s_2)$
            \\
            \hline
            $(++)$& 0\\
            \hline
            $(-+)$& 
            $\begin{array}{l|l}
                \left(k-k_1\right) \left(k-k_1+1\right) & k\geq
                  k_1 \\
                -k \left(k_1-k_2\right) \left(k_1-k_2+1\right)
                  \left(k_2+1\right) & k_2\geq k_1+1 \\
                \left(k_2-k_1\right) \left(-k_1+k_2+1\right)
                  \left(2 k_2 k+k+k_2+1\right) & k_2\geq k_1 \\
                -\left(k_2+1\right) \left(k-k_1+k_2\right)
                  \left(k-k_1+k_2+1\right) & k+k_2\geq k_1 \\
                -(k+1) k_2 \left(-k_1+k_2+1\right)
                  \left(-k_1+k_2+2\right) & k_2+1\geq k_1 \\
                k_2 \left(k-k_1+k_2+1\right)
                  \left(k-k_1+k_2+2\right) & k+k_2+1\geq k_1 \\
               \end{array}$
            \\
            \hline
            $(+-)$& 
            $\begin{array}{l|l}
                \left(k-k_2\right) \left(k-k_2+1\right)
                  \left(k_2+1\right) \phantom{ddddddddddddd} & k\geq k_2 \\
                -\left(k_1-k_2-1\right) \left(k_1-k_2\right)
                  \left(2 k_2 k+k+k_2\right) & k_1\geq k_2+1 \\
                (k+1) \left(k_1-k_2\right) \left(k_1-k_2+1\right)
                  \left(k_2+1\right) & k_1\geq k_2 \\
                \left(-k_2-1\right) \left(k+k_1-k_2\right)
                  \left(k+k_1-k_2+1\right) & k+k_1\geq k_2 \\
                -\left(k-k_2-1\right) \left(k-k_2\right) k_2 &
                  k\geq k_2+1 \\
                k \left(k_1-k_2-2\right) \left(k_1-k_2-1\right)
                  k_2 & k_1\geq k_2+2 \\
                \left(k+k_1-k_2-1\right) \left(k+k_1-k_2\right)
                  k_2 & k+k_1\geq k_2+1 \\
               \end{array}$
            \\
            \hline
            $(--)$& 
            $
            \begin{array}{l|l}
                -k (k+1) & \text{True} \\
                \left(k-k_1\right) \left(k-k_1+1\right) & k\geq
                  k_1 \\
                \left(k-k_2\right) \left(k-k_2+1\right)
                  \left(k_2+1\right) & k\geq k_2 \\
                -\left(k_2+1\right) \left(-k+k_1+k_2-1\right)
                  \left(-k+k_1+k_2\right) & k\geq k_1+k_2 \\
                -\left(k-k_2-1\right) \left(k-k_2\right) k_2 &
                  k\geq k_2+1 \\
                \left(k-k_1-k_2-1\right) \left(k-k_1-k_2\right)
                  k_2 & k\geq k_1+k_2+1 \\
               \end{array}$
            \\
            \hline
        \end{tabular}
    \end{align}
 The way to read this table is, for any $(s_1,s_2)$, go to the corresponding cell; then, add up every term in that cell for which the inequality holds true.  
    \item The elemental piece of the $g_{Y,2}$ factor is the following 
    \begin{align}
        &\oint\frac{dw}{2\pi i} \, g_k^{(H)}(v) (v^{-p\neq 0} - 1) \nn \\
        =& \sum_{m=1}^k \oint\frac{dw}{2\pi i} \, \frac{1}{m} (v^k-1)(v^{-p}-1) \nn \\
        =& -\sum_{m=1}^k \frac{1}{m} \min(m,p) \Theta(p) \nn \\
        =& -\pr{ k-\Theta(k-p) \sum_{m=1}^{k-p} \frac{m}{m+p} }\Theta(p) \nn \\
        = & -\pr{ k-\Sigma(k,p) }\Theta(p) \, ,
    \end{align}
    where $\Sigma(k,p) \equiv \Theta(k-p) \sum_{m=1}^{k-p} \frac{m}{m+p}$.
    This translates the $v$ powers  with the rule 
    \begin{align}
        v^{p} \mapsto -\pr{ k-\Sigma(k,-p) }\Theta(-p) \, ,
    \end{align}
    from the expansion of $\hat{g}_{k_1}^{(\phi)}(v^{s_1}) \hat{g}_{k_2}^{(\psi)} (v^{s_2})$
    \begin{align}
       -\hat{g}_{k_1}^{(\phi)}(v^{s_1}) \hat{g}_{k_2}^{(\psi)} (v^{s_2})
        =& -v^{k_1 s_1}-k_2 v^{k_2 s_2}-v^{k_2 s_2}+k_2 v^{k_1 s_1+k_2 s_2}+v^{k_1 s_1+k_2 s_2}\nn \\
        &+k_2 v^{k_2 s_2+s_2}-k_2 v^{k_1 s_1+k_2 s_2+s_2}+1
    \end{align}
    into a table
    \begin{align}\label{yukawa-gy2}
        \begin{tabular}{|c|l|}
            \hline
            $(s_1,s_2)$& $g_{Y,2}(k, k_1, k_2, s_1, s_2) $
            \\
            \hline
            $(++)$& 0\\
            \hline
            $(-+)$& 
            $\begin{array}{l|l}
                 k-\Sigma \left(k,k_1\right) & \text{True} \\
                 -\left(k_2+1\right) \left(k-\Sigma
                   \left(k,k_1-k_2\right)\right) \phantom{dddddd} \ \ & k_2\leq k_1 \\
                 k_2 \left(k-\Sigma
                   \left(k,k_1-k_2-1\right)\right) & k_2+1\leq k_1
                   \\
                \end{array}$
            \\
            \hline
            $(+-)$& 
            $\begin{array}{l|l}
                 k-\left(k_2+1\right) \Sigma
                   \left(k,k_2\right)+k_2 \Sigma
                   \left(k,k_2+1\right) & \text{True} \\
                 -\left(k_2+1\right) \left(k-\Sigma
                   \left(k,k_2-k_1\right)\right) & k_1\leq k_2 \\
                 k_2 \left(k-\Sigma
                   \left(k,-k_1+k_2+1\right)\right) & k_1\leq
                   k_2+1 \\
                \end{array}$
            \\
            \hline
            $(--)$& 
            $\begin{array}{l}
                -\Sigma \left(k,k_1\right)-\left(k_2+1\right)
                  \Sigma \left(k,k_2\right)+k_2 \Sigma
                  \left(k,k_2+1\right)\\+\left(k_2+1\right) \Sigma
                  \left(k,k_1+k_2\right)-k_2 \Sigma
                  \left(k,k_1+k_2+1\right)+k \\
               \end{array}$
            \\
            \hline
        \end{tabular}
    \end{align}
    \item The integral of $g_{Y,\log}$ uses the elemental log integral:  
    \be
        \CI(k) \equiv \oint \frac{dw}{2\pi i} \, v^k \log(-v) \cong k \pr{ 1 - \sum_{i=1}^{|k|} \frac{1}{i} } = k \pr{ 1 - H_{|k|} } \, .
    \ee
    to get 
    \begin{align}\label{yukawa-gylog}
       - g_{Y,\log}( k_1, k_2, s_1, s_2)
        =& \begin{cases}
        k_1 \left(H_{k_1}-H_{k_1+k_2}\right) & (s_1, s_2) = (+,+) \vspace{5px} \\ 
        \begin{array}{l}
        	k_1 H_{k_1}+\left(k_1-k_2-1\right) 
        	k_2 H_{-k_1+k_2+1} \\ 
        	+\left(k_2+1\right)
        	\left(k_2-k_1\right) 
        	H_{k_1-k_2}+k_2 
        	\end{array} & (s_1, s_2) = (+,-) \vspace{5px}\\
        \begin{array}{l} k_1 \left(-H_{k_1}\right)
        	+\left(k_1-k_2\right) 
        	\left(k_2+1\right) \\
        	H_{k_1-k_2}+
        	k_2\left(\left(-k_1+k_2+1\right) 
        	H_{-k_1+k_2+1}-1\right) 
        	\end{array} & (s_1, s_2) = (-,+) \vspace{5px}\\
        k_1 \left(H_{k_1+k_2}-H_{k_1}\right) & (s_1, s_2) = (-,-) \\
    \end{cases}
    \end{align}
\end{itemize}

Merging $g_{k_1}^{(\phi)}(v^{s_1}) g_{k_2}^{(\psi)} (v^{s_2})$ with (\ref{yukawa-nonlocal-general-final}) and applying our building blocks (\ref{yukawa-gy1-definition}) - (\ref{yukawa-gylog-definition}), we obtain the final formula for the quartic $\phi \psi \frac{1}{\p} \phi \psi$ Yukawa factor:
\begin{align}\label{yukawa-final-factor}
    -\hat{g}_{\phi\psi\frac{1}{\p}\phi\psi}(k_i, s_i) =& s_3 k_3 \bigg( 
        g_{Y,2}(\ell;k_1, k_2, s_{4}s_ks_1,s_{3}s_{4}s_2)
        - g_{Y,2}(k_3;k_1,k_2,s_{3}s_1,s_{3}s_2)
    \bigg)  \nn \\
    & -s_3 \, g_{Y,1}(k_3;k_1,k_2,s_{3}s_1,s_{3}s_2) 
     -s_4 \, g_{Y,1}(k_4;k_1,k_2,s_{4}s_1,s_{4}s_2) 
     \nn \\
    &
     +  s_{k}s_4 \, 
    g_{Y,2}(\ell;k_1,k_2,s_4s_ks_1,s_4s_ks_2)
    \nn \\
    &+ \begin{cases}
        k_3g_{Y,\log}(k_1,k_2,s_1,s_2) & s_3 s_4 = -1 \text{ and } k_3 \leq k_4 \\
        0 & \text{otherwise}
    \end{cases}
\end{align}
and formulas (\ref{yukawa-gy1}), (\ref{yukawa-gy2}) and (\ref{yukawa-gylog}) for $g_{Y,1}$, $g_{Y,2}$ and $g_{Y,\log}$ respectively.  The un-hatted $g_{\phi\psi\frac{1}{\p}\phi\psi}$ simply restores all the $\sqrt{k}$ and $\sqrt{k (k+1)}$ factors in the denominator from its constituent generating functions in (\ref{eq:PhiPsiPhiPsiTerm}). 

Finally, we also have to obtain the formula for the $\phi \psi \frac{1}{\p} \psi$ Yukawa factor.  Fortunately, by inspection, this factor is just the term $g_{Y,1}$ from (\ref{yukawa-gy1-definition}) that we have already evaluated in (\ref{yukawa-gy1})! That is,
\be
\hat{g}_{\phi \psi \frac{1}{\partial} \psi} (k_i, s_i) = g_{Y,1}(k_3, k_1, k_2, s_1 s_3, s_2 s_3),
\label{eq:yukawa-cubic-final-factor}
\ee
where $k_3$ is the external leg contracted with $\frac{1}{\partial \psi}$ from the interaction, $k_1$ is contracted with $\phi$, and $k_2$ is contracted with $\psi$. The sign $s_1 s_3$ is positive (negative) if $\phi$ is contracted in the same (opposite) direction as $\frac{1}{\partial} \psi$ is; the analogous statement holds for $s_2 s_3$ and the direction  $\psi$ is contracted.

\subsection{Supercharges $Q_+$ and $Q_-$ }

Finally, we apply our radial quantization methods to compute matrix elements of the supercharges $Q_+$ and $Q_-$, for the cases where the superpotential $W(\phi)$ has a quadratic $\phi^2$ mass term or a cubic $\phi^3$ interaction term.  The $Q_+$ supercharge is given by
\be
Q_+ = \sqrt{2} \int dx^-  W'(\phi) \psi ,
\ee
and so is $\sim \phi \psi$ or $\sim \phi^2 \psi$ for the mass or cubic term, respectively.  The $Q_-$ supercharge is unaffected by the deformation (just as $P_-$ is unaffected):
\be
Q_- = 2 \int dx^- (\p \phi) \psi.
\ee
In the Yukawa subsection above, we introduced notation that streamlines the derivation of generating functions.  Applying those results here, we find 
 
\paragraph{For $Q_+$:}
\be
	\frac{1}{N_{\rm FT}} \int dx dy dz e^{i (p x-p' z)} \Gcal_{\{ k_i \}, \{k'_j\}}^{(\phi \psi)}(x,y,z)  &=& \int_{\frac{1}{2} -i \infty}^{\frac{1}{2}+i \infty}  \frac{dw}{2\pi i}  g_\phi(k_1, v^{s_1}) g_\psi(k_2,v^{s_2}) \\
	&=& \frac{1}{\sqrt{k_1 k_2 (k_2+1)}} \int_{\frac{1}{2} -i \infty}^{\frac{1}{2}+i \infty} \frac{dw}{2\pi i} \hat g_\phi(k_1, v^{s_1}) \hat g_\psi(k_2,v^{s_2}) . \nn
\ee
Without loss of generality we can set $s_2 = +$. If $s_1 = +$ clearly the integral should be zero since both particles contracting to the right corresponds to zero mode and should vanish. If $s_1 = -$,
\begin{align}
	&\int_{\frac{1}{2} -i \infty}^{\frac{1}{2}+i \infty}  \frac{dw}{2\pi i} \hat g_\phi(k_1, v^{-1}) \hat g_\psi(k_2,v) \nn \\
	=& \oint \frac{dw}{2\pi i} \pr{v^{-k_1}-1} 
	\sqb{-k_2\pr{v^{k_2+1}-1} + (k_2 + 1)\pr{v^{k_2}-1} +1 } \nn \\
	=& \left\{ \begin{array}{lcl}
	- k_2 \pr{v^{-k_1}-1} \pr{v^{k_2+1}-1} 
		& \xrightarrow{\oint\frac{dw}{2\pi i}} 
		& -k_2 \min(k_1,k_2+1) \\
	+ (k_2+1) \pr{v^{-k_1}-1} \pr{v^{k_2}-1} 
		& \xrightarrow{\oint\frac{dw}{2\pi i}} 
		& + (k_2+1) \min(k_1,k_2) \\
	+ \pr{v^{-k_1}-1}
		& \xrightarrow{\oint\frac{dw}{2\pi i}} 
		& + 0
	\end{array}\right. \nn \\
	=& \begin{cases}
	k_1 & 0 < k_1 \leq k_2 \\
	0 & {\rm else} .
	\end{cases}
\end{align}
where $k_1 < 0$ can be seen as the case $s_1 = +$. The interaction term $\phi^2\psi$ can easily be reduced to a sum over mass terms:
\begin{align}
	\hat g_\phi(k_0, v^{s_0}) \hat g_\phi(k_1, v^{s_1}) \hat g_\psi(k_2,v) =
	&~ \hat g_\phi(-(s_0 k_0 + s_1 k_1), v^{-1}) \hat g_\psi(k_2,v) \nn \\
	&~- \hat g_\phi(-s_0 k_0, v^{-1}) g_\psi(k_2,v) \nn \\
	&~- \hat g_\phi(-s_1 k_1, v^{-1}) g_\psi(k_2,v)  .
\end{align}

\paragraph{For $Q_-$:}
\be
	\frac{1}{N_{\rm FT}} \int dx dy dz e^{i (p x-p' z)} \Gcal_{\{ k_i \}, \{k'_j\}}^{(\p \phi \psi)}(x,y,z)  &=& \int_{\frac{1}{2} -i \infty}^{\frac{1}{2}+i \infty}  \frac{dw}{2\pi i} g_{\d\phi}(k_1, v^{s_1}) g_\psi(k_2,v^{s_2}) \\
	&=& \frac{1}{\sqrt{k_1 k_2 (k_2+1)}}
	\int_{\frac{1}{2} -i \infty}^{\frac{1}{2}+i \infty}\frac{dw}{2\pi i} 
	\hat g_{\d\phi}(k_1, v^{s_1}) \hat g_\psi(k_2,v^{s_2}), \nn
\ee
where
\begin{align}
	\hat g_{\d\phi}(k_1, v^{-1}) \equiv \d_w \pr{v^{-k_1}-1}  
	= \frac{1}{w(w-1)}\pr{\frac{w-1}{w}}^{-k_1}\, .
\end{align}

Again without loss of generality set $s_2 = +$ and $s_1 = -$.  Using  the identity
\begin{align}
	\oint \frac{dw}{2\pi i} \frac{1}{w(w-1)}\pr{\frac{w-1}{w}}^{a}
	= \delta_{a,0}\, ,
\end{align}
we evaluate the integral above to be
\be
	\int_{\frac{1}{2} -i \infty}^{\frac{1}{2}+i \infty}\frac{dw}{2\pi i} \hat g_{\d \phi}(k_1, v^{-1}) \hat g_\psi(k_2,v)
	&=& \oint \frac{dw}{2\pi i} \d_w\pr{v^{-k_1}-1} 
	\sqb{-k_2\pr{v^{k_2+1}-1} + (k_2 + 1)\pr{v^{k_2}-1} +1 }  \nn\\
	&=&k_2(k_2+1) \times \begin{cases}
	1 & k_1 = k_2+1 \\
	-1 & k_1 = k_2 \\
	0 & {\rm else}
	\end{cases} .
\ee

\section{The Fate of Vertex Operators}
\label{app:VertexOps}

For a free massless scalar in 2D, there are two building blocks we can use to construct primary operators. First, we have the conserved current $J_\mu \equiv \p_\mu\phi$, which we've used throughout this work. However, there is also the infinite set of vertex operators
\be
V_\alpha(x) \equiv e^{i\alpha \phi(x)}.
\ee
In principle, these primary operators should be included in constructing our UV basis for free scalar field theory. However, we'll now demonstrate that in the presence of a mass term $\sim m^2 \phi^2$ the Hamiltonian matrix elements for these vertex operator states are all divergent, such that these states are lifted from the IR Hilbert space.

The inner product and matrix elements for vertex operators can be computed in terms of those of $\phi^n$, by writing each state as the sum
\be
|V_\alpha,p\> = \fr{1}{N_\alpha} \sum_n \fr{(i\alpha)^n}{n!} N_{\phi^n} |\phi^n,p\>.
\ee
To compute the norm of the $\phi^n$, we can use the Fock space method discussed in section~\ref{sec:FockSpace} to obtain
\be
N^2_{\phi^n} = \fr{1}{2p} \int dx \, e^{ipx} \<\phi^n(x) \phi^n(0)\> = \fr{n!}{2p} \int \fr{dp_1 \cdots dp_n}{(2\pi)^n 2p_1 \cdots 2p_n} (2\pi) \de(p-|p|_n).
\ee
Looking at the integrand, we see that this norm is actually logarithmically divergent, due to the $1/p_i$ singularities in the integration measure. We can regulate this divergence by placing a lower bound on the momentum of each individual particle,\footnote{This somewhat peculiar regulator was chosen to make the evaluation of these Fock space integrals much simpler, but the overall results will be the same with any other choice of regulator, such as imposing a more uniform cutoff on particle momentum or placing this system in finite volume.}
\be
\epsilon \leq \fr{p_i}{|p|_i} \leq 1-\epsilon \qquad (i=2,\ldots,n),
\ee
leading to the norm
\be
N^2_{\phi^n} = \fr{n! \log^{n-1} \fr{1}{\epsilon}}{4p^2 (2\pi)^{n-1}}.
\ee

Using this regulated monomial norm, we can then compute the normalization of the full vertex operator state
\be
N_\alpha^2 = \sum_n \fr{\alpha^{2n}}{(n!)^2} N^2_{\phi^n} = \fr{1}{4p^2} \sum_n \fr{\alpha^{2n}}{n!} \fr{ \log^{n-1} \fr{1}{\epsilon}}{(2\pi)^{n-1}} = \fr{\pi}{2p^2\epsilon^{\fr{\alpha^2}{2\pi}} \log\fr{1}{\epsilon}}.
\ee
The $\epsilon$-dependence in this norm is important for ensuring the orthogonality of distinct vertex operator states in the limit $\epsilon \ra 0$. For example, if we consider the inner product,
\be
\fr{\<V_\alpha,p|V_\beta,p'\>}{2p(2\pi)\de(p-p')} = \fr{1}{N_\alpha N_\beta} \sum_n \fr{(\alpha \beta)^n}{(n!)^2} N^2_{\phi^n} = \epsilon^{\fr{(\alpha-\beta)^2}{4\pi}},
\ee
we see that it vanishes as $\epsilon\ra 0$ unless $\alpha=\beta$, reproducing the familiar selection rule for two-point functions.

We can use the same approach to compute Hamiltonian matrix elements involving vertex operators. First, let's consider mixing between vertex operators and our basis built from $\p\phi$. For example, if we compute the matrix element between the $n$-particle monomial $(\p\phi)^n$ and an arbitrary vertex operator $V_\alpha$, we find that only the $\phi^n$ term in the expansion of $V_\alpha$ has a nonzero contribution, giving us the expression
\bq
\bal
\Mcal_{(\p\phi)^n,V_\alpha}^{(\phi^2)} &= \fr{(i\alpha)^n N_{\phi^n}}{n!N_\alpha} \Mcal_{(\p\phi)^n,\phi^n} = \fr{m^2(i\alpha)^n}{2 N_{(\p\phi)^n} N_\alpha} \int dx \, dz \, e^{ip(x-z)} \<(\p\phi)^n(x) \phi^2(0) \phi^n(z)\> \\
&= \fr{n!(i\alpha)^n}{N_{(\p\phi)^n} N_\alpha} \int \fr{dp_1 \cdots dp_n}{(2\pi)^n 2p_1 \cdots 2p_n} (2\pi)\de(p-|p|_n) \, p_1 \cdots p_n \sum_i \fr{m^2}{2p_i} \\
&= n(n-1)(i\alpha)^n\epsilon^{\fr{\alpha^2}{4\pi}}\log^{\fr{3}{2}}\tfrac{1}{\epsilon} \, \sqrt{ \fr{2\G(2n)}{(4\pi)^n \G(n+1)} }.
\eal
\eq
While the Fock space integral is logarithmically divergent, the normalization of the vertex operator causes this expression to vanish as $\epsilon \ra 0$ for $\alpha > 0$. This behavior holds for all matrix elements between states built from $\p\phi$ and those built from vertex operators, such that there is \emph{no} mixing between the $\alpha=0$ sector and vertex operators in the presence of a mass term (as well as higher $\phi^n$ interactions). We can therefore safely consider the states built from $\p\phi$ as an isolated system, with no effects due to vertex operators, as we have in this work.

We also can consider mass term matrix elements between vertex operators, which can be evaluated by first computing the $\phi^n$ matrix elements
\be
\Mcal^{(\phi^2)}_{\phi^n,\phi^n} = \fr{n!}{N_{\phi^n}^2} \int \fr{dp_1 \cdots dp_n}{(2\pi)^n 2p_1 \cdots 2p_n} (2\pi)\de(p-|p|_n) \sum_i \fr{m^2}{2p_i} = \fr{nm^2}{2\epsilon\log \fr{1}{\epsilon}}.
\ee
Using this monomial matrix element, we can then compute the full vertex operator matrix element
\be
\Mcal^{(\phi^2)}_{V_\alpha,V_\alpha} = \fr{1}{N_\alpha^2} \sum_n \fr{\alpha^{2n}}{(n!)^2} N^2_{\phi^n} \Mcal^{(\phi^2)}_{\phi^n,\phi^n} = \fr{\alpha^2 m^2}{4\pi\epsilon}.
\ee
As we can see, this matrix element diverges as $\epsilon \ra 0$, even after properly normalizing the external states. This behavior also holds for all states created by acting on $V_\alpha$ with factors of $\p\phi$. We therefore find that in every vertex operator sector (except $\alpha=0$) the mass term matrix elements are \emph{all} divergent as $\epsilon \ra 0$.

These divergent matrix elements lift all states created by vertex operators, removing them from the low-energy Hilbert space and leaving only states created by $\p\phi$, which is the set of states used in this work. This behavior is perhaps not too surprising, as vertex operators are all built from $\phi$, and the equation of motion $\p_+\p_-\phi = m^2\phi$ restricts $\phi$ to no longer be an independent degree of freedom. The removal of vertex operators from the massive scalar Hilbert space is analogous to the restriction to Dirichlet states for fermions.

It is important to note that vertex operators are only lifted from the Hilbert space because we are considering relevant deformations (i.e.~$\phi^2$) which completely break the shift symmetry $\phi \ra \phi + c$. However, if we instead considered a theory such as sine-Gordon, vertex operators with the appropriate periodicity would have finite matrix elements, such that they remain in the Hilbert space.

%%%%%%%%%%%%%%%%%%%%%%%%%%%%%%%%%%%%%%%%%%%%%%%%%%%%%%%%%%%%%%%%%%%%%%%%%%%%%
%%%%%%%%%%%%%%%%%%%%%%%%%%%%%%%%%%%%%%%%%%%%%%%%%%%%%%%%%%%%%%%%%%%%%%%%%%%%%

\section{Removable Phases in Matrix Elements}
\label{app:Phases}

When computing the Hamiltonian matrix elements, there are many factors of $i$ and $-1$ that arise at various steps in the calculation. While the resulting matrix must be Hermitian, this does not preclude the possibility of relative phases between distinct off-diagonal matrix elements, which suggests that one must be remarkably careful to obtain the correct relative phases for each matrix element. However, it is important to distinguish between overall phases which can simply be removed with a redefinition of the external states,
\be
|\Ocal,p\> \ra e^{i\phi} |\Ocal,p\>,
\ee
and the irreducible relative phase factors which affect the resulting Hamiltonian eigenvalues.

Throughout this work we have often used the notation $\doteq$ to indicate equations in which ``removable'' phases have been suppressed. In this appendix, we will now more carefully explain which phases can be removed with a redefinition of the basis states (and thus can be ignored). Because these removable phases have no effect on the final matrix elements, readers can therefore safely use any $\doteq$ equations in this work (i.e.~ignore the suppressed phases) in the context of lightcone conformal truncation.

The simplest context for understanding these overall phase factors is the Fock space method. Because lightcone momenta are manifestly real and positive, factors of $i$ can only originate from derivatives acting on the Fock space expansion of $\phi$. For example, consider the general $n$-particle monomial state
\bq
|\p^{\bk}\phi,p\> = \fr{1}{n! N_{\bk}} \int \fr{dp_1 \cdots dp_n}{(2\pi)^n 2p_1 \cdots 2p_n} (2\pi) \de(p-|p|_n) F_{\p^{\bk}\phi}(p) |p_1,\ldots,p_n\>.
\eq
Using the mode expansion for $\phi$ from eq.~\eqref{eq:phimodedecomp}, we can compute the wavefunction (now being careful to include factors of $i$),
\bq
\bal
F_{\p^{\bk}\phi}(p) &\equiv \<p_1,\ldots,p_n|\p^{\bk}\phi(0)\> \\
&= \int \fr{dp'_1 \cdots dp'_n}{(2\pi)^n \sqrt{2p'_1 \cdots 2p'_n}} (ip'_1)^{k_1} \cdots (ip'_n)^{k_n} \<p_1,\ldots,p_n|a^\dagger_{p'_1} \cdots a^\dagger_{p'_n}\> \\
&= i^\De\sum_{\bk' \in \textrm{perm}(\bk)} p_1^{k'_1} \cdots p_n^{k'_n}.
\eal
\eq
We thus see that the overall phase is set by the number of derivatives, or equivalently the \emph{scaling dimension} of the operator. Because primary operators are built from linear combinations of monomials with fixed scaling dimension, there are thus \emph{no relative phases} between individual monomials, only an overall factor of $i^\De$. However, this overall phase can be removed by redefining the basis state's normalization coefficient $N_\Ocal$,
\be
N_\Ocal \ra i^\De N_\Ocal,
\label{eq:CoeffRedef}
\ee
such that no relative phases arise when computing the inner products between basis states.

Crucially, because these phases arise solely due to the wavefunctions of the states, the overall phases of Hamiltonian matrix elements are also set by the scaling dimensions of the two external states. For example, if we consider the mass term matrix element between two $n$-particle monomials, we have
\bq
\bal
\Mcal_{\bk\bk'}^{(\phi^2)} &= \fr{1}{n! N^*_{\bk} N_{\bk'}} \int \fr{dp_1 \cdots dp_n}{(2\pi)^n 2p_1 \cdots 2p_n} (2\pi) \de(p-|p|_n) F^*_{\p^{\bk}\phi}(p) F_{\p^{\bk'}\phi}(p) \sum_{k=1}^n \fr{m^2}{2p_k} \\
&= \fr{i^{\De'-\De}}{n! N^*_{\bk} N_{\bk'}} \int \fr{dp_1 \cdots dp_n}{(2\pi)^n 2p_1 \cdots 2p_n} (2\pi) \de(p-|p|_n) \big|F_{\p^{\bk}\phi}(p)\big| \big|F_{\p^{\bk'}\phi}(p)\big| \sum_{k=1}^n \fr{m^2}{2p_k}.
\eal
\eq
We thus obtain an overall factor of $i^{\De'-\De}$ in \emph{every} mass term matrix element (as well as matrix elements for any other $\phi^n$ interaction). We can clearly remove this overall phase by the redefinition~\eqref{eq:CoeffRedef} of all normalization coefficients $N_\Ocal$, such that we can safely ignore it throughout the calculation. Note that once we remove this phase, the resulting matrix elements are all manifestly real.

In a nutshell, the $\doteq$ notation in this work simply indicates the suppression of any phase which contributes to the removable overall factor of $i^{\De'-\De}$ in every matrix element. However, this overall phase arises from very different contributions in each of the three calculational methods (Fock space, Wick contraction, and radial quantization).

In the Wick contraction method, there are two sources of overall phases: factors of $-1$ from derivatives acting on $\phi$ in the position space correlation function and factors of $i$ from the Fourier transform to momentum space. Of course, these results must simply reproduce the Fock space expressions, but let's briefly step through these contributions in the Wick contraction method.

First, we have the monomial two-point function in position space (now carefully including all minus signs),
\be
\<\p^{\bk}\phi(x) \p^{\bk'}\phi(0)\> = \fr{(-1)^{\De}}{(4\pi)^n x^{|\bk|+|\bk'|}} \sum_{\boldsymbol{\sigma} \in \textrm{perm}(\bk')} \G(k_1 + \sigma_1) \cdots \G(k_n + \sigma_n).
\label{eq:WickPhase}
\ee
The overall sign for this correlator is thus determined by the total number of derivatives (or equvalently the scaling dimension) of the \emph{left} operator.

Next, we Fourier transform this correlator to momentum space, including the resulting factors of $i$, to obtain the inner product
\bq
\bal
&\int dx \, e^{ipx} \<\p^{\bk}\phi(x) \p^{\bk'}\phi(0)\> \\
&= i^{\De'-\De} \fr{2\pi p^{|\bk|+|\bk'|-1}}{(4\pi)^n \G(|\bk|+|\bk'|)} \sum_{\boldsymbol{\sigma} \in \textrm{perm}(\bk')} \G(k_1 + \sigma_1) \cdots \G(k_n + \sigma_n).
\eal
\eq
Unsurprisingly, we obtain the same overall factor of $i^{\De'-\De}$ as the Fock space method, which can be removed with the state redefinition~\eqref{eq:CoeffRedef}. This same structure holds for matrix elements, as well. The overall sign of the position space three-point function $(-1)^\De$ is set by the scaling dimension of the left operator (i.e.~the bra state), and the Fourier transform contributes a factor of $i^{\De+\De_R+\De'}$. However, all of the scalar field deformations we consider have $\De_R=0$, such that we obtain the expected overall phase of $i^{\De'-\De}$.

In the radial quantization method, the phase structure is much more subtle. As a concrete example, let's consider the matrix element of a general $\phi^m$ interaction. The three-point function for primary operators is built from monomial correlators of the form
\be
G^{(\phi^m)}_{\bk\bk'}(x,y,z) = \<\p^{\bk}\phi(x) \phi^m(y) \p^{\bk'}\phi(z)\>.
\ee
Let's assume (without loss of generality) that $n \leq n'$, i.e.~that the number of particles in the bra state is less than or equal to the number in the ket state. We can then rewrite the power of $\phi^m$ as $m=2q+n'-n$, where $q$ is the number of particles in $\p^{\bk}\phi$ that contract with the $\phi^m$ interaction (and the remaining $q+n'-n$ particles in the interaction contract with $\p^{\bk'}\phi$).

As discussed in section~\ref{sec:RadialScalars}, our strategy for computing such correlation functions is to instead consider the higher-point correlation function
\be
G^{(\p\phi^m)}_{\bk\bk'}(x,y_i,z) = \<\p^{\bk}\phi(x) \p\phi(y_1) \cdots \p\phi(y_m) \p^{\bk'}\phi(z)\>,
\ee
integrate over the $y_i$ to eliminate the derivatives and reduce back to a correlator involving $\phi^m$, and finally Fourier transform to momentum space to obtain the contribution to a Hamiltonian matrix element. Let's now step through this procedure to identify all removable phase factors.

From the radial quantization mode expansion of $\p\phi$ in eq.~\eqref{eq:II:RadialPPhi}, we see that each insertion of $\p\phi$ in the correlator (both from the external states and the interaction) gives a factor of $i$, for a total of $i^{n+n'+m} = i^{2n'+2q}$. Each additional derivative in $\p^{\bk}\phi$ (the left monomial) also gives a factor of $-1$, leading to a factor of $(-1)^{\De-n}$. We thus have
\be
G^{(\p\phi^m)}_{\bk\bk'}(x,y_i,z) \sim (-1)^{\De-n+n'+q},
\ee
here $\sim$ indicates that we have dropped all other factors to focus on the removable phases in the expression.

Next, we must integrate with respect to the $y_i$ to reduce this to a correlator involving $\phi^m$. However, the integral for each $\p\phi(y_i)$ which contracts with $\p^{\bk'}\phi$ (i.e.~the ket state) gives a factor of $-1$. This contributes an additional factor of $(-1)^{q+n'-n}$, giving us
\be
G^{(\phi^m)}_{\bk\bk'}(x,y,z) \sim (-1)^{\De},
\ee
which matches the overall factor of $(-1)^\De$ obtained via the Wick contraction method, such as in eq.~\eqref{eq:WickPhase}. We then Fourier transform to momentum space, contributing a factor of $i^{\De+\De'}$, resulting in the familiar overall phase
\be
\int dx \, dz \, e^{ip(x-z)} G^{(\phi^m)}_{\bk\bk'}(x,0,z) \sim i^{\De'-\De}.
\ee

All of these phase contributions in the radial quantization method are removed by the redefinition~\eqref{eq:CoeffRedef}, and can therefore be ignored. However, it is important to note that there are \emph{additional} factors of $(-1)^{k}$ and $(-1)^{-k'}$ that arise in the radial quantization method which are \emph{not removed} by~\eqref{eq:CoeffRedef} and must therefore be included. These additional minus signs give rise to the factor of $(-1)^{\De-\De'}$ in the general matrix element given in eq.~\eqref{eq:II:ScalarPhiNFinal}.

To summarize, in every $\doteq$ equation we have suppressed any phase which contributes to the removable overall factor of $i^{\De'-\De}$ in the resulting matrix elements. The suppressed phases in each method are:
\begin{itemize}
\item \textbf{Fock Space:} a factor of $-i$ for each derivative acting on $\phi$ in the bra state, and a factor of $i$ for each derivative acting on $\phi$ in the ket state.
\item \textbf{Wick Contraction:} a factor of $-1$ for each derivative acting on $\phi$ in the bra state, and a factor of $i^{\De+\De'}$ from the Fourier transform to momentum space.
\item \textbf{Radial Quantization:} a factor of $i$ for each insertion of $\phi$ in the external states and interaction, a factor of $-1$ for each derivative acting on $\p\phi$ in the bra state, a factor of $-1$ for each $y_i$ integral which contracts with the ket state, and a factor of $i^{\De+\De'}$ from the Fourier transform to momentum space.
\end{itemize}

Turning to fermion matrix elements, we find that the phase structure is almost exactly the same as scalars, with two added complications: the inverse derivatives from integrating out $\chi$ and the fact that $\psi$ itself has nonzero scaling dimension $\De_\psi=\half$.

The overall phase structure is again simplest to see in the Fock space method, where we suppress a factor of $-i$ for each derivative in the bra state and a factor of $i$ for each derivative in the ket state, just like for scalars. However, unlike the scalar case, the total number of derivatives in a state is not the scaling dimension, but rather $|\bk| = \De-\fr{n}{2}$. The overall removable phase for fermions is thus $i^{\De'-\De-\fr{n'-n}{2}}$.

In both the Wick contraction method and the radial quantization method, the main difference between fermions and scalars is that the Lorentzian two-point function for $\psi$ has an overall factor of $e^{-i\pi\De_\psi}=-i$
\be
\<\psi(x) \psi(0)\> = \fr{-i}{4\pi x}.
\ee
For both methods, we thus have the same suppressed phase contributions as for scalars, plus a factor of $-i$ for each contraction of two fermions (or equivalently a factor of $(-i)^{1/2}$ for each insertion of $\psi$ in the external states and interaction), as well as a factor of $i$ coming from the inverse derivative in the interaction (both for the mass term and Yukawa interactions). Including all of these suppressed contributions in both methods, we recover the same removable phase of $i^{\De'-\De-\fr{n'-n}{2}}$ as in the Fock space method.

Note that the minus signs arising from anticommuting individual fermions do \emph{not} all cancel, and lead to needed relative minus signs in the computations of both inner products and matrix elements. Because of this, we have been careful to include all such factors of $-1$ in the $\doteq$ equations for fermions, only suppressing those phases which will eventually cancel.

%%%%%%%%%%%%%%%%%%%%%%%%%%%%%%%%%%%%%%%%%%%%%%%%%%%%%%%%%%%%%%%%%%%%%%%%%%%%%
%%%%%%%%%%%%%%%%%%%%%%%%%%%%%%%%%%%%%%%%%%%%%%%%%%%%%%%%%%%%%%%%%%%%%%%%%%%%%

\section{Efficient Techniques for Fock Space Method}
\label{app:FockSpaceTricks}

In this section, we present more efficient methods for dealing with Fock space LCT computations. While these techniques are less efficient than those introduced in Part II and are consequently not presented in our version of the LCT code, they may nevertheless be useful for two reasons. First, they provide a consistency check with other methods presented in this paper, and enable the usage of Fock space methods for moderately high $\Delta_{\textrm{max}}$. Furthermore, when starting out learning LCT, the most straightforward way to develop intuition for the structure of the matrix elements is usually to work directly with momentum space wavefunctions. For the interested reader, the technology developed in this section will greatly expedite that process. For brevity, we will restrict most of our discussion to scalars, but these techniques can easily be carried over to fermions.

\subsection{Symmetrizing Momentum Space Eigenfunctions}
\label{sec:symmetrizationtricks}
Let us recall from section \ref{sec:Jacobi} that in momentum space, Casimir eigenfunctions associated with LCT basis states can be written in terms of Jacobi polynomials $P_\ell^{(\alpha,\beta)}$: \begin{equation}
	\widehat{F}_{\Lvec}(p) \equiv  p_1 \dotsb p_n \prod_{i=1}^{n-1} |p|_{i+1}^{\ell_i} \widehat{P}_{\ell_i}^{(2|\ell|_{i-1}+2i-1,1)} \left(\frac{p_{i+1}-|p|_i}{|p|_{i+1}}\right), \label{eq:hattedCasimirEfuncs}
\end{equation} where \begin{equation}
	\begin{aligned}
			\widehat{P}_{\ell}^{(\alpha,\beta)}(x) &\equiv \mu_{\ell}^{(\alpha,\beta)} \, P_{\ell}^{(\alpha,\beta)}(x), \\
			\mu_\ell^{(\alpha,\beta)} &\equiv \sqrt{\frac{\Gamma(\ell+1)\Gamma(\ell+\alpha+\beta+1)\Gamma(2\ell+\alpha+\beta+2)}{\Gamma(\ell+\alpha+1)\Gamma(\ell+\beta+1)\Gamma(2\ell+\alpha+\beta+1)}},
		\end{aligned}	
\end{equation}The notation $\widehat{F}$ distinguishes \eqref{eq:hattedCasimirEfuncs} from \eqref{eq:JacobiBasis} since \eqref{eq:hattedCasimirEfuncs} has overall normalization factors, whose origin we will explain below. One way to arrive at \eqref{eq:hattedCasimirEfuncs} is to consider expanding LCT basis states in terms of  functions that are orthogonal with respect to the LCT inner product \eqref{eq:GeneralInner}, reproduced here for convenience:\footnote{Recall from section \ref{sec:Jacobi} that the last component of $\Lvec$ is zero ($\ell_n = 0 $).} \begin{equation}
	\begin{aligned}
		\frac{1}{n! 2p N_{\Lvec} N_{\Lvec'}^*}& \int \frac{dp_1 \dotsb dp_n}{(2\pi)^n 2p_1 \dotsb 2p_n} (2\pi) \delta(p-|p|_n) \widehat{F}_{\Lvec}(p) \widehat{F}_{\Lvec'}(p) \\
		&=  \frac{1}{n! 2p N_{\Lvec} N_{\Lvec'}^*} \frac{p^{2n+|\Lvec|+|\Lvec'|-1}}{(2\pi)^{n-1} 2^n}\int \frac{dx_1 \dotsb dx_n}{x_1 \dotsb x_n}  \delta(1-|x|_n) \widehat{F}_{\Lvec}(x) \widehat{F}_{\Lvec'}(x) \\
		&= \frac{1}{n! 2p | N_{\Lvec}|^2}  \frac{p^{2n+|\Lvec|+|\Lvec'|-1}}{(2\pi)^{n-1} 2^n} \cdot \delta_{\Lvec\Lvec'}, \label{eq:simplexorthogonality}
	\end{aligned}
\end{equation} where in the second line we changed variables $x_i = \frac{p_i}{p}$ and in the third line we used the orthogonality property of the (normalized) Jacobi polynomials defined in \eqref{eq:hattedCasimirEfuncs}. We can see that the eigenfunctions in \eqref{eq:hattedCasimirEfuncs} provide an orthogonal basis of wavefunctions for the LCT basis, with the overall normalization set by \begin{equation}
	N_{\Lvec} = \frac{p^{n+|\Lvec|-1} \pi^{(1-n)/2}}{2^n \sqrt{n!}}.
\end{equation} However, they are overcomplete, since these eigenfunctions correspond to states built from distinguishable particles, while our scalar LCT basis states are built out of indistinguishable $\phi$'s. We therefore need a linear combination of the eigenfunctions in \eqref{eq:hattedCasimirEfuncs} that is invariant under swapping any of the momenta $p_i \leftrightarrow p_j$.\footnote{For fermions, we would require antisymmetry under swapping any of the momenta.} \footnote{The space $\{\boldsymbol{x} \in \mathbb{R}^n \, : \, x_i \ge 0 \textrm{ and } |x|_n = 1\}$ in the second line of \eqref{eq:simplexorthogonality} is known as a \textit{simplex} \cite{dunkl_xu_2014,xu2017orthogonal}. Computing a complete, orthonormal basis of Fock space wavefunctions thus amounts to computing \textit{orthogonal, symmetric functions on the simplex}. To the best of our knowledge, this is an open problem in the mathematical literature.} We will now present an efficient brute-force method for this procedure.

First, note that we can map the momentum-space wavefunction $\widehat{F}_{\Lvec}(p) $ to a position space primary operator by taking $p_i^{k_i} \to \ptl^{k_i} \phi_i(x)$, where $\phi_i(x)$ is a \textit{distinguishable} particle indexed by $i$. If we write \eqref{eq:hattedCasimirEfuncs} as a sum over monomials \begin{equation}
	\widehat{F}_{\Lvec}(p) = \sum_{\boldsymbol{\sigma}} C_{\Lvec}^{\boldsymbol{\sigma}} p_1^{\sigma_1} \dotsb p_n^{\sigma_n},
\end{equation} where $C_{\Lvec}^{\boldsymbol{\sigma}}$ are the coefficients obtained from expanding out \eqref{eq:hattedCasimirEfuncs}, then this maps to an operator built out of distinguishable $\phi_i$'s: \begin{equation}
	\widehat{F}_{\Lvec}(p) \to \Ocal_{\Lvec}^{\textrm{dist}} (x) \doteq \sum_{\boldsymbol{\sigma}} C_{\Lvec}^{\boldsymbol{\sigma}} \ptl^{\sigma_1} \phi_1(x) \dotsb \ptl^{\sigma_n} \phi_n(x).
\end{equation} Phrased in this way, the basic idea to symmetrize over \eqref{eq:hattedCasimirEfuncs} is then simple: we identify all the $\phi_i$'s with a single $\phi$. The resulting object, guaranteed to be symmetric, can be written in the following way: \begin{equation}
	\Ocal_{\Lvec}^{\textrm{dist}} (x) \to \boxed{\Ocal_{\Lvec} (x) = \sum_{|\Kvec|=|\Lvec|+n} \Omega_{\Lvec}^{\boldsymbol{(1)}} (\Kvec-\boldsymbol{1}) \ptl^{\Kvec}\phi,} \label{eq:disttoindist}
\end{equation} where we have defined the coefficients \begin{equation}
	\Omega^{(\boldsymbol{1})}_{\Lvec}(\Kvec-\boldsymbol{1}) \equiv \sum_{\boldsymbol{\sigma}\in \textrm{perm}(\Kvec)}C_{\Lvec}^{\boldsymbol{\sigma}}.
\end{equation} The sum over permutations ensures that that the map \eqref{eq:disttoindist} identifies the $\phi_i$ with each other. The formula for the $\Omega$ coefficients can be determined from the expansion of Jacobi polynomials; it is given by \begin{equation}
	\Omega^{(\boldsymbol{\beta})}_{\Lvec}(\Kvec) \equiv (-1)^{|\Kvec|+\ell_n} \sqrt{\frac{\ell_n! \Gamma(\ell_n+\alpha_n+1)}{\prod_{i=1}^n k_i! \Gamma(k_i + \beta_i +1)}} \sum_{\boldsymbol{y}\in \textrm{perm}(\Kvec)} \mathcal{W}_{\Lvec}^{\boldsymbol{\beta}}(\boldsymbol{y}), \label{eq:OmegaDef1}
\end{equation} where \cite{Genest_2014} \begin{equation}
	\alpha_i \equiv 2|\Lvec|_{i-1} +|\boldsymbol{\beta}|_i + i - 1,
\end{equation} \begin{equation}
\label{W-Definiton-2}
\mathcal{W}_{\Lvec}^{(\boldsymbol{\beta})}(\mathbf{y})  = \prod_{i = 1}^{n-1} \hat{h}_{\ell_{i}}( |\mathbf{y}|_{i} - |\Lvec|_{i - 1}; \alpha_i, \beta_{i + 1}, |\mathbf{y}|_{i + 1} - |\Lvec|_{i - 1}),
\end{equation} and \begin{multline}
\hat{h}_{\ell}(y; \alpha, \beta, M) = \sqrt{\frac{(y+1)_\alpha(M-y+1)_\beta(\ell+1)_\alpha(\beta+\ell+1)_\alpha(2\ell+\alpha+\beta+1)}{\Gamma(\alpha+\beta+2+M+\ell)\Gamma(M-\ell+1)}} 
\\
\times 
(-1)^\ell \frac{\Gamma(M+1)}{\Gamma(\alpha+1)} {}_3F_{2}\left(\genfrac{}{}{0pt}{}{-\ell, \ell + \alpha + \beta +1, -y}{\alpha+1, -M}; 1\right).
\end{multline} In practice, it is much more efficient to use a recursion relation to determine the $\Omega$'s, starting from seed $\Omega$'s where the arguments are two-component vectors. Then, $n$-particle $\Omega$'s can be related to $(n-1)$-particle ones through the recurrence relation \begin{multline}
\label{Omega-Recurrence}
\Omega_{\Lvec}^{(\boldsymbol{\beta})}(\boldsymbol{k}) = (-1)^{\ell_n + \ell_{n-1}}\left[ \frac{\ell_{n-1}! \,\Gamma(\ell_{n-1} + \alpha_{n-1} + 1)}{\ell_{n}! \,\Gamma(\ell_n + \alpha_n + 1)}\right]^{-1/2}
\\
\times
\sum_{k_i \in \boldsymbol{k}}
\Omega_{\Lvec/\ell_n}^{(\boldsymbol{\beta})}(\boldsymbol{k}/k_i) \Bigg[(-1)^{k_i}[k_i ! \, \Gamma(k_i + \beta +1)]^{-1/2}
\\
\times \;\hat{h}_{\ell_{n-1}}(|\boldsymbol{k}|_n - k_i - |\Lvec|_{n-2}; \alpha_{n-1}, \beta, |\boldsymbol{k}|_n- |\Lvec|_{n-2}) \Bigg],
\end{multline}
where the sum is over all \textit{distinct} entries of $\bk$. \eqref{Omega-Recurrence} can be derived by observing that the  $\mathcal{W}$'s themselves satisfy a recurrence relation \begin{equation}
	\Wcal_{\Lvec}^{(\boldsymbol{\beta})}(\mathbf{y}) = \hat{h}_{\ell_{n-1}}(|\mathbf{y}|_{n-1}-|\Lvec|_{n-2}; \alpha_{n-1}, \beta_n, |\mathbf{y}|_n-|\Lvec|_{n-2}) \, \Wcal_{\Lvec/\ell_n}^{(\boldsymbol{\beta}/\beta_n)}(\mathbf{y}/y_n),
\end{equation} and then rewriting the sum over $\boldsymbol{y}$ that appears in \eqref{eq:OmegaDef1} as \begin{equation}
	\sum_{\yvec \in \textrm{perm}(\bk)} \Wcal^{(\Bvec)}_{\Lvec}(\yvec) = \sum_{\text{distinct }k_j\in\bk} \,\, \sum_{\yvec/y_n \in \textrm{perm}(\bk/k_j)}  \Wcal^{(\Bvec)}_{\Lvec}((y_1,\dots,y_{n-1},k_j)).
\end{equation}

We now have a method for computing manifestly symmetric position space primary operators starting from the momentum space eigenfunctions in \eqref{eq:hattedCasimirEfuncs}.\footnote{This map may be inverted, by writing monomials as linear combinations of Jacobi polynomials. See, e.g., \cite{Genest_2014}.}  All that remains is to orthogonalize them, which can be accomplished using any of the methods presented in Parts I and II.

For fermions, the modifications to the above discussion are straightforward; we can define analogous \textit{antisymmetrized} $\Omega$ coefficients \begin{equation}
	\widetilde{\Omega}^{(\boldsymbol{\beta})}_{\Lvec}(\Kvec) \equiv (-1)^{|\Kvec|+\ell_n} \sqrt{\frac{\ell_n! \Gamma(\ell_n+\alpha_n+1)}{\prod_{i=1}^n k_i! \Gamma(k_i + \beta_i +1)}} \sum_{\boldsymbol{y}\in \textrm{perm}(\Kvec)} (-1)^{N_{\textrm{perm}}^{(\boldsymbol{y}|\bk)}} \mathcal{W}_{\Lvec}^{\boldsymbol{\beta}}(\boldsymbol{y}),
\end{equation} where $N_{\textrm{perm}}^{(\boldsymbol{y}|\bk)}$ counts the permutations relating $\boldsymbol{y}$ to $\bk$. 

Finally, let us comment on the significance of the superscript $(\boldsymbol{\beta})$, which is closely related to the measure in \eqref{eq:simplexorthogonality}. The Jacobi polynomials defined in \eqref{eq:hattedCasimirEfuncs} are orthogonal with respect to the measure in \eqref{eq:simplexorthogonality}, but we can define a more general set of functions \begin{equation}
	\widehat{F}_{\Lvec}^{(\boldsymbol{\beta})}(p) \equiv  p_1 \dotsb p_n \prod_{i=1}^{n-1} |p|_{i+1}^{\ell_i} \widehat{P}_{\ell_i}^{(2|\ell|_{i-1}+|\beta|_i+i-1,\beta_{i+1})} \left(\frac{p_{i+1}-|p|_i}{|p|_{i+1}}\right), \label{eq:hattedCasimirEfuncs2}
\end{equation} that are orthogonal with respect to the measure \begin{equation}
	\begin{aligned}
		\int_0^\infty  \left( \prod_{k=1}^n dx_k x_k^{\beta_k-2}\right) \delta(1-|x|_n)  \widehat{F}_{\Lvec}^{(\boldsymbol{\beta})}(x) \widehat{F}^{(\boldsymbol{\beta})}_{\Lvec'}(x) \propto \delta_{\Lvec\Lvec'}.
	\end{aligned}
\end{equation} Then, the $\Omega$ (or $\widetilde{\Omega}$) coefficients inherit this measure and thus provide a map between functions orthgonal with respect to the general measure defined above and symmetric or antisymmetric position space operators. For fermions, this is especially useful as the distinction between Dirichlet $(\boldsymbol{\beta}=\boldsymbol{2})$ and non-Dirichlet $(\boldsymbol{\beta}=\boldsymbol{0})$ states can be phrased in terms of a modification to the inner product measure. On the other hand, as we have seen from the above discussion, scalars require $\boldsymbol{\beta}=\boldsymbol{1}$.

\subsection{Useful Jacobi Formulas} As a starting point, a reader may wish to work directly with Jacobi polynomials that are symmetrized without any of the tricks in section \ref{sec:symmetrizationtricks} (e.g.~by brute-force summing over all images $p_i \leftrightarrow p_j$ or by starting with symmetric functions and then expressing them in terms of Jacobis) and compute their matrix elements. Here we record some potentially useful formulas; many of these formulas can be found in \cite{book}.

It can often be useful to re-express Jacobi polynomials orthogonal with respect one measure as a linear combination of those that are orthogonal with respect to a different measure using the identity \begin{equation}
	\widehat{P}^{(\alpha,\beta)}_\ell (z) = \sum_{k=0}^\ell \Acal_{\ell k}^{(\alpha,\beta,\gamma,\delta)} \widehat{P}^{(\g,\de)}_k(z), \label{eq:JacobiMeasureIdentity}
\end{equation}
with the conversion coefficients given by
\be
\begin{split}
\Acal_{\ell k}^{(\a,\beta,\g,\de)} = &\fr{\mu^{(\a,\beta)}_\ell}{\mu^{(\g,\de)}_k} \fr{\G(k+\g+\de+1)\G(\ell+k+\a+\beta+1)\G(\ell+\a+1)}{\G(\ell+\a+\beta+1)\G(k+\a+1)\G(2k+\g+\de+1)\G(\ell-k+1)} \\
& \, \times \phantom{}_3F_2(k-\ell,\ell+k+\a+\beta+1,k+\g+1;k+\a+1,2k+\g+\de+2;1).
\end{split}
\ee The normalized Jacobi polynomials $\widehat{P}^{(\alpha,\beta)}_\ell(z)$ satisfy the orthogonality property \begin{equation}
	\int_{-1}^{1} dz (1-z)^\alpha (1+z)^\beta  \widehat{P}^{(\alpha,\beta)}_\ell(z)  \widehat{P}^{(\alpha,\beta)}_{\ell'}(z) = 2^{\alpha+\beta+1}\delta_{\ell \ell'}. \label{eq:HatJacobiOrtho}
\end{equation}

For evaluating integrals over the simplex defined in \eqref{eq:simplexorthogonality}, we find the following change of variables useful, which maps the simplex $|x| = 1$ to the hypercube $[-1,1]^{(n-1)}$ \begin{equation}
	z_i = \frac{x_{i+1}-|x|_i}{|x|_{i+1}}, \quad\quad\quad i = 1, \dots, n-1,
\end{equation} with the resulting Jacobian \begin{equation}
	\int_0^\infty dx_1 \dotsb dx_n \, \delta(1-|x|_n) \quad \rightarrow \quad \int_{-1}^1 \frac{(1-z_2)(1-z_3)^2 \dotsb (1-z_{n-1})^{n-2}}{2^{\frac{n(n-1)}{2}}} dz_1 \dotsb dz_{n-1}.
\end{equation} Combined with \eqref{eq:JacobiMeasureIdentity} and \eqref{eq:HatJacobiOrtho}, most matrix elements can be evaluated straightforwardly.

%%%%%%%%%%%%%%%%%%%%%%%%%%%%%%%%%%%%%%%%%%%%%%%%%%%%%%%%%%%%%%%%%%%%%%%%%%%%%
%%%%%%%%%%%%%%%%%%%%%%%%%%%%%%%%%%%%%%%%%%%%%%%%%%%%%%%%%%%%%%%%%%%%%%%%%%%%%
%%%%%%%%%%%%%%%%%%%%%%%%%%%%%%%%%%%%%%%%%%%%%%%%%%%%%%%%%%%%%%%%%%%%%%%%%%%%%

\bibliographystyle{utphys}
\bibliography{ManualBib}

\providecommand{\href}[2]{#2}\begingroup\raggedright\begin{thebibliography}{10}

\bibitem{doi:10.1142/0543}
L.~Polley and D.~E.~L. Pottinger, \href{http://dx.doi.org/10.1142/0543}{{\em
  Variational Calculations in Quantum Field Theory}}.
\newblock World Scientific, 1988.

\bibitem{Yurov:1989yu}
V.~P. Yurov and A.~B. Zamolodchikov, ``{Truncated conformal space approach to
  scaling Lee-Yang model},''
\href{http://dx.doi.org/10.1142/S0217751X9000218X}{{\em Int. J. Mod. Phys.}
  {\bfseries A5} (1990) 3221--3246}.
%%CITATION = IMPAE,A5,3221;%%.

\bibitem{Yurov:1991my}
V.~P. Yurov and A.~B. Zamolodchikov, ``{Truncated fermionic space approach to
  the critical 2-D Ising model with magnetic field},''
\href{http://dx.doi.org/10.1142/S0217751X91002161}{{\em Int. J. Mod. Phys.}
  {\bfseries A6} (1991) 4557--4578}.
%%CITATION = IMPAE,A6,4557;%%.

\bibitem{Coser:2014lla}
A.~Coser, M.~Beria, G.~P. Brandino, R.~M. Konik, and G.~Mussardo, ``{Truncated
  Conformal Space Approach for 2D Landau-Ginzburg Theories},''
  \href{http://dx.doi.org/10.1088/1742-5468/2014/12/P12010}{{\em J. Stat.
  Mech.} {\bfseries 1412} (2014) P12010},
\href{http://arxiv.org/abs/1409.1494}{{\ttfamily arXiv:1409.1494 [hep-th]}}.
%%CITATION = ARXIV:1409.1494;%%.

\bibitem{Rychkov:2014eea}
S.~Rychkov and L.~G. Vitale, ``{Hamiltonian truncation study of the $\phi^4$
  theory in two dimensions},''
  \href{http://dx.doi.org/10.1103/PhysRevD.91.085011}{{\em Phys. Rev.}
  {\bfseries D91} (2015) 085011},
\href{http://arxiv.org/abs/1412.3460}{{\ttfamily arXiv:1412.3460 [hep-th]}}.
%%CITATION = ARXIV:1412.3460;%%.

\bibitem{Rychkov:2015vap}
S.~Rychkov and L.~G. Vitale, ``{Hamiltonian truncation study of the $\phi^4$
  theory in two dimensions II. The $\mathbb Z_2$-broken phase and the Chang
  duality},'' \href{http://dx.doi.org/10.1103/PhysRevD.93.065014}{{\em Phys.
  Rev.} {\bfseries D93} no.~6, (2016) 065014},
\href{http://arxiv.org/abs/1512.00493}{{\ttfamily arXiv:1512.00493 [hep-th]}}.
%%CITATION = ARXIV:1512.00493;%%.

\bibitem{Bajnok:2015bgw}
Z.~Bajnok and M.~Lajer, ``{Truncated Hilbert space approach to the 2d
  $\phi^{4}$ theory},'' \href{http://dx.doi.org/10.1007/JHEP10(2016)050}{{\em
  JHEP} {\bfseries 10} (2016) 050},
\href{http://arxiv.org/abs/1512.06901}{{\ttfamily arXiv:1512.06901 [hep-th]}}.
%%CITATION = ARXIV:1512.06901;%%.

\bibitem{Gabai:2019ryw}
B.~Gabai and X.~Yin, ``{On The S-Matrix of Ising Field Theory in Two
  Dimensions},'' \href{http://arxiv.org/abs/1905.00710}{{\ttfamily
  arXiv:1905.00710 [hep-th]}}.

\bibitem{Rakovszky:2016ugs}
T.~Rakovszky, M.~Mesty\'{a}n, M.~Collura, M.~Kormos, and G.~Tak\'{a}cs,
  ``{Hamiltonian truncation approach to quenches in the Ising field theory},''
  \href{http://dx.doi.org/10.1016/j.nuclphysb.2016.08.024}{{\em Nucl. Phys. B}
  {\bfseries 911} (2016) 805--845},
  \href{http://arxiv.org/abs/1607.01068}{{\ttfamily arXiv:1607.01068
  [cond-mat.stat-mech]}}.

\bibitem{Hodsagi:2018sul}
K.~H\'{o}ds\'{a}gi, M.~Kormos, and G.~Tak\'{a}cs, ``{Quench dynamics of the
  Ising field theory in a magnetic field},''
  \href{http://dx.doi.org/10.21468/SciPostPhys.5.3.027}{{\em SciPost Phys.}
  {\bfseries 5} no.~3, (2018) 027},
  \href{http://arxiv.org/abs/1803.01158}{{\ttfamily arXiv:1803.01158
  [cond-mat.stat-mech]}}.

\bibitem{Elias-Miro:2015bqk}
J.~Elias~Mir\'{o}, M.~Montull, and M.~Riembau, ``{The renormalized Hamiltonian
  truncation method in the large $E_T$ expansion},''
  \href{http://dx.doi.org/10.1007/JHEP04(2016)144}{{\em JHEP} {\bfseries 04}
  (2016) 144},
\href{http://arxiv.org/abs/1512.05746}{{\ttfamily arXiv:1512.05746 [hep-th]}}.
%%CITATION = ARXIV:1512.05746;%%.

\bibitem{Elias-Miro:2017xxf}
J.~Elias~Mir\'{o}, S.~Rychkov, and L.~G. Vitale, ``{High-Precision Calculations
  in Strongly Coupled Quantum Field Theory with Next-to-Leading-Order
  Renormalized Hamiltonian Truncation},''
  \href{http://dx.doi.org/10.1007/JHEP10(2017)213}{{\em JHEP} {\bfseries 10}
  (2017) 213},
\href{http://arxiv.org/abs/1706.06121}{{\ttfamily arXiv:1706.06121 [hep-th]}}.
%%CITATION = ARXIV:1706.06121;%%.

\bibitem{Elias-Miro:2017tup}
J.~Elias~Mir\'{o}, S.~Rychkov, and L.~G. Vitale, ``{NLO Renormalization in the
  Hamiltonian Truncation},''
  \href{http://dx.doi.org/10.1103/PhysRevD.96.065024}{{\em Phys. Rev.}
  {\bfseries D96} no.~6, (2017) 065024},
\href{http://arxiv.org/abs/1706.09929}{{\ttfamily arXiv:1706.09929 [hep-th]}}.
%%CITATION = ARXIV:1706.09929;%%.

\bibitem{Lee:2000ac}
D.~Lee, N.~Salwen, and D.~Lee, ``{The Diagonalization of quantum field
  Hamiltonians},'' \href{http://dx.doi.org/10.1016/S0370-2693(01)00197-6}{{\em
  Phys. Lett. B} {\bfseries 503} (2001) 223--235},
  \href{http://arxiv.org/abs/hep-th/0002251}{{\ttfamily arXiv:hep-th/0002251}}.

\bibitem{Lee:2000xna}
D.~Lee, N.~Salwen, and M.~Windoloski, ``{Introduction to stochastic error
  correction methods},''
  \href{http://dx.doi.org/10.1016/S0370-2693(01)00198-8}{{\em Phys. Lett. B}
  {\bfseries 502} (2001) 329--337},
  \href{http://arxiv.org/abs/hep-lat/0010039}{{\ttfamily
  arXiv:hep-lat/0010039}}.

\bibitem{Hogervorst:2014rta}
M.~Hogervorst, S.~Rychkov, and B.~C. van Rees, ``{Truncated conformal space
  approach in d dimensions: A cheap alternative to lattice field theory?},''
  \href{http://dx.doi.org/10.1103/PhysRevD.91.025005}{{\em Phys. Rev.}
  {\bfseries D91} (2015) 025005},
\href{http://arxiv.org/abs/1409.1581}{{\ttfamily arXiv:1409.1581 [hep-th]}}.
%%CITATION = ARXIV:1409.1581;%%.

\bibitem{Hogervorst:2018otc}
M.~Hogervorst, ``{RG flows on $S^d$ and Hamiltonian truncation},''
  \href{http://arxiv.org/abs/1811.00528}{{\ttfamily arXiv:1811.00528
  [hep-th]}}.

\bibitem{EliasMiro:2020uvk}
J.~Elias~Mir\'{o} and E.~Hardy, ``{Exploring Hamiltonian Truncation in
  $d=2+1$},'' \href{http://arxiv.org/abs/2003.08405}{{\ttfamily
  arXiv:2003.08405 [hep-th]}}.

\bibitem{2018RPPh...81d6002J}
A.~J.~A. {James}, R.~M. {Konik}, P.~{Lecheminant}, N.~J. {Robinson}, and A.~M.
  {Tsvelik}, ``{Non-perturbative methodologies for low-dimensional
  strongly-correlated systems: From non-Abelian bosonization to truncated
  spectrum methods},'' \href{http://dx.doi.org/10.1088/1361-6633/aa91ea}{{\em
  Reports on Progress in Physics} {\bfseries 81} no.~4, (Apr., 2018) 046002},
  \href{http://arxiv.org/abs/1703.08421}{{\ttfamily arXiv:1703.08421
  [cond-mat.str-el]}}.

\bibitem{Katz:2013qua}
E.~Katz, G.~Marques~Tavares, and Y.~Xu, ``{Solving 2D QCD with an adjoint
  fermion analytically},''
  \href{http://dx.doi.org/10.1007/JHEP05(2014)143}{{\em JHEP} {\bfseries 05}
  (2014) 143},
\href{http://arxiv.org/abs/1308.4980}{{\ttfamily arXiv:1308.4980 [hep-th]}}.
%%CITATION = ARXIV:1308.4980;%%.

\bibitem{Katz:2014uoa}
E.~Katz, G.~Marques~Tavares, and Y.~Xu, ``{A solution of 2D QCD at Finite $N$
  using a conformal basis},''
\href{http://arxiv.org/abs/1405.6727}{{\ttfamily arXiv:1405.6727 [hep-th]}}.
%%CITATION = ARXIV:1405.6727;%%.

\bibitem{Katz:2016hxp}
E.~Katz, Z.~U. Khandker, and M.~T. Walters, ``{A Conformal Truncation Framework
  for Infinite-Volume Dynamics},''
  \href{http://dx.doi.org/10.1007/JHEP07(2016)140}{{\em JHEP} {\bfseries 07}
  (2016) 140},
\href{http://arxiv.org/abs/1604.01766}{{\ttfamily arXiv:1604.01766 [hep-th]}}.
%%CITATION = ARXIV:1604.01766;%%.

\bibitem{Anand:2017yij}
N.~Anand, V.~X. Genest, E.~Katz, Z.~U. Khandker, and M.~T. Walters, ``{RG flow
  from $\phi^4$ theory to the 2D Ising model},''
  \href{http://dx.doi.org/10.1007/JHEP08(2017)056}{{\em JHEP} {\bfseries 08}
  (2017) 056},
\href{http://arxiv.org/abs/1704.04500}{{\ttfamily arXiv:1704.04500 [hep-th]}}.
%%CITATION = ARXIV:1704.04500;%%.

\bibitem{Fitzpatrick:2018ttk}
A.~L. Fitzpatrick, J.~Kaplan, E.~Katz, L.~G. Vitale, and M.~T. Walters,
  ``{Lightcone effective Hamiltonians and RG flows},''
  \href{http://dx.doi.org/10.1007/JHEP08(2018)120}{{\em JHEP} {\bfseries 08}
  (2018) 120}, \href{http://arxiv.org/abs/1803.10793}{{\ttfamily
  arXiv:1803.10793 [hep-th]}}.

\bibitem{Delacretaz:2018xbn}
L.~V. Delacr\'{e}taz, A.~L. Fitzpatrick, E.~Katz, and L.~G. Vitale,
  ``{Conformal Truncation of Chern-Simons Theory at Large $N_f$},''
\href{http://arxiv.org/abs/1811.10612}{{\ttfamily arXiv:1811.10612 [hep-th]}}.
%%CITATION = ARXIV:1811.10612;%%.

\bibitem{Fitzpatrick:2018xlz}
A.~L. Fitzpatrick, E.~Katz, and M.~T. Walters, ``{Nonperturbative Matching
  Between Equal-Time and Lightcone Quantization},''
  \href{http://arxiv.org/abs/1812.08177}{{\ttfamily arXiv:1812.08177
  [hep-th]}}.

\bibitem{Anand:2019lkt}
N.~Anand, Z.~U. Khandker, and M.~T. Walters, ``{Momentum space CFT correlators
  for Hamiltonian truncation},''
  \href{http://arxiv.org/abs/1911.02573}{{\ttfamily arXiv:1911.02573
  [hep-th]}}.

\bibitem{Fitzpatrick:2019cif}
A.~L. Fitzpatrick, E.~Katz, M.~T. Walters, and Y.~Xin, ``{Solving the 2D SUSY
  Gross-Neveu-Yukawa Model with Conformal Truncation},''
  \href{http://arxiv.org/abs/1911.10220}{{\ttfamily arXiv:1911.10220
  [hep-th]}}.

\bibitem{Penedones:2010ue}
J.~Penedones, ``{Writing CFT correlation functions as AdS scattering
  amplitudes},'' \href{http://dx.doi.org/10.1007/JHEP03(2011)025}{{\em JHEP}
  {\bfseries 03} (2011) 025},
\href{http://arxiv.org/abs/1011.1485}{{\ttfamily arXiv:1011.1485 [hep-th]}}.
%%CITATION = ARXIV:1011.1485;%%.

\bibitem{reed1978iv}
M.~Reed and B.~Simon, {\em Methods of Modern Mathematical Physics IV: Analysis
  of Operators}.
\newblock Elsevier, 1978.

\bibitem{macdonald1933successive}
J.~MacDonald, ``{Successive approximations by the Rayleigh-Ritz variation
  method},'' \href{http://dx.doi.org/10.1103/PhysRev.43.830}{{\em Physical
  Review} {\bfseries 43} no.~10, (1933) 830}.

\bibitem{Dirac:1949cp}
P.~A. Dirac, ``{Forms of Relativistic Dynamics},''
  \href{http://dx.doi.org/10.1103/RevModPhys.21.392}{{\em Rev. Mod. Phys.}
  {\bfseries 21} (1949) 392--399}.

\bibitem{Weinberg:1966jm}
S.~Weinberg, ``{Dynamics at infinite momentum},''
\href{http://dx.doi.org/10.1103/PhysRev.150.1313}{{\em Phys. Rev.} {\bfseries
  150} (1966) 1313--1318}.
%%CITATION = PHRVA,150,1313;%%.

\bibitem{Bardakci:1969dv}
K.~Bardakci and M.~Halpern, ``{Theories at infinite momentum},''
  \href{http://dx.doi.org/10.1103/PhysRev.176.1686}{{\em Phys. Rev.} {\bfseries
  176} (1968) 1686--1699}.

\bibitem{Kogut:1969xa}
J.~B. Kogut and D.~E. Soper, ``{Quantum Electrodynamics in the Infinite
  Momentum Frame},'' \href{http://dx.doi.org/10.1103/PhysRevD.1.2901}{{\em
  Phys. Rev. D} {\bfseries 1} (1970) 2901--2913}.

\bibitem{Chang:1972xt}
S.-J. Chang, R.~G. Root, and T.-M. Yan, ``{Quantum field theories in the
  infinite momentum frame. 1. Quantization of scalar and Dirac fields},''
  \href{http://dx.doi.org/10.1103/PhysRevD.7.1133}{{\em Phys. Rev. D}
  {\bfseries 7} (1973) 1133--1148}.

\bibitem{Klauder:1969zz}
H.~Leutwyler, J.~R. Klauder, and L.~Streit, ``{Quantum field theory on
  lightlike slabs},''
\href{http://dx.doi.org/10.1007/BF02826338}{{\em Nuovo Cim.} {\bfseries A66}
  (1970) 536--554}.
%%CITATION = NUCIA,A66,536;%%.

\bibitem{Maskawa:1975ky}
T.~Maskawa and K.~Yamawaki, ``{The Problem of $P^+ = 0$ Mode in the Null Plane
  Field Theory and Dirac's Method of Quantization},''
\href{http://dx.doi.org/10.1143/PTP.56.270}{{\em Prog. Theor. Phys.} {\bfseries
  56} (1976) 270}.
%%CITATION = PTPKA,56,270;%%.

\bibitem{Brodsky:1997de}
S.~J. Brodsky, H.-C. Pauli, and S.~S. Pinsky, ``{Quantum chromodynamics and
  other field theories on the light cone},''
  \href{http://dx.doi.org/10.1016/S0370-1573(97)00089-6}{{\em Phys. Rept.}
  {\bfseries 301} (1998) 299--486},
\href{http://arxiv.org/abs/hep-ph/9705477}{{\ttfamily arXiv:hep-ph/9705477
  [hep-ph]}}.
%%CITATION = HEP-PH/9705477;%%.

\bibitem{Rutter:2018aog}
D.~Rutter and B.~C. van Rees, ``{Counterterms in Truncated Conformal
  Perturbation Theory},'' \href{http://arxiv.org/abs/1803.05798}{{\ttfamily
  arXiv:1803.05798 [hep-th]}}.

\bibitem{Anderson:1967zze}
P.~Anderson, ``{Infrared Catastrophe in Fermi Gases with Local Scattering
  Potentials},'' \href{http://dx.doi.org/10.1103/PhysRevLett.18.1049}{{\em
  Phys. Rev. Lett.} {\bfseries 18} (1967) 1049--1051}.

\bibitem{Chang:1968bh}
S.-J. Chang and S.-K. Ma, ``{Feynman rules and quantum electrodynamics at
  infinite momentum},''
\href{http://dx.doi.org/10.1103/PhysRev.180.1506}{{\em Phys. Rev.} {\bfseries
  180} (1969) 1506--1513}.
%%CITATION = PHRVA,180,1506;%%.

\bibitem{Yan:1973qg}
T.-M. Yan, ``{Quantum field theories in the infinite momentum frame IV.
  Scattering matrix of vector and Dirac fields and perturbation theory},''
\href{http://dx.doi.org/10.1103/PhysRevD.7.1780}{{\em Phys. Rev.} {\bfseries
  D7} (1973) 1780--1800}.
%%CITATION = PHRVA,D7,1780;%%.

\bibitem{Wilson:1994fk}
K.~G. Wilson, T.~S. Walhout, A.~Harindranath, W.-M. Zhang, R.~J. Perry, and
  S.~D. Glazek, ``{Nonperturbative QCD: A weak coupling treatment on the light
  front},'' \href{http://dx.doi.org/10.1103/PhysRevD.49.6720}{{\em Phys. Rev.}
  {\bfseries D49} (1994) 6720--6766},
\href{http://arxiv.org/abs/hep-th/9401153}{{\ttfamily arXiv:hep-th/9401153
  [hep-th]}}.
%%CITATION = HEP-TH/9401153;%%.

\bibitem{Tsujimaru:1997jt}
S.~Tsujimaru and K.~Yamawaki, ``{Zero mode and symmetry breaking on the light
  front},'' \href{http://dx.doi.org/10.1103/PhysRevD.57.4942}{{\em Phys. Rev.}
  {\bfseries D57} (1998) 4942--4964},
\href{http://arxiv.org/abs/hep-th/9704171}{{\ttfamily arXiv:hep-th/9704171
  [hep-th]}}.
%%CITATION = HEP-TH/9704171;%%.

\bibitem{Hellerman:1997yu}
S.~Hellerman and J.~Polchinski, ``{Compactification in the lightlike limit},''
  \href{http://dx.doi.org/10.1103/PhysRevD.59.125002}{{\em Phys. Rev. D}
  {\bfseries 59} (1999) 125002},
  \href{http://arxiv.org/abs/hep-th/9711037}{{\ttfamily arXiv:hep-th/9711037}}.

\bibitem{Burkardt2}
M.~Burkardt, ``{Much ado about nothing: Vacuum and renormalization on the light
  front},''
\href{http://arxiv.org/abs/hep-ph/9709421}{{\ttfamily arXiv:hep-ph/9709421
  [hep-ph]}}.
%%CITATION = HEP-PH/9709421;%%.

\bibitem{Yamawaki:1998cy}
K.~Yamawaki, ``{Zero mode problem on the light front},''
\href{http://arxiv.org/abs/hep-th/9802037}{{\ttfamily arXiv:hep-th/9802037
  [hep-th]}}.
%%CITATION = HEP-TH/9802037;%%.

\bibitem{Rozowsky:2000gy}
J.~S. Rozowsky and C.~B. Thorn, ``{Spontaneous symmetry breaking at infinite
  momentum without $P^+$ zero modes},''
  \href{http://dx.doi.org/10.1103/PhysRevLett.85.1614}{{\em Phys. Rev. Lett.}
  {\bfseries 85} (2000) 1614--1617},
  \href{http://arxiv.org/abs/hep-th/0003301}{{\ttfamily arXiv:hep-th/0003301}}.

\bibitem{Heinzl:2003jy}
T.~Heinzl, ``{Light cone zero modes revisited},''
\href{http://arxiv.org/abs/hep-th/0310165}{{\ttfamily arXiv:hep-th/0310165
  [hep-th]}}.
%%CITATION = HEP-TH/0310165;%%.

\bibitem{Beane:2013ksa}
S.~R. Beane, ``{Broken Chiral Symmetry on a Null Plane},''
  \href{http://dx.doi.org/10.1016/j.aop.2013.06.012}{{\em Annals Phys.}
  {\bfseries 337} (2013) 111--142},
\href{http://arxiv.org/abs/1302.1600}{{\ttfamily arXiv:1302.1600 [nucl-th]}}.
%%CITATION = ARXIV:1302.1600;%%.

\bibitem{Herrmann:2015dqa}
M.~Herrmann and W.~N. Polyzou, ``{Light-front vacuum},''
  \href{http://dx.doi.org/10.1103/PhysRevD.91.085043}{{\em Phys. Rev.}
  {\bfseries D91} no.~8, (2015) 085043},
\href{http://arxiv.org/abs/1502.01230}{{\ttfamily arXiv:1502.01230 [hep-th]}}.
%%CITATION = ARXIV:1502.01230;%%.

\bibitem{Hiller:2016itl}
J.~R. Hiller, ``{Nonperturbative light-front Hamiltonian methods},''
  \href{http://dx.doi.org/10.1016/j.ppnp.2016.06.002}{{\em Prog. Part. Nucl.
  Phys.} {\bfseries 90} (2016) 75--124},
\href{http://arxiv.org/abs/1606.08348}{{\ttfamily arXiv:1606.08348 [hep-ph]}}.
%%CITATION = ARXIV:1606.08348;%%.

\bibitem{Collins:2018aqt}
J.~Collins, ``{The non-triviality of the vacuum in light-front quantization: An
  elementary treatment},''
\href{http://arxiv.org/abs/1801.03960}{{\ttfamily arXiv:1801.03960 [hep-ph]}}.
%%CITATION = ARXIV:1801.03960;%%.

\bibitem{Henning:2019mcv}
B.~Henning and T.~Melia, ``{Conformal-helicity duality \& the Hilbert space of
  free CFTs},'' \href{http://arxiv.org/abs/1902.06747}{{\ttfamily
  arXiv:1902.06747 [hep-th]}}.

\bibitem{Mikhailov:2002bp}
A.~Mikhailov, ``{Notes on higher spin symmetries},''
\href{http://arxiv.org/abs/hep-th/0201019}{{\ttfamily arXiv:hep-th/0201019
  [hep-th]}}.
%%CITATION = HEP-TH/0201019;%%.

\bibitem{2dQCDToAppear}
N.~Anand, A.~L. Fitzpatrick, E.~Katz, and Y.~Xin, ``{Solving 2D QCD with
  Massive Quarks in Lightcone Conformal Truncation},'' {\em To appear} .

\bibitem{tHooft:1974pnl}
G.~'t~Hooft, ``{A Two-Dimensional Model for Mesons},''
\href{http://dx.doi.org/10.1016/0550-3213(74)90088-1}{{\em Nucl. Phys.}
  {\bfseries B75} (1974) 461--470}.
%%CITATION = NUPHA,B75,461;%%.

\bibitem{Coleman:1973ci}
S.~R. Coleman, ``{There are no Goldstone bosons in two-dimensions},''
  \href{http://dx.doi.org/10.1007/BF01646487}{{\em Commun. Math. Phys.}
  {\bfseries 31} (1973) 259--264}.

\bibitem{Ginsparg}
P.~H. Ginsparg, ``{Applied Conformal Field Theory},''
\href{http://arxiv.org/abs/hep-th/9108028}{{\ttfamily arXiv:hep-th/9108028
  [hep-th]}}.
%%CITATION = HEP-TH/9108028;%%.

\bibitem{Harindranath:1987db}
A.~Harindranath and J.~Vary, ``{Solving two-dimensional $\phi^4$ theory by
  discretized light front quantization},''
  \href{http://dx.doi.org/10.1103/PhysRevD.36.1141}{{\em Phys. Rev. D}
  {\bfseries 36} (1987) 1141--1147}.

\bibitem{Harindranath:1988zt}
A.~Harindranath and J.~Vary, ``{Stability of the Vacuum in Scalar Field Models
  in $1 + 1$ Dimensions},''
  \href{http://dx.doi.org/10.1103/PhysRevD.37.1076}{{\em Phys. Rev. D}
  {\bfseries 37} (1988) 1076--1078}.

\bibitem{Pauli:1985pv}
H.~C. Pauli and S.~J. Brodsky, ``{Solving Field Theory in One Space One Time
  Dimension},''
\href{http://dx.doi.org/10.1103/PhysRevD.32.1993}{{\em Phys. Rev.} {\bfseries
  D32} (1985) 1993}.
%%CITATION = PHRVA,D32,1993;%%.

\bibitem{Pauli:1985ps}
H.~C. Pauli and S.~J. Brodsky, ``{Discretized Light Cone Quantization: Solution
  to a Field Theory in One Space One Time Dimensions},''
\href{http://dx.doi.org/10.1103/PhysRevD.32.2001}{{\em Phys. Rev.} {\bfseries
  D32} (1985) 2001}.
%%CITATION = PHRVA,D32,2001;%%.

\bibitem{Burkardt:2016ffk}
M.~Burkardt, S.~S. Chabysheva, and J.~R. Hiller, ``{Two-dimensional light-front
  $\phi^4$ theory in a symmetric polynomial basis},''
  \href{http://dx.doi.org/10.1103/PhysRevD.94.065006}{{\em Phys. Rev.}
  {\bfseries D94} no.~6, (2016) 065006},
\href{http://arxiv.org/abs/1607.00026}{{\ttfamily arXiv:1607.00026 [hep-th]}}.
%%CITATION = ARXIV:1607.00026;%%.

\bibitem{Chabysheva:2015ynr}
S.~Chabysheva, ``{Light-front $\phi^4_{1+1}$ theory using a many-boson
  symmetric-polynomial basis},''
  \href{http://dx.doi.org/10.1007/s00601-016-1106-0}{{\em Few Body Syst.}
  {\bfseries 57} no.~8, (2016) 675--680},
  \href{http://arxiv.org/abs/1512.08770}{{\ttfamily arXiv:1512.08770
  [hep-ph]}}.

\bibitem{Sugihara:2004qr}
T.~Sugihara, ``{Density matrix renormalization group in a two-dimensional
  lambda phi4 Hamiltonian lattice model},''
  \href{http://dx.doi.org/10.1088/1126-6708/2004/05/007}{{\em JHEP} {\bfseries
  05} (2004) 007}, \href{http://arxiv.org/abs/hep-lat/0403008}{{\ttfamily
  arXiv:hep-lat/0403008}}.

\bibitem{Schaich:2009jk}
D.~Schaich and W.~Loinaz, ``{An improved lattice measurement of the critical
  coupling in $\phi_2^4$ theory},''
  \href{http://dx.doi.org/10.1103/PhysRevD.79.056008}{{\em Phys. Rev. D}
  {\bfseries 79} (2009) 056008},
  \href{http://arxiv.org/abs/0902.0045}{{\ttfamily arXiv:0902.0045 [hep-lat]}}.

\bibitem{Milsted:2013rxa}
A.~Milsted, J.~Haegeman, and T.~J. Osborne, ``{Matrix product states and
  variational methods applied to critical quantum field theory},''
  \href{http://dx.doi.org/10.1103/PhysRevD.88.085030}{{\em Phys. Rev. D}
  {\bfseries 88} (2013) 085030},
  \href{http://arxiv.org/abs/1302.5582}{{\ttfamily arXiv:1302.5582 [hep-lat]}}.

\bibitem{Bosetti:2015lsa}
P.~Bosetti, B.~De~Palma, and M.~Guagnelli, ``{Monte Carlo determination of the
  critical coupling in $\phi^4_2$ theory},''
  \href{http://dx.doi.org/10.1103/PhysRevD.92.034509}{{\em Phys. Rev. D}
  {\bfseries 92} no.~3, (2015) 034509},
  \href{http://arxiv.org/abs/1506.08587}{{\ttfamily arXiv:1506.08587
  [hep-lat]}}.

\bibitem{Zamolodchikov:1986gt}
A.~Zamolodchikov, ``{Irreversibility of the Flux of the Renormalization Group
  in a 2D Field Theory},'' {\em JETP Lett.} {\bfseries 43} (1986) 730--732.

\bibitem{Cappelli:1990yc}
A.~Cappelli, D.~Friedan, and J.~I. Latorre, ``{$C$-theorem and spectral
  representation},'' \href{http://dx.doi.org/10.1016/0550-3213(91)90102-4}{{\em
  Nucl. Phys. B} {\bfseries 352} (1991) 616--670}.

\bibitem{Mussardo:2010mgq}
G.~Mussardo, {\em {Statistical field theory}: {an introduction to exactly
  solved models in statistical physics}}.
\newblock Oxford Univ. Press, 2010.

\bibitem{Cappelli:1989yu}
A.~Cappelli and J.~I. Latorre, ``{Perturbation Theory of Higher Spin Conserved
  Currents Off Criticality},''
  \href{http://dx.doi.org/10.1016/0550-3213(90)90463-N}{{\em Nucl. Phys. B}
  {\bfseries 340} (1990) 659--691}.

\bibitem{Gukov1}
M.~Dedushenko and S.~Gukov, ``{IR duality in 2D $N=(0,2)$ gauge theory with
  noncompact dynamics},''
  \href{http://dx.doi.org/10.1103/PhysRevD.99.066005}{{\em Phys. Rev. D}
  {\bfseries 99} no.~6, (2019) 066005},
  \href{http://arxiv.org/abs/1712.07659}{{\ttfamily arXiv:1712.07659
  [hep-th]}}.

\bibitem{Gukov2}
S.~Gukov, D.~Pei, and P.~Putrov, ``{Trialities of minimally supersymmetric 2d
  gauge theories},'' \href{http://dx.doi.org/10.1007/JHEP04(2020)079}{{\em
  JHEP} {\bfseries 04} (2020) 079},
  \href{http://arxiv.org/abs/1910.13455}{{\ttfamily arXiv:1910.13455
  [hep-th]}}.

\bibitem{Antonuccio:1998zp}
F.~Antonuccio, O.~Lunin, and S.~Pinsky, ``{On exact supersymmetry in DLCQ},''
  \href{http://dx.doi.org/10.1016/S0370-2693(98)01274-X}{{\em Phys. Lett. B}
  {\bfseries 442} (1998) 173--179},
  \href{http://arxiv.org/abs/hep-th/9809165}{{\ttfamily arXiv:hep-th/9809165}}.

\bibitem{Kutasov:1993gq}
D.~Kutasov, ``{Two-dimensional QCD coupled to adjoint matter and string
  theory},'' \href{http://dx.doi.org/10.1016/0550-3213(94)90420-0}{{\em Nucl.
  Phys. B} {\bfseries 414} (1994) 33--52},
  \href{http://arxiv.org/abs/hep-th/9306013}{{\ttfamily arXiv:hep-th/9306013}}.

\bibitem{Bhanot:1993xp}
G.~Bhanot, K.~Demeterfi, and I.~R. Klebanov, ``{(1+1)-dimensional large N QCD
  coupled to adjoint fermions},''
  \href{http://dx.doi.org/10.1103/PhysRevD.48.4980}{{\em Phys. Rev.} {\bfseries
  D48} (1993) 4980--4990},
\href{http://arxiv.org/abs/hep-th/9307111}{{\ttfamily arXiv:hep-th/9307111
  [hep-th]}}.
%%CITATION = HEP-TH/9307111;%%.

\bibitem{Demeterfi:1993rs}
K.~Demeterfi, I.~R. Klebanov, and G.~Bhanot, ``{Glueball spectrum in a
  ($1+1$)-dimensional model for QCD},''
  \href{http://dx.doi.org/10.1016/0550-3213(94)90236-4}{{\em Nucl. Phys. B}
  {\bfseries 418} (1994) 15--29},
  \href{http://arxiv.org/abs/hep-th/9311015}{{\ttfamily arXiv:hep-th/9311015}}.

\bibitem{dalley1993string}
S.~Dalley and I.~R. Klebanov, ``{String spectrum of ($1+1$)-dimensional large-N
  QCD with adjoint matter},''
  \href{http://dx.doi.org/10.1103/PhysRevD.47.2517}{{\em Physical Review D}
  {\bfseries 47} no.~6, (1993) 2517}.

\bibitem{liu2020quantum}
J.~Liu and Y.~Xin, ``Quantum simulation of quantum field theories as quantum
  chemistry,'' \href{http://arxiv.org/abs/2004.13234}{{\ttfamily
  arXiv:2004.13234 [hep-th]}}.

\bibitem{Wetzel}
W.~Wetzel, ``{Two Loop Beta Function for the {Gross-Neveu} Model},''
  \href{http://dx.doi.org/10.1016/0370-2693(85)90551-9}{{\em Phys. Lett. B}
  {\bfseries 153} (1985) 297--299}.

\bibitem{fisher1972critical}
M.~E. Fisher, S.-k. Ma, and B.~Nickel, ``Critical exponents for long-range
  interactions,'' \href{http://dx.doi.org/10.1007/BF00398169}{{\em Physical
  Review Letters} {\bfseries 29} no.~14, (1972) 917}.

\bibitem{Behan:2017emf}
C.~Behan, L.~Rastelli, S.~Rychkov, and B.~Zan, ``{A scaling theory for the
  long-range to short-range crossover and an infrared duality},''
  \href{http://dx.doi.org/10.1088/1751-8121/aa8099}{{\em J. Phys. A} {\bfseries
  50} no.~35, (2017) 354002}, \href{http://arxiv.org/abs/1703.05325}{{\ttfamily
  arXiv:1703.05325 [hep-th]}}.

\bibitem{Burkardt}
M.~Burkardt, ``{Light front quantization of the Sine-Gordon model},''
\href{http://dx.doi.org/10.1103/PhysRevD.47.4628}{{\em Phys. Rev.} {\bfseries
  D47} (1993) 4628--4633}.
%%CITATION = PHRVA,D47,4628;%%.

\bibitem{dunkl_xu_2014}
C.~F. Dunkl and Y.~Xu, \href{http://dx.doi.org/10.1017/CBO9781107786134}{{\em
  Orthogonal Polynomials of Several Variables}}.
\newblock Cambridge University Press, 2~ed., 2014.

\bibitem{xu2017orthogonal}
Y.~Xu, ``{Orthogonal polynomials of several variables},''
  \href{http://arxiv.org/abs/1701.02709}{{\ttfamily arXiv:1701.02709
  [math.CA]}}.

\bibitem{Genest_2014}
V.~X. Genest and L.~Vinet, ``The multivariate hahn polynomials and the singular
  oscillator,'' \href{http://dx.doi.org/10.1088/1751-8113/47/45/455201}{{\em
  Journal of Physics A: Mathematical and Theoretical} {\bfseries 47} no.~45,
  (2014) 455201}.

\bibitem{book}
R.~Koekoek, P.~Lesky, and R.~Swarttouw,
  \href{http://dx.doi.org/10.1007/978-3-642-05014-5}{{\em Hypergeometric
  Orthogonal Polynomials and Their q-Analogues}}.
\newblock Springer, 2010.

\end{thebibliography}\endgroup

\end{document}